\documentclass[11pt]{article}

\usepackage{jheppub}

\input{epsf}
\usepackage{epsfig}
\usepackage{amssymb}
\usepackage{amsfonts}
\usepackage{amsbsy}
\usepackage[all]{xy}
\usepackage{amsmath}

\usepackage{amssymb,amscd}
\usepackage{mathrsfs}
\usepackage{amsmath,amsthm}
\usepackage{dsfont}
\def\be{ \begin{eqnarray} }
\def\ee{ \end{eqnarray}}

\def\Co0{{\rm Co}_0}

\def\det{{\rm det}}

\def\exp{{\rm exp}}

\def\I{{\rm i}}

\def\log{{\rm log}}
\def\mod{{\rm mod}}

\def\Tr{{\rm Tr}}

\def\half{\frac{1}{2}}
\def\p{\partial}

\def\one{{\hbox{ 1\kern-.8mm l}}}

\def\CB{{\cal B}}
\def\CC {{\cal C}}
\def\CD {{\cal D}}
\def\CE {{\cal E}}

\def\CH {{\cal H}}
\def\CI {{\cal I}}

\def\CL {{\cal L}}
\def\CM {{\cal M}}

\def\CO {{\cal O}}
\def\CP {{\cal P}}
\def\CR {{\cal R}}
\def\CV {{\cal V}}

\def\CX {{\cal X}}
\def\CO {{\cal O}}

\def\CE {{\cal E}}

\def\CH {{\cal H}}
\def\CI {{{\cal I}}}
\def\CB {{\cal B}}
\def\CQ {{\cal Q}}

\def\CT {{\cal T}}

\def\CX {{\cal X}}

\def\IC{\mathbb{C}}

\def\IF{\mathbb{Z}}

\def\IM{{\cal E}} 

\def\IP{\mathbb{P}}
\def\IQ{{\cal Q}} 
\def\IR{{\mathbb{R}}}

\def\IV{{\mathbb{V}}}
\def\IZ{{\mathbb{Z}}}

\def\fI{\mathfrak{I}}

\def\fS{\mathfrak{S}}

\def\rmk#1{\bigskip\noindent{\bf Remarks} }

\def\Pdeg{{\rm \textbf{P}} }

\usepackage{tikz}
\usetikzlibrary{arrows}
\usetikzlibrary{positioning}
\usetikzlibrary{shapes,snakes}
\usetikzlibrary{fit}

\def\lm{\limits}
\def\nn{\nonumber}
\def\X{{\cal X}}
\def\SS{\draw[ultra thick] (0,0) to[out=90,in=210] (0.5,0.5) to[out=30,in=270] (1,1) (1,0) to[out=90,in=330] (0.6,0.4) (0.4,0.6) to[out=150,in=270] (0,1);}
\def\S{\draw[ultra thick, purple] (0,0) to[out=90,in=210] (0.5,0.5) to[out=30,in=270] (1,1) (1,0) to[out=90,in=330] (0.6,0.4) (0.4,0.6) to[out=150,in=270] (0,1);}
\def\FINDANAME{instanton transplant\;}
\def\sqr{\filldraw[blue] (-0.1,0.1) -- (0.1,0.1) -- (0.1,-0.1) -- (-0.1,-0.1) -- (-0.1,0.1);}
\def\crl{\filldraw[white] (0,0) circle (0.1); \draw (0,0) circle (0.1);}
\def\dmd{\filldraw[red] (-0.15,0) -- (0,0.15) -- (0.15,0) -- (0,-0.15) -- (-0.15,0);}

	\title{Comments On The Two-Dimensional Landau-Ginzburg Approach To Link Homology}

	\author[a,b]{Dmitry Galakhov}
	\author[b]{ and Gregory W. Moore}
	\affiliation[a]
	{Institute for Information Transmission Problems, \\
		Bolshoy Karetny per. 19, build.1, Moscow 127051, Russia,}
	\affiliation[b]
	{NHETC and
		Department of Physics and Astronomy, Rutgers University \\
		126 Frelinghuysen Rd., Piscataway NJ 08855, USA}

	\emailAdd{galakhov@itep.ru}
	\emailAdd{galakhov@physics.rutgers.edu}
	\emailAdd{gmoore@physics.rutgers.edu}
	
	\abstract{We describe rules for computing a homology theory of knots and links in $\IR^3$. It is derived from the theory
of framed BPS states bound to domain walls separating two-dimensional Landau-Ginzburg
models with (2,2) supersymmetry. We illustrate the rules with some sample computations, obtaining results consistent
with Khovanov homology. We show that of the two Landau-Ginzburg models discussed in this context by Gaiotto and Witten one,
(the so-called Yang-Yang-Landau-Ginzburg model) does not lead to topological invariants
of links while the other, based on a model with target space equal to the universal cover of
 the moduli space of $SU(2)$ magnetic monopoles, will indeed produce a topologically invariant theory of knots and links.      
}

\begin{document}
	
\maketitle

\section{Introduction And Conclusion}\label{sec:Intro}

The physical interpretation of knot polynomials based on Chern-Simons gauge theory \cite{Witten:1988hf}
 has had wide-ranging
applications, both in physics and in mathematics. A subsequent very interesting development
in knot theory  was initiated by Khovanov's categorification of the Jones
polynomial \cite{Khovanov,BarNatan,Dolotin:2012sw}.  Given the profound impact of the
Chern-Simons interpretation of knot polynomials
one might hope that a physical interpretation of the more refined
knot homology theories will likewise lead to important advances,
both in physics and mathematics. This paper is concerned with
developing one of the physical approaches to the computation of
knot homology. (The entire program works for links with
any number of connected components, so henceforth we deprecate the
term ``knot'' in favor of ``link.'')

In searching for a physical generalization of the Chern-Simons interpretation
of link polynomials one might call to mind  the
relation between Chern-Simons theory and two-dimensional
conformal field theory \cite{Witten:1988hf,Elitzur:1989nr,Moore:1989vd}.
The generalization suitable for link homology
could then be attempted either   by generalizing the Chern-Simons theory
or by generalizing the conformal field theory. The approach we discuss involves both
generalizations. While many   interesting generalizations of the Chern-Simons side
have been proposed in the physics literature, this paper will focus on developing the approach of
Witten \cite{Witten:2011zz,WItten:2011pz,Witten:2014xwa}.
\footnote{
	Among the other physical approaches to the subject that are now available in the literature we would like to
mention
	\cite{Anokhina:2014hha,Arthamonov:2015rha,Aganagic:2011sg,Dolotin:2012sw,Gukov:2016gkn}. See also the references therein.
	The first interpretation of knot homology in terms of BPS states of which we are aware is \cite{Gukov:2004hz}.
	}
Witten's approach is
based on the six-dimensional (2,0) theory and its   KK reduction on a cigar
to five-dimensional SYM on a half-space.

Witten's approach uses equivariant Morse theory applied to an infinite-dimensional space of
fields, and the equations determining the Morse complex and the differential
on that complex are rather complicated partial differential equations
\cite{Haydys,Witten:2011zz} with subtle boundary conditions
\cite{Witten:2011zz,Mazzeo:2013zga}. This Morse complex can be interpreted
in terms of two-dimensional N=(2,2) Landau-Ginzburg (LG) theory, but it is a gauged LG
model with an infinite-dimensional target space (a space of complexified gauge potentials
on a three-manifold) with an infinite-dimensional gauge group (the group of unitary
gauge transformations) \cite{Gaiotto:2015aoa}. We will refer to this model as the CSLG model.
(For readers intimately familiar with \cite{Gaiotto:2015aoa} we mean the CSLG2 model.)
 From a computational viewpoint it was therefore an important
advance when Gaiotto and Witten proposed that one could replace the CSLG model by an equivalent \underline{ungauged} model with a
 \underline{finite-dimensional} target space  \cite{Gaiotto:2011nm}. Some related observations
 were also made in \cite{Cheng:2010yw}. One may view these
2d LG models with finite-dimensional targets as generalizations of rational conformal
field theories, the monodromy of whose conformal blocks leads to the formulation of link polynomials,
such as the Jones polynomial.
In this sense, the famous Chern-Simons gauge theory/RCFT correspondence is generalized.

While the  paper \cite{Gaiotto:2011nm} of Gaiotto and Witten demonstrated that one could use
certain 2d  ungauged LG models with finite-dimensional target space for the computation
of the Jones polynomial (and, presumably, its other Chern-Simons generalizations),
the paper did not explain how to compute the actual link homologies.
That is, it showed how to recover the Euler characters of the link homology, but not the homology
groups themselves.  One of the
 (several) motivations for the development of the web formalism described in
\cite{Gaiotto:2015zna,Gaiotto:2015aoa} was to provide a framework for the computation of these
homology groups using LG models. Section 3.6 of \cite{Gaiotto:2015zna} and Section 18.4 of
\cite{Gaiotto:2015aoa} sketch how the web formalism is expected to be related to the computation of link
homology. The present paper continues that line of development and aims to apply the web-formalism to
the  Landau-Ginzburg models of
\cite{Gaiotto:2011nm}. We will give semi-practical rules for the computation of link homology.
The reason for the modifier ``semi-'' is discussed in the final Section \ref{sec:FutureDirections}.

In fact the Gaiotto-Witten paper introduces \underline{two}  Landau-Ginzburg models with finite-dimensional target space.
We refer to them as  the
\emph{Yang-Yang-Landau-Ginzburg} (YYLG) model  and the \emph{Monopole-Landau-Ginzburg} (MLG) model. \footnote{The YYLG model is closely related to the work of Bigelow \cite{Bigelow}.}
The equations defining the generators of the Morse complex for the original CSLG model,
once deformed by ``symmetry-breaking at infinity,'' turn out to be
related to (the holomorphic part of) the equations defining moduli spaces of magnetic monopoles. This motivated Gaiotto and Witten
to introduce a Landau-Ginzburg model whose target space is the (finite dimensional)  moduli space of smooth $SU(2)$
monopoles. In order to capture crucial physical properties of the original CSLG model, Gaiotto and Witten also introduced
a superpotential on the monopole moduli space. (It has been generalized to other gauge groups in \cite{Braverman:2014ysa}.)
This superpotential is most easily expressed, for the case of gauge group $SU(2)$, using
the parametrization of magnetic monopoles in terms of rational maps. See Section \ref{subsec:MLG-Model} below.

The regions at infinity of monopole moduli space are well-described by collections of widely separated
singly-charged monopoles. When the superpotential confines monopoles to such regions one
can replace the MLG model by a slightly simpler model that we call
the   \emph{Naive-Monopole-Landau-Ginzburg} (NMLG) model. The NMLG model is described in Section \ref{subsec:NMLG-Model}.
In this model one chooses a direction in $\IR^3$, to be thought of as the $y$-direction and introduces
chiral superfields $Y_i$ associated with the $y$-positions of the monopole centers. Another set of chiral superfields
$w_i$ is associated with the projections of the monopole  centers onto the plane orthogonal to the $y$-axis.
The NMLG model is  singular when monopole centers collide. In the NMLG model, if no two centers are on the same
line parallel to the  $y$-axis then the $Y_i$ superfields are heavy and can be integrated out.
The result is the
YYLG model of Section \ref{subsec:YYLG}.  The YYLG  model is in many ways the easiest to work with, and it has the most
direct connection to conformal field theory. For example, the holomorphic brane amplitudes associated to Lefshetz
thimbles are just free-field representations of conformal blocks.

We can summarize the logical connections between
 Chern-Simons theory, conformal blocks in 2d CFT, 5d SYM, YYLG, NMLG and MLG
 in the following schematic diagram:

\begin{center}
	\begin{tikzpicture}
	\node(A) at (-2,0) {Chern-Simons};
	\node(B) at (2,0) {Conformal Blocks};
	\node(C) at (4,-1) {YYLG};
	\node(D) at (2,-2) {NMLG};
	\node(E) at (-2,-2) {MLG};
	\node(F) at (-4,-1) {5d SYM};
	\path (A) edge[<->] (B) (B) edge[<-] (C) (C) edge[<->] (D) (D) edge[<->] (E) (E) edge[<->] (F) (F) edge[->] (A);
	\draw[rounded corners, dashed] (-3.7,0) -- (-3.7,0.3) -- (3.7,0.3) -- (3.7,-0.2) -- (-3.7,-0.2) -- (-3.7,0);
	\draw[rounded corners, dashed] (-5,-1.5) -- (-5,-0.7) -- (5,-0.7) -- (5,-2.2) -- (-5,-2.2) -- (-5,-1.5);
	\node at (7.2,0) {Jones polynomials};
	\node at (7.2,-1.5) {Khovanov polynomials};
	\end{tikzpicture}
\end{center}

We now outline the paper and sketch some of the results.

This paper assumes some familiarity with the (curved) web formalism of \cite{Gaiotto:2015aoa}, including the
proposed application to link homology in Section 18.4 of that paper. For a relatively brief introduction
to the material the reader might wish to consult \cite{Gaiotto:2015zna}.  In particular we will use the
notation of those papers.  We give a lightning review in
Section \ref{sec:Review}.

After briefly reviewing three relevant LG models in Section \ref{sec:ThreeLGModels},
 we give some simple rules for computing link homologies in  Section \ref{sec:KnotHomRules}.
 The curved web formalism,   reviewed in Section \ref{sec:CurvedWebs},  applies to all three classes of models.
The solitons and their (curved) boosted trajectories
- from which one constructs the generators of the link complex and important ingredients for determining its differential -
just depend on
the central charge function - and hence since the three kinds of models are closely related to each other, the
relevant curved webs used to construct the differential will look the same: There is a one-one correspondence of the relevant vacua and solitons
and edges of curved webs in the three theories.  What is not the same are the vertices of the graphs, i.e. the  ``interior amplitudes,''
to use the language of \cite{Gaiotto:2015aoa}. Thus the differential and hence the link complex is \underline{not} the same for the
YYLG, NMLG, and MLG models.

In fact, after illustrating how the formalism works with a number of examples in Section \ref{sec:Examples}, we will find a surprise:
In Section \ref{sec:R-Invariance} we investigate the important issue of whether the link homology complexes constructed using our formalism
really do have three-dimensional general covariance up to quasi-isomorphism. This is not obvious since our rules depend on a presentation
of the link as a tangle - that is, as an evolution of a collection of points in the complex plane as a function of a real parameter.
The complex depends as well  on a projection of the tangle
into a plane. Thus the complexes \emph{per se} are most definitely not objects that can be intrinsically associated to the topology of a link in
$\IR^3$.  What we would
like to show is that, nevertheless, the \underline{homology}  of the complexes is intrinsically three-dimensional. This is true
if and only if  two complexes associated to link projections related by the five Reidemeister moves are quasi-isomorphic.
 In Section \ref{sec:R-Invariance} we show how to check invariance up to quasi-isomorphism
for the Reidemeister moves. The demonstration relies on making
  certain assumptions about the existence of suitable vertices in the web formalism. Then, in Section
\ref{sec:Obstruction}, we show that in the YYLG and NMLG models there can be obstructions to the existence of these necessary vertices.
There are two kinds of things that can go wrong. First, a wall of marginal stability can interfere with the existence of certain solitons
entering the vertex. This is described in Section \ref{subsec:Problem-RI}. Second there can be simple topological obstructions to the existence of
solutions of the $\zeta$-instanton equation used to construct the vertex. The topological obstructions arise from
the fact that in the YYLG and NMLG models the target space is not simply connected.
 In all three models the most natural description of the model begins with a
multi-valued superpotential $W$ on a target space $\bar X$ such that $dW$ is a well-defined one-form on $\bar X$.  ($W$ has its origin
in a Chern-Simons term in the gauge theory, and the Chern-Simons term is multivalued under large gauge
transformations. This is the source of multi-valuedness.)  One therefore
defines the model on a cyclic cover $X\to \bar X$ where $W$ is single-valued. We will find that, in the YYLG and NMLG
models the relevant cyclic cover $X$ is not simply connected, and the existence of the ``$\zeta$-instantons''
needed to define a suitable differential   is obstructed, simply because the boundary conditions for such $\zeta$-instantons define
homotopically nontrivial loops in $X$.

In Section \ref{sec:ResolveObstruction} we argue that for the   MLG model both of the obstructions identified above vanish.
The first obstruction (which already vanishes in the NMLG model) disappears because there is no singularity
when the points $w_i$ collide with the positions $z_a$ of the link. The second obstruction vanishes in the
monopole model because the cyclic cover of monopole moduli space coincides with the simply connected
universal cover.  Our main conclusion is therefore that the link complexes
for the YYLG and NMLG models do not lead to well-defined link homology groups. That is,  the link
homology groups depend on the link projection and are not topological invariants of links. But, given our
check of invariance under Reidemeister moves up to quasi-isomorphism, the homology of the complex for the MLG model
indeed defines a topological invariant of links in $\IR^3$.

In Appendix \ref{app:FermionTTstar} we discuss the difficult question of how to compute the
Fermion number grading, a.k.a. the homological grading, of the complexes. This requires an
excursion - interesting in its own right - into $tt^*$ equations \cite{Cecotti:1992rm} for a generalization
of the CFIV index \cite{Cecotti:1992qh} to the case of LG interfaces. In Appendix \ref{app:Signs} we discuss
the even more difficult, but quite necessary, task of how to determine the signs of the
matrix elements of the differential on the complex.

There is independent, but unpublished, work that overlaps with our results.
Some time ago C. Manolescu observed that there are obstructions to using what
we call the YYLG model for defining knot homology. Moreover, in some work in progress,
M. Abouzaid and I. Smith are using somewhat different methods to construct a knot homology theory
that is closely related to what we call the MLG model. 

The present paper is a summary and extension of the
PhD thesis of the first author \cite{GalakhovPhD}. Some further details, omitted here, can be found in this thesis.

\section*{Acknowledgements}

GM would like to thank D.~Gaiotto and E.~Witten for many discussions about link homology
over the past few years, and for much collaboration on closely related material. DG would like to
thank S.~Arthamonov, A.~Mironov, A.~Morozov, S.~Shakirov for fruitful discussions of some aspects of the knot theory. We also thank
M. ~Abouzaid, M.~Aganagic,  J.~Clingempeel, T.~Dimofte, and I.~Smith for correspondence and discussions on related matters.
The work of DG and GM is supported by the DOE under grant DE-SC0007897. GM gratefully acknowledges the hospitality of the 
Aspen Center for Physics (under NSF Grant No. PHY-1066293) during the completion of this paper. 
The work of DG is partly supported by RFBR grants 15-31-20832-mol-a-ved, 15-52-50041-YaF, 16-01-00291.

\section{Lightning review}\label{sec:Review}

In this section we review some essential points from \cite{Gaiotto:2015aoa,Gaiotto:2015zna}.

\subsection{Interfaces From Tangles}\label{subsec:Tangles}

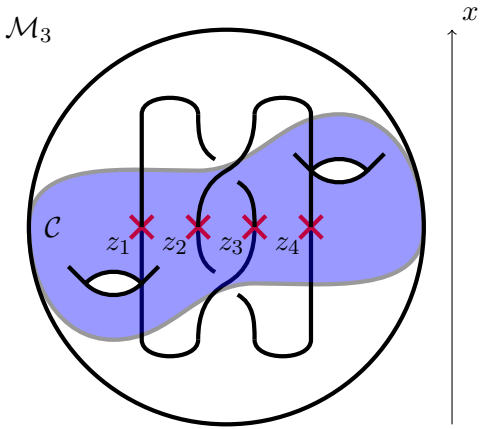
\begin{figure}[h]
	\begin{center}
		\begin{tikzpicture}
		\begin{scope}[scale=1.5]
		\begin{scope}[shift={(0,-1)}]
		\draw [ultra thick] (-0.25,-0.5) to [out=90,in=210] (0,0) to [out=30, in=270] (0.25,0.5) (-0.25,0.5) to [out=270, in=150] (-0.1,0.1) (0.1,-0.1) to [out=330, in=90] (0.25,-0.5);
		\end{scope}
		\begin{scope}[shift={(0.25,0)}]
		\draw [ultra thick] (-1,-0.5)-- (-1,-1.5) to [out=270, in=270] (-0.5,-1.5);
		\end{scope}
		\begin{scope}[shift={(-0.25,0)}]
		\begin{scope}[xscale=-1]
		\draw [ultra thick] (-1,-0.5)-- (-1,-1.5) to [out=270, in=270] (-0.5,-1.5);
		\end{scope}
		\end{scope}
		\begin{scope}[shift={(-1.75,-0.5)}]
		\draw[ultra thick, fill=blue, opacity=0.4] (0,0) to[out=90,in=180] (1.5,0.5) to[out=0,in=180] (2.75,1) to[out=0,in=90] (3.5,0) to[out=270,in=0] (2,-0.5) to [out=180,in=0] (0.75,-1) to[out=180,in=270] (0,0);
		\begin{scope}[shift={(0.5,-0.5)}]
		\draw[ultra thick, fill=white] (0,0) to[out=45,in=135] (0.5,0) to[out=225,in=315] (0,0);
		\draw[ultra thick] (0,0) -- (-0.15,0.15) (0.5,0) -- (0.65,0.15);
		\end{scope}
		\begin{scope}[shift={(2.5,0.5)}]
		\draw[ultra thick, fill=white] (0,0) to[out=45,in=135] (0.5,0) to[out=225,in=315] (0,0);
		\draw[ultra thick] (0,0) -- (-0.15,0.15) (0.5,0) -- (0.65,0.15);
		\end{scope}
		\begin{scope}[shift={(1,0)}]\draw[ultra thick, purple] (-0.1,-0.1)--(0.1,0.1) (0.1,-0.1)--(-0.1,0.1); \end{scope}
		\begin{scope}[shift={(1.5,0)}]\draw[ultra thick, purple] (-0.1,-0.1)--(0.1,0.1) (0.1,-0.1)--(-0.1,0.1); \end{scope}
		\begin{scope}[shift={(2,0)}]\draw[ultra thick, purple] (-0.1,-0.1)--(0.1,0.1) (0.1,-0.1)--(-0.1,0.1); \end{scope}
		\begin{scope}[shift={(2.5,0)}]\draw[ultra thick, purple] (-0.1,-0.1)--(0.1,0.1) (0.1,-0.1)--(-0.1,0.1); \end{scope}
		\node[below left] at (1,0) {$z_1$};
		\node[below left] at (1.5,0) {$z_2$};
		\node[below left] at (2,0) {$z_3$};
		\node[below left] at (2.5,0) {$z_4$};
		\end{scope}
		\draw [ultra thick] (-0.25,-0.5) to [out=90,in=210] (0,0) to [out=30, in=270] (0.25,0.5) (-0.25,0.5) to [out=270, in=150] (-0.1,0.1) (0.1,-0.1) to [out=330, in=90] (0.25,-0.5);
		\begin{scope}[shift={(0.25,0)}]
		\draw [ultra thick] (-0.5,0.5) to [out=90, in=90] (-1,0.5) --(-1,-0.5);
		\end{scope}
		\begin{scope}[shift={(-0.25,0)}]
		\begin{scope}[xscale=-1]
		\draw [ultra thick] (-0.5,0.5) to [out=90, in=90] (-1,0.5) --(-1,-0.5);
		\end{scope}
		\end{scope}
		\draw[ultra thick](-1.75,-0.5) to[out=270,in=180] (0,-2.25) to[out=0,in=270] (1.75,-0.5) to[out=90,in=0] (0,1.25) to[out=180,in=90] (-1.75,-0.5);
		\node [right] at (-1.7,-0.5) {$\cal C$};
		\node at (-1.75,1.25){${\cal M}_3$};
		\draw[->] (2,-2.25) -- (2,1.25);
		\node[above right] at (2,1.25) {$x$};
		\end{scope}
		\end{tikzpicture}
	\end{center}
	\caption{A tangle embedded in a three-manifold ${\cal M}_3$. Transverse slices to the $x$-plane are a Riemann surface $C$ so the
tangle is described by an evolving set of points $z_a(x)$ in $C$. In this paper we will take $C$ to be the complex plane so that
$\CM_3$ is just $\IC \times \IR = \IR^3$.  } \label{fig:tangle}
\end{figure}

We begin by recalling a few points from Section 18.4 of  \cite{Gaiotto:2015aoa}.
The link $L$ whose link-homology we wish to formulate will be presented as a tangle, that is, as
set of distinct points   $z_a$ in the complex plane continuously evolving with a parameter $x$. The points are allowed to be
  created or annihilated in pairs at critical values of $x$ but otherwise cannot collide\footnote{In principle one could consider punctures on a general Riemann surface $C$, and hence discuss links in
$C\times \IR$. However, in this paper will will strictly limit ourselves to the case where $C$ is the complex plane.}. See Fig.\ref{fig:tangle}.

In link-homology the strands of the tangle, and hence the points $z_a$ are labeled by
irreducible representations of some compact simple Lie group. In this paper we consider
$G$ to be $SU(2)$, so the points will be labeled by positive integers $k_a$, to be thought
of as the dimension of an irreducible representation of $SU(2)$ minus one (i.e. $k_a$ is just
twice the spin), and indeed the Chern-Simons
invariant associated to the Euler character of the link homology will be that for gauge
group $SU(2)$ with Wilson lines in the $k_a$-dimensional irreducible representations.

In the Landau-Ginzburg approach to link homology the collection  $\{ z_a, k_a\} $ parameterizes a family of
massive $1+1$ dimensional quantum field  theories with $(2,2)$ supersymmetry, $\CT(z_a, k_a)$. The spatial
axis of the ``worldsheet'' on which the quantum field theory is defined is to be identified
with the $x$ axis in Fig.\ref{fig:tangle}, and the time axis is not shown in that figure.
The transverse space $C = \IC$ is a parameter space of the family of quantum field theories.
In a tangle the $z_a$ vary as a function of $x$, and therefore the parameters of the quantum field
theory vary as a function of $x$. This defines what is known as a ``Janus'' or an ``interface.''
If $\wp$ denotes a particular tangle $\{ z_a(x) \}$, or equivalently a path of theories, then
reference \cite{Gaiotto:2015aoa} explains that one can define an object $\fI[\wp]$ in an
$A_\infty$-category of Interfaces between the theories associated with the initial and final
points of the path. For composable paths $\wp_1, \wp_2$ one can moreover define a product $\boxtimes$ of
interfaces so that
\be\label{eq:ComposeInterfaces}
\fI[\wp_1 \circ \wp_2 ] \sim  \fI[\wp_1] \boxtimes \fI[\wp_2]
\ee
where $\sim$ denotes homotopy equivalence of Interfaces - a concept whose details we will not need here.
The key point for us will be that part of the data of an Interface is its \emph{Chan-Paton data}, $\CE(\fI)$.
This is a matrix of $\IZ$-graded complexes, the positions of matrix elements being labeled by an ordered pair of massive vacua of the theory.
We will write $\CE(\fI[\wp])_{ij'}$ where $i$ is a vacuum in the initial theory and $j'$ is a vacuum in the final theory.
The Chan-Paton complexes have the property that if $\fI[\wp_1] \sim \fI[\wp_2]$ are homotopy equivalent then for each matrix element $ij'$
the chain complexes $\CE(\fI[\wp_1])_{ij'}$ and $\CE(\fI[\wp_1])_{ij'}$  are quasi-isomorphic. Moreover,
\be\label{eq:ComposeChanPatonMatrices}
\CE(\fI[\wp_1] \boxtimes \fI[\wp_2] ) = \CE(\fI[\wp_1])   \CE(\fI[\wp_2])
\ee
where the product on the RHS is defined by matrix multiplication together with tensor product and
direct sum of $\IZ$-graded complexes.
Finally, the only data in an Interface between the empty theory and itself is a single chain complex.
Referring again to Fig.\ref{fig:tangle}  we see that for large and small values of $x$ we have the
empty theory. The associated chain complex is the link-homology complex defined by the family
of massive quantum field theories.

As shown in \cite{Gaiotto:2015aoa}, interfaces define a notion of parallel transport of the category
of branes over the parameter space of the quantum field theory. Moreover, suitably interpreted, the
parallel transport is given by a ``flat connection.'' Therefore we can hope that homotopic deformations
of the path of parameters $\{ z_a(x) \}$ will lead to chain homotopic complexes. Therefore, this is a
natural setting for the formulation of link homology.

Thanks to the multiplicative and homotopy properties of
Interfaces we have just reviewed one can break up the Interface into a product of several elementary
Interfaces associated with braiding of two strands and creation and annihilation of two strands.
In this way one can hope to make explicit computations of link homology Interfaces.
Explaining how to do such explicit computations is precisely the goal
of this paper, for a specific family of massive Landau-Ginzburg models.

We now explain how these general concepts can be realized in certain families of Landau-Ginzburg models.

\subsection{Supersymmetric Landau-Ginzburg Models And BPS Solitons}

Let us briefly recall the basics  of $1+1$ dimensional ${\cal N}=(2,2)$ supersymmetric Landau-Ginzburg model. The target space of this model is a complex $N$-dimensional K\"ahler manifold $X$ equipped with a holomorphic function $W:\;X\to \IC$ called the superpotetnial. If $\phi^I$ are complex local coordinates on $X$
then we introduce corresponding chiral superfields $\Phi^I = \phi^I + \cdots$ and write the action
\be
S=\int d^2x \left[\int d^4\theta\; K\left(\Phi ,\bar{\Phi} \right)+\frac{1}{2}\left(\int d^2 \theta\; W\left(\Phi \right)+\int d^2 \bar\theta\; \bar W\left(\bar\Phi \right)\right)\right]
\ee
The 4 supercharges in this theory have certain chirality and fermion number, and we denote them accordingly:
\footnote{Notice that our choice of  fermion number convention is opposite to that in \cite{Gaiotto:2015aoa}. This change of convention will be convenient in
some definitions below. The $\CR$-interfaces we will define   are represented by \emph{right} helices and have an expansion $\CR\sim S\oplus X$ where $S$ is an interface representing the classical $R$-matrix and having   fermion number $0$, while $X$ is a ``higher order correction''. It is convenient to have a convention in
which  the corrections $X$ include contributions carrying fermion number $1$ and higher.}
\be\label{SUSY_conv}
\begin{array}{c|c|c|c|c}
 & Q_+ & \bar Q_+ & Q_-& \bar Q_-\\
 \hline
 \chi & +1 & +1 & -1 & -1 \\
 \hline
 {\rm \textbf{F}} & +1 & -1 &-1 & +1\\
\end{array}
\ee
They satisfy the superalgebra
\be\label{SUSY}
\begin{split}
\left\{Q_{\pm},\bar Q_{\pm}\right\}=H\pm P\\
\left\{Q_+, Q_-\right\}=\bar Z,\quad  \left\{\bar Q_+,\bar Q_-\right\}=Z
\end{split}
\ee
Where $H$ and $P$ are Hamiltonian and momentum operators correspondingly, and $Z$ is the central charge.

The vacua of the theory on the real line are in one-one correspondence with the critical points
of $W$, denoted by $\phi_i$ with $i\in \IV$. The masses of particle excitations above the vacuum
are determined by the eigenvalues of the  Hessian $\p_{IJ}^2 W$. We assume that $W$ is Morse so the
vacua are all massive.

The superalgebra (\ref{SUSY}) has a a family of subalgebras parameterized by a phase $\zeta$:
\be\label{eq:su_charges}
\CQ_{\zeta}=\bar Q_--\zeta Q_+,\quad \bar \CQ_{\zeta}=Q_--\zeta^{-1}\bar Q_+
\ee
so that $\CQ_{\zeta}^2=\bar \CQ_{\zeta}^2=0$ and the Hamiltonian is almost $\CQ_{\zeta}$-exact:
\be\label{Hamiltonian}
H=\frac{1}{2}\left\{\CQ_{\zeta},\bar\CQ_{\zeta} \right\}+{\rm Re}\left[\zeta^{-1}Z\right]
\ee

Any  stationary finite energy field configuration on the real line
should approach one of the vacuum values at spatial infinity:   $\phi(-\infty)=\phi_i$, $\phi(+\infty)=\phi_j$. The central charge in the this topologivally non-trivial background becomes non-zero: $Z_{ij}=\I (W_i-W_j)$ where $W_i$ denotes the critical value of the superpotential in the vacuum $i$.
 Finite energy field configurations satisfy a Bogomolny-Prasad-Sommerfeld (BPS) bound $E\geq |Z_{ij}|$.

Field configurations  minimizing the Hamiltonian (\ref{Hamiltonian}) and therefore saturating the BPS bound on the classical level are called BPS solitons.
They can be determined as solutions to the following equation:
\be\label{soliton}
\p_x\phi^I=\frac{\I\zeta_{ij}}{2}g^{I\bar J}{\frac{\p \bar W}{\p\bar \phi^{\bar J}}},\quad \lim\lm_{x\to -\infty}\phi(x)=\phi_i,\; \lim\lm_{x\to +\infty}\phi(x)=\phi_j
\ee
Where $g^{I\bar J}$ is inverse K\"ahler metric on $X$ and $\zeta_{ij}=\frac{Z_{ij}}{|Z_{ij}|}$.

Solitons interpolating between vacua $i$ and $j$ can be thought of as domain walls separating theories in vacua $i$ and $j$; to exponentially good accuracy in the
length scale $\ell_W$ of the theory, the fields are nearly constant except for a small region of ``width'' $\ell_W$. The length scale $\ell_W$ is
determined by the inverse of the smallest absolute value of the eigenvalues of the Hessian at the critical points.

\subsection{Half-Supersymmetric Landau-Ginzburg Interfaces}\label{sec:HSLGInterfaces}

The interfaces discussed in Section \ref{subsec:Tangles} above can be easily realized in the present setting.
We consider a family of superpotentials $W(\phi^I; z_a)$ and allow the parameters $z_a$ to be spatially varying.
When punctures are created or annihilated at a critical value $x_*$ we must view the interface at $x_*$
as one between theories with different numbers
of parameters.  We follow the
general philosophy of \cite{Gaiotto:2015aoa} and view the LG model as a special case of
a supersymmetric quantum mechanics with target space $M =   {\rm Map}(\IR\to X)$. According to \cite{Witten:1982im}
we can view the whole construction in terms of Morse theory.   The Morse function in this case is given by the following expression:
\be\label{eq:InterfaceMorseFunction}
h=-\int\lm_D \left(\lambda-\frac{1}{2}\;{\rm Re}\;\left[\zeta^{-1}W\left(\phi^I(x)|z_a(x)\right)\right]\; dx\right)
\ee
where the 1-form $\lambda$ trivializes the K\"ahler form  $d\lambda=\frac{i}{2}g_{I\bar J}\; d\phi^I\wedge d\bar\phi^{\bar J}$. An interface defined in this way preserves the subalgebra of $\CQ_{\zeta}$ and $\bar\CQ_{\zeta}$ supercharges.

Following the Morse theory interpretation we are interested in the perturbative vacua defined by the critical points on $M$ where $\delta h=0$.
These correspond to solutions of the   \emph{forced} soliton equation:
\be
\label{forced_soliton}
\p_x\phi^I(x)=\frac{\I\zeta}{2}g^{I\bar J}\overline{\p_J W\left(\phi^I(x)|z_a(x)\right)}
\ee
The Morse-Smale-Witten complex of critical points will define the Chan-Paton data of the interface.
The position of the matrix elements is labeled by a pair of a vacuum of the theory on the far left and a
vacuum of the theory on the far right. This pair of vacua serves to define boundary conditions for
equation \eqref{forced_soliton}.

We now introduce a convenient diagrammatic notation for discussing the MSW complexes.
We work in the adiabatic limit where the rate of change of parameters is very small  $dz_a/dx\ll \ell_W^{-1}$, where $\ell_W$ is the maximal soliton width.
In this situation we can construct approximate solutions to the forced $\zeta$-soliton equation (\ref{forced_soliton})
 from solutions to the ordinary $\zeta$-soliton equation (\ref{soliton}) as follows. In the adiabatic limit a class of solutions - the
 \emph{hovering solutions} - are simply described by fields which, at each $x$ are critical points for $W(\phi;z(x))$ for the same value of $x$.
 In equations:  If $\phi^I_i(x)$ is a critical point:
\be
dW(\phi^I|z_a(x))\Big|_{\phi_i^I(x)}=0
\ee
(where the exterior derivative only takes derivatives with respect to $\phi^I$)
then, since the $z_a(x)$ vary continuously once we have a chosen a critical point $i$ at some $x_0$ the field
$\phi^I_i(x)$ will evolve continuously. (Here we use the assumption that we have a family of Morse superpotentials
so that critical points never collide.)

Suppose we consider an interface $\fI_{x_1,x_2}$ where the parameters $z_a(x)$ are changing only for  $x$ between $x_1$ and  $x_2$. We diagrammatically
denote the hovering solutions of type $i$ by a straight line with the index $i$ at either end. The contribution of the hovering solution in
vacuum $i$ to the Chan-Paton complex is just $\IZ$ in degree zero in matrix element $ii$. We denote this graphically by:
\be
\CE[\fI_{x_1,x_2}]^{\rm hovering} =\bigoplus\lm_i \begin{array}{c}
	\begin{tikzpicture}
	\draw[ultra thick] (-1,0) -- (1,0);
	\node[above] at (-1,0) {$i$}; \node[above] at (1,0) {$i$};
	\end{tikzpicture}
\end{array}
\ee
In addition to the hovering solutions there are also solutions where solitons are localized near \emph{binding points}. The binding points are
points   $x_c$ where the phase of some central charge, say, $Z_{ij}(x_c)$ of the superpotential at $z(x_c)$,  coincides with the phase  $\zeta$ entering the equation \eqref{forced_soliton}.
At a binding point of type $ij$ one can glue a soliton of type $ij$ for the theory with parameter $z(x_c)$ to a hovering solution of type $i$ on the
left and hovering solution of type $j$ on the right. In our diagrammatic notation we will denote such solitons as

\begin{center}
	\begin{tikzpicture}
	\draw[ultra thick] (-1,0) -- (1,0);
	\begin{scope}[shift={(0,0)}]
	\filldraw [red] (-0.1,0.1) -- (0.1,0.1) -- (0.1,-0.1) -- (-0.1,-0.1) -- (-0.1,0.1);
	\end{scope}
	\node[above] at (-1,0) {$i$}; \node[above] at (1,0) {$j$};
	\end{tikzpicture}
\end{center}

If there is only one binding point of type $ij$ in the domain $[x_1, x_2]$ the Chan-Paton complex is denoted by:
\be
\CE[\fI_{x_1,x_2}]=\left(\bigoplus\lm_i \begin{array}{c}
	\begin{tikzpicture}
	\draw[ultra thick] (-1,0) -- (1,0);
	\node[above] at (-1,0) {$i$}; \node[above] at (1,0) {$i$};
	\end{tikzpicture}
\end{array}\right)\oplus \begin{array}{c}
\begin{tikzpicture}
\draw[ultra thick] (-1,0) -- (1,0);
\begin{scope}[shift={(0,0)}]
\filldraw [red] (-0.1,0.1) -- (0.1,0.1) -- (0.1,-0.1) -- (-0.1,-0.1) -- (-0.1,0.1);
\end{scope}
\node[above] at (-1,0) {$i$}; \node[above] at (1,0) {$j$};
\end{tikzpicture}
\end{array}
\ee
Of course in a given interval there can be many binding points and hence we must sum over all possible insertions of solitons
at the binding points. A convenient way to do this is to divide the interval into segments, each containing one binding point
and use the general property \eqref{eq:ComposeChanPatonMatrices} of composition of interfaces. For example, we
can diagrammatically glue two interfaces $\fI_{x_1,x_2}$ and $\fI_{x_2,x_3}$, each with a single binding point of type $ij$
and $jk$ respectively,   to produce an interface
with Chan-Paton data:
\be\label{eq:CP-Tensor1}
\begin{split}
	\left(\left(\bigoplus\lm_I \begin{array}{c}
		\begin{tikzpicture}
		\draw[ultra thick] (-0.5,0) -- (0.5,0);
		\node[above] at (-0.5,0) {$i$}; \node[above] at (0.5,0) {$i$};
		\end{tikzpicture}
	\end{array}\right)\oplus \begin{array}{c}
	\begin{tikzpicture}
	\draw[ultra thick] (-0.5,0) -- (0.5,0);
	\begin{scope}[shift={(0,0)}]
	\filldraw [red] (-0.1,0.1) -- (0.1,0.1) -- (0.1,-0.1) -- (-0.1,-0.1) -- (-0.1,0.1);
	\end{scope}
	\node[above] at (-0.5,0) {$i$}; \node[above] at (0.5,0) {$j$};
	\end{tikzpicture}
\end{array}\right)\otimes
\left(\left(\bigoplus\lm_i \begin{array}{c}
	\begin{tikzpicture}
	\draw[ultra thick] (-0.5,0) -- (0.5,0);
	\node[above] at (-0.5,0) {$i$}; \node[above] at (0.5,0) {$i$};
	\end{tikzpicture}
\end{array}\right)\oplus \begin{array}{c}
\begin{tikzpicture}
\draw[ultra thick] (-0.5,0) -- (0.5,0);
\begin{scope}[shift={(0,0)}]
\filldraw [blue] (-0.1,0.1) -- (0.1,0.1) -- (0.1,-0.1) -- (-0.1,-0.1) -- (-0.1,0.1);
\end{scope}
\node[above] at (-0.5,0) {$j$}; \node[above] at (0.5,0) {$k$};
\end{tikzpicture}
\end{array}\right)=\\ =
\left(\bigoplus\lm_i \begin{array}{c}
	\begin{tikzpicture}
	\draw[ultra thick] (-0.5,0) -- (0.5,0);
	\node[above] at (-0.5,0) {$i$}; \node[above] at (0.5,0) {$i$};
	\end{tikzpicture}
\end{array}\right)\oplus \begin{array}{c}
\begin{tikzpicture}
\draw[ultra thick] (-0.5,0) -- (0.5,0);
\begin{scope}[shift={(0,0)}]
\filldraw [red] (-0.1,0.1) -- (0.1,0.1) -- (0.1,-0.1) -- (-0.1,-0.1) -- (-0.1,0.1);
\end{scope}
\node[above] at (-0.5,0) {$i$}; \node[above] at (0.5,0) {$j$};
\end{tikzpicture}
\end{array}\oplus \begin{array}{c}
\begin{tikzpicture}
\draw[ultra thick] (-0.5,0) -- (0.5,0);
\begin{scope}[shift={(0,0)}]
\filldraw [blue] (-0.1,0.1) -- (0.1,0.1) -- (0.1,-0.1) -- (-0.1,-0.1) -- (-0.1,0.1);
\end{scope}
\node[above] at (-0.5,0) {$j$}; \node[above] at (0.5,0) {$k$};
\end{tikzpicture}
\end{array}\oplus \begin{array}{c}
\begin{tikzpicture}
\draw[ultra thick] (-1,0) -- (1,0);
\begin{scope}[shift={(-0.5,0)}]
\filldraw [red] (-0.1,0.1) -- (0.1,0.1) -- (0.1,-0.1) -- (-0.1,-0.1) -- (-0.1,0.1);
\end{scope}
\begin{scope}[shift={(0.5,0)}]
\filldraw [blue] (-0.1,0.1) -- (0.1,0.1) -- (0.1,-0.1) -- (-0.1,-0.1) -- (-0.1,0.1);
\end{scope}
\node[above] at (-1,0) {$i$}; \node[above] at (0,0) {$j$}; \node[above] at (1,0) {$k$};
\end{tikzpicture}
\end{array}
\end{split}
\ee
In general we can take  an ordered product over all the binding points of interfaces each containing only one binding point.
We order the product from left to right by increasing value of the $x$-position of the binding point.

The next step in defining the MSW complex is to describe the grading. As explained in \cite{Gaiotto:2015aoa} the solutions
to \eqref{forced_soliton} define generators whose grading is the fermion number ${\rm \textbf{F}}$ of the corresponding
semiclassical groundstate. This fermion number is the APS $\eta$-invariant of a certain Dirac operator obtained by linearizing
the soliton equation. Unfortunately, the computation of such $\eta$-invariants is rather difficult. Fortunately, for fixed boundary
conditions $\phi_i$ at $x \to -\infty$ and $\phi_{j'}$ at $x \to + \infty$ the different soliton solutions will all have
fermion numbers differing by an integer. That is, the MSW complex is graded by a $\IZ$-torsor. To prove this we first
note that the fermion number of a solution to \eqref{forced_soliton} that is described as a collection of solitons
localized near binding points is the sum of the fermion numbers of those solitons. (Indeed this statement was implicit
in \eqref{eq:CP-Tensor1}.) Now, again for a soliton in a LG model $W(\phi;z)$ where $z$ is now independent of $x$
an $ij$ type soliton has a fermion number of the form $f_{ij} + n_{ij}$ where \cite{Cecotti:1992qh,Cecotti:1992rm,Gaiotto:2015aoa}:
\be
f_{ij}\;\mod\; \IZ = \frac{1}{2\pi\I} \left( \log\; \det\; W''(\phi_j) - \log\; \det\; W''(\phi_i) \right)\; \mod \;\IZ
\ee
Therefore, $f_{ij} + f_{jk} = f_{ik} + m_{ik}$, where $m_{ik}$ is an integer. The fermion number becomes
the ``homological grading'' of the link homology. In this paper we will only determine this homological grading up to
an overall constant, but the relative integers are important in defining the link homology. Since $\eta$ invariants
can be difficult to compute we describe another method by which they can be computed in Appendix \ref{app:FermionTTstar}.

\subsection{Zeta Instantons And Curved Webs}\label{sec:CurvedWebs}

It now remains to describe the differential on the Chan-Paton complexes $\CE(\fI[\wp])_{ij'}$ associated with paths in families of
LG  superpotentials. By general principles of Morse theory the differential
should be constructed from the  solutions of the instanton equation interpolating between two critical
points of $h$ which are ``rigid,'' that is whose only modulus is that of time translation. The condition on the moduli of the instanton
is required because the differential is defined by the supercharge $\CQ_{\zeta}$ and we
want the  matrix element of the differential between perturbative groundstates $\Psi_1$ and $\Psi_2$,
denoted by $\langle \Psi_2|\CQ_{\zeta}|\Psi_1\rangle$, to be nonzero. The number of moduli correspond to the number of
fermion zeromodes, and we should therefore have precisely one fermion zeromode.

 For the Morse function $h$ in equation \eqref{eq:InterfaceMorseFunction}
above the SQM instanton equation is the \emph{$\zeta$-instanton equation} on $\IR^2$:
\footnote{Notice that due to our choice of fermion number convention \eqref{SUSY_conv} we have a minus sign in front of $\I\p_{\tau}$ in comparison to \cite{Gaiotto:2015aoa}. So field configurations we call \emph{instantons} are \emph{anti-instantons} in the sense of \cite{Gaiotto:2015aoa}.}
\be\label{instanton}
\left(\p_x-\I \p_{\tau}\right)\phi^I(x,\tau)=\frac{\I\zeta}{2}g^{I\bar J}\overline{\p_J W\left(\phi^I(x,\tau)|z_a(x)\right)}
\ee
The instanton is meant to interpolate between two solutions of  (\ref{forced_soliton}) in the far past and the far future
whose fermion number differs by $1$. This defines the boundary conditions on \eqref{instanton}.

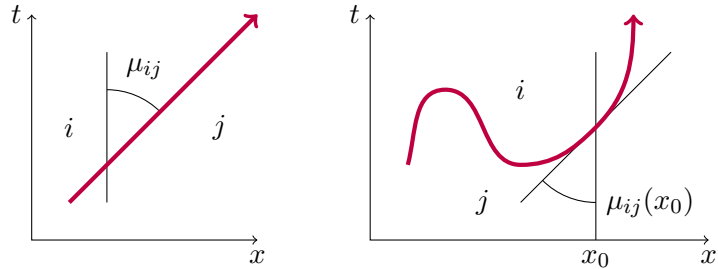
\begin{figure}
	\begin{center}
		\begin{tikzpicture}
		\draw[<->] (0,3) -- (0,0) -- (3,0);
		\draw (1,0.5) -- (1,2.5);
		\draw (1,2) arc (90:45:1);
		\draw[ultra thick, purple, ->] (0.5,0.5) -- (3,3);
		\node[above] at (1.5,2) {$\mu_{ij}$};
		\node[below] at (3,0) {$x$};
		\node[left] at (0,3) {$t$};
		\node at (0.5,1.5) {$i$};
		\node at (2.5,1.5) {$j$};
		\begin{scope}[shift={(4.5,0)}]
		\draw[<->] (0,3) -- (0,0) -- (4.5,0);
		\node[below] at (4.5,0) {$x$};
		\node[left] at (0,3) {$t$};
		\draw (3,0) -- (3,2.5) (2,0.5) -- (4,2.5);
		\draw[ultra thick, purple, ->] (0.5,1) to [out=75, in=180] (1,2) to [out=0,in=180] (2,1) to [out=0, in=225] (3,1.5) to[out=45, in=270] (3.5,3);
		\node[below] at (3,0) {$x_0$};
		\draw (3,0.5) arc (270:225:1);
		\node[right] at (3,0.5) {$\mu_{ij}(x_0)$};
		\node at (1.5,0.5) {$j$};
		\node at (2,2) {$i$};
		\end{scope}
		\end{tikzpicture}
	\end{center}
	\caption{(a): The worldline of a boosted soliton. To exponentially good accuracy the field $\phi$ is in the vacuum $i$ and $j$ away
from a small region of width $\ell_W$ around the straight line we have indicated. (b) The worldline of a curved boosted soliton associated
to an interface of LG models.  }\label{fig:curved_web1}
\end{figure}

We now describe good approximations to   $\zeta$-instantons in terms of \emph{curved webs}. These will give us
a formalism for constructing the   differential on the MSW complexes. To begin, we first consider a theory with
a fixed superpotential. (That is, we consider the theory with $z_a$ independent of $x$.) We recall the
\emph{boosted soliton} solution to the $\zeta$-instanton equation in such theories.
 This is a particular solution
obtained by ``boosting'' the soliton so that the preserved supersymmetry is $\CQ_\zeta, \bar \CQ_{\zeta}$.
Since we are working in Euclidean space a ``boost'' is really a rotation:
\be
\left(
\begin{array}{c}
 x'\\
 \tau'
\end{array}\right)=\left(\begin{array}{cc}
\cos \mu & -\sin \mu \\
\sin \mu & \cos \mu
\end{array}\right)\left(
\begin{array}{c}
	x\\
	\tau
\end{array}\right)
\ee
Under this ``boost'' the $\zeta$-instanton equation transforms as
\be\label{boosted-instanton}
\left(\p_{x'}-\I\p_{\tau'}\right)\phi^I=\frac{\I \zeta e^{\I\mu}}{2}g^{I\bar J}{\frac{\p \bar W}{\p\bar \phi^{\bar J}}}
\ee
Therefore if we consider an $ij$ soliton as a function of $x'$ and choose the rotation angle $\mu_{ij}=-{\rm Arg}[\zeta^{-1}Z_{ij}]$
then $\phi_{ij}(x')$ will be a solution of the $\zeta$-instanton equation. We call this a boosted soliton. To exponentially good
accuracy the field is nearly constant in the vacua $\phi\cong \phi_i$ and $\phi \cong \phi_j$ away from the ``worldline'' of
the soliton shown in Figure \ref{fig:curved_web1}(a).

Let us now consider an interface of LG models defined by a path of parameters  $z_a(x) $. In the adiabatic approximation we can consider
boosted soliton solutions smoothly changing slope with $x$. This leads to edges of a curved web such as
is shown in Figure  \ref{fig:curved_web1}(b).

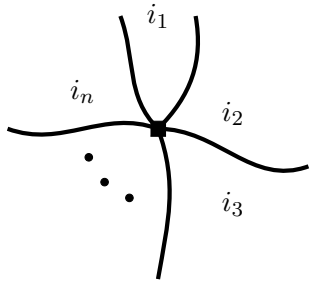
\begin{figure}[h!]
\begin{center}
	\begin{tikzpicture}
	\draw[ultra thick] (0,0) to[out=45,in=280] (0.5,1.5);
	\draw[ultra thick] (0,0) to[out=135,in=290] (-0.5,1.5);
	\draw[ultra thick,] (0,0) to[out=160,in=340] (-2,0);
	\draw[ultra thick] (0,0) to[out=290,in=80] (0,-2);
	\draw[ultra thick] (0,0) to[out=0,in=200] (2,-0.5);
	\node at (-1,0.5) {$i_n$};
	\node at (0,1.5) {$i_1$};
	\node at (1,0.2) {$i_2$};
	\node at (1,-1) {$i_3$};
	\filldraw [black] (-0.71,-0.71) circle (0.05);
	\filldraw [black] (-0.92,-0.38) circle (0.05);
	\filldraw [black] (-0.38,-0.92) circle (0.05);
	\filldraw [black] (-0.1,0.1) -- (0.1,0.1) -- (0.1,-0.1) -- (-0.1,-0.1);
	\end{tikzpicture}
\end{center}
\caption{A bulk vertex in a curved web.}\label{fig:bulk_vertex}
\end{figure}

In equations, a $\zeta$-instanton for the interface, generalizing the boosted soliton, will have a worldline in the  $(x,\tau)$ space satisfying:
\be\label{domain_wall}
\frac{dx(s)}{ds}=-\frac{{\rm Im}\;\left[\zeta^{-1}Z_{ij}\right]}{|Z_{ij}|},\quad \frac{d\tau(s)}{ds}=\frac{{\rm Re}\;\left[\zeta^{-1}Z_{ij}\right]}{|Z_{ij}|}
\ee
where we have chosen the parameter $s$ so that the tangent vector has unit norm. We will call these
\emph{curved boosted solitons}.   As in the discussion of the generators of the MSW complex we must allow for the existence of more general solutions of the $\zeta$-instanton
solution. These can be approximated by curved boosted solitons intersecting at vertices as shown in Figure \ref{fig:bulk_vertex}.

The central charges of the boosted solitons at such a vertex satisfy a natural conservation law:
\be\label{eq:cons_law}
Z_{i_1 i_2}+Z_{i_2 i_3}+\ldots+Z_{i_n i_1}=0
\ee
(where the central charges are evaluated in the theory at the appropriate value of $x$).
These are the $\zeta$-vertices of \cite{Gaiotto:2015aoa}. The path integral associates certain quantities to these vertices known
as \emph{interior amplitudes}. They are denoted by   $\beta_{i_1,\ldots,i_n}$ and define solutions to an $L_\infty$ Maurer-Cartan
equation. Fortunately we will not need to know their precise values for our specific computations - we will just need to know if they are
zero or not, and this information can sometimes be deduced from consistency conditions associated with the flatness of the parallel
transport associated with $\fI[\wp]$. (One such consistency condition is, for example the
Cecotti-Vafa wall-crossing formula.) Of course, for arbitrarily complicated knots one might well need
further information to determine relevant interior amplitudes.

\section{Three Landau-Ginzburg Models }\label{sec:ThreeLGModels}

In this section we define the  three Landau-Ginzburg models relevant to this paper.

\subsection{Yang-Yang-Landau-Ginzburg model}\label{subsec:YYLG}

Let $P_s(\{z_a\})$ denote the configuration space of $m$ ordered and \underline{distinct} points in the punctured complex
plane $\IC - \{ z_a \}$. Thus a point in $P_m(\{z_a\})$ is an ordered $m$-tuple $(w_1, \dots, w_s)$ with $w_i \not= z_a$
for all $i,a$ and $w_i \not= w_j$ for $i\not=j$. The notation is also meant to indicate that we have chosen an
ordering of the points $z_a$. Let $B_m(\{z_a\}) = P_m(\{ z_a \})/S_m$ be the set of unordered distinct
points on $\IC - \{ z_a \}$.
 The target space $X$ of the YYLG model will be a certain cyclic cover of $B_m(\{ z_a \})$.
We may take the K\"ahler metric inherited from the flat metric on the
complex plane. To define the cyclic cover we introduce the multi-valued superpotential on $B_m(\{ z_a \})$:
\be\label{eq:YY-superpotential}
W_{\rm YY}(w_i|z_a)=\sum\lm_{a=1}^n\sum\lm_{i=1}^m k_a\log(w_i-z_a)-2\sum\lm_{1\leq i<j\leq m}\log(w_i-w_j)+c\sum\lm_i w_i
\ee
and we take $X$ to be the cyclic cover on which $W$ is single-valued. That is $X =  \widetilde{ B_m(\{ z_a \})}/H$ where
$H \subset \pi_1 (B_m(\{ z_a \}))$ is the kernel of $\oint dW$.   It will be of some importance below that $\pi_1(X)$ is nontrivial.

Following \cite{Gaiotto:2011nm} we note that the equations determining the vacua of the theory are the   Bethe ansatz equations
for the Gaudin model with irregular singularity. Moreover the vacua are easily described in the large $c$ limit in the case
$m\leq n$ and $k_a = 1$ for all $a$: For each $i$, $1\leq i \leq m$
there is a unique $a$, $1\leq a \leq n$  such that
\be
w_i = z_{a(i)} - \frac{1}{c} + \CO(c^{-2})
\ee
Moreover, for $i\not= j$ we have $a(i) \not= a(j)$. That is,  there is an ``exclusion principle''
that prevents more than one LG field $w_i$ from taking a value $z_{a(i)} - \frac{1}{c} + \CO(c^{-2})$. (Otherwise $(w_i-w_j)^{-1}$ would
contribute $\CO(c^2)$ to the Bethe ansatz equation.)
Thus, the projection of the critical points from $X$ into $B_m(\{ z_a \})$  may be denoted by a map $\vec \epsilon: \{ z_a \} \to \{\pm \}$,
where a $+$ means that the point $z_{a} - \frac{1}{c} + \CO(c^{-2})$ near $z_a$ is ``occupied'' by some $w_i$ and $-$ means it is unoccupied.
Of course, the true critical points are on the cyclic cover $X$ so the critical points can be
labeled -   noncanonically - by $(\vec \epsilon, n)$, $n \in \IZ$.
If we order the points $\{ z_a \}$ and the fields $w_i$
(thus choosing a noncanonical lift to $P_m(\{ z_ a\})$) then we can encode $\vec \epsilon$ by  an ordered tuple
of $\pm$:
\be
\vec \epsilon = \{ \epsilon_1, \dots, \epsilon_n \}
\ee
and this is how we will in fact denote vacua in explicit computations below.

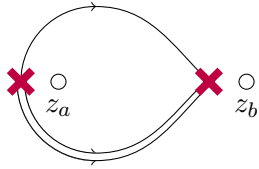
\begin{figure}[h!]
	\begin{center}
		\begin{tikzpicture}
		\draw (0,0) circle (0.1);
		\draw (2.5,0) circle (0.1);
		\node[below] at (0,-0.1) {$z_a$};
		\node[below] at (2.5,-0.1) {$z_b$};
		\draw[->] (-0.5,0) to [out=90,in=180] (0.5,1);
		\draw (0.5,1) to[out=0,in=135] (2,0);
		\draw[->] (-0.55,0) to [out=270,in=180] (0.5,-1.05);
		\draw[->] (-0.45,0) to [out=270,in=180] (0.5,-0.95);
		\filldraw[white] (-0.55,0) to [out=270,in=180] (0.5,-1.05) to [out=0,in=225] (2.05,0) -- (1.95,0) to [out=225,in=0] (0.5,-0.95) to [out=180,in=270] (-0.45,0);
		\draw (-0.55,0) to [out=270,in=180] (0.5,-1.05) to [out=0,in=225] (2.05,0) -- (1.95,0) to [out=225,in=0] (0.5,-0.95) to [out=180,in=270] (-0.45,0);
		\begin{scope}[shift={(-0.5,0)}]
		\draw[line width=3, purple] (-0.15,-0.15) -- (0.15,0.15) (-0.15,0.15) -- (0.15,-0.15);
		\end{scope}
		\begin{scope}[shift={(2,0)}]
		\draw[line width=3, purple] (-0.15,-0.15) -- (0.15,0.15) (-0.15,0.15) -- (0.15,-0.15);
		\end{scope}
		\end{tikzpicture}
	\end{center}
	\caption{Two punctures $z_a$ and $z_b$ are denoted by open circles. The nearby points
$z_a - c^{-1} + \CO(c^{-2})$ are denoted by crosses. (Here, and in all subsequent figures
we will take $c$ to be positive and $\I \zeta =1$. ) There are two distinct solitons illustrated here. The curve on
top is one soliton where some $w_i$ travels from $z_a$ to $z_b$, while the doubled curve on bottom is a second soliton
where $w_i$ travels from $z_a$ to $z_b$. Note that the double line on the bottom curve is a notation for the second
soliton. It is \underline{not} meant to indicate that there are two solitons!  }\label{fig:2_solitons}
\end{figure}

The generic soliton in the theory  will connect vacua in which one point $z_a$ is occupied and another $z_b$ is unoccupied
with the reverse situation where $z_a$ is unoccupied and $z_b$ is occupied, while the occupation of all other points
$z_c$ with $c\not= a,b$ is unchanged. This may be visualized as follows:  A single field $w_i$ will travel from
$z_a$ to $z_b$ as shown in Figure \ref{fig:2_solitons}   while the other fields $w_j$, $j \not=i$ will ``hover'' near
their respective singularities. Note that for large $c$ it follows from the $\zeta$-soliton equation that the soliton
generally moves parallel to $\I \zeta \bar c$ for increasing $x$. The relative fermion number of these two solitons
is discussed in Appendix \ref{app:FermionTTstar}.

\bigskip
\noindent
\textbf{Remarks}:

\begin{enumerate}

\item If we add a suitable function of the $z_a$ to $W$ then
 integrals of $\exp (b^{2} W)$ or $\exp (b^{-2} W)$ over the Lefshetz thimbles associated to the above vacua give a
basis of conformal blocks of Feigin Fuks theory with $Q= b+b^{-1} $. (For the free field representation in the Liouville field theory see \cite{DF},
for the free field representation in the WZW model see \cite{GMOMS}).
This is a key point in deriving link polynomials from the monodromy of these conformal blocks \cite{Ruth_Lawrence,Gaiotto:2011nm}.

\item As we have mentioned, the superpotential is multivalued on $B_m(\{z_a\})$ and we must pass to
a cyclic cover $X$ to define the target space of the YYLG model. Therefore, each Bethe root, denoted by $i$,
corresponds to a point on   $B_m(\{z_a\})$ and lifts to a collection $(i,n)$ of vacua on $X$. The central charge   of a soliton
from $(i,n)$ to $(j,m)$ is
\be\label{eq:CenChargDiff}
\I ( W_i - W_j + 2\pi \I (n-m) )
\ee
Therefore, in principle,  there could be infinitely many solitons on $X$ associated with each of the
two curves shown in Figure \ref{fig:2_solitons}. In fact this does not occur. Explicit numerical
solution of the $\zeta$-soliton equation (for values of $\zeta$ suitable to the slopes implied
by \eqref{eq:CenChargDiff}) shows that each of the curves shown in Figure  \ref{fig:2_solitons}
lifts to a \underline{unique} soliton on $X$  (up to simultaneous shift of $n,m$).
In other words amongst the possible slopes of central charges shown in in Figure
\ref{fig:CC} only one supports a true $\zeta$-soliton. In the language
of \cite{Gaiotto:2015aoa}, for each pair $i,j$ of Bethe roots,
 the MSW spaces $R_{(i,n),(j,m)}$ are nonvanishing
only for a single value of $n-m$.
\begin{figure}[h]
\begin{center}
\begin{tikzpicture}
\draw (0,0) to[out=180,in=0] (-0.25,0.5) to[out=0,in=180] (0,1);
\node[left] at (-0.25,0.5) {$2\pi\I$};
\begin{scope}[shift={(2,-0.5)},xscale=-1]
\draw (0,0) to[out=180,in=0] (-0.25,0.5) to[out=0,in=180] (0,1);
\node[right] at (-0.25,0.5) {$2\pi\I$};
\end{scope}
\node at (-1,2.5) {$W_i$};
\node at (3,2) {$W_j$};
\foreach \i in {0,1,3}
{
	\begin{scope}[shift={(0,1)}]
	\draw[gray] (0,0) -- (2,-1.5+\i);
	\begin{scope}[scale=0.6]
	\draw[gray,->] (0,0) -- (2,-1.5+\i);
	\end{scope}
	\end{scope}
}
\foreach \i in {2}
{
	\begin{scope}[shift={(0,1)}]
	\draw[red, ultra thick] (0,0) -- (2,-1.5+\i);
	\begin{scope}[scale=0.6]
	\draw[red, ultra thick,->] (0,0) -- (2,-1.5+\i);
	\end{scope}
	\end{scope}
}
\foreach \i in {0,...,3}
{
	\begin{scope}[shift={(0,\i)}]
		\draw[ultra thick] (-0.1,-0.1) -- (0.1,0.1) (-0.1,0.1) -- (0.1,-0.1);
	\end{scope}
	\begin{scope}[shift={(2,\i-0.5)}]
	\draw[ultra thick] (-0.1,-0.1) -- (0.1,0.1) (-0.1,0.1) -- (0.1,-0.1);
	\end{scope}
}
\end{tikzpicture}
\qquad
\begin{tikzpicture}
\draw (0,0) to[out=180,in=0] (-0.25,0.5) to[out=0,in=180] (0,1);
\node[left] at (-0.25,0.5) {$2\pi\I$};
\begin{scope}[shift={(2,-0.5)},xscale=-1]
\draw (0,0) to[out=180,in=0] (-0.25,0.5) to[out=0,in=180] (0,1);
\node[right] at (-0.25,0.5) {$2\pi\I$};
\end{scope}
\node at (-1,2.5) {$W_i$};
\node at (3,2) {$W_j$};
\foreach \i in {0,2,3}
{
	\begin{scope}[shift={(0,1)}]
	\draw[gray] (0,0) -- (2,-1.5+\i);
	\begin{scope}[scale=0.6]
	\draw[gray,->] (0,0) -- (2,-1.5+\i);
	\end{scope}
	\end{scope}
}
\foreach \i in {1}
{
	\begin{scope}[shift={(0,1)}]
	\draw[blue, ultra thick] (0,0) -- (2,-1.5+\i);
	\begin{scope}[scale=0.6]
	\draw[blue, ultra thick,->] (0,0) -- (2,-1.5+\i);
	\end{scope}
	\end{scope}
}
\foreach \i in {0,...,3}
{
	\begin{scope}[shift={(0,\i)}]
	\draw[ultra thick] (-0.1,-0.1) -- (0.1,0.1) (-0.1,0.1) -- (0.1,-0.1);
	\end{scope}
	\begin{scope}[shift={(2,\i-0.5)}]
	\draw[ultra thick] (-0.1,-0.1) -- (0.1,0.1) (-0.1,0.1) -- (0.1,-0.1);
	\end{scope}
}
\end{tikzpicture}
\end{center}
\caption{Here we show the slopes in the complex plane of $W$ illustrating potential lifts to $X$ of the solitons illustrated in Figure \ref{fig:2_solitons}. In fact, amongst all the potential lifts, only one straight line corresponds to a true solution on $X$. On the left we show the projection to the $W$-plane of the lift of the lower soliton in Figure \ref{fig:2_solitons} and on the right the projection of the upper soliton. That is:  $\color{red} \blacksquare$: $\Longrightarrow$, $\color{blue} \blacksquare$: $\longrightarrow$.}\label{fig:CC}
\end{figure}
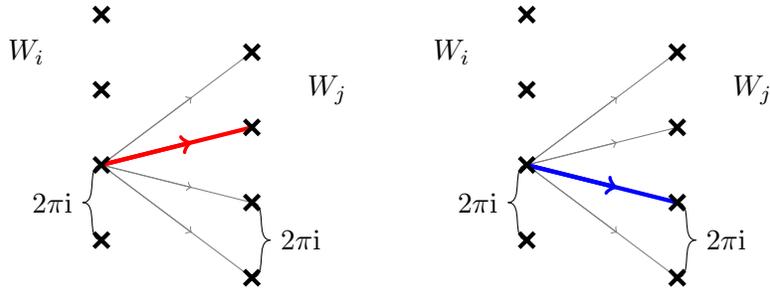

\item In addition to the above solitons we must mention that there are also ``purely flavour solitons''
connecting vacua $(i,n)$ to $(i,m)$ where one LG field $w_i(x)$ near $z_{a(i)}$ loops around
$z_{a(i)}$. If the $z_a$ are constant the solution can only have winding number $\pm 1$ but if
$z_a(x)$ vary arbitrary winding numbers can occur. These solutions always have central
charge given by $Z = 2\pi (m-n)$. Marginal stability phenomena associated with these flavour
solitons will prove to be problematical for verifying the RI move in the YYLG model, as discussed
in Section \ref{sec:Obstruction} below.

\end{enumerate}

\subsection{Naive Monopole Model}\label{subsec:NMLG-Model}

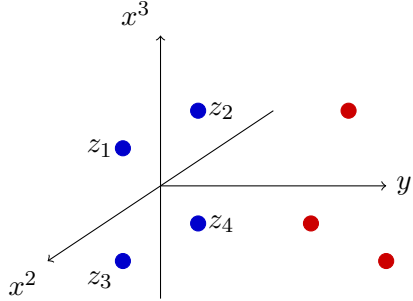
\begin{figure}[h!]
	\begin{center}
		\begin{tikzpicture}
		\draw[->] (0,-1.5) -- (0,2);
		\draw[->] (1.5,1) -- (-1.5,-1);
		\draw[->] (0,0) -- (3,0);
		\filldraw[blue!80!black] (-0.5,-1) circle (0.1) (-0.5,0.5) circle (0.1) (0.5,-0.5) circle (0.1) (0.5,1) circle (0.1);
		\node[above left] at (0,2) {$x^3$};
		\node[below left] at (-1.5,-1) {$x^2$};
		\node[right] at (3,0) {$y$};
		\node[left] at (-0.5,0.5) {$z_1$};
		\node[right] at (0.5,1) {$z_2$};
		\node[below left] at (-0.5,-1) {$z_3$};
		\node[right] at (0.5,-0.5) {$z_4$};
		\filldraw[red!80!black] (2,-0.5) circle (0.1) (3,-1) circle (0.1) (2.5,1) circle (0.1);
		\end{tikzpicture}
	\end{center}
	\caption{An illustration of the Gaiotto-Witten monopole model for link homology. The red dots, localized for $y\gg 1$
represent smooth $SU(2)$ monopoles of monopole charge $1$ (when the roots $w_i$ of $Q$ are all distinct).
  The behavior of the superpotential  at $w_i =z_a$  is meant to
take into account the effects of the 't Hooft-loop boundary conditions of the underlying 5d SYM theory.  } \label{fig:mono}
\end{figure}

Gaiotto and Witten introduced a second LG model, which we call the MLG model. One can obtain the YYLG model from the MLG model by a
procedure of ``integrating out fields,'' but it will be convenient to introduce here a kind of intermediate model that we call the
``Naive Monopole Landau-Ginzburg Model'' (NMLG model). The NMLG model is based on the fact that regions at infinity of monopole
moduli space are well-approximated by the configuration space of $m$ charge one monopoles.
 We promote each of the chiral superfields $w_i$ of the YYLG to a pair of chiral superfields
$(w_i, Y_i)$ where $e^{Y_i} \in \IC^*$ and now there is no restriction forbidding $w_i = z_a$, although we continue to
require $w_i \not= w_j$ for $i\not=j$. The idea is that $(w_i, e^{Y_i})$, $i=1,\dots, m$  can be
considered as coordinates near infinity of a moduli space of smooth magnetic monopoles on $\IR^3 = \IR \times \IC$ of total magnetic charge $m$.
For example, for a single monopole we could write: $w = x^2 + \I x^3$ and $Y= \mu y + \I \vartheta$ where $(x,y,z)$ denote
the center of mass of a monopole and $\vartheta$ denotes the electric phase. Here $\mu$ is a dimensional scale controlling the
monopole size \cite{Gaiotto:2011nm}.

More invariantly, the target space is taken to be a cyclic cover of $(\IC \times \IC^*)^m/S_m $.  The metric on the $w_i$ is
again induced from the flat metric on the plane and that on the $Y_i$ is induced from $\vert dY_i \vert^2$.

The superpotential of the model is
\be
W_{\rm naive}(w_i,Y_i|z_a)=c\sum\lm_i w_i+\sum\lm_i Y_i+\sum\lm_i \frac{\prod\lm_a (w_i-z_a)^{k_a}}{\prod\lm_{j\neq i}(w_i-w_j)}e^{-Y_i}
\ee
and again we take the target space $X$ to be the cyclic cover of $(\IC \times \IC^*)^m/S_m $ on which $W$ is single-valued.

The NMLG model can be related to the YYLG as follows. Define
\be\label{eq:subs}
\tilde Y_i :=  Y_i -\sum\lm_{a=1}^n k_a\log(w_i-z_a)+\sum\lm_{j\neq i}\log(w_i-w_j)
\ee

Then we have:
\be
W_{\rm NMLG} (w_i,Y_i|z_a)=W_{YY}(w_i|z_a)+\sum\lm_i\left(\tilde{Y}_i+ e^{-\tilde Y_i}\right)
\ee
From this representation it is easy to describe the vacua. For each of the vacua $(\vec \epsilon, n)$ of the
YYLG we have a corresponding vacuum of the NMLG model with $e^{\tilde Y_i} = 1$. So there is
 a one-one correspondence between the vacua and solitons of
the YYLG and NMLG models.
Nevertheless, the solitons in the NMLG model differ from those in the YYLG model
in an important way. That will be the subject of the discussion of Section \ref{sec:Res_RI}.

\bigskip
\noindent
\textbf{Remark}: A proper derivation of the YYLG model from the NMLG model would also take into account
kinetic terms and one-loop determinants. We expect this will not introduce any problems, but we have
not investigated this issue in detail.

\subsection{Smooth Monopole Landau-Ginzburg Model}\label{subsec:MLG-Model}

Finally, we come to the MLG model of \cite{Gaiotto:2011nm}. The target space of this
model is the universal cover of the moduli space of $m$ magnetic monopoles in $SU(2)$
Yang-Mills-Higgs theory on $\IR^3$. The moduli space has the form
$\IR^4 \times \CM_0$ where $\CM_0$ is the simply-connected strongly-centered moduli space.
The most natural metric would be the hyperk\"ahler metric, although fortunately we will
not need to know any details of this metric.

In order to describe the superpotential we use the identification of the moduli space
with a space of rational maps $\IC \IP^1 \to \IC \IP^1$ of degree $m$ and preserving the point at infinity
\cite{Donaldson,Atiyah:1988jp,Hurtubise}.
Thus the rational map can be taken to be $ u \mapsto P(u)/Q(u)$ where $Q$ is a monic polynomial of degree $m$
and $P$ has degree $m-1$ with nonvanishing leading coefficient. The polynomials $P,Q$ are relatively prime
so $P$ never vanishes at the roots of $Q$.  We will
denote those roots (which need not be distinct)  by $w_i$:
\be
Q(u) = \prod_{i=1}^m(u-w_i).
\ee
The superpotential is
\be\label{eq:MLG-SUPERPOT}
W_{\rm MLG}(P,Q|z_a)=\sum\lm_{i=1}^m \Biggl\{
\mathop{\rm Res}\lm_{u=w_i} \left[K(u) \frac{ P(u)}{Q(u)} du \right]
-     \log  P(w_i) +c  w_i \Biggr\}
\ee
where $K(u):=\prod\lm_{a=1}^n (u-z_a)^{k_a}$.
This superpotential is multivalued on the monopole moduli space but single valued on a cyclic cover.
This cyclic cover coincides with the simply connected cover of monopole moduli space.

In order to relate the MLG model to the NMLG model we note that, if the roots $w_i$ of $Q$ are all distinct and we define   $e^{Y_i}:= P(w_i)$
then evaluation of the residue in \eqref{eq:MLG-SUPERPOT} reproduces $W_{\rm NMLG}$.  However, a crucial difference from the
NMLG model is that now, not only is it possible to have $w_i = z_a$ but the superpotential is also smooth when roots $w_i$ and $w_j$
coincide. For example, if $w_1 \to w_2$ but all other $w_i$ are distinct then the sum of the residues at $w_1$ and $w_2$ is just $F'(w_2)$ where
\be
F(w) = \frac{ K(w) P(w) }{ \prod_{j=3}^m (w - w_j) }.
\ee

Another way to make the smoothness manifest is the following remark, due to D. Gaiotto. The 
sum of the residues of the first term is the residue at infinity. For the second term 
we note that $\prod_i P(w_i) = R(Q,P)$ is the resultant of polynomials $Q$ and $P$. 
The resultant can be expressed as a polynomial in the coefficients of $P$ and $Q$, 
and since $P$ and $Q$ are relatively prime it never vanishes. Finally, denoting the 
coefficients of $Q$ by $Q(w) = w^m + Q_{m-1} w^{m-1} + \cdots $ the last term 
is proportional to $Q_{m-1}$. Altogether then, 
\be\label{eq:MLG-SUPERPOT-alt}
W_{\rm MLG}(P,Q|z_a)= -  
\mathop{\rm Res}\lm_{u=\infty} \left[K(u) \frac{ P(u)}{Q(u)} du \right]
-     \log R(Q,P) - c Q_{m-1}
\ee
makes no reference to the zeros $w_i$, and is manifestly smooth on the cover of moduli space.

\section{The Rules For Knot Homology}\label{sec:KnotHomRules}

\subsection{Decomposing Interfaces Into Basic Products}

From the  general properties
\eqref{eq:ComposeInterfaces} and \eqref{eq:ComposeChanPatonMatrices} of interfaces it follows that
the chain complex of the link homology, at least as a bi-graded vector space (or $\IZ$-module)
can be rather easily computed from products of elementary factors. In this section
we describe those elementary factors. The differential on the complex will be described at
the end of this section, and is much harder to compute.

We have described the complexes as bi-graded. As we have discussed, one grading is by the
fermion number operator ${\rm \textbf{F}}$. The other grading, which we write as
$q^{{\rm \textbf{P}} } $ in the Poincar\'e polynomial arises from the periods of
the superpotential $dW$. For example, when the $z_a$ are independent of space a $\zeta$-soliton projects to a
closed path $\gamma: \IR \to B_m(\{ z_a \})$ in the YYLG model and would have $q$-degree given by
\be\label{eq:P-degree}
{\rm \textbf{P} } = -\frac{1}{\pi \I } \oint_{\gamma} d W
\ee
so that ${\rm \textbf{P} }= +1$ for the upper soliton and ${\rm \textbf{P} }= -1$
 for the lower soliton in Figure \ref{fig:2_solitons}.

When we consider a braiding interface with a closed path $z_a(x)$ in $B_n(\IC)$
the ${\rm \textbf{P} }$-degree of the $\zeta$-solitons solving \eqref{forced_soliton}
is determined as follows. Notice that there is a natural mapping
\be
P_{n+m}(\IC) \to P_n(\IC)
\ee
whose fibers are $P_m(\IC - \{z_a \}$). Choosing an ordering of the $z_a$'s,
an interface defines a path $\bar \gamma$ in $P_n(\IC)$ and a forced
$\zeta$-soliton for this interface then describes a path in $\gamma: \IR \to P_{n+m}(\IC)$
that covers $\bar\gamma$. We then again take the  $\rm \bf P$-degree to be given
by equation \eqref{eq:P-degree} (for the projection with unordered points).

Since ${\rm \textbf{P} }$ commutes with $\CQ_{\zeta}$ we can
decompose the Chan-Paton factors as a direct sum of complexes
$\CE_\alpha(\fI)$ with fixed grading $\alpha$ under ${\rm \textbf{P} }$.
We will write such a   decomposition as:
\be
\CE(\fI) = \oplus_{\alpha} q^\alpha  \CE_\alpha(\fI)
\ee
where the pre-factor $q^\alpha$ is merely a convenient book-keeping device.

A crucial point is that, while we generally describe the rules in the framework of the YYLG
model, the same rules can be adapted to the NMLG and MLG models, thanks to the one-one
correspondence of vacua, solitons, and edges of curved webs.

{\bf Remark:} It is well-known that quantum invariants of links such as Chern-Simons invariants are in fact invariants of embeddings of ribbons in $\IR^3$ rather than circles. This will, of course, also be a feature of the link homologies. The net result of this is that the overall $(\Pdeg,{\bf F})$-bidegree of the link complex can be shifted by a change of the framing of the link. Correspondingly, our basic interfaces are defined up to an overall (but universal) shift of $(\Pdeg,{\bf F})$-bidegree, and in verifying some of the Reidemeister moves we will need to account for a shift of $(\Pdeg,{\bf F})$-bidegree via a framing anomaly.

\subsection{The Braiding Interface}\label{sec:Brading}

Let us now describe the basic braiding interfaces. The interface is defined by taking the
points   $z_a$ to be  $x$-independent for $x \leq x_1$ and $x \geq x_1 + L$
while for for $x_1 \leq x \leq x_1 + L$, all but two points, say $z_{a_1}$ and $z_{a_2}$
remain constant while
\be\label{eq:braiding-a}
\begin{split}
z_{a_1}(x) & = \half(z_{a_1} + z_{a_2}) + e^{  \I \pi (x-x_1)/L } \half(z_{a_1} - z_{a_2}) \\
z_{a_2}(x) & = \half(z_{a_1} + z_{a_2}) - e^{  \I \pi (x-x_1)/L } \half(z_{a_1} - z_{a_2}) \\
\end{split}
\ee
We will denote the interface associated with such a braiding by  $\CR_{a_1, a_2}$,
or just by $\CR$ when the strands are understood.

We can take $L$ to be very large and we can take $z_{a_1}, z_{a_2}$ to be very far from all the
other punctures. Therefore,   we can describe the contribution to the Chan-Paton
interfaces by focusing on a tensor factor associated with the theory with one LG field $w$ and two punctures
$z_{a_1}$ and $z_{a_2}$. The Chan-Paton matrix for this subtheory is a $4\times 4$ matrix with matrix elements specified by
$(\epsilon_{a_1}, \epsilon_{a_2}; \epsilon_{a_2}, \epsilon_{a_1})$. We write the complex diagrammatically as follows:

	\be\label{rules_a}
	\begin{split}
		\CE\left(\begin{array}{c}
			\begin{tikzpicture}
			\begin{scope}[scale=0.8]
			\draw[ultra thick] (0,0) -- (0,0.5) to[out=90,in=210] (0.5,1) to[out=30,in=270] (1,1.5) -- (1,2) (1,0) -- (1,0.5) to[out=90,in=330] (0.6,0.9) (0.4,1.1) to[out=150,in=270] (0,1.5) -- (0,2);
			\end{scope}
			\end{tikzpicture}
		\end{array}\right)= q^{\frac{1}{2}} \begin{array}{c}
		\begin{tikzpicture}
		\begin{scope}[scale=0.8]
		\node[above] at (0,2) {$+$};
		\node[above] at (1,2) {$+$};
		\node[below] at (0,0) {$+$};
		\node[below] at (1,0) {$+$};
		\draw[ultra thick, purple] (0,0) -- (0,0.5) to[out=90,in=210] (0.5,1) to[out=30,in=270] (1,1.5) -- (1,2) (1,0) -- (1,0.5) to[out=90,in=330] (0.6,0.9) (0.4,1.1) to[out=150,in=270] (0,1.5) -- (0,2);
		\end{scope}
		\end{tikzpicture}
	\end{array}\oplus q^{\frac{1}{2}}\begin{array}{c}
	\begin{tikzpicture}
	\begin{scope}[scale=0.8]
	\node[above] at (0,2) {$-$};
	\node[above] at (1,2) {$-$};
	\node[below] at (0,0) {$-$};
	\node[below] at (1,0) {$-$};
	\draw[ultra thick, purple] (0,0) -- (0,0.5) to[out=90,in=210] (0.5,1) to[out=30,in=270] (1,1.5) -- (1,2) (1,0) -- (1,0.5) to[out=90,in=330] (0.6,0.9) (0.4,1.1) to[out=150,in=270] (0,1.5) -- (0,2);
	\end{scope}
	\end{tikzpicture}
\end{array}\oplus q^{-\frac{1}{2}}\begin{array}{c}
\begin{tikzpicture}
\begin{scope}[scale=0.8]
\node[above] at (0,2) {$-$};
\node[above] at (1,2) {$+$};
\node[below] at (0,0) {$+$};
\node[below] at (1,0) {$-$};
\draw[ultra thick, purple] (0,0) -- (0,0.5) to[out=90,in=210] (0.5,1) to[out=30,in=270] (1,1.5) -- (1,2) (1,0) -- (1,0.5) to[out=90,in=330] (0.6,0.9) (0.4,1.1) to[out=150,in=270] (0,1.5) -- (0,2);
\end{scope}
\end{tikzpicture}
\end{array}\oplus\\ \oplus q^{-\frac{1}{2}}\begin{array}{c}
\begin{tikzpicture}
\begin{scope}[scale=0.8]
\node[above] at (0,2) {$+$};
\node[above] at (1,2) {$-$};
\node[below] at (0,0) {$-$};
\node[below] at (1,0) {$+$};
\draw[ultra thick, purple] (0,0) -- (0,0.5) to[out=90,in=210] (0.5,1) to[out=30,in=270] (1,1.5) -- (1,2) (1,0) -- (1,0.5) to[out=90,in=330] (0.6,0.9) (0.4,1.1) to[out=150,in=270] (0,1.5) -- (0,2);
\end{scope}
\end{tikzpicture}
\end{array}\oplus q^{\frac{1}{2}}\begin{array}{c}
\begin{tikzpicture}
\begin{scope}[scale=0.8]
\node[above] at (0,2) {$+$};
\node[above] at (1,2) {$-$};
\node[below] at (0,0) {$+$};
\node[below] at (1,0) {$-$};
\draw (0,0.45) -- (1,0.45);
\draw[ultra thick, purple] (0,0) -- (0,0.5) to[out=90,in=210] (0.5,1) to[out=30,in=270] (1,1.5) -- (1,2) (1,0) -- (1,0.5) to[out=90,in=330] (0.6,0.9) (0.4,1.1) to[out=150,in=270] (0,1.5) -- (0,2);
\end{scope}
\end{tikzpicture}
\end{array}\oplus q^{-\frac{3}{2}}\begin{array}{c}
\begin{tikzpicture}
\begin{scope}[scale=0.8]
\node[above] at (0,2) {$+$};
\node[above] at (1,2) {$-$};
\node[below] at (0,0) {$+$};
\node[below] at (1,0) {$-$};
\draw (0,0.5) -- (1,0.5) (0,0.4) -- (1,0.4);
\filldraw[black] (0.5,0.45) circle (0.1);
\draw[ultra thick, purple] (0,0) -- (0,0.5) to[out=90,in=210] (0.5,1) to[out=30,in=270] (1,1.5) -- (1,2) (1,0) -- (1,0.5) to[out=90,in=330] (0.6,0.9) (0.4,1.1) to[out=150,in=270] (0,1.5) -- (0,2);
\end{scope}
\end{tikzpicture}
\end{array}
\end{split}\ee

In this figure the $x$ axis runs vertically oriented upwards and the braiding of $z_{a_1}(x)$ and $z_{a_2}(x)$ is indicated
on the LHS so that if we look down on the plane from the top the braiding is counterclockwise.
 On the RHS $\pm$ signs at the bottom and top  indicate the  initial and final vacuum, respectively.\footnote{More precisely the sequence of $\pm$ signs is the projection of the vacuum to $B_s(\{z_a\})$ then lifted to $P_s(\{z_a\})$.}
%
%
Note that there are only six terms out a possible sixteen
matrix elements.  The other matrix elements are the zero complex. All of the six nonzero complexes are one dimensional.
Each one-dimensional complex will be denoted below by $\IZ[\Psi]$ where $\Psi$ is a generator corresponding to
a solution of the relevant $\zeta$-soliton
equation.
\footnote{Note that, although the quantum-mechanical interpretation of Morse theory naturally leads to an MSW
complex made from complex vector spaces, in fact the formalism we describe makes sense for complexes defined
over the integers. This is a familiar oddity of Witten's approach to Morse theory.}
 The first four complexes  are generated
by the hovering solutions to \eqref{forced_soliton}. The last two complexes are also one dimensional but a
binding point has been occupied by a soliton where $w$ travels from $z_{a_1}(x_0)$ to $z_{a_2}(x_0)$
(and $x_0$ is the binding point).   The single and double lines refer to the type
of the solitons shown in figure \ref{fig:2_solitons} above. We use the convention
 that solitons depicted in the diagrams flow from left to right. This is justified by
 the large $c$ approximation, as mentioned above.

Similarly, for the clockwise braiding
\be\label{eq:braiding-b}
\begin{split}
z_{a_1}(x) & = \half(z_{a_1} + z_{a_2}) + e^{ - \I \pi (x-x_1)/L } \half(z_{a_1} - z_{a_2}) \\
z_{a_2}(x) & = \half(z_{a_1} + z_{a_2}) - e^{-  \I \pi (x-x_1)/L } \half(z_{a_1} - z_{a_2}) \\
\end{split}
\ee
we have the Chan-Paton data:
\be
\label{rules_b}
\begin{split}
	\CE\left(\begin{array}{c}
		\begin{tikzpicture}
		\begin{scope}[scale=0.8, xscale=-1]
		\draw[ultra thick] (0,0) -- (0,0.5) to[out=90,in=210] (0.5,1) to[out=30,in=270] (1,1.5) -- (1,2) (1,0) -- (1,0.5) to[out=90,in=330] (0.6,0.9) (0.4,1.1) to[out=150,in=270] (0,1.5) -- (0,2);
		\end{scope}
		\end{tikzpicture}
	\end{array}\right)= q^{-\frac{1}{2}} \begin{array}{c}
	\begin{tikzpicture}
	\begin{scope}[scale=0.8, xscale=-1]
	\node[above] at (0,2) {$+$};
	\node[above] at (1,2) {$+$};
	\node[below] at (0,0) {$+$};
	\node[below] at (1,0) {$+$};
	\draw[ultra thick, purple] (0,0) -- (0,0.5) to[out=90,in=210] (0.5,1) to[out=30,in=270] (1,1.5) -- (1,2) (1,0) -- (1,0.5) to[out=90,in=330] (0.6,0.9) (0.4,1.1) to[out=150,in=270] (0,1.5) -- (0,2);
	\end{scope}
	\end{tikzpicture}
\end{array}\oplus q^{-\frac{1}{2}}\begin{array}{c}
\begin{tikzpicture}
\begin{scope}[scale=0.8, xscale=-1]
\node[above] at (0,2) {$-$};
\node[above] at (1,2) {$-$};
\node[below] at (0,0) {$-$};
\node[below] at (1,0) {$-$};
\draw[ultra thick, purple] (0,0) -- (0,0.5) to[out=90,in=210] (0.5,1) to[out=30,in=270] (1,1.5) -- (1,2) (1,0) -- (1,0.5) to[out=90,in=330] (0.6,0.9) (0.4,1.1) to[out=150,in=270] (0,1.5) -- (0,2);
\end{scope}
\end{tikzpicture}
\end{array}\oplus q^{\frac{1}{2}}\begin{array}{c}
\begin{tikzpicture}
\begin{scope}[scale=0.8, xscale=-1]
\node[above] at (0,2) {$-$};
\node[above] at (1,2) {$+$};
\node[below] at (0,0) {$+$};
\node[below] at (1,0) {$-$};
\draw[ultra thick, purple] (0,0) -- (0,0.5) to[out=90,in=210] (0.5,1) to[out=30,in=270] (1,1.5) -- (1,2) (1,0) -- (1,0.5) to[out=90,in=330] (0.6,0.9) (0.4,1.1) to[out=150,in=270] (0,1.5) -- (0,2);
\end{scope}
\end{tikzpicture}
\end{array}\oplus\\ \oplus q^{\frac{1}{2}}\begin{array}{c}
\begin{tikzpicture}
\begin{scope}[scale=0.8, xscale=-1]
\node[above] at (0,2) {$+$};
\node[above] at (1,2) {$-$};
\node[below] at (0,0) {$-$};
\node[below] at (1,0) {$+$};
\draw[ultra thick, purple] (0,0) -- (0,0.5) to[out=90,in=210] (0.5,1) to[out=30,in=270] (1,1.5) -- (1,2) (1,0) -- (1,0.5) to[out=90,in=330] (0.6,0.9) (0.4,1.1) to[out=150,in=270] (0,1.5) -- (0,2);
\end{scope}
\end{tikzpicture}
\end{array}\oplus q^{\frac{3}{2}}\begin{array}{c}
\begin{tikzpicture}
\begin{scope}[scale=0.8, xscale=-1]
\node[above] at (0,2) {$+$};
\node[above] at (1,2) {$-$};
\node[below] at (0,0) {$+$};
\node[below] at (1,0) {$-$};
\draw (0,1.55) -- (1,1.55);
\filldraw[white] (0.5,1.55) circle (0.1);
\draw (0.5,1.55) circle (0.1);
\draw[ultra thick, purple] (0,0) -- (0,0.5) to[out=90,in=210] (0.5,1) to[out=30,in=270] (1,1.5) -- (1,2) (1,0) -- (1,0.5) to[out=90,in=330] (0.6,0.9) (0.4,1.1) to[out=150,in=270] (0,1.5) -- (0,2);
\end{scope}
\end{tikzpicture}
\end{array}\oplus q^{-\frac{1}{2}}\begin{array}{c}
\begin{tikzpicture}
\begin{scope}[scale=0.8, xscale=-1]
\node[above] at (0,2) {$+$};
\node[above] at (1,2) {$-$};
\node[below] at (0,0) {$+$};
\node[below] at (1,0) {$-$};
\draw (0,1.5) -- (1,1.5) (0,1.6) -- (1,1.6);
\draw[ultra thick, purple] (0,0) -- (0,0.5) to[out=90,in=210] (0.5,1) to[out=30,in=270] (1,1.5) -- (1,2) (1,0) -- (1,0.5) to[out=90,in=330] (0.6,0.9) (0.4,1.1) to[out=150,in=270] (0,1.5) -- (0,2);
\end{scope}
\end{tikzpicture}
\end{array}
\end{split}\ee
%

We will denote  the interface associated with \eqref{rules_a} by $\CR^{-1}$ below, or by $\CR^{-1}_{a_1, a_2}$ when
there are several strands. Of course, we expect \eqref{rules_a} and \eqref{rules_b} to be inverse to each other
so that $\CR \boxtimes \CR^{-1}$ is homotopic to the identity interface. This is the second Reidemeister move,
and, at the level  of Chan-Paton data it will be demonstrated in Section \ref{subsec:R2} below.

\bigskip
\noindent
\textbf{Remarks}

\begin{enumerate}

\item Notice we have switched the notion of the fermion number to the opposite in comparison to \cite{Gaiotto:2015aoa} and this
has important implications for keeping track of past- and future-stable binding points.
\footnote{See \cite{Gaiotto:2015aoa}, Section 7.4.1 for the definition of past- and future- stability of binding points.}
For \underline{future stable}  binding points as
$x$ moves from $x_0-\epsilon $ to $x_0 + \epsilon$ the vacuum weights
$z_i(x) - z_j(x)$ rotate in the \underline{counter-clockwise} direction through the
imaginary axis, while for \underline{\underline{past stable}} binding points
$z_i(x) - z_j(x)$ rotates \underline{\underline{clockwise}}.  With our definition of fermion numbers the vacuum weights in a LG model
are given by $z_i = -\zeta^{-1}  W_i$ and in the large $c$ limit
the $+-$ soliton has vacuum weights
\be
z_{ij} \cong -\zeta^{-1}  c(z_{a(i)} - z_{a(j)} )
\ee
We therefore conclude that the binding points in the
in the last two terms of \eqref{rules_a} are \underline{future-stable}
while those in the last
two terms of \eqref{rules_b} are \underline{\underline{past-stable}}.
%
%
%

\item One can compute the \Pdeg-degree of the forced solitons
represented in \eqref{rules_a} and \eqref{rules_b} using
\eqref{eq:P-degree}. However, to do this it is necessary to
modify the superpotential \eqref{eq:YY-superpotential}
with an additional term
\be
\Delta W = - \half \sum_{a<b} k_a k_b \log( z_a - z_b)
\ee
This term does not affect the dynamics of the LG fields $w_i$
but is needed in the free-field representations of conformal
blocks whose monodromy defines link invariants. Indeed in
order to produce the standard R-matrix of quantum group
theory from the Euler character of the Chan-Paton data of
the interface $\CR$ the overall factor of $q^{1/2}$ in
\eqref{rules_a} and of $q^{-1/2}$ in \eqref{rules_b}
is necessary. In this way we compute the $\Pdeg$-degree
of the complexes in \eqref{rules_a} and \eqref{rules_b}.

\item  Computing the homological grading, or Fermion number, of the
complexes above is much more subtle. A procedure for computing this
Fermion number - of some interest in its own right - is
explained in Appendix \ref{app:FermionTTstar}. The result is that
we can make a ``gauge choice'' so that the fermion numbers of the
first five complexes in \eqref{rules_a} are zero, while the solid dot
  ($\begin{tikzpicture}
\filldraw[black] (0,0) circle (0.1);
\end{tikzpicture}$) on the horizontal line on the soliton in
the last complex of \eqref{rules_a} indicates that the fermion number of the
soliton has been shifted by $+1$. Similarly we can take the Fermion number
of all but the fifth complex in \eqref{rules_b} to be zero, and the open circle  ($\begin{tikzpicture}
\draw (0,0) circle (0.1);
\end{tikzpicture}$) on the soliton in the fifth complex indicates that the fermion number
has been shifted by $-1$.

\item To recover the standard calculation of the Jones polynomial from the $SU(2)_q$ $R$-matrix one simply
computes the Euler characters of the above complexes. The Chan-Paton matrix of complexes is reduced in this way
to  matrices with matrix elements given by taking the trace with  $q^{\bf P}(-1)^{\bf F}$. The result is:
\be
\begin{split}
\chi\left(\CR\right)=q^{\frac{1}{2}}\begin{array}{c}
\begin{tikzpicture}
\begin{scope}[scale=0.6]
\draw[ultra thick] (-0.5,-0.5) -- (0.5,0.5) (0.5,-0.5) -- (0.1,-0.1) (-0.1,0.1) -- (-0.5,0.5);
\node[below] at (-0.5,-0.5) {$+$}; \node[below] at (0.5,-0.5) {$+$};
\node[above] at (-0.5,0.5) {$+$}; \node[above] at (0.5,0.5) {$+$};
\end{scope}
\end{tikzpicture}
\end{array}\mkern-10mu +q^{\frac{1}{2}}\begin{array}{c}
\begin{tikzpicture}
\begin{scope}[scale=0.6]
\draw[ultra thick] (-0.5,-0.5) -- (0.5,0.5) (0.5,-0.5) -- (0.1,-0.1) (-0.1,0.1) -- (-0.5,0.5);
\node[below] at (-0.5,-0.5) {$-$}; \node[below] at (0.5,-0.5) {$-$};
\node[above] at (-0.5,0.5) {$-$}; \node[above] at (0.5,0.5) {$-$};
\end{scope}
\end{tikzpicture}
\end{array} \mkern-10mu +q^{-\frac{1}{2}}\begin{array}{c}
\begin{tikzpicture}
\begin{scope}[scale=0.7]
\draw[ultra thick] (-0.5,-0.5) -- (0.5,0.5) (0.5,-0.5) -- (0.1,-0.1) (-0.1,0.1) -- (-0.5,0.5);
\node[below] at (-0.5,-0.5) {$+$}; \node[below] at (0.5,-0.5) {$-$};
\node[above] at (-0.5,0.5) {$-$}; \node[above] at (0.5,0.5) {$+$};
\end{scope}
\end{tikzpicture}
\end{array} \mkern-10mu +q^{-\frac{1}{2}}\begin{array}{c}
\begin{tikzpicture}
\begin{scope}[scale=0.7]
\draw[ultra thick] (-0.5,-0.5) -- (0.5,0.5) (0.5,-0.5) -- (0.1,-0.1) (-0.1,0.1) -- (-0.5,0.5);
\node[below] at (-0.5,-0.5) {$-$}; \node[below] at (0.5,-0.5) {$+$};
\node[above] at (-0.5,0.5) {$+$}; \node[above] at (0.5,0.5) {$-$};
\end{scope}
\end{tikzpicture}
\end{array} \mkern-10mu +\left(q^{\frac{1}{2}}-q^{-\frac{3}{2}}\right) \mkern-10mu \begin{array}{c}
\begin{tikzpicture}
\begin{scope}[scale=0.7]
\draw[ultra thick] (-0.5,-0.5) -- (0.5,0.5) (0.5,-0.5) -- (0.1,-0.1) (-0.1,0.1) -- (-0.5,0.5);
\node[below] at (-0.5,-0.5) {$+$}; \node[below] at (0.5,-0.5) {$-$};
\node[above] at (-0.5,0.5) {$+$}; \node[above] at (0.5,0.5) {$-$};
\end{scope}
\end{tikzpicture}
\end{array}
\end{split}
\ee
This expression is just Drinfeld's universal $R$-matrix for $SU(2)_q$ \cite{Drinfeld} intertwining two 2-dimensional irreducible representations of spin $\frac{1}{2}$. To get it explicitly one should identify our parametrization of the vacua with  vectors in the tensor product of two spin $\frac{1}{2}$ representations. We simply
identify $|+\rangle=\left|-\frac{1}{2},\frac{1}{2}\right\rangle$ and $|-\rangle=\left|+\frac{1}{2},\frac{1}{2}\right\rangle$. Physically, we may think of $SU(2)_q$ as a representation of a tight binding model for LG fields: vacua are determined by filling numbers of vacua near nodes (strands) where fields are bound, and expressions like $E\otimes F$ represent hopping operators.
Of course, a similar statement holds for the  $\CR^{-1}$ and the creation/annihilation interfaces discussed in
Section \ref{subsec:CrAn-Interfaces} below. This also fits well with the approach to knot polynomials via vertex state models.
(See for example   \cite[Figure 2]{Witten:2014xwa}.)

\item At this stage we can compare the link complexes following form LG interfaces with the well-known mathematical construction due to Khovanov \cite{Khovanov,BarNatan,Dolotin:2012sw} (and other generalizations of the hypercube model, see e.g. \cite{Anokhina:2014hha,Morozov:2015iha}). Khovanov's complex is generated by vertices of a hypercube. Each self-intersection in the link diagram has two possible resolutions and
    these resolutions generate the edges of a hypercube connecting the vertices:
\be
\begin{array}{c}
\begin{tikzpicture}
\begin{scope}[scale=0.6]
\draw[ultra thick] (-0.5,-0.5) -- (0.5,0.5) (-0.5,0.5) -- (-0.1,0.1) (0.1,-0.1) -- (0.5,-0.5);
\end{scope}
\end{tikzpicture}
\end{array}\longrightarrow
\left[
\begin{array}{c}
	\begin{tikzpicture}
	\begin{scope}[scale=0.6,rotate=90]
	\draw[ultra thick] (-0.5,-0.5) to[out=45,in=135] (0.5,-0.5) (-0.5,0.5) to[out=315,in=225] (0.5,0.5);
	\end{scope}
	\end{tikzpicture}
\end{array} \mathop{\rule[1.2mm]{15mm}{.1pt}}^{\rm edge}
\begin{array}{c}
	\begin{tikzpicture}
	\begin{scope}[scale=0.6]
	\draw[ultra thick] (-0.5,-0.5) to[out=45,in=135] (0.5,-0.5) (-0.5,0.5) to[out=315,in=225] (0.5,0.5);
	\end{scope}
	\end{tikzpicture}
\end{array}
\right]
\ee
The resolution implies a substitution of an intersection by 2 non-intersecting neighbouring cycles. To compare this complex to one we have presented in the field theory we would like to construct ``Chan-Paton data for Khovanov's complex''. These data can be guessed by  associating to   a cycle in the resolved diagram a 2-dimensional graded vector space: we assign $+$ and $-$ to vectors of different degree. So the Chan-Paton data associated to the $\CR$-twist in Khovanov's complex contains:
\be
\underbrace{2}_{\substack{\rm \#\; vertices \\ \rm in\; one\; resolution}} \times \underbrace{2}_{\substack{\rm \#\; neighbouring\\ \rm cycles}}\times
\underbrace{2}_{\substack{\rm dim.\; of\; space\\ \rm accoc.\; to \; 1 \; cycle}}=8
\ee
terms as opposed to the $6$ terms of \eqref{rules_a}. A more detailed analysis in \cite{GalakhovPhD} shows that the two ``missing'' terms have the same $\Pdeg$-degree and $\bf F$-number differing by $1$, so they cancel in the Euler characteristic. Thus, generically, Khovanov's complex is \underline{not isomorphic} to that derived from the LG model. However as we see in the examples in Section \ref{sec:Examples} Khovanov's complex can be conjectured to be \underline{quasi-isomorphic} to one following from the LG model.

\end{enumerate}

\subsection{The Creation And Annihilation Interfaces}\label{subsec:CrAn-Interfaces}

In order to produce a link, rather than a braid it is necessary to have creation and annihilation
interfaces between theories with $n$ punctures and $n\pm 2$ punctures.
 We will denote the fusion interface annihilating $z_{a_1}$ and $z_{a_2}$ by $\cap_{a_1, a_2}$ and the creation interface by $\cup_{a_1,a_2}$. We will also refer to these
interfaces as the \emph{cap} and \emph{cup} respectively. Since the tangle
begins at large negative $x$ with no punctures, and no LG fields $w$ each creation adds one LG field
and each annihilation removes one LG field. Therefore, the total number of LG fields at any value of $x$
will be $m=n/2$ if there are $n$ punctures at evolution parameter $x$ for the tangle.

The Chan-Paton data for the basic annihilation and creation interfaces can be illustrated in
our diagrammatic formalism as:
\begin{subequations}
	\be
	\CE\left(\begin{array}{c}
		\begin{tikzpicture}
		\begin{scope}[scale=0.8,yscale=-1]
		\draw[ultra thick] (0,1) to[out=270,in=180] (0.5,0) to[out=0,in=270] (1,1);
		\end{scope}
		\end{tikzpicture}
	\end{array}\right)=q^{\frac{1}{2}}\begin{array}{c}
	\begin{tikzpicture}
	\begin{scope}[scale=0.8]
	\draw[ultra thick, purple] (-0.25,1) -- (1.25,1) (0,0) -- (0,1) (1,0) -- (1,1);
	\node[below] at (0,0) {$-$};
	\node[above] at (0,1) {$-$};
	\node[below] at (1,0) {$+$};
	\node[above] at (1,1) {$+$};
	\end{scope}
	\end{tikzpicture}
\end{array}\oplus q^{-\frac{1}{2}}\begin{array}{c}
\begin{tikzpicture}
\begin{scope}[scale=0.8]
\draw (0,0.55) -- (1,0.55);
\draw (0,0.45) -- (1,0.45);
\filldraw[black] (0.5,0.5) circle (0.1);
\draw[ultra thick, purple] (-0.25,1) -- (1.25,1) (0,0) -- (0,1) (1,0) -- (1,1);
\node[below] at (0,0) {$+$};
\node[above] at (0,1) {$-$};
\node[below] at (1,0) {$-$};
\node[above] at (1,1) {$+$};
\end{scope}
\end{tikzpicture}
\end{array}\label{rules_c}\ee
\be\label{rules_d}
\CE\left(\begin{array}{c}
	\begin{tikzpicture}
	\begin{scope}[scale=0.8]
	\draw[ultra thick] (0,1) to[out=270,in=180] (0.5,0) to[out=0,in=270] (1,1);
	\end{scope}
	\end{tikzpicture}
\end{array}\right)=q^{\frac{1}{2}}\begin{array}{c}
\begin{tikzpicture}
\begin{scope}[scale=0.8]
\draw (0,0.5) -- (1,0.5);
\filldraw[white] (0.5,0.5) circle (0.1);
\draw (0.5,0.5) circle (0.1);
\draw[ultra thick, purple] (-0.25,0) -- (1.25,0) (0,0) -- (0,1) (1,0) -- (1,1);
\node[below] at (0,0) {$+$};
\node[above] at (0,1) {$-$};
\node[below] at (1,0) {$-$};
\node[above] at (1,1) {$+$};
\end{scope}
\end{tikzpicture}
\end{array}\oplus q^{-\frac{1}{2}}\begin{array}{c}
\begin{tikzpicture}
\begin{scope}[scale=0.8]
\draw[ultra thick, purple] (-0.25,0) -- (1.25,0) (0,0) -- (0,1) (1,0) -- (1,1);
\node[below] at (0,0) {$+$};
\node[above] at (0,1) {$+$};
\node[below] at (1,0) {$-$};
\node[above] at (1,1) {$-$};
\end{scope}
\end{tikzpicture}
\end{array}
\ee
\end{subequations}

Each of the diagrams on the RHS of   \eqref{rules_c} and \eqref{rules_d} is used to define a rank one
complex. The $\Pdeg$-degree in the first term of \eqref{rules_c} and the second term of \eqref{rules_d} is a choice, motivated from the quantum group rules for Chern-Simons invariants. Given these, the $\Pdeg$-degrees of the second term of \eqref{rules_c} and the first term of \eqref{rules_d} follow from the period of $dW$. The fermion number of the first term in \eqref{rules_c} is zero and second term of \eqref{rules_d} is zero. In order to construct the differential using curved webs we will require some interpretation of the solitons we have indicated. The boldface
 solid horizontal line indicates that lines terminate and will be related to $\CQ_\zeta$-supersymmetry
 preserving branes in the LG theory in a way we now describe.

\begin{figure}[h!]
\begin{center}
\begin{tikzpicture}
\begin{scope}[scale=0.7,xscale=-1]
\draw[ultra thick, dashed] (-1,-1) to[out=270,in=225] (0,-1) to[out=45,in=270] (1,1);
\filldraw (0,-1) circle (0.1);
\filldraw[blue!40!black, opacity=0.5] (-3.5,-1.5) -- (0.5,-1.5) -- (3.5,1.5) -- (-0.5,1.5) -- (-3.5,-1.5);
\draw[<-] (-3,0) -- (3,0);
\draw[<-] (-2,-2) -- (2,2);
\draw[->] (0,-2) -- (0,6);
\node[above] at (0,6) {$x$};
\node at (2,-1) {$z$};
\node[below right] at (-0.5,-1.5) {$\CL_{(+-)}$};
\draw[ultra thick,red!70!black] (-1,3) -- (1,3);
\draw[ultra thick,red!70!black,-<] (-1,3) -- (0,3);
\draw[ultra thick] (-1,-1) to[out=90,in=225] (0,2) to (1,3) to[out=45,in=270] (1.5,4) (1,1) to[out=90,in=315] (0.1,1.9) (-0.1,2.1) to (-1,3) to[out=135,in=270] (-1.5,4);
\begin{scope}[shift={(0,4)}]
\filldraw[green!40!black, opacity=0.3] (-3.5,-1.5) -- (0.5,-1.5) -- (3.5,1.5) -- (-0.5,1.5) -- (-3.5,-1.5);
\draw (-3,0) -- (3,0);
\draw (-2,-2) -- (2,2);
\end{scope}
\draw[ultra thick] (-1.5,4) -- (-1.5,5) (1.5,4) -- (1.5,5);
\node(A) at (-3,5) {$x_{\cup}^{PT}$};
\path (A) edge[->] (0,4);
\node(B) at (-5,2) {$x_{\cup}^s$};
\path (B) edge[->] (0,3);
\node(C) at (-4,0.5) {$x_{\cup}^{\CL}$};
\path (C) edge[->] (0,0);
\node(C) at (-5,-1.3) {$x_{\cup}$};
\path (C) edge[->] (0,-1);
\end{scope}
\end{tikzpicture}
\quad
\begin{tikzpicture}
	\draw[->] (-1,0) -- (4,0);
	\draw[dashed] (-1,0) -- (-1,2.5);
	\draw[dashed] (2,0) -- (2,2.5) (2,0) to[out=90,in=210] (4,2.5);
	\node[right] at (4,0) {$x$};
	\draw[ultra thick, blue!40!black] (0,0) -- (0,2.5);
	\draw[ultra thick, green!60!black] (3,0) -- (3,2.5);
	\node[below] at (-1,0) {$x_{\cup}$};
	\node[below] at (0,0) {$x_{\cup}^{\CL}$};
	\node[below] at (2,-0.1) {$x_{\cup}^{s}$};
	\node[below] at (3,0) {$x_{\cup}^{PT}$};
	\begin{scope}[shift={(2,0)}]
	\filldraw[red!40!black] (-0.1,0.1) -- (0.1,0.1) -- (0.1,-0.1) -- (-0.1,-0.1) -- (-0.1,0.1);
	\end{scope}
	\node[right] at (0,2) {$\CL_{(+-)}$};
	\node[below] at (2,-0.8) {$\tau=-\infty$};
\end{tikzpicture}
\end{center}
\caption{Cup interface in details.}\label{fig:cup}
\end{figure}
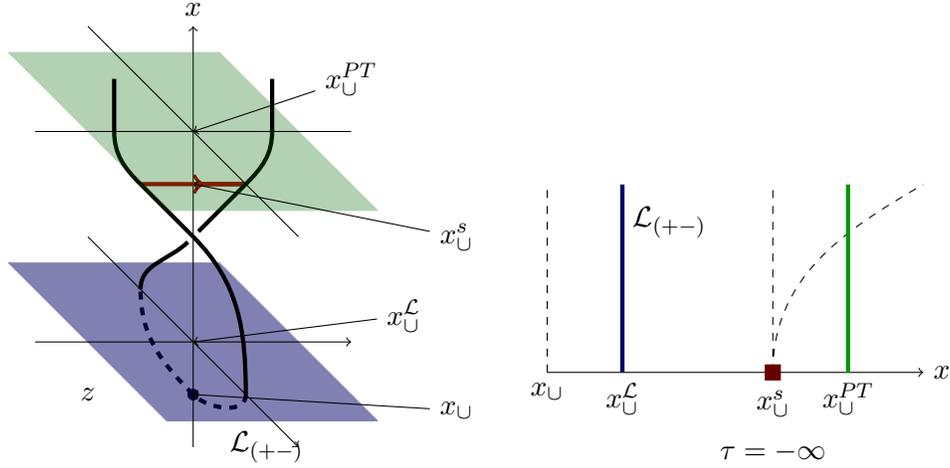

To begin, we focus locally
on a theory with $2$ punctures, in which case the cup and cap interfaces are  just boundary conditions in a theory
with $2$ punctures and a single LG field $w$.
Consider, for example, the cup interface where the strands are created at a value $x_{\cup}$. Imagine that
at some suitable $x_{\cup}^{PT}>x_{\cup}$ instead of fusing the two strands as $x$ decreases
we rather twist the strands in the clockwise or counterclockwise direction. Before the strands have twisted
by a full rotation of angle $\pi$ a binding point will appear. We choose the sense of the twist so that a \emph{past-stable} binding point $x_{\cup}^{s}$ ($x_{\cup}<x_{\cup}^{s}<x_{\cup}^{PT}$) appears. See Figure \ref{fig:cup}(a).
We then choose some $x_{\cup}^{\CL}$ ($x_{\cup}<x_{\cup}^{\CL}< x_{\cup}^{s}$)  and put a boundary condition on the theory at $x_{\cup}^{\CL}$ by inserting
a Lefshetz thimble $\CL_{(+-)}$ corresponding to the vacuum $+-$.
%
%
There are then two possible spacetime trajectories of solitons in this situation, as shown in
Figure \ref{fig:cup}(b). Note that, locally,  three possible trajectories could emanate from $(x=x_{\cup}^{s}, \tau = -\infty)$ since
it is a past stable binding point, but the trajectory that bends to the left would be incompatible with the
boundary condition at $x_{\cup}^{\CL}$. The two dashed curves in Figure \ref{fig:cup}(b) represent two $\zeta$-instantons
consistent with the initial condition in the first term in \eqref{rules_d} whereas the Lefshetz thimble at $x_{\cup}^{\CL}$
without the soliton at $x_{\cup}^{s}$ represents the second term in \eqref{rules_d}.

\bigskip
\noindent
\textbf{Remarks}

\begin{enumerate}

\item The intuition for the cup and cap interfaces comes from the $R$-matrix of the quantum group $SU(2)_q$.
This is a matrix between representations  $(\half) \otimes (\half ) \rightarrow (\half) \otimes (\half)$
and in the isotypical decomposition $(\half) \otimes (\half) = (0) \oplus (1)$ it is diagonal. A strand
with a spin 0 representation is invisible. Therefore, if we can project the R-matrix onto the spin zero
channel we will have ``annihilated'' two strands. For example for the cup the projection is obtained by
taking the third and fifth terms in \eqref{rules_b} and shifting the $\Pdeg$-grading by multiplying with
$q^{-1}$.

\item The cup interface can be understood in terms of D-brane boundary conditions for a suitable brane
inserted at   $x_{\cup}^{PT}$. To do this  we count all the possible solutions to the $\zeta$-instanton equation in the half-space $(x_{\cup}^{\CL},\infty)$ and present a ``hologram'' at $x_{\cup}^{PT}$ delivering the same solutions. In the case of $n=2$, $m=1$ there are 3 possible boundary conditions to consider
illustrated by the Lefshetz thimble at $x^{\CL}_{\cup}$ and the two dashed lines emitted from $x^s_{\cup}$. These can be imposed by the
direct sum of branes:
\be\label{eq:ThreeThimle}
\CE_{ij}(V)=\left(\begin{array}{c}
\begin{tikzpicture}
\begin{scope}[scale=0.8]
\draw[ultra thick] (0,0) -- (0,2);
\node[left] at (0,1.5) {$\CL_{(-+)}$};
\end{scope}
\end{tikzpicture}
\end{array}\right)\oplus \left(\begin{array}{c}
\begin{tikzpicture}
\begin{scope}[scale=0.8]
\draw[ultra thick] (0,0) -- (0,2);
\node[left] at (0,1.5) {$\CL_{(+-)}$};
\end{scope}
\end{tikzpicture}
\end{array}\right)\oplus\left[t q^{-1}\right]\left(\begin{array}{c}
\begin{tikzpicture}
\begin{scope}[scale=0.8]
\draw[ultra thick] (0,0) -- (0,2);
\node[left] at (0,1.5) {$\CL_{(+-)}$};
\node[left] at (0,0.5) {$\CL_{(-+)}$};
\draw[->] (0,1) -- (0.5,1.3);
\begin{scope}[shift={(0,1)}]
\filldraw[red] (-0.1,0.1) -- (0.1,0.1) -- (0.1,-0.1) -- (-0.1,-0.1) -- (-0.1,0.1);
\end{scope}
\end{scope}
\end{tikzpicture}
\end{array}\right)
\ee
Notice there are three branes. The first Lefshetz thimble $\CL_{(-+)}$ corresponds to the vertical dashed line on the right in Figure \ref{fig:cup}. The second Lefshetz thimble is the one indicated at $x_{\cup}^{\CL}$ in figure \ref{fig:cup}, and the third brane involves a boundary changing operator corresponding to the bent dashed curve in figure \ref{fig:cup}. The third brane has a shift of quantum numbers following from the relative gradings of the two terms in \eqref{rules_c}. In the case $n\geq 4$ a new boundary condition may appear: A binding point may appear at infinity. For example, it may appear in the following process:
\be
\begin{array}{c}
\begin{tikzpicture}
\draw (0,0.3) -- (2,0.3);
\draw[purple, ultra thick] (0,0) -- (0,2) (0.5,1) -- (0.5,2) (1.5,1) -- (1.5,2) (2,0) -- (2,2) (0.4,1) -- (1.6,1);
\node[below] at (0,0) {$+$}; \node[below] at (2,0) {$-$};
\node[below] at (0.5,1) {$+$}; \node[below] at (1.5,1) {$-$};
\node[above] at (0,2) {$-$}; \node[above] at (0.5,2) {$+$}; \node[above] at (1.5,2) {$-$}; \node[above] at (2,2) {$+$};
\node[above] at (1,0.3) {$1$};
\end{tikzpicture}
\end{array}\mathop{\longrightarrow}^{\CQ_{\zeta}}
\begin{array}{c}
	\begin{tikzpicture}
	\draw (0.5,1.3) -- (1.5,1.3) (0,1.7) -- (0.5,1.7) (1.5,1.7) -- (2,1.7);
	\draw[purple, ultra thick] (0,0) -- (0,2) (0.5,1) -- (0.5,2) (1.5,1) -- (1.5,2) (2,0) -- (2,2) (0.4,1) -- (1.6,1);
	\node[below] at (0,0) {$+$}; \node[below] at (2,0) {$-$};
	\node[below] at (0.5,1) {$+$}; \node[below] at (1.5,1) {$-$};
	\node[above] at (0,2) {$-$}; \node[above] at (0.5,2) {$+$}; \node[above] at (1.5,2) {$-$}; \node[above] at (2,2) {$+$};
	\node[left] at (0,1.7) {$3$}; \node[right] at (2,1.7) {$4$};
	\node[above] at (1,1.3) {$2$};
	\end{tikzpicture}
\end{array}\qquad
\begin{array}{c}
\begin{tikzpicture}
\draw[<->] (0,2) -- (0,0) -- (3,0);
\draw (0,1.8) -- (2.5,1.8);
\node[right] at (3,0) {$x$}; \node[above] at (0,2) {$\tau$};
\draw[ultra thick] (0.5,0.5) -- (1.5,0.75) -- (2.5,0.5) (1.5,1.8) -- (1.5,0.75) -- (2.5,1.25);
\begin{scope}[shift={(1.5,0.75)}]
\filldraw[black] (-0.1,0.1) -- (0.1,0.1) -- (0.1,-0.1) -- (-0.1,-0.1) -- (-0.1,0.1);
\end{scope}
\begin{scope}[shift={(1.5,1.8)}]
\filldraw[red] (-0.1,0.1) -- (0.1,0.1) -- (0.1,-0.1) -- (-0.1,-0.1) -- (-0.1,0.1);
\end{scope}
\node [above] at (1.5,1.9) {$2$};
\end{tikzpicture}
\end{array}
\ee
Where solitons 1, 3, 4 are at binding points not shown on the right.
However since $x_2$ is a past binding point the only $\zeta$-instantons is a straight vertical line emanating from some interior amplitude. Altogether we have an effective brane:
\be
\begin{split}
\CE_{ij}(V)=\left(\begin{array}{c}
	\begin{tikzpicture}
	\begin{scope}[scale=0.8]
	\draw[ultra thick] (0,0) -- (0,2);
	\node[left] at (0,1.5) {$\CL_{(-+)}$};
	\end{scope}
	\end{tikzpicture}
\end{array}\right)\oplus\left[t q^{-1}\right]\left(\begin{array}{c}
\begin{tikzpicture}
\begin{scope}[scale=0.8]
\draw[ultra thick] (0,0) -- (0,2);
\node[left] at (0,1.5) {$\CL_{(+-)}$};
\node[left] at (0,0.5) {$\CL_{(-+)}$};
\draw[->] (0,1) -- (0.5,1.3);
\begin{scope}[shift={(0,1)}]
\filldraw[red] (-0.1,0.1) -- (0.1,0.1) -- (0.1,-0.1) -- (-0.1,-0.1) -- (-0.1,0.1);
\end{scope}
\end{scope}
\end{tikzpicture}
\end{array}\right)\oplus \\ \oplus\left(\begin{array}{c}
\begin{tikzpicture}
\begin{scope}[scale=0.8]
\draw[ultra thick] (0,0) -- (0,2);
\node[left] at (0,1.5) {$\CL_{(+-)}$};
\end{scope}
\end{tikzpicture}
\end{array}\right)\oplus\left[t^{-1} q\right]\left(\begin{array}{c}
\begin{tikzpicture}
\begin{scope}[scale=0.8]
\draw[ultra thick] (0,0) -- (0,2);
\node[left] at (0,1.5) {$\CL_{(+-)}$};
\node[left] at (0,0.5) {$\CL_{(-+)}$};
\draw[dashed,->] (-0.5,0.7) -- (0.5,1.3);
\begin{scope}[shift={(0,1)}]
\filldraw[red] (-0.1,0.1) -- (0.1,0.1) -- (0.1,-0.1) -- (-0.1,-0.1) -- (-0.1,0.1);
\end{scope}
\end{scope}
\end{tikzpicture}
\end{array}\right)
\end{split}
\ee
Here the dashed line indicates that the boundary operator should be coupled to an interior amplitude.

\item In a tangle with many strands we can always consider a link projection such that all the creation
events take place at a single large negative $x_{\rm begin}$ and all the annihilation events take
place at a single large positive $x_{\rm end}$:
\begin{center}
\begin{tikzpicture}
\begin{scope}[scale=0.7]
\draw[ultra thick] (0,0) -- (7,0) -- (7,1.5) -- (0,1.5) -- (0,0);
\foreach \i in {1,3,6}
{
	\begin{scope}[shift={(0,1.5)}]
	\begin{scope}[shift={(-0.5+\i,0)},yscale=-1]
	\draw[ultra thick] (0,0) -- (0,-0.5) to[out=270,in=270] (1,-0.5) -- (1,0);
	\end{scope}
	\end{scope}
	\begin{scope}[shift={(-0.5+\i,0)}]
	\draw[ultra thick] (0,0) -- (0,-0.5) to[out=270,in=270] (1,-0.5) -- (1,0);
	\end{scope}
}
\begin{scope}[shift={(4,-0.5)}]
\foreach \i in {0,1,2} {\filldraw (0.5*\i,0) circle (0.05);}
\end{scope}
\begin{scope}[shift={(4,2)}]
\foreach \i in {0,1,2} {\filldraw (0.5*\i,0) circle (0.05);}
\end{scope}
\node at (3.5,0.75) {\bf BRAID};
\end{scope}
\end{tikzpicture}
\end{center}
Examples shown below are the link projections for the Hopf link
and trefoil, but  \underline{not} for the figure eight. In this case we can view the interface is
equivalent to studying the LG model with $m=n/2$ fields $w_i$ on the
interval $[x_{\rm begin},x_{\rm end}]$. Our rules above can be interpreted as defining suitable boundary
conditions at $x_{\rm begin}$ and $x_{\rm end}$. We can describe these boundary conditions
as follows. Suppose that $\I \zeta \bar c > 0$, so that solitons flow horizontally towards positive $x$
and Lefshetz thimbles stretch in the direction of positive real part for $w$.
With that choice suppose   the singularities are created in widely-separated pairs all sitting on a
vertical line in the complex plane. For each pair we have two possible Lefshetz thimbles $\CL_{+-}$
and $\CL_{-+}$:
\be
\CL_{+-}=\begin{array}{c}
	\begin{tikzpicture}
	\draw (0,0) circle (0.1) (0,1) circle (0.1);
	\begin{scope}[shift={(0,1)}]
	\draw[thick] (2,-0.25) -- (0,-0.25) to[out=180,in=270] (-0.5,0) to[out=90,in=180] (0,0.25) -- (2,0.25);
		\end{scope}
	\begin{scope}[shift={(-0.5,0)}]
	\draw[purple,ultra thick] (-0.1,-0.1) -- (0.1,0.1) (0.1,-0.1) -- (-0.1,0.1);
	\end{scope}
	\begin{scope}[shift={(-0.5,1)}]
	\draw[purple,ultra thick] (-0.1,-0.1) -- (0.1,0.1) (0.1,-0.1) -- (-0.1,0.1);
	\end{scope}
	\end{tikzpicture}
\end{array},\qquad
\CL_{-+}=\begin{array}{c}
\begin{tikzpicture}
\draw (0,0) circle (0.1) (0,1) circle (0.1);
\draw[thick] (2,-0.25) -- (0,-0.25) to[out=180,in=270] (-0.5,0) to[out=90,in=180] (0,0.25) -- (2,0.25);
\begin{scope}[shift={(-0.5,0)}]
\draw[purple,ultra thick] (-0.1,-0.1) -- (0.1,0.1) (0.1,-0.1) -- (-0.1,0.1);
\end{scope}
\begin{scope}[shift={(-0.5,1)}]
\draw[purple,ultra thick] (-0.1,-0.1) -- (0.1,0.1) (0.1,-0.1) -- (-0.1,0.1);
\end{scope}
\end{tikzpicture}
\end{array}
\ee
If the pairs are widely separated the terms $\sim \CO( (w_i - w_j)^{-1} )$ in the flow equations
will be suppressed and the full Lefshetz thimble in the $m$-dimensional space $P_m( \IC- \{ z_a \})$
is approximately
\be
\CL_{\vec \epsilon} \cong \CL_{\epsilon_1\epsilon_2} \times \CL_{\epsilon_3\epsilon_4} \times \cdots \times \CL_{\epsilon_{2m-1}\epsilon_{2m} }
\ee
Now from our discussion above we have seen that for a single LG field $w$ with two singularities the
boundary condition corresponds to a sum of three branes:
\be
\CL_{-+} \oplus \CL_{-+}[\CO] \oplus \CL_{+-}
\ee
for a suitable boundary-changing operator $\CO \in {\rm Hop}(\CL_{+-}, \CL_{-+})$. (We have not described this
operator very explicitly, although we have indicated its effect on the curved webs.) The notation
 $\CL_{-+}[\CO]$ means that $\CO$ has been inserted on the boundary as in the third term of
 \eqref{eq:ThreeThimle}. Let us define
\be
\CL_{\epsilon}[e^{\CO}]
:= \begin{cases}
\CL_{-+} \oplus \CL_{-+}[\CO] & \epsilon = -+ \\
\CL_{+-}   & \epsilon =+- \\
\end{cases}
\ee
The creation of the $m$ pairs of singularities, arranged vertically in widely-separated pairs
then corresponds to the D-brane
\be
\mathfrak{B} = \bigoplus\lm_{\vec \epsilon}  \CL_{\vec \epsilon} [e^{\hat\CO}]
\ee
where $\hat \CO$ is the sum of operators $\sum_{i=1}^s \CO^{(i)}$ with $\CO^{(i)}$ being
the operator $\CO$ for the $i^{th}$ LG field $w_i$.

\end{enumerate}

\subsection{The Differential}

Using the above rules one can construct a chain complex for a link described as a tangle as a bi-graded vector space (or $\IZ$-module).
However, to describe the link homology it is also necessary to describe the differential. The differential for  the Chan-Paton data of a
product of
interfaces $\fI_1\boxtimes \fI_2$ is not a simple sum of differentials for $\CE(\fI_1)$ and $\CE(\fI_2)$. Rather, in the web formalism it is defined by
contraction with the taut element of the composite webs made by combining in interfaces as described in Sections 6.2 of \cite{Gaiotto:2015aoa}. In the LG incarnation of the web-formalism it is described by
counting (with signs) solutions to the $\zeta$-instanton equation with boundary conditions specified by initial and final $\zeta$-solitons. As we have discussed,
these can be described using the curved web formalism. In the next section we will show that the vertices can be determined by simple consistency conditions,
at least for simple links.

\section{Some Examples}\label{sec:Examples}

We now illustrate the rules of Section \ref{sec:KnotHomRules} with several examples.
Of course, it is important to demonstrate that the complexes we construct are invariant
under a change of the link projection by the Reidemeister moves. We will give general
arguments for this invariance in Section \ref{sec:R-Invariance} below.

Also it is natural to compare the resulting Poincar\'e polynomials to known examples of Khovanov homology \cite{Khovanov, BarNatan, Dolotin:2012sw}.
In the examples discussed below they coincide up to a simple change of variables: Khovanov polynomials $\mathscr{K}(q,t)$
turn out to be related to Poincar\'e polynomials of LG link cohomology $\CP(q,t)$ by the following relation:
\be
\mathscr{K}(q,t)=\CP(qt,t).
\ee
One way to motivate this relation is to compare the polynomials for the unknot. With our conventions
the Euler character is $-(q + 1/q)$ whereas Khovanov postulates that it should be $q+1/q$ and a natural
way to account for this is by such a shift of Fermion number grading.

\subsection{The Unknot}

The unknot is the easiest calculation. Using our rules we get the following MSW complex:
\be
\IM\left(\begin{array}{c}
	\begin{tikzpicture}
	\draw[ultra thick] (0,0) circle (0.3);
	\end{tikzpicture}
\end{array}\right)=q \underbrace{\begin{array}{c}
	\begin{tikzpicture}
	\begin{scope}[scale=0.8]
	\draw (0,0.5) -- (1,0.5);
	\filldraw[white] (0.5,0.5) circle (0.1);
	\draw (0.5,0.5) circle (0.1);
	\draw[ultra thick, purple] (-0.25,0) -- (1.25,0) (0,0) -- (0,1) (1,0) -- (1,1) (-0.25,1) -- (1.25,1);
	\node[below] at (0,0) {$+$};
	\node[above] at (0,1) {$-$};
	\node[below] at (1,0) {$-$};
	\node[above] at (1,1) {$+$};
	\end{scope}
	\end{tikzpicture}
\end{array}}_{\IF[\Psi_1]}\oplus \; q^{-1}\underbrace{\begin{array}{c}
\begin{tikzpicture}
\begin{scope}[scale=0.8]
\draw (0,0.55) -- (1,0.55);
\draw (0,0.45) -- (1,0.45);
\filldraw[black] (0.5,0.5) circle (0.1);
\draw[ultra thick, purple] (-0.25,0) -- (1.25,0) (-0.25,1) -- (1.25,1) (0,0) -- (0,1) (1,0) -- (1,1);
\node[below] at (0,0) {$+$};
\node[above] at (0,1) {$-$};
\node[below] at (1,0) {$-$};
\node[above] at (1,1) {$+$};
\end{scope}
\end{tikzpicture}
\end{array}}_{\IF[\Psi_2]}
\ee

The $q$-grading splits this complex in two one-dimensional subcomplexes
$$
\IM_{1}=0\mathop{\to}\lm^{\IQ_{\zeta}}\IF[\Psi_1]\mathop{\to}\lm^{\IQ_{\zeta}} 0,\quad \IM_{-1}=0\mathop{\to}\lm^{\IQ_{\zeta}}\IF[\Psi_2]\mathop{\to}\lm^{\IQ_{\zeta}} 0,
$$
and in each of them the differential acts trivially. The fermion numbers are
$$
{\rm \textbf{F}}(\Psi_1)=-1,\quad {\rm \textbf{F}}(\Psi_2)=+1.
$$
Accordingly the link homology is rank $2$ and may be denoted
\be
H^{\bullet, \bullet}({\rm Unknot}) = t^{-1} q \IZ \oplus t^{} q^{-1} \IZ
\ee
We will generally summarize this kind of data by specifying the Poincar\'e polynomial:
\be
{\cal P}(q,t|{\rm Unknot})= \frac{q}{t}+\frac{t}{q} .
\ee

In Section \ref{sec:R-Invariance} we will give general arguments that the complexes we construct only
depend on the link-projection up to quasi-isomorphism, by considering the Reidemeister moves. However,
as an illustration of the general principles, let us  add a twist to the unknot and calculate its polynomial again:
\be\label{tw_unknot}
\IM\left(\begin{array}{c}
	\begin{tikzpicture}
	\begin{scope}[scale=0.5]
	\draw [ultra thick] (-0.5,-0.5) -- (0.5,0.5) to [out=45,in=135] (-0.5,0.5) -- (-0.2,0.2) (0.2,-0.2) -- (0.5,-0.5) to [out=315,in=225] (-0.5,-0.5);
	\end{scope}
	\end{tikzpicture}
\end{array}\right)=q^{-\frac{1}{2}}\left(
\begin{array}{c}
	\begin{tikzpicture}
	\begin{scope}[scale=0.8]
	\draw (0,0.25) -- (1,0.25) (0,2.3) -- (1,2.3) (0,2.2) -- (1,2.2);
	\filldraw[white] (0.5,0.25) circle (0.1);
	\draw (0.5,0.25) circle (0.1);
	\filldraw[black] (0.5,2.25) circle (0.1);
	\node[above] at (0,2.5) {$-$};
	\node[above] at (1,2.5) {$+$};
	\node[below] at (0,0) {$+$};
	\node[below] at (1,0) {$-$};
	\node[left] at (0,0.25) {$x_{\cup}^s$};
	\node[left] at (0,2.25) {$x_{\cap}^s$};
	\draw[purple, ultra thick] (0,0) -- (0,1) to[out=90,in=210] (0.5,1.5) to[out=30, in=270] (1,2) -- (1,2.5)  (1,0) -- (1,1) to [out=90,in=330] (0.6,1.4) (0.4,1.6) to [out=150,in=270] (0,2) -- (0,2.5) (-0.25,0) -- (1.25,0) (-0.25,2.5) -- (1.25,2.5);
	\end{scope}
	\end{tikzpicture}
\end{array}\oplus
\begin{array}{c}
	\begin{tikzpicture}
	\begin{scope}[scale=0.8]
	\node[above] at (0,2.5) {$-$};
	\node[above] at (1,2.5) {$+$};
	\node[below] at (0,0) {$+$};
	\node[below] at (1,0) {$-$};
	\draw[purple, ultra thick] (0,0) -- (0,1) to[out=90,in=210] (0.5,1.5) to[out=30, in=270] (1,2) -- (1,2.5)  (1,0) -- (1,1) to [out=90,in=330] (0.6,1.4) (0.4,1.6) to [out=150,in=270] (0,2) -- (0,2.5) (-0.25,0) -- (1.25,0) (-0.25,2.5) -- (1.25,2.5);
	\end{scope}
	\end{tikzpicture}
\end{array}\oplus \begin{array}{c}
\begin{tikzpicture}
\begin{scope}[scale=0.8]
\node[left] at (0,2.25) {$x_{\cap}^s$};
\node[left] at (0,1) {$x_1$};
\draw (0,1) -- (1,1) (0,2.3) -- (1,2.3) (0,2.2) -- (1,2.2);
\filldraw[black] (0.5,2.25) circle (0.1);
\node[above] at (0,2.5) {$-$};
\node[above] at (1,2.5) {$+$};
\node[below] at (0,0) {$+$};
\node[below] at (1,0) {$-$};
\draw[purple, ultra thick] (0,0) -- (0,1) to[out=90,in=210] (0.5,1.5) to[out=30, in=270] (1,2) -- (1,2.5)  (1,0) -- (1,1) to [out=90,in=330] (0.6,1.4) (0.4,1.6) to [out=150,in=270] (0,2) -- (0,2.5) (-0.25,0) -- (1.25,0) (-0.25,2.5) -- (1.25,2.5);
\end{scope}
\end{tikzpicture}
\end{array}
\right)\oplus q^{-\frac{5}{2}}
\begin{array}{c}
	\begin{tikzpicture}
	\begin{scope}[scale=0.8]
	\node[left] at (0,2.25) {$x_{\cap}^s$};
	\node[left] at (0,1) {$x_2$};
	\draw (0,1) -- (1,1) (0,0.9) -- (1,0.9) (0,2.3) -- (1,2.3) (0,2.2) -- (1,2.2);
	\filldraw[black] (0.5,2.25) circle (0.1);
	\filldraw[black] (0.5,0.95) circle (0.1);
	\node[above] at (0,2.5) {$-$};
	\node[above] at (1,2.5) {$+$};
	\node[below] at (0,0) {$+$};
	\node[below] at (1,0) {$-$};
	\draw[purple, ultra thick] (0,0) -- (0,1) to[out=90,in=210] (0.5,1.5) to[out=30, in=270] (1,2) -- (1,2.5)  (1,0) -- (1,1) to [out=90,in=330] (0.6,1.4) (0.4,1.6) to [out=150,in=270] (0,2) -- (0,2.5) (-0.25,0) -- (1.25,0) (-0.25,2.5) -- (1.25,2.5);
	\end{scope}
	\end{tikzpicture}
\end{array}
\ee
In this equation each of the complexes is one-dimensional and denoted by $\IZ[\Psi_\alpha]$, $\alpha=1,\dots, 4$, reading from left to right.
Taking into account the $q$ and $t$-gradings we have
\be
\IM_{-\frac{1}{2}}=\left(0\to (\IF[\Psi_1]\oplus\IF[\Psi_2])\to\IF[\Psi_3]\to 0\right),\quad \IM_{-\frac{5}{2}}=\left(0\to\IF[\Psi_4]\to 0\right)
\ee

The cohomology of $\IM_{-\frac{5}{2}}$ is easily computed since $\CQ_{\zeta}$ must vanish.
To calculate the cohomology of $\IM_{-\frac{1}{2}}$ we need to construct the differential $\IQ_{\zeta}$  between solitons with
fermion number differing by one.   We will draw explicit curved webs with a single time-translation modulus:
\be
\langle \Psi_3|\IQ_{\zeta}|\Psi_1\rangle =1\sim\begin{array}{c}
	\begin{tikzpicture}
	\draw [<->] (0,2.5) -- (0,0) -- (3.5,0);
	\draw (0,2) -- (3,2);
	\draw[ultra thick] (0.5,0) to [out=90, in=195] (1,1) to [out=15,in=270] (1.5,2) (2.5,0) -- (2.5,2);
	\begin{scope}[shift={(2.5,0)}]
	\filldraw[red] (-0.1,0.1) -- (0.1,0.1) -- (0.1,-0.1) -- (-0.1,-0.1);
	\end{scope}
	\node[below] at (2.5,-0.1) {$x_{\cap}^s$};
	\begin{scope}[shift={(2.5,2)}]
	\filldraw[red] (-0.1,0.1) -- (0.1,0.1) -- (0.1,-0.1) -- (-0.1,-0.1);
	\end{scope}
	\node[above] at (2.5,2.1) {$x_{\cap}^s$};
	\begin{scope}[shift={(0.5,0)}]
	\filldraw[red] (-0.1,0.1) -- (0.1,0.1) -- (0.1,-0.1) -- (-0.1,-0.1);
	\end{scope}
	\node[below] at (0.5,-0.1) {$x_{\cup}^s$};
	\begin{scope}[shift={(1.5,2)}]
	\filldraw[red] (-0.1,0.1) -- (0.1,0.1) -- (0.1,-0.1) -- (-0.1,-0.1);
	\end{scope}
	\node[above] at (1.5,2.1) {$x_1$};
	\node[left] at (0,1) {$T_0$};
	\draw[dashed] (0,1) -- (1,1);
	\node[left] at (0,2.5) {$\tau$};
	\node[below] at (3.5,0) {$x$};
	\end{tikzpicture}
\end{array}, \quad
\langle \Psi_3|\IQ_{\zeta}|\Psi_2\rangle =1\sim\begin{array}{c}
	\begin{tikzpicture}
	\draw [<->] (0,2.5) -- (0,0) -- (3.5,0);
	\draw (0,2) -- (3,2);
	\draw[ultra thick] (2.5,2) to [out=270, in=0] (2,1) to[out=180,in=270] (1.5,2);
	\begin{scope}[shift={(1.5,2)}]
	\filldraw[red] (-0.1,0.1) -- (0.1,0.1) -- (0.1,-0.1) -- (-0.1,-0.1);
	\end{scope}
	\node[above] at (1.5,2.1) {$x_1$};
	\begin{scope}[shift={(2.5,2)}]
	\filldraw[red] (-0.1,0.1) -- (0.1,0.1) -- (0.1,-0.1) -- (-0.1,-0.1);
	\end{scope}
	\node[above] at (2.5,2.1) {$x_{\cap}^s$};
	\node[left] at (0,1) {$T_0$};
	\draw[dashed] (0,1) -- (2,1);
	\node[left] at (0,2.5) {$\tau$};
	\node[below] at (3.5,0) {$x$};
	\end{tikzpicture}
\end{array}
\ee
In  these pictures the future binding point $x_1$ corresponds to a soliton coming from the $\cal R$-interface.

Thus, in the basis $\Psi_\alpha$, the differential  $\IQ_{\zeta}$ takes the form
\be
\IQ_{\zeta}=\left(\begin{array}{c}
	1\\ 1
\end{array}\right)
\ee
and hence the  cohomology of the subcomplex is one-dimensional and can be represented by
\be
H^{0}(\IM_{-\frac{1}{2}},\IQ_{\zeta})=\IF[\Psi_1-\Psi_2]
\ee
Thus the Poincar\'e   polynomial is
\be
{\cal P}\left(q,t\Big|\begin{array}{c}
	\begin{tikzpicture}
	\begin{scope}[scale=0.5]
	\draw [ultra thick] (-0.5,-0.5) -- (0.5,0.5) to [out=45,in=135] (-0.5,0.5) -- (-0.2,0.2) (0.2,-0.2) -- (0.5,-0.5) to [out=315,in=225] (-0.5,-0.5);
	\end{scope}
	\end{tikzpicture}
\end{array}\right)=q^{-\frac{3}{2}}t \left(\frac{q}{t}+\frac{t}{q}\right)
\ee
This
differs from ${\cal P}(q,t|{\rm Unknot})$ only by an overall monomial factor that can be attributed to a change of link framing.
%
%

\subsection{The Hopf Link}\label{sec:Hopf}

We now calculate the link homology of a Hopf link
$$
\begin{tikzpicture}
\draw [ultra thick] (-0.5,-0.5) to [out=90,in=210] (0,0) to [out=30, in=270] (0.5,0.5) (-0.5,0.5) to [out=270, in=150] (-0.1,0.1) (0.1,-0.1) to [out=330, in=90] (0.5,-0.5);
\begin{scope}[shift={(0,-1)}]
\draw [ultra thick] (-0.5,-0.5) to [out=90,in=210] (0,0) to [out=30, in=270] (0.5,0.5) (-0.5,0.5) to [out=270, in=150] (-0.1,0.1) (0.1,-0.1) to [out=330, in=90] (0.5,-0.5);
\end{scope}
\draw [ultra thick] (-0.5,0.5) to [out=90, in=90] (-1,0.5) -- (-1,-1.5) to [out=270, in=270] (-0.5,-1.5);
\begin{scope}[xscale=-1]
\draw [ultra thick] (-0.5,0.5) to [out=90, in=90] (-1,0.5) -- (-1,-1.5) to [out=270, in=270] (-0.5,-1.5);
\end{scope}
\end{tikzpicture}
$$

\def\Hopf{
	\draw[purple, ultra thick] (0,0) -- (0,1) (1,0) -- (1,1) (-0.25,0) -- (1.25,0);
	\begin{scope} [shift={(2,0)}]
		\draw[purple, ultra thick] (0,0) -- (0,1) (1,0) -- (1,1) (-0.25,0) -- (1.25,0);
	\end{scope}
	\begin{scope}[yscale=-1,shift={(0,-7.5)}]
		\draw[purple, ultra thick] (0,0) -- (0,1) (1,0) -- (1,1) (-0.25,0) -- (1.25,0);
		\begin{scope} [shift={(2,0)}]
			\draw[purple, ultra thick] (0,0) -- (0,1) (1,0) -- (1,1) (-0.25,0) -- (1.25,0);
		\end{scope}
	\end{scope}
	\begin{scope}[shift={(1,1.5)}]
		\draw[purple, ultra thick] (0,0) -- (0,0.5) to[out=90, in=210] (0.5,1) to[out=30,in=270] (1,1.5) -- (1,2) (1,0) -- (1,0.5) to[out=90,in=330] (0.6,0.9) (0.4,1.1) to [out=150,in=270] (0,1.5) -- (0,2);
	\end{scope}
	\begin{scope}[shift={(1,4)}]
		\draw[purple, ultra thick] (0,0) -- (0,0.5) to[out=90, in=210] (0.5,1) to[out=30,in=270] (1,1.5) -- (1,2) (1,0) -- (1,0.5) to[out=90,in=330] (0.6,0.9) (0.4,1.1) to [out=150,in=270] (0,1.5) -- (0,2);
	\end{scope}
	\draw[purple, ultra thick] (0,1.5) -- (0,6) (3,1.5) -- (3,6);
	\node[below] at (0,0) {$+$}; \node[below] at (1,0) {$-$}; \node[below] at (2,0) {$+$}; 	\node[below] at (3,0) {$-$};
	\node[above] at (0,7.5) {$-$}; \node[above] at (1,7.5) {$+$}; \node[above] at (2,7.5) {$-$}; 	\node[above] at (3,7.5) {$+$};
}

The calculation gives us four  subcomplexes of definite $\Pdeg$-degree:
\be
\nn
\IM({\rm Hopf})=q^{-3}\begin{array}{c}
	\begin{tikzpicture}
	\begin{scope}[scale=0.6]
	\Hopf
	\node at (0,1.25) {$+$};\node at (1,1.25) {$-$};\node at (2,1.25) {$+$};\node at (3,1.25) {$+$};
	\node at (1,3.75) {$+$};\node at (2,3.75) {$-$};
	\node at (0,6.25) {$+$};\node at (1,6.25) {$-$};\node at (2,6.25) {$+$};\node at (3,6.25) {$+$};
	\draw (0,7.05) -- (1,7.05) (0,6.95) -- (1,6.95) (2,7.05) -- (3,7.05) (2,6.95) -- (3,6.95);
	\filldraw[black] (0.5,7) circle (0.1) (2.5,7) circle (0.1);
	\end{scope}
	\end{tikzpicture}
\end{array}\oplus q^{-1} \begin{array}{c}
\begin{tikzpicture}
\begin{scope}[scale=0.6]
\Hopf
\node at (0,1.25) {$-$};\node at (1,1.25) {$+$};\node at (2,1.25) {$-$};\node at (3,1.25) {$+$};
\node at (1,3.75) {$+$};\node at (2,3.75) {$-$};
\node at (0,6.25) {$-$};\node at (1,6.25) {$+$};\node at (2,6.25) {$-$};\node at (3,6.25) {$+$};
\draw (0,0.5) -- (1,0.5) (2,0.5) -- (3,0.5) (1,2.05) -- (2,2.05) (1,1.95) -- (2,1.95) (1,4.55) -- (2,4.55) (1,4.45) -- (2,4.45);
\filldraw[white] (0.5,0.5) circle (0.1) (2.5,0.5) circle (0.1);
\draw (0.5,0.5) circle (0.1) (2.5,0.5) circle (0.1);
\filldraw[black] (1.5,2) circle (0.1) (1.5,4.5) circle (0.1);
\end{scope}
\end{tikzpicture}
\end{array}\oplus q^3 \begin{array}{c}
\begin{tikzpicture}
\begin{scope}[scale=0.6]
\Hopf
\node at (0,1.25) {$-$};\node at (1,1.25) {$+$};\node at (2,1.25) {$-$};\node at (3,1.25) {$+$};
\node at (1,3.75) {$+$};\node at (2,3.75) {$-$};
\node at (0,6.25) {$-$};\node at (1,6.25) {$+$};\node at (2,6.25) {$-$};\node at (3,6.25) {$+$};
\draw (0,0.5) -- (1,0.5) (2,0.5) -- (3,0.5) (1,2) -- (2,2) (1,4.5) -- (2,4.5);
\filldraw[white] (0.5,0.5) circle (0.1) (2.5,0.5) circle (0.1);
\draw (0.5,0.5) circle (0.1) (2.5,0.5) circle (0.1);
\end{scope}
\end{tikzpicture}
\end{array}\oplus\ee
\be\label{eq:E_Hopf}
\oplus q\left(
\begin{array}{c}
	\begin{tikzpicture}
	\begin{scope}[scale=0.6]
	\Hopf
	\node at (0,1.25) {$-$};\node at (1,1.25) {$+$};\node at (2,1.25) {$-$};\node at (3,1.25) {$+$};
	\node at (1,3.75) {$-$};\node at (2,3.75) {$+$};
	\node at (0,6.25) {$-$};\node at (1,6.25) {$+$};\node at (2,6.25) {$-$};\node at (3,6.25) {$+$};
	\draw (0,0.5) -- (1,0.5) (2,0.5) -- (3,0.5);
	\filldraw[white] (0.5,0.5) circle (0.1) (2.5,0.5) circle (0.1);
	\draw (0.5,0.5) circle (0.1) (2.5,0.5) circle (0.1);
	\end{scope}
	\end{tikzpicture}
\end{array}\oplus
\begin{array}{c}
	\begin{tikzpicture}
	\begin{scope}[scale=0.6]
	\Hopf
	\node at (0,1.25) {$-$};\node at (1,1.25) {$+$};\node at (2,1.25) {$-$};\node at (3,1.25) {$+$};
	\node at (1,3.75) {$+$};\node at (2,3.75) {$-$};
	\node at (0,6.25) {$-$};\node at (1,6.25) {$+$};\node at (2,6.25) {$-$};\node at (3,6.25) {$+$};
	\draw[orange] (0,0.5) -- (1,0.5) (2,0.5) -- (3,0.5);
	\draw[red] (1,2.05) -- (2,2.05) (1,1.95) -- (2,1.95);
	\draw[blue] (1,4.5) -- (2,4.5);
	\filldraw[white] (0.5,0.5) circle (0.1) (2.5,0.5) circle (0.1);
	\draw (0.5,0.5) circle (0.1) (2.5,0.5) circle (0.1);
	\filldraw[black] (1.5,2) circle (0.1);
	\end{scope}
	\end{tikzpicture}
\end{array}\oplus  \begin{array}{c}
\begin{tikzpicture}
\begin{scope}[scale=0.6]
\Hopf
\node at (0,1.25) {$-$};\node at (1,1.25) {$+$};\node at (2,1.25) {$-$};\node at (3,1.25) {$+$};
\node at (1,3.75) {$+$};\node at (2,3.75) {$-$};
\node at (0,6.25) {$-$};\node at (1,6.25) {$+$};\node at (2,6.25) {$-$};\node at (3,6.25) {$+$};
\draw[orange] (0,0.5) -- (1,0.5) (2,0.5) -- (3,0.5);
\draw[blue] (1,2) -- (2,2);
\draw[red] (1,4.55) -- (2,4.55) (1,4.45) -- (2,4.45);
\filldraw[white] (0.5,0.5) circle (0.1) (2.5,0.5) circle (0.1);
\draw (0.5,0.5) circle (0.1) (2.5,0.5) circle (0.1);
\filldraw[black] (1.5,4.5) circle (0.1);
\end{scope}
\end{tikzpicture}
\end{array}\oplus \right. \ee \be \left.\oplus
\begin{array}{c}
	\begin{tikzpicture}
	\begin{scope}[scale=0.6]
	\Hopf
	\node at (0,1.25) {$-$};\node at (1,1.25) {$+$};\node at (2,1.25) {$+$};\node at (3,1.25) {$-$};
	\node at (1,3.75) {$+$};\node at (2,3.75) {$+$};
	\node at (0,6.25) {$-$};\node at (1,6.25) {$+$};\node at (2,6.25) {$+$};\node at (3,6.25) {$-$};
	\draw (0,0.5) -- (1,0.5);
	\filldraw[white] (0.5,0.5) circle (0.1);
	\draw (0.5,0.5) circle (0.1);
	\draw[green] (2,7.05) -- (3,7.05) (2,6.95) -- (3,6.95);
	\filldraw[black] (2.5,7) circle (0.1);
	\end{scope}
	\end{tikzpicture}
\end{array}\oplus \begin{array}{c}
\begin{tikzpicture}
\begin{scope}[scale=0.6]
\Hopf
\node at (0,1.25) {$+$};\node at (1,1.25) {$-$};\node at (2,1.25) {$-$};\node at (3,1.25) {$+$};
\node at (1,3.75) {$-$};\node at (2,3.75) {$-$};
\node at (0,6.25) {$+$};\node at (1,6.25) {$-$};\node at (2,6.25) {$-$};\node at (3,6.25) {$+$};
\begin{scope}[shift={(2,0)}]
\draw (0,0.5) -- (1,0.5);
\filldraw[white] (0.5,0.5) circle (0.1);
\draw (0.5,0.5) circle (0.1);
\end{scope}
\begin{scope}[shift={(-2,0)}]
\draw[green] (2,7.05) -- (3,7.05) (2,6.95) -- (3,6.95);
\filldraw[black] (2.5,7) circle (0.1);
\end{scope}
\end{scope}
\end{tikzpicture}
\end{array}
\right)\nn
\ee

The only non-trivial contribution is $\IM_1$, we denote generators of this subcomplex by $ \Psi_i $, $i=1,\dots, 5$, reading from
left to right.
Then the complex has entries at gradings
$-2$, $-1$ and $0$:
\be
\IM_1=\left(0\to \IF[\Psi_1]\to \IF[\Psi_2]\oplus\IF[\Psi_3]\to \IF[\Psi_4]\oplus\IF[\Psi_5]\right)
\ee

As before, in order to compute the differential we must count $\zeta$-instantons with one modulus. Henceforth we will omit stationary solitons in our curved webs to keep the figures simple and readable.
%
%
To compute the differential from fermion number $-2$ to $-1$ we have first
\be
\langle\Psi_2|\IQ_{\zeta}|\Psi_1\rangle =1\sim\begin{array}{c}
	\begin{tikzpicture}
	\draw[<->] (0,2) -- (0,0) -- (2.5,0);
	\draw (0,1.5) -- (2,1.5);
	\draw[dashed] (0,0.5) -- (1,0.5) (2,0) -- (2,1.5);
	\draw[ultra thick] (0.5,1.5) -- (0.5,1.2) to [out=270,in=180] (1,0.5) to[out=0,in=270] (1.5,1.2) -- (1.5,1.5);
	\begin{scope}[shift={(1.5,1.5)}]
	\filldraw[blue] (-0.1,0.1) -- (0.1,0.1) -- (0.1,-0.1) -- (-0.1,-0.1);
	\end{scope}
	\begin{scope}[shift={(0.5,1.5)}]
	\filldraw[red] (-0.1,0.1) -- (0.1,0.1) -- (0.1,-0.1) -- (-0.1,-0.1);
	\end{scope}
	\node[below] at (2.5,0) {$x$};
	\node[left] at (0,2) {$\tau$};
	\node[left] at (0,0.5) {$T_0$};
	\node[below] at (2,0) {$+\infty$};
	\end{tikzpicture}
\end{array}
\ee
where the colors of the solitons in the link diagram $\Psi_2$ match the colors of the binding points.
Similarly, for the other matrix element with   permuted vertices:
\be
\langle\Psi_3|\IQ_{\zeta}|\Psi_1\rangle =1\sim\begin{array}{c}
	\begin{tikzpicture}
	\draw[<->] (0,2) -- (0,0) -- (2.5,0);
	\draw (0,1.5) -- (2,1.5);
	\draw[dashed] (0,0.5) -- (1,0.5) (2,0) -- (2,1.5);
	\draw[ultra thick] (0.5,1.5) -- (0.5,1.2) to [out=270,in=180] (1,0.5) to[out=0,in=270] (1.5,1.2) -- (1.5,1.5);
	\begin{scope}[shift={(1.5,1.5)}]
	\filldraw[red] (-0.1,0.1) -- (0.1,0.1) -- (0.1,-0.1) -- (-0.1,-0.1);
	\end{scope}
	\begin{scope}[shift={(0.5,1.5)}]
	\filldraw[blue] (-0.1,0.1) -- (0.1,0.1) -- (0.1,-0.1) -- (-0.1,-0.1);
	\end{scope}
	\node[below] at (2.5,0) {$x$};
	\node[left] at (0,2) {$\tau$};
	\node[left] at (0,0.5) {$T_0$};
	\end{tikzpicture}
\end{array}
\ee

To construct the differential from grading $-1$ to $0$
we construct elements $\langle\Psi_4|\IQ_{\zeta}|\Psi_2\rangle$, $\langle\Psi_5|\IQ_{\zeta}|\Psi_2\rangle$:
\be
\langle\Psi_4|\IQ_{\zeta}|\Psi_2\rangle=\langle\Psi_5|\IQ_{\zeta}|\Psi_2\rangle=1\sim \begin{array}{c}
	\begin{tikzpicture}
	\draw[<->] (-0.5,2) -- (-0.5,0) -- (3,0);
	\draw[ultra thick] (0,0) to[out=90,in=180] (0.5,1);
	\draw[ultra thick] (1.5,1) to[out=0,in=270] (2,1.5);
	\draw[ultra thick] (0.5,0) -- (0.5,1) -- (1.5,1) -- (1.5,0);
	\draw[dashed] (-0.5,1) -- (0.5,1) (2.5,1.5) -- (2.5,0);
	\draw (-0.5,1.5) -- (2.5,1.5);
	\begin{scope}[shift={(0.5,1)}]
	\filldraw[black] (-0.1,0.1) -- (0.1,0.1) -- (0.1,-0.1) -- (-0.1,-0.1);
	\end{scope}
	\begin{scope}[shift={(1.5,1)}]
	\filldraw[black] (-0.1,0.1) -- (0.1,0.1) -- (0.1,-0.1) -- (-0.1,-0.1);
	\end{scope}
	\begin{scope}[shift={(0,0)}]
	\filldraw[orange] (-0.1,0.1) -- (0.1,0.1) -- (0.1,-0.1) -- (-0.1,-0.1);
	\end{scope}
	\begin{scope}[shift={(2,1.5)}]
	\filldraw[green] (-0.1,0.1) -- (0.1,0.1) -- (0.1,-0.1) -- (-0.1,-0.1);
	\end{scope}
	\begin{scope}[shift={(0.5,0)}]
	\filldraw[red] (-0.1,0.1) -- (0.1,0.1) -- (0.1,-0.1) -- (-0.1,-0.1);
	\end{scope}
	\begin{scope}[shift={(1.5,0)}]
	\filldraw[blue] (-0.1,0.1) -- (0.1,0.1) -- (0.1,-0.1) -- (-0.1,-0.1);
	\end{scope}
	\node[left] at (-0.5,2){$\tau$};
	\node[below] at (3,0){$x$};
	\node[left] at (-0.5,1) {$T_0$};
	\end{tikzpicture}
\end{array}
\ee
Notice that the binding points from the $\cal R$-interface
(denoted by the blue and red boxes in the figure) are future-stable binding points in the infinite past, so the vertical trajectories are exactly straight.

Analogous instantons saturate the two remaining matrix elements
%
%
\be
\langle\Psi_4|\IQ_{\zeta}|\Psi_3\rangle=\langle\Psi_5|\IQ_{\zeta}|\Psi_3\rangle=-1\sim \begin{array}{c}
	\begin{tikzpicture}
	\draw[<->] (-0.5,2) -- (-0.5,0) -- (3,0);
	\draw[ultra thick] (0,0) to[out=90,in=180] (0.5,1);
	\draw[ultra thick] (1.5,1) to[out=0,in=270] (2,1.5);
	\draw[ultra thick] (0.5,0) -- (0.5,1) -- (1.5,1) -- (1.5,0);
	\draw[dashed] (-0.5,1) -- (0.5,1) (2.5,1.5) -- (2.5,0);
	\draw (-0.5,1.5) -- (2.5,1.5);
	\begin{scope}[shift={(0.5,1)}]
	\filldraw[black] (-0.1,0.1) -- (0.1,0.1) -- (0.1,-0.1) -- (-0.1,-0.1);
	\end{scope}
	\begin{scope}[shift={(1.5,1)}]
	\filldraw[black] (-0.1,0.1) -- (0.1,0.1) -- (0.1,-0.1) -- (-0.1,-0.1);
	\end{scope}
	\begin{scope}[shift={(0,0)}]
	\filldraw[orange] (-0.1,0.1) -- (0.1,0.1) -- (0.1,-0.1) -- (-0.1,-0.1);
	\end{scope}
	\begin{scope}[shift={(2,1.5)}]
	\filldraw[green] (-0.1,0.1) -- (0.1,0.1) -- (0.1,-0.1) -- (-0.1,-0.1);
	\end{scope}
	\begin{scope}[shift={(0.5,0)}]
	\filldraw[blue] (-0.1,0.1) -- (0.1,0.1) -- (0.1,-0.1) -- (-0.1,-0.1);
	\end{scope}
	\begin{scope}[shift={(1.5,0)}]
	\filldraw[red] (-0.1,0.1) -- (0.1,0.1) -- (0.1,-0.1) -- (-0.1,-0.1);
	\end{scope}
	\node[left] at (-0.5,2){$\tau$};
	\node[below] at (3,0){$x$};
	\node[left] at (-0.5,1) {$T_0$};
	\end{tikzpicture}
\end{array}
\ee
We have assigned $-1$ to these elements according to the chosen sign rule discussed in Appendix \ref{app:Signs}.
One can also deduce the signs in this case just from $\CQ_\zeta^2=0$.

All in all, we have derived  the complex:
\be
\IM_1=\left(0\longrightarrow \IF[\Psi_1]\mathop{\longrightarrow}^{\IQ_{\zeta}^{(21)}}\IF[\Psi_2]\oplus\IF[\Psi_3]\mathop{\longrightarrow}^{\IQ_{\zeta}^{(32)}} \IF[\Psi_4]\oplus\IF[\Psi_5]\mathop{\longrightarrow} 0\right)
\ee
where
\be
\IQ_{\zeta}^{(21)}=\left(\begin{array}{c}
	1\\ 1\\
\end{array}\right),\quad \IQ_{\zeta}^{(32)}=\left(\begin{array}{cc}
1& -1\\
1& -1\\
\end{array}\right)
\ee
Obviously, $\IQ_{\zeta}^{(32)}\IQ_{\zeta}^{(21)}=0$ and the cohomology can be represented by
\be
H^{\bullet}(\IM_1,\IQ_{\zeta})=\IF[\Psi_4-\Psi_5]
\ee

The link polynomial of the Hopf link reads:
\be
{\cal P}(q,t|{\rm Hopf})=q^{-3}t^2+q^{-1}+ q+q^3 t^{-2}=\left(\frac{q}{t}+\frac{t}{q}\right)\left(\frac{q^2}{t}+\frac{t}{q^2}\right)
\ee

\def\Trefoil{
	\begin{scope}[shift={(1,1)}]
		\S;
	\end{scope}
	\begin{scope}[shift={(1,2.5)}]
		\S;
	\end{scope}
	\begin{scope}[shift={(1,4)}]
		\S;
	\end{scope}
	\draw[ultra thick, purple] (0,0) -- (0,0.5) (1,0) -- (1,0.5) (-0.25,0) -- (1.25,0);
	\begin{scope}[shift={(2,0)}]
		\draw[ultra thick, purple] (0,0) -- (0,0.5) (1,0) -- (1,0.5) (-0.25,0) -- (1.25,0);
	\end{scope}
	\begin{scope}[shift={(0,5.5)}]
		\draw[ultra thick, purple] (0,0) -- (0,0.5) (1,0) -- (1,0.5) (-0.25,0.5) -- (1.25,0.5);
	\end{scope}
	\begin{scope}[shift={(2,5.5)}]
		\draw[ultra thick, purple] (0,0) -- (0,0.5) (1,0) -- (1,0.5) (-0.25,0.5) -- (1.25,0.5);
	\end{scope}
	\draw[ultra thick, purple] (0,1) -- (0,5) (3,1) -- (3,5);
	\node[below] at (0,0) {$+$};\node[below] at (1,0) {$-$};\node[below] at (2,0) {$+$};\node[below] at (3,0) {$-$};
	\node[above] at (0,6) {$-$};\node[above] at (1,6) {$+$};\node[above] at (2,6) {$-$};\node[above] at (3,6) {$+$};
}

\subsection{The Trefoil}\label{subsec:Trefoil}

We now consider the trefoil knot. (This knot is denoted as ${\bf 3_1}$ in Rolfsen knot table \cite{Rolfsen}. The 3 indicates the
number of intersections.)

$$
\begin{tikzpicture}
\begin{scope}[scale=0.7]
\draw [ultra thick] (-0.5,-0.5) to [out=90,in=210] (0,0) to [out=30, in=270] (0.5,0.5) (-0.5,0.5) to [out=270, in=150] (-0.1,0.1) (0.1,-0.1) to [out=330, in=90] (0.5,-0.5);
\begin{scope}[shift={(0,-1)}]
\draw [ultra thick] (-0.5,-0.5) to [out=90,in=210] (0,0) to [out=30, in=270] (0.5,0.5) (-0.5,0.5) to [out=270, in=150] (-0.1,0.1) (0.1,-0.1) to [out=330, in=90] (0.5,-0.5);
\end{scope}
\begin{scope}[shift={(0,-2)}]
\draw [ultra thick] (-0.5,-0.5) to [out=90,in=210] (0,0) to [out=30, in=270] (0.5,0.5) (-0.5,0.5) to [out=270, in=150] (-0.1,0.1) (0.1,-0.1) to [out=330, in=90] (0.5,-0.5);
\end{scope}
\draw [ultra thick] (-0.5,0.5) to [out=90, in=90] (-1,0.5) -- (-1,-2.5) to [out=270, in=270] (-0.5,-2.5);
\begin{scope}[xscale=-1]
\draw [ultra thick] (-0.5,0.5) to [out=90, in=90] (-1,0.5) -- (-1,-2.5) to [out=270, in=270] (-0.5,-2.5);
\end{scope}
\end{scope}
\end{tikzpicture}
$$

Applying our interfaces we get an MSW complex of the form:  
\be
\IM({\bf 3_1})=\bigoplus\lm_{i=-2}^2 q^{2i}\IM_{2i}.
\ee
The complexes $\CE_{-4}$ and $\CE_{4}$ are one-dimensional, so the cohomology is readily computed. 

Let us now  consider the complex $\CE_2$. It is 7-dimensional and is concentrated in three Fermion degrees:
\be
\CE_2=\left(0\to \CC_{-2}\mathop{\longrightarrow}^{\CQ_{\zeta}^{(-2)}} \CC_{-1}\mathop{\longrightarrow}^{\CQ_{\zeta}^{(-1)}} \CC_0\to 0\right)
\ee
Where
\be
\CC_{-2}=
\begin{array}{c}
\begin{tikzpicture}
\begin{scope}[scale=0.6]
\draw (0,0.25) -- (1,0.25);
\filldraw [white] (0.5,0.25) circle (0.1);
\draw (0.5,0.25) circle (0.1);
\begin{scope}[shift={(2,0)}]
\draw (0,0.25) -- (1,0.25);
\filldraw [white] (0.5,0.25) circle (0.1);
\draw (0.5,0.25) circle (0.1);
\end{scope}
\begin{scope}[shift={(1,4)}]
\draw (0.1,0.2) -- (0.9,0.2);
\end{scope}
\Trefoil
\node at (0,0.75) {$-$};\node at (1,0.75) {$+$};\node at (2,0.75) {$-$};\node at (3,0.75) {$+$};
\node at (1,2.25) {$+$};\node at (2,2.25) {$-$};
\node at (1,3.75) {$-$};\node at (2,3.75) {$+$};
\node at (0,5.25) {$-$};\node at (1,5.25) {$+$};\node at (2,5.25) {$-$};\node at (3,5.25) {$+$};
\end{scope}
\end{tikzpicture}
\end{array}\oplus
\begin{array}{c}
	\begin{tikzpicture}
	\begin{scope}[scale=0.6]
	\draw (0,0.25) -- (1,0.25);
	\filldraw [white] (0.5,0.25) circle (0.1);
	\draw (0.5,0.25) circle (0.1);
	\begin{scope}[shift={(2,0)}]
	\draw (0,0.25) -- (1,0.25);
	\filldraw [white] (0.5,0.25) circle (0.1);
	\draw (0.5,0.25) circle (0.1);
	\end{scope}
	\begin{scope}[shift={(1,1)}]
	\draw (0.1,0.2) -- (0.9,0.2);
	\end{scope}
	\Trefoil
	\node at (0,0.75) {$-$};\node at (1,0.75) {$+$};\node at (2,0.75) {$-$};\node at (3,0.75) {$+$};
	\node at (1,2.25) {$+$};\node at (2,2.25) {$-$};
	\node at (1,3.75) {$-$};\node at (2,3.75) {$+$};
	\node at (0,5.25) {$-$};\node at (1,5.25) {$+$};\node at (2,5.25) {$-$};\node at (3,5.25) {$+$};
	\end{scope}
	\end{tikzpicture}
\end{array}
\ee
\be
\CC_{-1}=\begin{array}{c}
	\begin{tikzpicture}
	\begin{scope}[scale=0.6]
	\draw (0,0.25) -- (1,0.25);
	\filldraw [white] (0.5,0.25) circle (0.1);
	\draw (0.5,0.25) circle (0.1);
	\begin{scope}[shift={(2,0)}]
	\draw (0,0.25) -- (1,0.25);
	\filldraw [white] (0.5,0.25) circle (0.1);
	\draw (0.5,0.25) circle (0.1);
	\end{scope}
	\begin{scope}[shift={(1,1)}]
	\draw (0,0.1) -- (1,0.1) (0.1,0.2) -- (0.9,0.2);
	\filldraw (0.5,0.15) circle (0.1);
	\end{scope}
	\begin{scope}[shift={(1,2.5)}]
	\draw (0.1,0.2) -- (0.9,0.2);
	\end{scope}
	\begin{scope}[shift={(1,4)}]
	\draw (0.1,0.2) -- (0.9,0.2);
	\end{scope}
	\Trefoil
	\node at (0,0.75) {$-$};\node at (1,0.75) {$+$};\node at (2,0.75) {$-$};\node at (3,0.75) {$+$};
	\node at (1,2.25) {$+$};\node at (2,2.25) {$-$};
	\node at (1,3.75) {$+$};\node at (2,3.75) {$-$};
	\node at (0,5.25) {$-$};\node at (1,5.25) {$+$};\node at (2,5.25) {$-$};\node at (3,5.25) {$+$};
	\end{scope}
	\end{tikzpicture}
\end{array}\oplus
\begin{array}{c}
	\begin{tikzpicture}
	\begin{scope}[scale=0.6]
	\draw (0,0.25) -- (1,0.25);
	\filldraw [white] (0.5,0.25) circle (0.1);
	\draw (0.5,0.25) circle (0.1);
	\begin{scope}[shift={(2,0)}]
	\draw (0,0.25) -- (1,0.25);
	\filldraw [white] (0.5,0.25) circle (0.1);
	\draw (0.5,0.25) circle (0.1);
	\end{scope}
	\begin{scope}[shift={(1,1)}]
	\draw (0.1,0.2) -- (0.9,0.2);
	\end{scope}
	\begin{scope}[shift={(1,2.5)}]
	\draw (0,0.1) -- (1,0.1) (0.1,0.2) -- (0.9,0.2);
	\filldraw (0.5,0.15) circle (0.1);
	\end{scope}
	\begin{scope}[shift={(1,4)}]
	\draw (0.1,0.2) -- (0.9,0.2);
	\end{scope}
	\Trefoil
	\node at (0,0.75) {$-$};\node at (1,0.75) {$+$};\node at (2,0.75) {$-$};\node at (3,0.75) {$+$};
	\node at (1,2.25) {$+$};\node at (2,2.25) {$-$};
	\node at (1,3.75) {$+$};\node at (2,3.75) {$-$};
	\node at (0,5.25) {$-$};\node at (1,5.25) {$+$};\node at (2,5.25) {$-$};\node at (3,5.25) {$+$};
	\end{scope}
	\end{tikzpicture}
\end{array}\oplus
\begin{array}{c}
	\begin{tikzpicture}
	\begin{scope}[scale=0.6]
	\draw (0,0.25) -- (1,0.25);
	\filldraw [white] (0.5,0.25) circle (0.1);
	\draw (0.5,0.25) circle (0.1);
	\begin{scope}[shift={(2,0)}]
	\draw (0,0.25) -- (1,0.25);
	\filldraw [white] (0.5,0.25) circle (0.1);
	\draw (0.5,0.25) circle (0.1);
	\end{scope}
	\begin{scope}[shift={(1,1)}]
	\draw (0.1,0.2) -- (0.9,0.2);
	\end{scope}
	\begin{scope}[shift={(1,2.5)}]
	\draw (0.1,0.2) -- (0.9,0.2);
	\end{scope}
	\begin{scope}[shift={(1,4)}]
	\draw (0,0.1) -- (1,0.1) (0.1,0.2) -- (0.9,0.2);
	\filldraw (0.5,0.15) circle (0.1);
	\end{scope}
	\Trefoil
	\node at (0,0.75) {$-$};\node at (1,0.75) {$+$};\node at (2,0.75) {$-$};\node at (3,0.75) {$+$};
	\node at (1,2.25) {$+$};\node at (2,2.25) {$-$};
	\node at (1,3.75) {$+$};\node at (2,3.75) {$-$};
	\node at (0,5.25) {$-$};\node at (1,5.25) {$+$};\node at (2,5.25) {$-$};\node at (3,5.25) {$+$};
	\end{scope}
	\end{tikzpicture}
\end{array}
\ee
\be
\CC_0=\begin{array}{c}
	\begin{tikzpicture}
	\begin{scope}[scale=0.6]
	\draw (0,0.25) -- (1,0.25);
	\filldraw [white] (0.5,0.25) circle (0.1);
	\draw (0.5,0.25) circle (0.1);
	\begin{scope}[shift={(2,5.5)}]
	\draw (0,0.2) -- (1,0.2) (0,0.3) -- (1,0.3);
	\filldraw (0.5,0.25) circle (0.1);
	\end{scope}
	\Trefoil
	\node at (0,0.75) {$-$};\node at (1,0.75) {$+$};\node at (2,0.75) {$+$};\node at (3,0.75) {$-$};
	\node at (1,2.25) {$+$};\node at (2,2.25) {$+$};
	\node at (1,3.75) {$+$};\node at (2,3.75) {$+$};
	\node at (0,5.25) {$-$};\node at (1,5.25) {$+$};\node at (2,5.25) {$+$};\node at (3,5.25) {$-$};
	\end{scope}
	\end{tikzpicture}
\end{array}\oplus \begin{array}{c}
\begin{tikzpicture}
\begin{scope}[scale=0.6]
\begin{scope}[shift={(2,0)}]
\draw (0,0.25) -- (1,0.25);
\filldraw [white] (0.5,0.25) circle (0.1);
\draw (0.5,0.25) circle (0.1);
\end{scope}
\begin{scope}[shift={(0,5.5)}]
\draw (0,0.2) -- (1,0.2) (0,0.3) -- (1,0.3);
\filldraw (0.5,0.25) circle (0.1);
\end{scope}
\Trefoil
\node at (0,0.75) {$+$};\node at (1,0.75) {$-$};\node at (2,0.75) {$-$};\node at (3,0.75) {$+$};
\node at (1,2.25) {$-$};\node at (2,2.25) {$-$};
\node at (1,3.75) {$-$};\node at (2,3.75) {$-$};
\node at (0,5.25) {$+$};\node at (1,5.25) {$-$};\node at (2,5.25) {$-$};\node at (3,5.25) {$+$};
\end{scope}
\end{tikzpicture}
\end{array}
\ee
Recall that each of the diagrams on the right hand side of the above three equations is one-dimensional. It is useful to 
 denote generators of these one-dimensional complexes by $\Psi_{ij}$ where the index $j$ refers to the  order in which it appears in the above expansion
 for $\CC_i$. 
Using the graphical notation of Appendix \ref{app:Signs} we construct a diagram representing the $\IQ_{\zeta}$-matrix elements (arrows go from degree $[f]$ to degree $[f+1]$, solid arrows correspond to matrix element $+1$, dashed arrows correspond to matrix elements $-1$):
\be
\begin{array}{c}
\begin{tikzpicture}
	\node (A) at (-2,0.5) {\begin{tikzpicture} \draw circle (0.1); \end{tikzpicture}};
	\node (B) at (-2,-0.5) {\begin{tikzpicture} \draw circle (0.1); \end{tikzpicture}};
	\node (C) at (0,1) {\begin{tikzpicture} \draw circle (0.1); \end{tikzpicture}};
	\node (D) at (0,0) {\begin{tikzpicture} \draw circle (0.1); \end{tikzpicture}};
	\node (E) at (0,-1) {\begin{tikzpicture} \draw circle (0.1); \end{tikzpicture}};
	\node (F) at (2,0.5) {\begin{tikzpicture} \draw circle (0.1); \end{tikzpicture}};
	\node (G) at (2,-0.5) {\begin{tikzpicture} \draw circle (0.1); \end{tikzpicture}};
	\path (A) edge[->] (C) (A) edge[->] (D) (B) edge[->] (D) (B) edge[->] (E) (C) edge[->] (F) (D) edge[->,dashed] (F) (E) edge[->] (F) (C) edge[->] (G) (D) edge[->,dashed] (G) (E) edge[->] (G);
	\node[left] at (-2.1,0.5) {$\Psi_{-21}$}; 	\node[left] at (-2.1,-0.5) {$\Psi_{-22}$};
	\node[above] at (0,1.1) {$\Psi_{-11}$}; \node[above] at (0,0.1) {$\Psi_{-12}$}; 	\node[above] at (0,-0.9) {$\Psi_{-13}$};
	\node[right] at (2.1,0.5) {$\Psi_{01}$}; 	\node[right] at (2.1,-0.5) {$\Psi_{02}$};
\end{tikzpicture}
\end{array}\label{eq:CG_1}
\ee
All the matrix elements presented in this diagram are close analogs of ones encountered in the example of the Hopf link. However we should comment on the absence of two matrix elements: $\langle\Psi_{-13}|\CQ_{\zeta}|\Psi_{-21}\rangle=0$ and $\langle\Psi_{-11}|\CQ_{\zeta}|\Psi_{-22}\rangle=0$. Let us start with the matrix element $\langle\Psi_{-13}|\CQ_{\zeta}|\Psi_{-21}\rangle$. The only curved web approximately representing an instanton saturating this matrix element has the following form:
\be
\begin{array}{c}
\begin{tikzpicture}
\draw[<->] (0,2) -- (0,0) -- (3.5,0); \draw (0,1.5) -- (3.5,1.5);
\node[left] at (0,2) {$\tau$};\node[right] at (3.5,0) {$x$};
\draw[ultra thick] (1,1.5) -- (1,0.75) to[out=300,in=270] (2,1.5) (2.5,1.5) to[out=270,in=120] (3,0.75) -- (3,0);
\draw[thick, snake=bumps] (1,0.75) -- (3,0.75);
\begin{scope}[shift={(1,0.75)}]
\filldraw[black] (-0.1,0.1) -- (0.1,0.1) -- (0.1,-0.1) -- (-0.1,-0.1);
\end{scope}
\begin{scope}[shift={(3,0.75)}]
\filldraw[black] (-0.1,0.1) -- (0.1,0.1) -- (0.1,-0.1) -- (-0.1,-0.1);
\end{scope}
\node[left] at (0.9,0.75) {$\beta_a$}; \node[right] at (3.1,0.75) {$\beta_a'$};
\begin{scope}[shift={(1,1.5)}]
\sqr
\end{scope}
\begin{scope}[shift={(2,1.5)}]
\sqr
\end{scope}
\begin{scope}[shift={(2.5,1.5)}]
\dmd
\end{scope}
\begin{scope}[shift={(3,0)}]
\sqr
\end{scope}
\end{tikzpicture}
\end{array}
\ee
Here the blue squares ($\begin{array}{c}
\begin{tikzpicture}
\sqr
\end{tikzpicture}
\end{array}$) 
denote binding points corresponding to single line solitons in Figure \ref{fig:2_solitons} and red diamonds
($\begin{array}{c}
\begin{tikzpicture}
\dmd
\end{tikzpicture}
\end{array}$)
denote double line solitons. The horizontal curvy line denotes the  ``purely flavour soliton'' we have mentioned in Section \ref{subsec:YYLG}, Remark 3. 
These ``purely flavour solitons''   will play an important role in  Section \ref{sec:Obstruction}. The contribution of this diagram is proportional to a product of interior amplitudes $\beta_a\beta_a'$. These interior amplitudes are zero, since incoming solitons do not satisfy the ``conservation law'' \eqref{eq:cons_law}.

Similarly we can construct a curved web representing an instanton saturating
 $\langle\Psi_{-11}|\CQ_{\zeta}|\Psi_{-22}\rangle$:
\be
\begin{array}{c}
	\begin{tikzpicture}
	\draw[<->] (0,2) -- (0,0) -- (3.5,0); \draw (0,1.5) -- (3.5,1.5);
	\node[left] at (0,2) {$\tau$};\node[right] at (3.5,0) {$x$};
	\draw[thick,snake=bumps] (1,0.75) -- (2,0.75);
	\begin{scope}[shift={(1,0)}]
	\draw[ultra thick] (1,1.5) -- (1,0.75) to[out=300,in=270] (2,1.5);
	\begin{scope}[shift={(1,0.75)}]
	\filldraw[black] (-0.1,0.1) -- (0.1,0.1) -- (0.1,-0.1) -- (-0.1,-0.1);
	\end{scope}
	\begin{scope}[shift={(1,1.5)}]
	\sqr
	\end{scope}
	\begin{scope}[shift={(2,1.5)}]
	\sqr
	\end{scope}
	\end{scope}
	\begin{scope}[shift={(-2,0)}]
	\draw[ultra thick] (2.5,1.5) to[out=270,in=120] (3,0.75) -- (3,0);
	\begin{scope}[shift={(3,0.75)}]
	\filldraw[black] (-0.1,0.1) -- (0.1,0.1) -- (0.1,-0.1) -- (-0.1,-0.1);
	\end{scope}
	\begin{scope}[shift={(2.5,1.5)}]
	\dmd
	\end{scope}
	\begin{scope}[shift={(3,0)}]
	\sqr
	\end{scope}
	\end{scope}
	\node[left] at (0.9,0.75) {$\beta_b$}; \node[below] at (2,0.7) {$\beta_b'$};
	\end{tikzpicture}
\end{array}
\ee
In this case we assume the matrix element is proportional to $\beta_b\beta_b'$.
The amplitudes $\beta_b$ and $\beta_b'$ (and similarly $\beta_a$ and $\beta_a'$) are related by a $PT$-transform, that is, by a  
rotation (i.e. a ``Euclidean boost'')  by  $\pi$. They always appear in such pairs. We can consider the boost as a transformation of $\zeta$ and denote these amplitudes as $\beta(\zeta)$ and $\beta(-\zeta)$. We do not calculate these amplitudes here; however, by the path integral description of $\beta$ 
(see \cite{Gaiotto:2015aoa}, section 14) we expect that interior amplitudes are $PT$-invariant and therefore $\beta_b(\zeta)=\beta_b(-\zeta)=\beta_b$. Since
$\beta_b$ satisfies the conservation law \eqref{eq:cons_law},   it might be non-zero. However, if it were non-zero then the condition $\CQ_{\zeta}^2=0$ 
would be violated. Indeed the sum of the three paths $\Psi_{-22}\to \Psi_{-1 i}\to \Psi_{01}$ for $i=1,2,3$ in the cohomology graph \eqref{eq:CG_1} 
is proportional to $\beta_b^2$ and therefore  $\CQ_{\zeta}^2=0$ implies $\beta_b=0$. So from now on we \underline{assume}:
\be
\beta\left[\begin{array}{c}
\begin{tikzpicture}
\draw[ultra thick,->] (0,0) -- (-0.5,1);
\draw[ultra thick,->] (0,0) -- (-0.5,-1);
\draw[thick, snake=bumps] (0,0) -- (1,0); \draw[thick, ->] (1,0) -- (1.2,0);
\filldraw[black] (-0.1,0.1) -- (0.1,0.1) -- (0.1,-0.1) -- (-0.1,-0.1);
\begin{scope}[shift={(-0.25,-0.5)}]
\sqr
\end{scope}
\begin{scope}[shift={(-0.25,0.5)}]
\dmd
\end{scope}
\end{tikzpicture}
\end{array}\right]=0
\ee
Thus we easily read off values of the $\CQ$-matrix elements from \eqref{eq:CG_1}:
\be
\CQ_{\zeta}^{(-2)}=\left(\begin{array}{cc}
	1& 0\\
	1& 1\\
	0& 1\\
\end{array}\right),\quad  \CQ_{\zeta}^{(-1)}=\left(\begin{array}{ccc}
1& -1 & 1\\
1& -1 & 1\\
\end{array}\right)
\ee
The cohomology is concentrated in degree $\textbf{F}=0$. The image of $\CQ_{\zeta}^{(-1)}$ is the 
one-dimensional subspace generated by $\Psi_{01}+\Psi_{02}$ and so a representative of a generator 
of the cohomology is the vector $\Psi_{01}-\Psi_{02}$. We write this as 
\be
H^{\bullet}(\CE_2,\CQ_{\zeta})=\IZ[\Psi_{01}-\Psi_{02}].
\ee

In a similar way we consider subcomplex $\CE_0$. It is 5-dimensional and concentrated in Fermion 
degrees $-1$ and $0$: 
\be
\CE_0=\left(0 \to \CC_{-1}\to \CC_0\to 0\right)
\ee
Where
\be
\CC_{-1}=\begin{array}{c}
	\begin{tikzpicture}
	\begin{scope}[scale=0.6]
	\draw (0,0.25) -- (1,0.25);
	\filldraw [white] (0.5,0.25) circle (0.1);
	\draw (0.5,0.25) circle (0.1);
	\begin{scope}[shift={(2,0)}]
	\draw (0,0.25) -- (1,0.25);
	\filldraw [white] (0.5,0.25) circle (0.1);
	\draw (0.5,0.25) circle (0.1);
	\end{scope}
	\begin{scope}[shift={(1,4)}]
	\draw (0,0.1) -- (1,0.1) (0.1,0.2) -- (0.9,0.2);
	\filldraw (0.5,0.15) circle (0.1);
	\end{scope}
	\Trefoil
	\node at (0,0.75) {$-$};\node at (1,0.75) {$+$};\node at (2,0.75) {$-$};\node at (3,0.75) {$+$};
	\node at (1,2.25) {$-$};\node at (2,2.25) {$+$};
	\node at (1,3.75) {$+$};\node at (2,3.75) {$-$};
	\node at (0,5.25) {$-$};\node at (1,5.25) {$+$};\node at (2,5.25) {$-$};\node at (3,5.25) {$+$};
	\end{scope}
	\end{tikzpicture}
\end{array}\oplus
\begin{array}{c}
	\begin{tikzpicture}
	\begin{scope}[scale=0.6]
	\draw (0,0.25) -- (1,0.25);
	\filldraw [white] (0.5,0.25) circle (0.1);
	\draw (0.5,0.25) circle (0.1);
	\begin{scope}[shift={(2,0)}]
	\draw (0,0.25) -- (1,0.25);
	\filldraw [white] (0.5,0.25) circle (0.1);
	\draw (0.5,0.25) circle (0.1);
	\end{scope}
	\begin{scope}[shift={(1,1)}]
	\draw (0,0.1) -- (1,0.1) (0.1,0.2) -- (0.9,0.2);
	\filldraw (0.5,0.15) circle (0.1);
	\end{scope}
	\Trefoil
	\node at (0,0.75) {$-$};\node at (1,0.75) {$+$};\node at (2,0.75) {$-$};\node at (3,0.75) {$+$};
	\node at (1,2.25) {$+$};\node at (2,2.25) {$-$};
	\node at (1,3.75) {$-$};\node at (2,3.75) {$+$};
	\node at (0,5.25) {$-$};\node at (1,5.25) {$+$};\node at (2,5.25) {$-$};\node at (3,5.25) {$+$};
	\end{scope}
	\end{tikzpicture}
\end{array}
\ee
\be
\CC_0=\begin{array}{c}
	\begin{tikzpicture}
	\begin{scope}[scale=0.6]
	\draw (0,0.25) -- (1,0.25);
	\filldraw [white] (0.5,0.25) circle (0.1);
	\draw (0.5,0.25) circle (0.1);
	\begin{scope}[shift={(2,0)}]
	\draw (0,0.25) -- (1,0.25);
	\filldraw [white] (0.5,0.25) circle (0.1);
	\draw (0.5,0.25) circle (0.1);
	\end{scope}
	\begin{scope}[shift={(1,1)}]
	\draw (0.1,0.2) -- (0.9,0.2);
	\end{scope}
	\begin{scope}[shift={(1,2.5)}]
	\draw (0,0.1) -- (1,0.1) (0.1,0.2) -- (0.9,0.2);
	\filldraw (0.5,0.15) circle (0.1);
	\end{scope}
	\begin{scope}[shift={(1,4)}]
	\draw (0,0.1) -- (1,0.1) (0.1,0.2) -- (0.9,0.2);
	\filldraw (0.5,0.15) circle (0.1);
	\end{scope}
	\Trefoil
	\node at (0,0.75) {$-$};\node at (1,0.75) {$+$};\node at (2,0.75) {$-$};\node at (3,0.75) {$+$};
	\node at (1,2.25) {$+$};\node at (2,2.25) {$-$};
	\node at (1,3.75) {$+$};\node at (2,3.75) {$-$};
	\node at (0,5.25) {$-$};\node at (1,5.25) {$+$};\node at (2,5.25) {$-$};\node at (3,5.25) {$+$};
	\end{scope}
	\end{tikzpicture}
\end{array}\oplus
\begin{array}{c}
	\begin{tikzpicture}
	\begin{scope}[scale=0.6]
	\draw (0,0.25) -- (1,0.25);
	\filldraw [white] (0.5,0.25) circle (0.1);
	\draw (0.5,0.25) circle (0.1);
	\begin{scope}[shift={(2,0)}]
	\draw (0,0.25) -- (1,0.25);
	\filldraw [white] (0.5,0.25) circle (0.1);
	\draw (0.5,0.25) circle (0.1);
	\end{scope}
	\begin{scope}[shift={(1,1)}]
	\draw (0,0.1) -- (1,0.1) (0.1,0.2) -- (0.9,0.2);
	\filldraw (0.5,0.15) circle (0.1);
	\end{scope}
	\begin{scope}[shift={(1,2.5)}]
	\draw (0.1,0.2) -- (0.9,0.2);
	\end{scope}
	\begin{scope}[shift={(1,4)}]
	\draw (0,0.1) -- (1,0.1) (0.1,0.2) -- (0.9,0.2);
	\filldraw (0.5,0.15) circle (0.1);
	\end{scope}
	\Trefoil
	\node at (0,0.75) {$-$};\node at (1,0.75) {$+$};\node at (2,0.75) {$-$};\node at (3,0.75) {$+$};
	\node at (1,2.25) {$+$};\node at (2,2.25) {$-$};
	\node at (1,3.75) {$+$};\node at (2,3.75) {$-$};
	\node at (0,5.25) {$-$};\node at (1,5.25) {$+$};\node at (2,5.25) {$-$};\node at (3,5.25) {$+$};
	\end{scope}
	\end{tikzpicture}
\end{array}\oplus
\begin{array}{c}
	\begin{tikzpicture}
	\begin{scope}[scale=0.6]
	\draw (0,0.25) -- (1,0.25);
	\filldraw [white] (0.5,0.25) circle (0.1);
	\draw (0.5,0.25) circle (0.1);
	\begin{scope}[shift={(2,0)}]
	\draw (0,0.25) -- (1,0.25);
	\filldraw [white] (0.5,0.25) circle (0.1);
	\draw (0.5,0.25) circle (0.1);
	\end{scope}
	\begin{scope}[shift={(1,1)}]
	\draw (0,0.1) -- (1,0.1) (0.1,0.2) -- (0.9,0.2);
	\filldraw (0.5,0.15) circle (0.1);
	\end{scope}
	\begin{scope}[shift={(1,2.5)}]
	\draw (0,0.1) -- (1,0.1) (0.1,0.2) -- (0.9,0.2);
	\filldraw (0.5,0.15) circle (0.1);
	\end{scope}
	\begin{scope}[shift={(1,4)}]
	\draw (0.1,0.2) -- (0.9,0.2);
	\end{scope}
	\Trefoil
	\node at (0,0.75) {$-$};\node at (1,0.75) {$+$};\node at (2,0.75) {$-$};\node at (3,0.75) {$+$};
	\node at (1,2.25) {$+$};\node at (2,2.25) {$-$};
	\node at (1,3.75) {$+$};\node at (2,3.75) {$-$};
	\node at (0,5.25) {$-$};\node at (1,5.25) {$+$};\node at (2,5.25) {$-$};\node at (3,5.25) {$+$};
	\end{scope}
	\end{tikzpicture}
\end{array}
\ee
As in the case of $\CE_2$ we denote generators of the various rank one subspaces by   $\Psi_{ij}$. It is easy to calculate the corresponding matrix elements
and we eventually get the  graph:
\be
\begin{array}{c}
	\begin{tikzpicture}
	\node (A) at (-2,0.5) {\begin{tikzpicture} \draw circle (0.1); \end{tikzpicture}};
	\node (B) at (-2,-0.5) {\begin{tikzpicture} \draw circle (0.1); \end{tikzpicture}};
	\node (C) at (0,1) {\begin{tikzpicture} \draw circle (0.1); \end{tikzpicture}};
	\node (D) at (0,0) {\begin{tikzpicture} \draw circle (0.1); \end{tikzpicture}};
	\node (E) at (0,-1) {\begin{tikzpicture} \draw circle (0.1); \end{tikzpicture}};
	\path (A) edge[->] (C) (A) edge[->] (D) (B) edge[->] (D) (B) edge[->] (E);
	\node[left] at (-2.1,0.5) {$\Psi_{-11}$}; 	\node[left] at (-2.1,-0.5) {$\Psi_{-12}$};
	\node[right] at (0.1,1) {$\Psi_{01}$}; \node[right] at (0.1,0) {$\Psi_{02}$}; 	\node[right] at (0.1,-1) {$\Psi_{03}$};
\end{tikzpicture}
\end{array}
\ee
or, equivalently, the matrix representation of $\CQ_{\zeta}$:
\be
\CQ_{\zeta}=\left(\begin{array}{cc}
1& 0\\
1& 1\\
0& 1\\
\end{array}\right)
\ee
This matrix has no nontrivial kernel and the image is rank two. The vector $\Psi_{02}$ is never in 
the image so  we conclude
\be
H^{\bullet}(\CE_0,\CQ_{\zeta})=\IZ[\Psi_{02}].
\ee

Now we consider subcomplex $\CE_{-2}$:
\be\label{tref_term}
\IM_{-2}=\begin{array}{c}
	\begin{tikzpicture}
	\begin{scope}[scale=0.6]
	\draw (0,0.25) -- (1,0.25);
	\filldraw [white] (0.5,0.25) circle (0.1);
	\draw (0.5,0.25) circle (0.1);
	\begin{scope}[shift={(2,0)}]
	\draw (0,0.25) -- (1,0.25);
	\filldraw [white] (0.5,0.25) circle (0.1);
	\draw (0.5,0.25) circle (0.1);
	\end{scope}
	\begin{scope}[shift={(1,1)}]
	\draw (0,0.1) -- (1,0.1) (0.1,0.2) -- (0.9,0.2);
	\filldraw (0.5,0.15) circle (0.1);
	\end{scope}
	\begin{scope}[shift={(1,2.5)}]
	\draw (0,0.1) -- (1,0.1) (0.1,0.2) -- (0.9,0.2);
	\filldraw (0.5,0.15) circle (0.1);
	\end{scope}
	\begin{scope}[shift={(1,4)}]
	\draw (0,0.1) -- (1,0.1) (0.1,0.2) -- (0.9,0.2);
	\filldraw (0.5,0.15) circle (0.1);
	\end{scope}
	\Trefoil
	\node at (0,0.75) {$-$};\node at (1,0.75) {$+$};\node at (2,0.75) {$-$};\node at (3,0.75) {$+$};
	\node at (1,2.25) {$+$};\node at (2,2.25) {$-$};
	\node at (1,3.75) {$+$};\node at (2,3.75) {$-$};
	\node at (0,5.25) {$-$};\node at (1,5.25) {$+$};\node at (2,5.25) {$-$};\node at (3,5.25) {$+$};
	\node[left] at (0,0.25) {1}; \node[right] at (3,0.25) {2};
	\node[left] at (1,1.5) {3}; \node[left] at (1,3) {4}; \node[left] at (1,4.5) {5};
	\end{scope}
	\end{tikzpicture}
\end{array} \oplus
\begin{array}{c}
	\begin{tikzpicture}
	\begin{scope}[scale=0.6]
	\begin{scope}[shift={(1,2.5)}]
	\draw (0,0.1) -- (1,0.1);
	\end{scope}
	\begin{scope}[shift={(0,5.5)}]
	\draw (0,0.2) -- (1,0.2) (0,0.3) -- (1,0.3);
	\filldraw (0.5,0.25) circle (0.1);
	\end{scope}
	\begin{scope}[shift={(2,5.5)}]
	\draw (0,0.2) -- (1,0.2) (0,0.3) -- (1,0.3);
	\filldraw (0.5,0.25) circle (0.1);
	\end{scope}
	\Trefoil
	\node at (0,0.75) {$+$};\node at (1,0.75) {$-$};\node at (2,0.75) {$+$};\node at (3,0.75) {$-$};
	\node at (1,2.25) {$+$};\node at (2,2.25) {$-$};
	\node at (1,3.75) {$+$};\node at (2,3.75) {$-$};
	\node at (0,5.25) {$+$};\node at (1,5.25) {$-$};\node at (2,5.25) {$+$};\node at (3,5.25) {$-$};
	\node[left] at (1,3) {6};
	\node[left] at (0,5.75) {7};
	\node[right] at (3,5.75) {8};
	\end{scope}
	\end{tikzpicture}
\end{array}
\ee

This subcomplex is two dimensional: one generator has homological degree [+1] and the other has homological degree [+2]:
\be
\IM_{-2}=\left(0\longrightarrow\IF[\Psi_1]\longrightarrow\IF[\Psi_2]\longrightarrow 0\right)
\ee

The matrix element $\langle\Psi_2|\IQ_{\zeta}|\Psi_1\rangle$ is saturated by the following field configuration:
\be\label{eq:tref_diag}
\begin{array}{c}
	\begin{tikzpicture}
	\node[left] at (0,1) {$T_0$};
	\draw[dashed] (0,1) -- (1.5,1) (3,2) -- (3,1.25);
	\draw[<->] (0,2.5) -- (0,0) -- (5.5,0);
	\node[left] at (0,2.5) {$\tau$}; \node[right] at (5.5,0) {$x$};
	\draw (0,2) -- (5.5,2);
	\draw[ultra thick] (0.5,0) to[out=90,in=135] (1.5,1) to[out=225,in=90] (1,0) (1.5,0) -- (1.5,1) -- (2.5,1) -- (2.5,0) (3,2) to[out=270,in=60] (2.5,1) -- (4,1) -- (4,0) (4.5,2) to[out=270,in=45] (4,1) to[out=315,in=270] (5,2);
	\begin{scope}[shift={(0.5,0)}]
	\filldraw[red] (-0.1,0.1) -- (0.1,0.1) -- (0.1,-0.1) -- (-0.1,-0.1);
	\end{scope}
	\begin{scope}[shift={(1,0)}]
	\filldraw[red] (-0.1,0.1) -- (0.1,0.1) -- (0.1,-0.1) -- (-0.1,-0.1);
	\end{scope}
	\begin{scope}[shift={(1.5,0)}]
	\filldraw[red] (-0.1,0.1) -- (0.1,0.1) -- (0.1,-0.1) -- (-0.1,-0.1);
	\end{scope}
	\begin{scope}[shift={(2.5,0)}]
	\filldraw[red] (-0.1,0.1) -- (0.1,0.1) -- (0.1,-0.1) -- (-0.1,-0.1);
	\end{scope}
	\begin{scope}[shift={(4,0)}]
	\filldraw[red] (-0.1,0.1) -- (0.1,0.1) -- (0.1,-0.1) -- (-0.1,-0.1);
	\end{scope}
	\begin{scope}[shift={(3,2)}]
	\filldraw[red] (-0.1,0.1) -- (0.1,0.1) -- (0.1,-0.1) -- (-0.1,-0.1);
	\end{scope}
	\begin{scope}[shift={(4.5,2)}]
	\filldraw[red] (-0.1,0.1) -- (0.1,0.1) -- (0.1,-0.1) -- (-0.1,-0.1);
	\end{scope}
	\begin{scope}[shift={(5,2)}]
	\filldraw[red] (-0.1,0.1) -- (0.1,0.1) -- (0.1,-0.1) -- (-0.1,-0.1);
	\end{scope}
	\begin{scope}[shift={(1.5,1)}]
	\filldraw[blue] (-0.1,0.1) -- (0.1,0.1) -- (0.1,-0.1) -- (-0.1,-0.1);
	\end{scope}
	\begin{scope}[shift={(4,1)}]
	\filldraw[blue] (-0.1,0.1) -- (0.1,0.1) -- (0.1,-0.1) -- (-0.1,-0.1);
	\end{scope}
	\begin{scope}[shift={(2.5,1)}]
	\filldraw[green!80!black] (-0.1,0.1) -- (0.1,0.1) -- (0.1,-0.1) -- (-0.1,-0.1);
	\end{scope}
	\node[below] at (0.5,-0.1) {$1$}; \node[below] at (1,-0.1) {$2$}; \node[below] at (1.5,-0.1) {$3$}; \node[below] at (2.5,-0.1) {$4$}; \node[below] at (4,-0.1) {$5$};
	\node[above] at (3,2.1) {$6$}; \node[above] at (4.5,2.1) {$7$}; \node[above] at (5,2.1) {$8$};
	\node[above] at (2,1) {$a$}; \node[above] at (3.25,1) {$b$};
	\end{tikzpicture}
\end{array}
\ee
Here we have marked binding points by corresponding numbers in equation \eqref{tref_term}. 

Notice here the bulk vertices are 4-valent. The choice of 4-valent vertex ($\color{blue} \blacksquare$) marked by blue is dictated by necessity. Indeed consider a collision point of solitons 1, 2 and 3. For a transition associated to soliton 3 to be possible a vacant vacuum near the second strand from left 
(see the RHS of \eqref{tref_term}) should be filled and one near the third strand should be unoccupied, this is done exactly by solitons 1 and 2. So neither soliton 1 nor 2 can cross the vertical line of soliton 3 trajectory, otherwise the fan of vacua for such an intersection vertex would be inconsistent. Hence the trajectories of solitons 1 and 2 should simultaneously hit the trajectory of soliton 3 creating a new soliton. Similar reasoning holds for a collision of solitons $5$, $7$ and  $8$.

The bulk vertex marked by green ($\color{green!80!black} \blacksquare$) is different.
The boosted solitons marked $a$ and $b$ on this diagram describe a LG field interpolating from the left-most strand to the   right-most strand 
in \eqref{tref_term} while solitons $4$ and $6$ interpolate between strands in the middle. In principle they should not interact. We can move side strands far away from the strands in the middle in the direction perpendicular to the knot diagram plane without changing the knot diagram. So, generically, these solitons should not interfere. What happens in this vertex is that, for both pairs of solitons $a-b$ and $4-6$, the central charge changes simultaneously by a flavour contribution $2\pi$ (the flavour charge is not conserved in the MLG model as we will see in Section \ref{sec:Obstruction}) so that the overall central charge is conserved.

Thus we have
\be
H^{\bullet} (\IM_{-2},\IQ_{\zeta})=\{ 0 \}
\ee

Summarizing the cohomologies of the other subcomplexes  we finally get the following Poincar\'e polynomial for the trefoil:
\be
{\cal P}(q,t|{\bf 3_1})=q^{-4} t^3+1+q^2+q^4 t^{-2}.
\ee

\subsection{The Figure-eight Knot}\label{sec:figure-eight}

Now we consider the figure-eight knot (denoted by  ${\bf 4_1}$ in the Rolfsen knot table \cite{Rolfsen})
$$
\begin{array}{c}
\begin{tikzpicture}
\begin{scope}[scale=0.6]
\SS;
\begin{scope}[shift={(0,1)}]
\SS;
\end{scope}
\begin{scope}[xscale=-1, shift={(-2,2)}]
\SS;
\end{scope}
\begin{scope}[xscale=-1, shift={(-2,3)}]
\SS;
\end{scope}
\draw[ultra thick] (2,2) -- (2,0) to[out=270,in=270] (1,0) (0,0) to[out=270,in=270] (-0.5,0) -- (-0.5,4) to[out=90,in=90] (2,4) (0,2) -- (0,4) to[out=90,in=90] (1,4);
\end{scope}
\end{tikzpicture}
\end{array}
$$
The MSW complex contains six subcomplexes:
\be
\IM({\bf 4_1})=\bigoplus\lm_{i=0}^{5}q^{2i-5} \IM_{2i-5}
\ee
As usual the subcomplex $\IM_{-5}$ is one dimensional:
\be
\IM_{-5}=\begin{array}{c}
	\begin{tikzpicture}
	\begin{scope}[scale=0.6]
	\begin{scope}[shift={(0,2.5)}]
	\draw (0,0.1) -- (1,0.1);
	\draw (0.1,0.2) -- (0.9,0.2);
	\filldraw (0.5,0.15) circle (0.1);
	\end{scope}
	\begin{scope}[shift={(0,7)}]
	\draw (0,0.2) -- (1,0.2) (0,0.3) -- (1,0.3);
	\filldraw (0.5,0.25) circle (0.1);
	\end{scope}
	\begin{scope}[shift={(0,8.5)}]
	\draw (0,0.2) -- (1,0.2) (0,0.3) -- (1,0.3);
	\filldraw (0.5,0.25) circle (0.1);
	\end{scope}
	\begin{scope}[shift={(-1,0)}]
	\draw[ultra thick,purple] (-0.25,0) -- (1.25,0) (0,0) -- (0,0.5) (1,0) -- (1,0.5);
	\end{scope}
	\begin{scope}[shift={(1,0)}]
	\draw[ultra thick,purple] (-0.25,0) -- (1.25,0) (0,0) -- (0,0.5) (1,0) -- (1,0.5);
	\end{scope}
	\begin{scope}[shift={(0,1)}]
	\S;
	\end{scope}
	\begin{scope}[shift={(0,2.5)}]
	\S;
	\end{scope}
	\begin{scope}[shift={(1,4)}]
	\begin{scope}[yscale=-1, shift={(0,-1)}]
	\S;
	\end{scope}
	\end{scope}
	\begin{scope}[shift={(1,5.5)}]
	\begin{scope}[yscale=-1, shift={(0,-1)}]
	\S;
	\end{scope}
	\end{scope}
	\begin{scope}[yscale=-1,shift={(0,-7.5)}]
	\draw[ultra thick,purple] (-0.25,0) -- (1.25,0) (0,0) -- (0,0.5) (1,0) -- (1,0.5);
	\end{scope}
	\begin{scope}[yscale=-1,shift={(0,-9)}]
	\draw[ultra thick,purple] (-0.25,0) -- (1.25,0) (0,0) -- (0,0.5) (1,0) -- (1,0.5);
	\end{scope}
	\draw[ultra thick, purple] (-1,1) -- (-1,7.5) to [out=90,in=270] (0,8) (0,4) -- (0,6.5) (2,1) -- (2,3.5) (2,7) -- (2,7.5) to[out=90,in=270] (1,8);
	\node[below] at (-1,0) {$+$}; \node[below] at (0,0) {$-$}; \node[below] at (1,0) {$+$}; \node[below] at (2,0) {$-$};
	\node at (-1,0.75) {$+$}; \node at (0,0.75) {$-$}; \node at (1,0.75) {$+$}; \node at (2,0.75) {$-$};
	\node at (0,2.25) {$+$}; \node at (1,2.25) {$-$};
	\node at (0,3.75) {$+$}; \node at (1,3.75) {$-$}; \node at (2,3.75) {$-$};
	\node at (1,5.25) {$-$}; \node at (2,5.25) {$-$};
	\node at (0,6.75) {$+$}; \node at (1,6.75) {$-$}; \node at (2,6.75) {$-$};
	\node at (0,8.25) {$+$}; \node at (1,8.25) {$-$};
	\end{scope}
	\end{tikzpicture}
\end{array}
\ee
Now we consider the subcomplex $\IM_{-3}$:
\be
\IM_{-3}=\begin{array}{c}
	\begin{tikzpicture}
	\begin{scope}[scale=0.6]
	\draw (-1,0.25) -- (0,0.25) (1,0.25) -- (2,0.25);
	\filldraw[white] (-0.5,0.25) circle (0.1) (1.5,0.25) circle (0.1);
	\draw (-0.5,0.25) circle (0.1) (1.5,0.25) circle (0.1);
	\begin{scope}[shift={(0,1)}]
	\draw (0,0.1) -- (1,0.1) (0.1,0.2) -- (0.9,0.2);
	\filldraw (0.5,0.15) circle (0.1);
	\end{scope}
	\begin{scope}[shift={(0,2.5)}]
	\draw (0,0.1) -- (1,0.1) (0.1,0.2) -- (0.9,0.2);
	\filldraw (0.5,0.15) circle (0.1);
	\end{scope}
	\begin{scope}[shift={(1,4)}]
	\draw (0,0.9) -- (1,0.9) (0.1,0.8) -- (0.9,0.8);
	\end{scope}
	\begin{scope}[shift={(1,5.5)}]
	\draw (0,0.9) -- (1,0.9) (0.1,0.8) -- (0.9,0.8);
	\end{scope}
	\begin{scope}[shift={(0,7)}]
	\draw (0,0.2) -- (1,0.2) (0,0.3) -- (1,0.3);
	\filldraw (0.5,0.25) circle (0.1);
	\end{scope}
	\begin{scope}[shift={(-1,0)}]
	\draw[ultra thick,purple] (-0.25,0) -- (1.25,0) (0,0) -- (0,0.5) (1,0) -- (1,0.5);
	\end{scope}
	\begin{scope}[shift={(1,0)}]
	\draw[ultra thick,purple] (-0.25,0) -- (1.25,0) (0,0) -- (0,0.5) (1,0) -- (1,0.5);
	\end{scope}
	\begin{scope}[shift={(0,1)}]
	\S;
	\end{scope}
	\begin{scope}[shift={(0,2.5)}]
	\S;
	\end{scope}
	\begin{scope}[shift={(1,4)}]
	\begin{scope}[yscale=-1, shift={(0,-1)}]
	\S;
	\end{scope}
	\end{scope}
	\begin{scope}[shift={(1,5.5)}]
	\begin{scope}[yscale=-1, shift={(0,-1)}]
	\S;
	\end{scope}
	\end{scope}
	\begin{scope}[yscale=-1,shift={(0,-7.5)}]
	\draw[ultra thick,purple] (-0.25,0) -- (1.25,0) (0,0) -- (0,0.5) (1,0) -- (1,0.5);
	\end{scope}
	\begin{scope}[yscale=-1,shift={(0,-9)}]
	\draw[ultra thick,purple] (-0.25,0) -- (1.25,0) (0,0) -- (0,0.5) (1,0) -- (1,0.5);
	\end{scope}
	\draw[ultra thick, purple] (-1,1) -- (-1,7.5) to [out=90,in=270] (0,8) (0,4) -- (0,6.5) (2,1) -- (2,3.5) (2,7) -- (2,7.5) to[out=90,in=270] (1,8);
	\node[below] at (-1,0) {$+$}; \node[below] at (0,0) {$-$}; \node[below] at (1,0) {$+$}; \node[below] at (2,0) {$-$};
	\node at (-1,0.75) {$-$}; \node at (0,0.75) {$+$}; \node at (1,0.75) {$-$}; \node at (2,0.75) {$+$};
	\node at (0,2.25) {$+$}; \node at (1,2.25) {$-$};
	\node at (0,3.75) {$+$}; \node at (1,3.75) {$-$}; \node at (2,3.75) {$+$};
	\node at (1,5.25) {$-$}; \node at (2,5.25) {$+$};
	\node at (0,6.75) {$+$}; \node at (1,6.75) {$-$}; \node at (2,6.75) {$+$};
	\node at (0,8.25) {$-$}; \node at (1,8.25) {$+$};
	\node[left] at (-1,0.25) {1};
	\node[right] at (2,0.25) {2};
	\node[left] at (0,1.5) {3};
	\node[left] at (0,3) {4};
	\node[left] at (1,4.5) {5};
	\node[left] at (1,6) {6};
	\end{scope}
	\end{tikzpicture}
\end{array}\oplus \begin{array}{c}
\begin{tikzpicture}
\begin{scope}[scale=0.6]
\begin{scope}[shift={(0,2.5)}]
\draw (0,0.1) -- (1,0.1);
\end{scope}
\begin{scope}[shift={(0,7)}]
\draw (0,0.2) -- (1,0.2) (0,0.3) -- (1,0.3);
\filldraw (0.5,0.25) circle (0.1);
\end{scope}
\begin{scope}[shift={(0,8.5)}]
\draw (0,0.2) -- (1,0.2) (0,0.3) -- (1,0.3);
\filldraw (0.5,0.25) circle (0.1);
\end{scope}
\begin{scope}[shift={(-1,0)}]
\draw[ultra thick,purple] (-0.25,0) -- (1.25,0) (0,0) -- (0,0.5) (1,0) -- (1,0.5);
\end{scope}
\begin{scope}[shift={(1,0)}]
\draw[ultra thick,purple] (-0.25,0) -- (1.25,0) (0,0) -- (0,0.5) (1,0) -- (1,0.5);
\end{scope}
\begin{scope}[shift={(0,1)}]
\S;
\end{scope}
\begin{scope}[shift={(0,2.5)}]
\S;
\end{scope}
\begin{scope}[shift={(1,4)}]
\begin{scope}[yscale=-1, shift={(0,-1)}]
\S;
\end{scope}
\end{scope}
\begin{scope}[shift={(1,5.5)}]
\begin{scope}[yscale=-1, shift={(0,-1)}]
\S;
\end{scope}
\end{scope}
\begin{scope}[yscale=-1,shift={(0,-7.5)}]
\draw[ultra thick,purple] (-0.25,0) -- (1.25,0) (0,0) -- (0,0.5) (1,0) -- (1,0.5);
\end{scope}
\begin{scope}[yscale=-1,shift={(0,-9)}]
\draw[ultra thick,purple] (-0.25,0) -- (1.25,0) (0,0) -- (0,0.5) (1,0) -- (1,0.5);
\end{scope}
\draw[ultra thick, purple] (-1,1) -- (-1,7.5) to [out=90,in=270] (0,8) (0,4) -- (0,6.5) (2,1) -- (2,3.5) (2,7) -- (2,7.5) to[out=90,in=270] (1,8);
\node[below] at (-1,0) {$+$}; \node[below] at (0,0) {$-$}; \node[below] at (1,0) {$+$}; \node[below] at (2,0) {$-$};
\node at (-1,0.75) {$+$}; \node at (0,0.75) {$-$}; \node at (1,0.75) {$+$}; \node at (2,0.75) {$-$};
\node at (0,2.25) {$+$}; \node at (1,2.25) {$-$};
\node at (0,3.75) {$+$}; \node at (1,3.75) {$-$}; \node at (2,3.75) {$-$};
\node at (1,5.25) {$-$}; \node at (2,5.25) {$-$};
\node at (0,6.75) {$+$}; \node at (1,6.75) {$-$}; \node at (2,6.75) {$-$};
\node at (0,8.25) {$+$}; \node at (1,8.25) {$-$};
\node[left] at (0,3) {7};
\node[left] at (0,8.25) {8};
\end{scope}
\end{tikzpicture}
\end{array}
\ee
It is easy to construct the curved web  interpolating between these states, it is analogous to the 
web in diagram \eqref{eq:tref_diag}. We have denoted the binding points by numbers corresponding to their
location in the diagram for the $\zeta$-instanton:
\be
\begin{array}{c}
	\begin{tikzpicture}
	\node[left] at (0,1) {$T_0$};
	\draw[dashed] (0,1) -- (1.5,1);
	\draw[<->] (0,2.5) -- (0,0) -- (6,0);
	\node[left] at (0,2.5) {$\tau$}; \node[right] at (6,0) {$x$};
	\draw (0,2) -- (6,2);
	\draw[ultra thick] (0.5,0) to[out=90,in=135] (1.5,1) to[out=225,in=90] (1,0) (1.5,0) -- (1.5,1) -- (2.5,1) -- (2.5,0) (3,2) to[out=270,in=60] (2.5,1) -- (4,1) -- (4,0) (4,1) -- (5,1) -- (5,0) (5.5,2) to[out=270,in=0] (5,1);
	\begin{scope}[shift={(0.5,0)}]
	\filldraw[red] (-0.1,0.1) -- (0.1,0.1) -- (0.1,-0.1) -- (-0.1,-0.1);
	\end{scope}
	\begin{scope}[shift={(1,0)}]
	\filldraw[red] (-0.1,0.1) -- (0.1,0.1) -- (0.1,-0.1) -- (-0.1,-0.1);
	\end{scope}
	\begin{scope}[shift={(1.5,0)}]
	\filldraw[red] (-0.1,0.1) -- (0.1,0.1) -- (0.1,-0.1) -- (-0.1,-0.1);
	\end{scope}
	\begin{scope}[shift={(2.5,0)}]
	\filldraw[red] (-0.1,0.1) -- (0.1,0.1) -- (0.1,-0.1) -- (-0.1,-0.1);
	\end{scope}
	\begin{scope}[shift={(4,0)}]
	\filldraw[red] (-0.1,0.1) -- (0.1,0.1) -- (0.1,-0.1) -- (-0.1,-0.1);
	\end{scope}
	\begin{scope}[shift={(3,2)}]
	\filldraw[red] (-0.1,0.1) -- (0.1,0.1) -- (0.1,-0.1) -- (-0.1,-0.1);
	\end{scope}
	\begin{scope}[shift={(5,0)}]
	\filldraw[red] (-0.1,0.1) -- (0.1,0.1) -- (0.1,-0.1) -- (-0.1,-0.1);
	\end{scope}
	\begin{scope}[shift={(5.5,2)}]
	\filldraw[red] (-0.1,0.1) -- (0.1,0.1) -- (0.1,-0.1) -- (-0.1,-0.1);
	\end{scope}
	\begin{scope}[shift={(1.5,1)}]
	\filldraw[blue] (-0.1,0.1) -- (0.1,0.1) -- (0.1,-0.1) -- (-0.1,-0.1);
	\end{scope}
	\begin{scope}[shift={(4,1)}]
	\filldraw[black] (-0.1,0.1) -- (0.1,0.1) -- (0.1,-0.1) -- (-0.1,-0.1);
	\end{scope}
	\begin{scope}[shift={(5,1)}]
	\filldraw[black] (-0.1,0.1) -- (0.1,0.1) -- (0.1,-0.1) -- (-0.1,-0.1);
	\end{scope}
	\begin{scope}[shift={(2.5,1)}]
	\filldraw[green!80!black] (-0.1,0.1) -- (0.1,0.1) -- (0.1,-0.1) -- (-0.1,-0.1);
	\end{scope}
	\node[below] at (0.5,-0.1) {$1$}; \node[below] at (1,-0.1) {$2$}; \node[below] at (1.5,-0.1) {$3$}; \node[below] at (2.5,-0.1) {$4$}; \node[below] at (4,-0.1) {$5$};
	\node[below] at (5,-0.1) {$6$}; \node[above] at (3,2.1) {$7$}; \node[above] at (5.5,2.1) {$8$};
	\end{tikzpicture}
\end{array}
\ee
So we conclude
\be
\langle\Psi_2|\IQ_{\zeta}|\Psi_1\rangle=1
\ee
Thus the cohomology group is zero:
\be
H^{\bullet}(\IM_{-3},\IQ_{\zeta})=\{ 0 \} . 
\ee

The complex $\IM_{-1}$ has nontrivial entries for three fermion numbers:
\be
\IM_{-1}=\left(0\longrightarrow {\cal C}^{(-1)}\mathop{\longrightarrow}^{\CQ_{\zeta}^{(-1)}} {\cal C}^{(0)}\mathop{\longrightarrow}^{\CQ_{\zeta}^{(0)}} {\cal C}^{(1)}\longrightarrow 0\right)
\ee
Where
\begin{subequations}
\be\label{eq:fig8-A}
{\cal C}^{(-1)}=\begin{array}{c}
	\begin{tikzpicture}
	\begin{scope}[scale=0.6]
	\draw (-1,0.25) -- (0,0.25) (1,0.25) -- (2,0.25);
	\filldraw[white] (-0.5,0.25) circle (0.1) (1.5,0.25) circle (0.1);
	\draw (-0.5,0.25) circle (0.1) (1.5,0.25) circle (0.1);
	\begin{scope}[shift={(1,4)}]
	\draw (0,0.9) -- (1,0.9) (0.1,0.8) -- (0.9,0.8);
	\end{scope}
	\begin{scope}[shift={(1,5.5)}]
	\draw (0,0.9) -- (1,0.9) (0.1,0.8) -- (0.9,0.8);
	\end{scope}
	\begin{scope}[shift={(0,7)}]
	\draw (0,0.2) -- (1,0.2) (0,0.3) -- (1,0.3);
	\filldraw (0.5,0.25) circle (0.1);
	\end{scope}
	\begin{scope}[shift={(-1,0)}]
	\draw[ultra thick,purple] (-0.25,0) -- (1.25,0) (0,0) -- (0,0.5) (1,0) -- (1,0.5);
	\end{scope}
	\begin{scope}[shift={(1,0)}]
	\draw[ultra thick,purple] (-0.25,0) -- (1.25,0) (0,0) -- (0,0.5) (1,0) -- (1,0.5);
	\end{scope}
	\begin{scope}[shift={(0,1)}]
	\S;
	\end{scope}
	\begin{scope}[shift={(0,2.5)}]
	\S;
	\end{scope}
	\begin{scope}[shift={(1,4)}]
	\begin{scope}[yscale=-1, shift={(0,-1)}]
	\S;
	\end{scope}
	\end{scope}
	\begin{scope}[shift={(1,5.5)}]
	\begin{scope}[yscale=-1, shift={(0,-1)}]
	\S;
	\end{scope}
	\end{scope}
	\begin{scope}[yscale=-1,shift={(0,-7.5)}]
	\draw[ultra thick,purple] (-0.25,0) -- (1.25,0) (0,0) -- (0,0.5) (1,0) -- (1,0.5);
	\end{scope}
	\begin{scope}[yscale=-1,shift={(0,-9)}]
	\draw[ultra thick,purple] (-0.25,0) -- (1.25,0) (0,0) -- (0,0.5) (1,0) -- (1,0.5);
	\end{scope}
	\draw[ultra thick, purple] (-1,1) -- (-1,7.5) to [out=90,in=270] (0,8) (0,4) -- (0,6.5) (2,1) -- (2,3.5) (2,7) -- (2,7.5) to[out=90,in=270] (1,8);
	\node[below] at (-1,0) {$+$}; \node[below] at (0,0) {$-$}; \node[below] at (1,0) {$+$}; \node[below] at (2,0) {$-$};
	\node at (-1,0.75) {$-$}; \node at (0,0.75) {$+$}; \node at (1,0.75) {$-$}; \node at (2,0.75) {$+$};
	\node at (0,2.25) {$-$}; \node at (1,2.25) {$+$};
	\node at (0,3.75) {$+$}; \node at (1,3.75) {$-$}; \node at (2,3.75) {$+$};
	\node at (1,5.25) {$-$}; \node at (2,5.25) {$+$};
	\node at (0,6.75) {$+$}; \node at (1,6.75) {$-$}; \node at (2,6.75) {$+$};
	\node at (0,8.25) {$-$}; \node at (1,8.25) {$+$};
	\end{scope}
	\end{tikzpicture}
\end{array}\oplus \begin{array}{c}
\begin{tikzpicture}
\begin{scope}[scale=0.6]
\draw (-1,0.25) -- (0,0.25) (1,0.25) -- (2,0.25);
\filldraw[white] (-0.5,0.25) circle (0.1) (1.5,0.25) circle (0.1);
\draw (-0.5,0.25) circle (0.1) (1.5,0.25) circle (0.1);
\begin{scope}[shift={(0,1)}]
\draw (0,0.1) -- (1,0.1) (0.1,0.2) -- (0.9,0.2);
\filldraw (0.5,0.15) circle (0.1);
\end{scope}
\begin{scope}[shift={(-1,0)}]
\draw[ultra thick,purple] (-0.25,0) -- (1.25,0) (0,0) -- (0,0.5) (1,0) -- (1,0.5);
\end{scope}
\begin{scope}[shift={(1,0)}]
\draw[ultra thick,purple] (-0.25,0) -- (1.25,0) (0,0) -- (0,0.5) (1,0) -- (1,0.5);
\end{scope}
\begin{scope}[shift={(0,1)}]
\S;
\end{scope}
\begin{scope}[shift={(0,2.5)}]
\S;
\end{scope}
\begin{scope}[shift={(1,4)}]
\begin{scope}[yscale=-1, shift={(0,-1)}]
\S;
\end{scope}
\end{scope}
\begin{scope}[shift={(1,5.5)}]
\begin{scope}[yscale=-1, shift={(0,-1)}]
\S;
\end{scope}
\end{scope}
\begin{scope}[yscale=-1,shift={(0,-7.5)}]
\draw[ultra thick,purple] (-0.25,0) -- (1.25,0) (0,0) -- (0,0.5) (1,0) -- (1,0.5);
\end{scope}
\begin{scope}[yscale=-1,shift={(0,-9)}]
\draw[ultra thick,purple] (-0.25,0) -- (1.25,0) (0,0) -- (0,0.5) (1,0) -- (1,0.5);
\end{scope}
\draw[ultra thick, purple] (-1,1) -- (-1,7.5) to [out=90,in=270] (0,8) (0,4) -- (0,6.5) (2,1) -- (2,3.5) (2,7) -- (2,7.5) to[out=90,in=270] (1,8);
\node[below] at (-1,0) {$+$}; \node[below] at (0,0) {$-$}; \node[below] at (1,0) {$+$}; \node[below] at (2,0) {$-$};
\node at (-1,0.75) {$-$}; \node at (0,0.75) {$+$}; \node at (1,0.75) {$-$}; \node at (2,0.75) {$+$};
\node at (0,2.25) {$+$}; \node at (1,2.25) {$-$};
\node at (0,3.75) {$-$}; \node at (1,3.75) {$+$}; \node at (2,3.75) {$+$};
\node at (1,5.25) {$+$}; \node at (2,5.25) {$+$};
\node at (0,6.75) {$-$}; \node at (1,6.75) {$+$}; \node at (2,6.75) {$+$};
\node at (0,8.25) {$-$}; \node at (1,8.25) {$+$};
\end{scope}
\end{tikzpicture}
\end{array}
\ee
\be\label{eq:fig8-B}
{\cal C}^{(0)}=\begin{array}{c}
	\begin{tikzpicture}
	\begin{scope}[scale=0.6]
	\draw (-1,0.25) -- (0,0.25) (1,0.25) -- (2,0.25);
	\filldraw[white] (-0.5,0.25) circle (0.1) (1.5,0.25) circle (0.1);
	\draw (-0.5,0.25) circle (0.1) (1.5,0.25) circle (0.1);
	\begin{scope}[shift={(0,1)}]
	\draw (0,0.1) -- (1,0.1);
	\end{scope}
	\begin{scope}[shift={(0,2.5)}]
	\draw (0,0.1) -- (1,0.1) (0.1,0.2) -- (0.9,0.2);
	\filldraw (0.5,0.15) circle (0.1);
	\end{scope}
	\begin{scope}[shift={(1,4)}]
	\draw (0,0.9) -- (1,0.9) (0.1,0.8) -- (0.9,0.8);
	\end{scope}
	\begin{scope}[shift={(1,5.5)}]
	\draw (0,0.9) -- (1,0.9) (0.1,0.8) -- (0.9,0.8);
	\end{scope}
	\begin{scope}[shift={(0,7)}]
	\draw (0,0.2) -- (1,0.2) (0,0.3) -- (1,0.3);
	\filldraw (0.5,0.25) circle (0.1);
	\end{scope}
	\begin{scope}[shift={(-1,0)}]
	\draw[ultra thick,purple] (-0.25,0) -- (1.25,0) (0,0) -- (0,0.5) (1,0) -- (1,0.5);
	\end{scope}
	\begin{scope}[shift={(1,0)}]
	\draw[ultra thick,purple] (-0.25,0) -- (1.25,0) (0,0) -- (0,0.5) (1,0) -- (1,0.5);
	\end{scope}
	\begin{scope}[shift={(0,1)}]
	\S;
	\end{scope}
	\begin{scope}[shift={(0,2.5)}]
	\S;
	\end{scope}
	\begin{scope}[shift={(1,4)}]
	\begin{scope}[yscale=-1, shift={(0,-1)}]
	\S;
	\end{scope}
	\end{scope}
	\begin{scope}[shift={(1,5.5)}]
	\begin{scope}[yscale=-1, shift={(0,-1)}]
	\S;
	\end{scope}
	\end{scope}
	\begin{scope}[yscale=-1,shift={(0,-7.5)}]
	\draw[ultra thick,purple] (-0.25,0) -- (1.25,0) (0,0) -- (0,0.5) (1,0) -- (1,0.5);
	\end{scope}
	\begin{scope}[yscale=-1,shift={(0,-9)}]
	\draw[ultra thick,purple] (-0.25,0) -- (1.25,0) (0,0) -- (0,0.5) (1,0) -- (1,0.5);
	\end{scope}
	\draw[ultra thick, purple] (-1,1) -- (-1,7.5) to [out=90,in=270] (0,8) (0,4) -- (0,6.5) (2,1) -- (2,3.5) (2,7) -- (2,7.5) to[out=90,in=270] (1,8);
	\node[below] at (-1,0) {$+$}; \node[below] at (0,0) {$-$}; \node[below] at (1,0) {$+$}; \node[below] at (2,0) {$-$};
	\node at (-1,0.75) {$-$}; \node at (0,0.75) {$+$}; \node at (1,0.75) {$-$}; \node at (2,0.75) {$+$};
	\node at (0,2.25) {$+$}; \node at (1,2.25) {$-$};
	\node at (0,3.75) {$+$}; \node at (1,3.75) {$-$}; \node at (2,3.75) {$+$};
	\node at (1,5.25) {$-$}; \node at (2,5.25) {$+$};
	\node at (0,6.75) {$+$}; \node at (1,6.75) {$-$}; \node at (2,6.75) {$+$};
	\node at (0,8.25) {$-$}; \node at (1,8.25) {$+$};
	\end{scope}
	\end{tikzpicture}
\end{array}\oplus \begin{array}{c}
\begin{tikzpicture}
\begin{scope}[scale=0.6]
\draw (-1,0.25) -- (0,0.25) (1,0.25) -- (2,0.25);
\filldraw[white] (-0.5,0.25) circle (0.1) (1.5,0.25) circle (0.1);
\draw (-0.5,0.25) circle (0.1) (1.5,0.25) circle (0.1);
\begin{scope}[shift={(0,1)}]
\draw (0,0.1) -- (1,0.1) (0.1,0.2) -- (0.9,0.2);
\filldraw (0.5,0.15) circle (0.1);
\end{scope}
\begin{scope}[shift={(0,2.5)}]
\draw (0,0.1) -- (1,0.1);
\end{scope}
\begin{scope}[shift={(1,4)}]
\draw (0,0.9) -- (1,0.9) (0.1,0.8) -- (0.9,0.8);
\end{scope}
\begin{scope}[shift={(1,5.5)}]
\draw (0,0.9) -- (1,0.9) (0.1,0.8) -- (0.9,0.8);
\end{scope}
\begin{scope}[shift={(0,7)}]
\draw (0,0.2) -- (1,0.2) (0,0.3) -- (1,0.3);
\filldraw (0.5,0.25) circle (0.1);
\end{scope}
\begin{scope}[shift={(-1,0)}]
\draw[ultra thick,purple] (-0.25,0) -- (1.25,0) (0,0) -- (0,0.5) (1,0) -- (1,0.5);
\end{scope}
\begin{scope}[shift={(1,0)}]
\draw[ultra thick,purple] (-0.25,0) -- (1.25,0) (0,0) -- (0,0.5) (1,0) -- (1,0.5);
\end{scope}
\begin{scope}[shift={(0,1)}]
\S;
\end{scope}
\begin{scope}[shift={(0,2.5)}]
\S;
\end{scope}
\begin{scope}[shift={(1,4)}]
\begin{scope}[yscale=-1, shift={(0,-1)}]
\S;
\end{scope}
\end{scope}
\begin{scope}[shift={(1,5.5)}]
\begin{scope}[yscale=-1, shift={(0,-1)}]
\S;
\end{scope}
\end{scope}
\begin{scope}[yscale=-1,shift={(0,-7.5)}]
\draw[ultra thick,purple] (-0.25,0) -- (1.25,0) (0,0) -- (0,0.5) (1,0) -- (1,0.5);
\end{scope}
\begin{scope}[yscale=-1,shift={(0,-9)}]
\draw[ultra thick,purple] (-0.25,0) -- (1.25,0) (0,0) -- (0,0.5) (1,0) -- (1,0.5);
\end{scope}
\draw[ultra thick, purple] (-1,1) -- (-1,7.5) to [out=90,in=270] (0,8) (0,4) -- (0,6.5) (2,1) -- (2,3.5) (2,7) -- (2,7.5) to[out=90,in=270] (1,8);
\node[below] at (-1,0) {$+$}; \node[below] at (0,0) {$-$}; \node[below] at (1,0) {$+$}; \node[below] at (2,0) {$-$};
\node at (-1,0.75) {$-$}; \node at (0,0.75) {$+$}; \node at (1,0.75) {$-$}; \node at (2,0.75) {$+$};
\node at (0,2.25) {$+$}; \node at (1,2.25) {$-$};
\node at (0,3.75) {$+$}; \node at (1,3.75) {$-$}; \node at (2,3.75) {$+$};
\node at (1,5.25) {$-$}; \node at (2,5.25) {$+$};
\node at (0,6.75) {$+$}; \node at (1,6.75) {$-$}; \node at (2,6.75) {$+$};
\node at (0,8.25) {$-$}; \node at (1,8.25) {$+$};
\end{scope}
\end{tikzpicture}
\end{array}\oplus
\begin{array}{c}
	\begin{tikzpicture}
	\begin{scope}[scale=0.6]
	\draw (-1,0.25) -- (0,0.25) (1,0.25) -- (2,0.25);
	\filldraw[white] (-0.5,0.25) circle (0.1) (1.5,0.25) circle (0.1);
	\draw (-0.5,0.25) circle (0.1) (1.5,0.25) circle (0.1);
	\begin{scope}[shift={(0,1)}]
	\draw (0,0.1) -- (1,0.1) (0.1,0.2) -- (0.9,0.2);
	\filldraw (0.5,0.15) circle (0.1);
	\end{scope}
	\begin{scope}[shift={(0,2.5)}]
	\draw (0,0.1) -- (1,0.1) (0.1,0.2) -- (0.9,0.2);
	\filldraw (0.5,0.15) circle (0.1);
	\end{scope}
	\begin{scope}[shift={(1,4)}]
	\draw (0,0.9) -- (1,0.9);
	\filldraw[white] (0.5,0.9) circle (0.1);
	\draw (0.5,0.9) circle (0.1);
	\end{scope}
	\begin{scope}[shift={(1,5.5)}]
	\draw (0,0.9) -- (1,0.9) (0.1,0.8) -- (0.9,0.8);
	\end{scope}
	\begin{scope}[shift={(0,7)}]
	\draw (0,0.2) -- (1,0.2) (0,0.3) -- (1,0.3);
	\filldraw (0.5,0.25) circle (0.1);
	\end{scope}
	\begin{scope}[shift={(-1,0)}]
	\draw[ultra thick,purple] (-0.25,0) -- (1.25,0) (0,0) -- (0,0.5) (1,0) -- (1,0.5);
	\end{scope}
	\begin{scope}[shift={(1,0)}]
	\draw[ultra thick,purple] (-0.25,0) -- (1.25,0) (0,0) -- (0,0.5) (1,0) -- (1,0.5);
	\end{scope}
	\begin{scope}[shift={(0,1)}]
	\S;
	\end{scope}
	\begin{scope}[shift={(0,2.5)}]
	\S;
	\end{scope}
	\begin{scope}[shift={(1,4)}]
	\begin{scope}[yscale=-1, shift={(0,-1)}]
	\S;
	\end{scope}
	\end{scope}
	\begin{scope}[shift={(1,5.5)}]
	\begin{scope}[yscale=-1, shift={(0,-1)}]
	\S;
	\end{scope}
	\end{scope}
	\begin{scope}[yscale=-1,shift={(0,-7.5)}]
	\draw[ultra thick,purple] (-0.25,0) -- (1.25,0) (0,0) -- (0,0.5) (1,0) -- (1,0.5);
	\end{scope}
	\begin{scope}[yscale=-1,shift={(0,-9)}]
	\draw[ultra thick,purple] (-0.25,0) -- (1.25,0) (0,0) -- (0,0.5) (1,0) -- (1,0.5);
	\end{scope}
	\draw[ultra thick, purple] (-1,1) -- (-1,7.5) to [out=90,in=270] (0,8) (0,4) -- (0,6.5) (2,1) -- (2,3.5) (2,7) -- (2,7.5) to[out=90,in=270] (1,8);
	\node[below] at (-1,0) {$+$}; \node[below] at (0,0) {$-$}; \node[below] at (1,0) {$+$}; \node[below] at (2,0) {$-$};
	\node at (-1,0.75) {$-$}; \node at (0,0.75) {$+$}; \node at (1,0.75) {$-$}; \node at (2,0.75) {$+$};
	\node at (0,2.25) {$+$}; \node at (1,2.25) {$-$};
	\node at (0,3.75) {$+$}; \node at (1,3.75) {$-$}; \node at (2,3.75) {$+$};
	\node at (1,5.25) {$-$}; \node at (2,5.25) {$+$};
	\node at (0,6.75) {$+$}; \node at (1,6.75) {$-$}; \node at (2,6.75) {$+$};
	\node at (0,8.25) {$-$}; \node at (1,8.25) {$+$};
	\end{scope}
	\end{tikzpicture}
\end{array}\oplus
\begin{array}{c}
	\begin{tikzpicture}
	\begin{scope}[scale=0.6]
	\draw (-1,0.25) -- (0,0.25) (1,0.25) -- (2,0.25);
	\filldraw[white] (-0.5,0.25) circle (0.1) (1.5,0.25) circle (0.1);
	\draw (-0.5,0.25) circle (0.1) (1.5,0.25) circle (0.1);
	\begin{scope}[shift={(0,1)}]
	\draw (0,0.1) -- (1,0.1) (0.1,0.2) -- (0.9,0.2);
	\filldraw (0.5,0.15) circle (0.1);
	\end{scope}
	\begin{scope}[shift={(0,2.5)}]
	\draw (0,0.1) -- (1,0.1) (0.1,0.2) -- (0.9,0.2);
	\filldraw (0.5,0.15) circle (0.1);
	\end{scope}
	\begin{scope}[shift={(1,4)}]
	\draw (0,0.9) -- (1,0.9) (0.1,0.8) -- (0.9,0.8);
	\end{scope}
	\begin{scope}[shift={(1,5.5)}]
	\draw (0,0.9) -- (1,0.9);
	\filldraw[white] (0.5,0.9) circle (0.1);
	\draw (0.5,0.9) circle (0.1);
	\end{scope}
	\begin{scope}[shift={(0,7)}]
	\draw (0,0.2) -- (1,0.2) (0,0.3) -- (1,0.3);
	\filldraw (0.5,0.25) circle (0.1);
	\end{scope}
	\begin{scope}[shift={(-1,0)}]
	\draw[ultra thick,purple] (-0.25,0) -- (1.25,0) (0,0) -- (0,0.5) (1,0) -- (1,0.5);
	\end{scope}
	\begin{scope}[shift={(1,0)}]
	\draw[ultra thick,purple] (-0.25,0) -- (1.25,0) (0,0) -- (0,0.5) (1,0) -- (1,0.5);
	\end{scope}
	\begin{scope}[shift={(0,1)}]
	\S;
	\end{scope}
	\begin{scope}[shift={(0,2.5)}]
	\S;
	\end{scope}
	\begin{scope}[shift={(1,4)}]
	\begin{scope}[yscale=-1, shift={(0,-1)}]
	\S;
	\end{scope}
	\end{scope}
	\begin{scope}[shift={(1,5.5)}]
	\begin{scope}[yscale=-1, shift={(0,-1)}]
	\S;
	\end{scope}
	\end{scope}
	\begin{scope}[yscale=-1,shift={(0,-7.5)}]
	\draw[ultra thick,purple] (-0.25,0) -- (1.25,0) (0,0) -- (0,0.5) (1,0) -- (1,0.5);
	\end{scope}
	\begin{scope}[yscale=-1,shift={(0,-9)}]
	\draw[ultra thick,purple] (-0.25,0) -- (1.25,0) (0,0) -- (0,0.5) (1,0) -- (1,0.5);
	\end{scope}
	\draw[ultra thick, purple] (-1,1) -- (-1,7.5) to [out=90,in=270] (0,8) (0,4) -- (0,6.5) (2,1) -- (2,3.5) (2,7) -- (2,7.5) to[out=90,in=270] (1,8);
	\node[below] at (-1,0) {$+$}; \node[below] at (0,0) {$-$}; \node[below] at (1,0) {$+$}; \node[below] at (2,0) {$-$};
	\node at (-1,0.75) {$-$}; \node at (0,0.75) {$+$}; \node at (1,0.75) {$-$}; \node at (2,0.75) {$+$};
	\node at (0,2.25) {$+$}; \node at (1,2.25) {$-$};
	\node at (0,3.75) {$+$}; \node at (1,3.75) {$-$}; \node at (2,3.75) {$+$};
	\node at (1,5.25) {$-$}; \node at (2,5.25) {$+$};
	\node at (0,6.75) {$+$}; \node at (1,6.75) {$-$}; \node at (2,6.75) {$+$};
	\node at (0,8.25) {$-$}; \node at (1,8.25) {$+$};
	\end{scope}
	\end{tikzpicture}
\end{array}
\ee

\be\label{eq:fig8-C}
{\cal C}^{(1)}=\begin{array}{c}
	\begin{tikzpicture}
	\begin{scope}[scale=0.6]
	\begin{scope}[shift={(0,8.5)}]
	\draw (0,0.3) -- (1,0.3) (0,0.2) -- (1,0.2);
	\filldraw (0.5,0.25) circle (0.1);
	\end{scope}
	\begin{scope}[shift={(-1,0)}]
	\draw[ultra thick,purple] (-0.25,0) -- (1.25,0) (0,0) -- (0,0.5) (1,0) -- (1,0.5);
	\end{scope}
	\begin{scope}[shift={(1,0)}]
	\draw[ultra thick,purple] (-0.25,0) -- (1.25,0) (0,0) -- (0,0.5) (1,0) -- (1,0.5);
	\end{scope}
	\begin{scope}[shift={(0,1)}]
	\S;
	\end{scope}
	\begin{scope}[shift={(0,2.5)}]
	\S;
	\end{scope}
	\begin{scope}[shift={(1,4)}]
	\begin{scope}[yscale=-1, shift={(0,-1)}]
	\S;
	\end{scope}
	\end{scope}
	\begin{scope}[shift={(1,5.5)}]
	\begin{scope}[yscale=-1, shift={(0,-1)}]
	\S;
	\end{scope}
	\end{scope}
	\begin{scope}[yscale=-1,shift={(0,-7.5)}]
	\draw[ultra thick,purple] (-0.25,0) -- (1.25,0) (0,0) -- (0,0.5) (1,0) -- (1,0.5);
	\end{scope}
	\begin{scope}[yscale=-1,shift={(0,-9)}]
	\draw[ultra thick,purple] (-0.25,0) -- (1.25,0) (0,0) -- (0,0.5) (1,0) -- (1,0.5);
	\end{scope}
	\draw[ultra thick, purple] (-1,1) -- (-1,7.5) to [out=90,in=270] (0,8) (0,4) -- (0,6.5) (2,1) -- (2,3.5) (2,7) -- (2,7.5) to[out=90,in=270] (1,8);
	\node[below] at (-1,0) {$+$}; \node[below] at (0,0) {$-$}; \node[below] at (1,0) {$+$}; \node[below] at (2,0) {$-$};
	\node at (-1,0.75) {$+$}; \node at (0,0.75) {$-$}; \node at (1,0.75) {$+$}; \node at (2,0.75) {$-$};
	\node at (0,2.25) {$+$}; \node at (1,2.25) {$-$};
	\node at (0,3.75) {$-$}; \node at (1,3.75) {$+$}; \node at (2,3.75) {$-$};
	\node at (1,5.25) {$-$}; \node at (2,5.25) {$+$};
	\node at (0,6.75) {$-$}; \node at (1,6.75) {$+$}; \node at (2,6.75) {$-$};
	\node at (0,8.25) {$+$}; \node at (1,8.25) {$-$};
	\end{scope}
	\end{tikzpicture}
\end{array}\oplus \begin{array}{c}
\begin{tikzpicture}
\begin{scope}[scale=0.6]
\draw (-1,0.25) -- (0,0.25) (1,0.25) -- (2,0.25);
\filldraw[white] (-0.5,0.25) circle (0.1) (1.5,0.25) circle (0.1);
\draw (-0.5,0.25) circle (0.1) (1.5,0.25) circle (0.1);
\begin{scope}[shift={(0,1)}]
\draw (0,0.1) -- (1,0.1) (0.1,0.2) -- (0.9,0.2);
\filldraw (0.5,0.15) circle (0.1);
\end{scope}
\begin{scope}[shift={(0,2.5)}]
\draw (0,0.1) -- (1,0.1) (0.1,0.2) -- (0.9,0.2);
\filldraw (0.5,0.15) circle (0.1);
\end{scope}
\begin{scope}[shift={(0,7)}]
\draw (0,0.2) -- (1,0.2) (0,0.3) -- (1,0.3);
\filldraw (0.5,0.25) circle (0.1);
\end{scope}
\begin{scope}[shift={(-1,0)}]
\draw[ultra thick,purple] (-0.25,0) -- (1.25,0) (0,0) -- (0,0.5) (1,0) -- (1,0.5);
\end{scope}
\begin{scope}[shift={(1,0)}]
\draw[ultra thick,purple] (-0.25,0) -- (1.25,0) (0,0) -- (0,0.5) (1,0) -- (1,0.5);
\end{scope}
\begin{scope}[shift={(0,1)}]
\S;
\end{scope}
\begin{scope}[shift={(0,2.5)}]
\S;
\end{scope}
\begin{scope}[shift={(1,4)}]
\begin{scope}[yscale=-1, shift={(0,-1)}]
\S;
\end{scope}
\end{scope}
\begin{scope}[shift={(1,5.5)}]
\begin{scope}[yscale=-1, shift={(0,-1)}]
\S;
\end{scope}
\end{scope}
\begin{scope}[yscale=-1,shift={(0,-7.5)}]
\draw[ultra thick,purple] (-0.25,0) -- (1.25,0) (0,0) -- (0,0.5) (1,0) -- (1,0.5);
\end{scope}
\begin{scope}[yscale=-1,shift={(0,-9)}]
\draw[ultra thick,purple] (-0.25,0) -- (1.25,0) (0,0) -- (0,0.5) (1,0) -- (1,0.5);
\end{scope}
\draw[ultra thick, purple] (-1,1) -- (-1,7.5) to [out=90,in=270] (0,8) (0,4) -- (0,6.5) (2,1) -- (2,3.5) (2,7) -- (2,7.5) to[out=90,in=270] (1,8);
\node[below] at (-1,0) {$+$}; \node[below] at (0,0) {$-$}; \node[below] at (1,0) {$+$}; \node[below] at (2,0) {$-$};
\node at (-1,0.75) {$-$}; \node at (0,0.75) {$+$}; \node at (1,0.75) {$-$}; \node at (2,0.75) {$+$};
\node at (0,2.25) {$+$}; \node at (1,2.25) {$-$};
\node at (0,3.75) {$+$}; \node at (1,3.75) {$-$}; \node at (2,3.75) {$+$};
\node at (1,5.25) {$+$}; \node at (2,5.25) {$-$};
\node at (0,6.75) {$+$}; \node at (1,6.75) {$-$}; \node at (2,6.75) {$+$};
\node at (0,8.25) {$-$}; \node at (1,8.25) {$+$};
\end{scope}
\end{tikzpicture}
\end{array}
\ee
\end{subequations}

As in our previous examples,  denote   generators of the above complex by
 $\Psi_{ij}$ where $i$ is homological degree and $j$ is order number of the diagram as it appears in the above expansion.

The diagram of $\IQ_{\zeta}$ matrix elements is
%
%
\be
\begin{array}{c}
	\begin{tikzpicture}
	\node (A) at (-2,1) {\begin{tikzpicture} \draw circle (0.1); \end{tikzpicture}};
	\node (B) at (-2,-1) {\begin{tikzpicture} \draw circle (0.1); \end{tikzpicture}};
	\node (C) at (0,1.5) {\begin{tikzpicture} \draw circle (0.1); \end{tikzpicture}};
	\node (D) at (0,0.5) {\begin{tikzpicture} \draw circle (0.1); \end{tikzpicture}};
	\node (E) at (0,-0.5) {\begin{tikzpicture} \draw circle (0.1); \end{tikzpicture}};
	\node (F) at (0,-1.5) {\begin{tikzpicture} \draw circle (0.1); \end{tikzpicture}};
	\node (G) at (2,1) {\begin{tikzpicture} \draw circle (0.1); \end{tikzpicture}};
	\node (H) at (2,-1) {\begin{tikzpicture} \draw circle (0.1); \end{tikzpicture}};
	\path (A) edge[->] (C) (A) edge[->] (D)  (B) edge[->] (E) (B) edge[->] (F) (E) edge[->,dashed] (H) (F) edge[->] (H) (E) edge[->,dashed] (G) (F) edge[->] (G);
	\node[left] at (-2.1,1) {$\Psi_{-11}$};
	\node[left] at (-2.1,-1) {$\Psi_{-12}$};
	\node[right] at (2.1,1) {$\Psi_{11}$};
	\node[right] at (2.1,-1) {$\Psi_{12}$};
	\node[above] at (0,1.6) {$\Psi_{01}$};
	\node[above] at (0,0.6) {$\Psi_{02}$};
	\node[above] at (0,-0.4) {$\Psi_{03}$};
	\node[above] at (0,-1.4) {$\Psi_{04}$};
\end{tikzpicture}
\end{array}\label{Fig8_sgn_diag}
\ee
The curved webs corresponding to all the nonzero matrix elements indicated here are similar to those we have already encountered.  
 Matrix elements such as $\langle\Psi_{01}|\CQ_{\zeta}|\Psi_{-12}\rangle$ are zero since they are proportional to interior amplitudes $\beta_{a,b}$ 
  which were shown to vanish in our discussion of  the trefoil example (see Section \ref{subsec:Trefoil}).

A new argument is required to   demonstrate that  $\langle\Psi_{11}|\CQ_{\zeta}|\Psi_{01}\rangle$ and $\langle\Psi_{11}|\CQ_{\zeta}|\Psi_{02}\rangle$ are zero. Let us divide the spatial interval in two pieces along the red dashed line marked on the following diagram:
$$
\begin{array}{c}
\begin{tikzpicture}
\begin{scope}[scale=0.6]
\draw[dashed, ultra thick, red] (-1,2) -- (2.5,2);
\SS;
\begin{scope}[shift={(0,1)}]
\SS;
\end{scope}
\begin{scope}[xscale=-1, shift={(-2,2)}]
\SS;
\end{scope}
\begin{scope}[xscale=-1, shift={(-2,3)}]
\SS;
\end{scope}
\draw[ultra thick] (2,2) -- (2,0) to[out=270,in=270] (1,0) (0,0) to[out=270,in=270] (-0.5,0) -- (-0.5,4) to[out=90,in=90] (2,4) (0,2) -- (0,4) to[out=90,in=90] (1,4);
\end{scope}
\end{tikzpicture}
\end{array}
$$
 We construct curved webs for boundary conditions of each piece:
\be
\begin{array}{c}
\begin{tikzpicture}
\draw[<-] (0,2) -- (0,0) -- (3.5,0);
\draw[->] (4.5,0) -- (7,0);
\draw (0,1.5) -- (3.5,1.5) (4.5,1.5) -- (7,1.5);
\node[left] at (0,2) {$\tau$};
\node[right] at (7,0) {$x$};
\draw[snake=bumps] (3.5,1.5) -- (3.5,0) (4.5,0) -- (4.5,1.5);
\draw[ultra thick] (0.5,0) to[out=90,in=135] (1.5,0.75) to[out=225,in=90] (1,0) (1.5,0) -- (1.5,0.75) -- (2.5,0.75) -- (2.5,0) (3.5,0.25) -- (2.5,0.75) -- (3.5,1.25) (5,0) to[out=90,in=0] (4.5,0.5) (4.5,1) to[out=0,in=135] (6,0.75) to[out=225,in=90] (5.5,0) (6,0.75) to[out=0,in=270] (6.5,1.5);
\begin{scope}[shift={(0.5,0)}]
\filldraw[red](-0.1,-0.1) -- (-0.1,0.1) -- (0.1,0.1) -- (0.1,-0.1) -- (-0.1,-0.1);
\end{scope}
\begin{scope}[shift={(1,0)}]
\filldraw[red](-0.1,-0.1) -- (-0.1,0.1) -- (0.1,0.1) -- (0.1,-0.1) -- (-0.1,-0.1);
\end{scope}
\begin{scope}[shift={(1.5,0)}]
\filldraw[red](-0.1,-0.1) -- (-0.1,0.1) -- (0.1,0.1) -- (0.1,-0.1) -- (-0.1,-0.1);
\end{scope}
\begin{scope}[shift={(2.5,0)}]
\filldraw[red](-0.1,-0.1) -- (-0.1,0.1) -- (0.1,0.1) -- (0.1,-0.1) -- (-0.1,-0.1);
\end{scope}
\begin{scope}[shift={(5,0)}]
\filldraw[red](-0.1,-0.1) -- (-0.1,0.1) -- (0.1,0.1) -- (0.1,-0.1) -- (-0.1,-0.1);
\end{scope}
\begin{scope}[shift={(5.5,0)}]
\filldraw[red](-0.1,-0.1) -- (-0.1,0.1) -- (0.1,0.1) -- (0.1,-0.1) -- (-0.1,-0.1);
\end{scope}
\begin{scope}[shift={(6.5,1.5)}]
\filldraw[red](-0.1,-0.1) -- (-0.1,0.1) -- (0.1,0.1) -- (0.1,-0.1) -- (-0.1,-0.1);
\end{scope}
\begin{scope}[shift={(1.5,0.75)}]
\filldraw[black](-0.1,-0.1) -- (-0.1,0.1) -- (0.1,0.1) -- (0.1,-0.1) -- (-0.1,-0.1);
\end{scope}
\begin{scope}[shift={(2.5,0.75)}]
\filldraw[black](-0.1,-0.1) -- (-0.1,0.1) -- (0.1,0.1) -- (0.1,-0.1) -- (-0.1,-0.1);
\end{scope}
\begin{scope}[shift={(6,0.75)}]
\filldraw[black](-0.1,-0.1) -- (-0.1,0.1) -- (0.1,0.1) -- (0.1,-0.1) -- (-0.1,-0.1);
\end{scope}
\node[right] at (3.5,0.25) {$1$};
\node[right] at (3.5,1.25) {$2$};
\node[left] at (4.5,0.45) {$3$};
\node[left] at (4.5,1.05) {$4$};
\end{tikzpicture}
\end{array}
\ee
Now we should try to glue these parts along the curvy line. We claim that, actually, it is impossible to join these boundary 
conditions so that the corresponding $\zeta$-instanton does not, in fact, exist. To understand this, 
let  us describe in more detail the solitons $1$, $2$, $3$ and $4$ that are supposed to be joined at the cut. It will be important to 
recall the distinction between single- and double-line solitons depicted in Figure \ref{fig:2_solitons}.  Solitons that interfere in the scattering process of the cup solitons in $1$ and $2$ ``cancel'' each other. (We already noticed a similar effect  when discussing the   Hopf link \eqref{eq:E_Hopf}.) Notice that cup single line solitons are mapped to double line solitons. Therefore we conclude that solitons $1$ and $2$ are double line solitons.   
On the other hand soliton $3$ is a continuation of a trajectory emanated by a binding point of a double line soliton. This trajectory goes thtough a twist. And 
after crossing the twist the double line soliton becomes  a single line soliton. Therefore  we are unable to connect trajectories of solitons $1$ and $3$, and hence there is no curved web connecting the boundary conditions implied by states $\Psi_{11}$ and $\Psi_{01}$. A similar argument holds for the matrix element connecting $\Psi_{11}$ and $\Psi_{02}$.

The resulting matrix elements of the supercharge are:
\be
\CQ_{\zeta}^{(-1)}=\left(\begin{array}{cc}
	1 & 0 \\
	1 & 0 \\
	0 & 1\\
	0 & 1\\
\end{array}\right),\quad \CQ_{\zeta}^{(0)}=\left(
\begin{array}{cccc}
	0 & 0 & -1 &1\\
	0 & 0 & -1 &1\\
\end{array}
\right)
\ee
And we check easily $\IQ_{\zeta}^{(0)}\IQ_{\zeta}^{(-1)}=0$. It is now easy to calculate the cohomology of this complex:
\be
H^{\bullet} (\IM_{-1},\IQ_{\zeta})\cong\IF[\Psi_{01}-\Psi_{02}]\oplus\IF[\Psi_{11}-\Psi_{12}]
\ee

The figure-eight knot is known to be achiral \cite{achiral}, so it is ambient isotopic to its mirror image. To go from a knot to its mirror image it is enough to map the chosen evolution direction $x$ to $-x$. Under this map we would expect the following map of the degrees $\Pdeg$ and $\bf F$:
\be
\Pdeg\to -\Pdeg,\quad {\bf F}\to -{\bf F}
\ee
Using this fact we can restore the rest of the Poincar\'e polynomial from this symmetry:
\be
{\cal P}(q,t|{\bf 4_1})={\cal P}(q^{-1},t^{-1}|{\bf 4_1})
\ee
Thus, we conclude that the Poincar\'e polynomial for the homology of this knot is:
\be
{\cal P}(q,t|{\bf 4_1})=q^{-5}t^3+q^{-1}(1+t)+q(1+t^{-1})+q^5 t^{-3}.
\ee

\section{Invariance Under Reidemeister Moves}\label{sec:R-Invariance}

The rules of Section \ref{sec:KnotHomRules} depend on both a link-projection into a plane and on a choice of
``time''-direction through which the tangle evolves. For example,
in the   tangle presentation we gave above if we write $z = x^2 + \I  x^3$ then the link projection
could be in the $(x^1, x^2)$ plane, with ``time'' in the $x^1$ direction. But of course
we could make other choices.  A good link homology theory will define
complexes which are quasi-isomorphic if we make changes of the link-projection and of the evolution
direction $x^1$.

Reidemeister classified equivalence of link projections onto the plane and showed that two projections
of the same link can always be related by a series of three basic moves known as the Reidemeister moves
$RI, RII, RIII$. When we take into account the extra data of a choice of   ``time'' direction
in the plane and analogous theorem
requires two more moves (up to time reversal) \cite{FY,RT}. We will call them ``Reidemeister 0 moves.''
Therefore, it suffices to show that our rules give quasi-isomorphic complexes under changes given by
the Reidemeister moves $R0$ through $RIII$.  In this section we explain how that works in detail.

The general strategy we use follows the discussion in Section 10.7 of \cite{Gaiotto:2015aoa}.
Let $\sigma$ be a continuous parameter for the interpolation of one link-projection into another.
Then we introduce an analog of the $\zeta$-instanton equation, but now in $(x,\sigma)$ space:
\be\label{Hmtpy}
\left(\p_x-\I\p_{\sigma}\right)\phi^I(x,\sigma)=\frac{\I\zeta}{2}g^{I\bar J}\overline{\p_J W\left(\phi^I(x,\sigma)|z_a(x,\sigma)\right)}
\ee
Solutions to this equation will be called ``$\sigma \zeta$-instantons'' for brevity.

Let $\CE^{(1)}$ and $\CE^{(2)}$ be two complexes constructed using two link projections related by a
homotopy in $\sigma$. As explained in \cite{Gaiotto:2015aoa}, counting \underline{rigid} $\sigma\zeta$-instantons
produces a map
\be
U : \CE^{(1)} \rightarrow \CE^{(2)}
\ee
that preserves bi-gradings up to an overall shift (related to the framing anomaly) and moreover (anti)-commutes with the differentials
\be
U \CQ_{\zeta}^{(1)} = \pm \CQ_{\zeta}^{(2)} U
\ee
To establish an isomorphism of cohomologies $H(\CE^{(1)},\CQ_{\zeta}^{(1)}) \cong H(\CE^{(2)},\CQ_{\zeta}^{(2)})$ one must find a chain map $U'$ in the opposite direction:
\be
U' : \CE^{(2)} \rightarrow \CE^{(1)}
\ee
that graded commutes with the differentials and moreover satisfies
\be
UU' = {\rm \bf Id} + T_1 Q^{(2)} + Q^{(2)} T_2  \qquad \quad U'U = {\rm \bf Id} + T_3 Q^{(1)} + Q^{(1)} T_4
\ee
for some maps $T_i$.

When implementing this idea it is very useful to construct the analog of the hovering $\zeta$-solitons
for the case of the $\sigma\zeta$-instanton equation. We work in the limit of very slow variation of the homotopy parameter $\sigma$ ($|dz_a/d\sigma|\ll 1$)
and take
\be
\phi^I(x,\sigma)=\varphi^I(x,\sigma)+O\left(\left|\frac{dz_a}{d\sigma}\right|\right)
\ee
where $\varphi^I(x,\sigma)$ is a continuously evolving family of critical points of the Morse height function $h$ \eqref{eq:InterfaceMorseFunction}:
\be\
\p_x\varphi^I(x,\sigma)=\frac{\I\zeta}{2}g^{I\bar J}\overline{\p_J W\left(\varphi^I(x,\sigma)|z_a(x,\sigma)\right)}
\ee
This solution (if it exists) is invertible and can be used to define matrix elements of $U$ and $U'$.

\begin{figure}[h!]
\begin{center}
\includegraphics[scale=0.5]{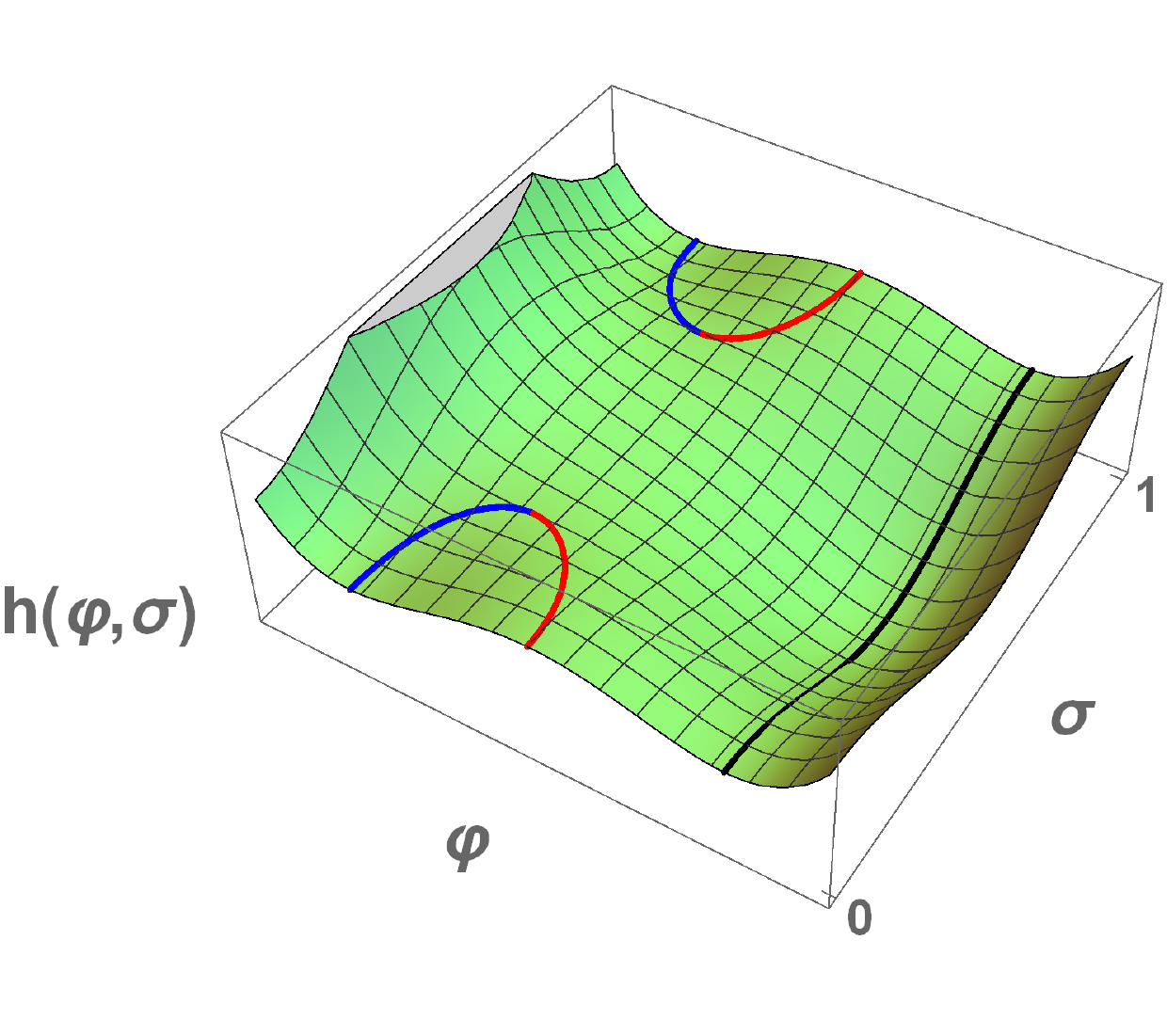}
\end{center}
\caption{Homotopic evolution of the Morse height function. }\label{fig:Morse}
\end{figure}

There are several types of evolution of the critical points of the Morse height function $h$ with the homotopy flow preserving the homology of the Morse complex \cite[Section 10.7]{Gaiotto:2015aoa}: field configurations corresponding to critical points of $h$ vary smoothly one to another, or merge, or emerge by pairs (see Figure \ref{fig:Morse}). Merging (or emerging) critical points correspond to ``fake'' ground states that form a subcomplex quasi-isomorphic to a null complex and are lifted non-perturbatively. In more complicated case there are possible moves depicted in \cite[Figure 123(d,e,f)]{Gaiotto:2015aoa} however we will not encounter them.

A major advantage of this approach is that local homotopies of the tangle modify solutions to the forced $\zeta$-soliton equation defined in Section \ref{sec:HSLGInterfaces} only locally as well. This, in turn, simplifies the task of checking quasi-isomorphism under homotopic deformation of the tangle drastically, in comparison to the calculation of the actual cohomology groups. The reason the computation of the cohomology groups themselves is much harder is that the $\zeta$-instanton equation has a highly non-local behavior, depending on contributions from the binding points over all the tangle.
 A related general remark is that the existence of a  $\zeta$-instanton saturating a matrix element $\langle\Psi_2|\CQ_{\zeta}|\Psi_1\rangle$ is \underline{necessary} for   $\IZ[\Psi_1]\oplus\IZ[\Psi_2]$ to be equivalent to zero,
 however it is not, in general,  \underline{sufficient} due to the non-locality of the differential. The states  $\IZ[\Psi_1]$ and $\IZ[\Psi_2]$ can be connected
 by a differential to other states and hence will not, in general, form a   well-defined subcomplex of the entire knot complex.

During the evolution with the homotopy flow critical points of $h$ given by the forced $\zeta$-soliton solutions may vary in such a way that the binding points travel along the interface and sometimes ``scatter'' through the $n$-valent interior amplitude vertices discussed in Section \ref{sec:CurvedWebs}. So we can remark that to prove quasi-isomorphism of complexes corresponding to homotopic superpotentials and to calculate the differential we need the same amount of data, in particular, interior amplitudes.

\subsection{Reidemiester 0}\label{subsec:R0}

The ``Reidemeister 0'' condition of \cite{FY,RT} really consists of two sub-conditions,
illustrated in \eqref{ex_move_a} and \eqref{ex_move_b} below.

The first Reidemeister 0 condition allows us to add new strands through humps:
\be\label{ex_move_a}
\begin{split}
\CI_{13} \sim \cup_{12}\boxtimes \cap_{23}\\
\begin{array}{c}
	\begin{tikzpicture}
	\draw[ultra thick] (0,0) -- (0,1);
	\end{tikzpicture}
\end{array}\sim \begin{array}{c}
\begin{tikzpicture}
\draw[ultra thick] (0,1) -- (0,0.5) to[out=270, in=270] (0.5,0.5) to[out=90,in=90] (1,0.5) -- (1,0);
\end{tikzpicture}
\end{array}
\end{split}
\ee

Using our rules we decompose both sides:
\be
\begin{array}{c}
	\begin{tikzpicture}
	\draw[ultra thick] (0,0) -- (0,1);
	\end{tikzpicture}
\end{array}=\begin{array}{c}
\begin{tikzpicture}
\draw[purple, ultra thick] (0,0) -- (0,1);
\node[below] at (0,0) {$-$}; \node[above] at (0,1) {$-$};
\end{tikzpicture}
\end{array}\oplus \begin{array}{c}
\begin{tikzpicture}
\draw[purple, ultra thick] (0,0) -- (0,1);
\node[below] at (0,0) {$+$}; \node[above] at (0,1) {$+$};
\end{tikzpicture}
\end{array},\quad
\begin{array}{c}
	\begin{tikzpicture}
	\draw[ultra thick] (0,1) -- (0,0.5) to[out=270, in=270] (0.5,0.5) to[out=90,in=90] (1,0.5) -- (1,0);
	\end{tikzpicture}
\end{array}=\begin{array}{c}
\begin{tikzpicture}
\begin{scope}[scale=0.7]
\draw[purple, ultra thick] (0,2) -- (0,0.5) (1,1.5) -- (1,0.5) (2,1.5) -- (2,0) (-0.1,0.5) -- (1.1,0.5) (0.9,1.5) -- (2.1,1.5);
\node[below] at (0,0.5) {$+$}; \node[below] at (1,0.5) {$-$}; \node[below] at (2,0) {$+$};
\node[above] at (0,2) {$+$}; \node[above] at (1,1.5) {$-$}; \node[above] at (2,1.5) {$+$};
\end{scope}
\end{tikzpicture}
\end{array}\oplus  \begin{array}{c}
\begin{tikzpicture}
\begin{scope}[scale=0.7]
\draw (0,0.75) -- (1,0.75) (1,1.2) -- (2,1.2) (1,1.3) -- (2,1.3);
\filldraw[white] (0.5,0.75) circle (0.1);
\draw (0.5,0.75) circle (0.1);
\filldraw[black] (1.5,1.25) circle (0.1);
\draw[purple, ultra thick] (0,2) -- (0,0.5) (1,1.5) -- (1,0.5) (2,1.5) -- (2,0) (-0.1,0.5) -- (1.1,0.5) (0.9,1.5) -- (2.1,1.5);
\node[below] at (0,0.5) {$+$}; \node[below] at (1,0.5) {$-$}; \node[below] at (2,0) {$-$};
\node[above] at (0,2) {$-$}; \node[above] at (1,1.5) {$-$}; \node[above] at (2,1.5) {$+$};
\end{scope}
\end{tikzpicture}
\end{array}
\ee

In this situation we cannot use the construction of $U$ given by counting $\sigma\zeta$ instantons and we must
present another argument. We will call it the  \FINDANAME argument.

When this interface is glued into a larger interface we will continue to have an isomorphism $U$ defined by rigid $\sigma\zeta$-instantons:
\be
U: \CE_{ij}\left(\underbrace{\begin{array}{c}
\begin{tikzpicture}
\begin{scope}[scale=0.7]
\filldraw[violet!50!white] (0,0) circle (1.5);
\filldraw[white] (0,0) circle (1);
\draw[black] (0,0) circle (1);
\draw[ultra thick] (0,1) -- (0,-1);
\end{scope}
\end{tikzpicture}
\end{array}}_{\mathscr{V}}\right)  \to \CE_{ij}\left(\underbrace{\begin{array}{c}
\begin{tikzpicture}
\begin{scope}[scale=0.7]
\filldraw[violet!50!white] (0,0) circle (1.5);
\filldraw[white] (0,0) circle (1);
\draw[black] (0,0) circle (1);
\draw[ultra thick] (0,1) to[out=270,in=180] (-0.25,-0.25) to[out=0,in=180] (0.25,0.25) to[out=0,in=90]  (0,-1);
\end{scope}
\end{tikzpicture}
\end{array}}_{\mathscr{H}}\right)
\ee

We need to show that this isomorphism commutes with the differential:
\be
  U \CQ_{\zeta}^{\mathscr{V}} = \pm  \CQ_{\zeta}^{\mathscr{H}} U
\ee
 We note that a typical $\zeta$-instanton contributing to the
differential on the complex $\CE_{ij}(\mathscr{V}) $
 is represented by a boosted soliton $\X$ interpolating between the strand in question and some other strand (denoted by a dashed line)
 with the soliton moving, say, towards the positive $x$-direction. We can illustrate this as:
\be\label{eq:identity}
\begin{array}{c}
	\begin{tikzpicture}
	\begin{scope}[scale=0.6]
	\draw[->] (-0.5,0) -- (-0.5,2);
	\node[above] at (-0.5,2) {$x$};
	\draw[ultra thick, purple] (0,0) -- (0,2);
	\node[below] at (0,0) {$+$};
	\node[right] at (0,1) {$-$};
	\node[above] at (0,2) {$-$};
	\draw[->] (0,0.5) -- (1,0.5);
	\draw[ultra thick, purple, dashed] (1,0) -- (1,2);
	\node[right] at (1,0.5) {$\X(\tau_0)$};
	\end{scope}
	\end{tikzpicture}
\end{array}\quad\mathop{\longrightarrow}^{\IQ_{\zeta}}\quad
\begin{array}{c}
	\begin{tikzpicture}
	\begin{scope}[scale=0.6]
	\draw[->] (-0.5,0) -- (-0.5,2);
	\node[above] at (-0.5,2) {$x$};
	\draw[ultra thick, purple] (0,0) -- (0,2);
	\node[below] at (0,0) {$+$};
	\node[right] at (0,1) {$+$};
	\node[above] at (0,2) {$-$};
	\draw[->] (0,1.5) -- (1,1.5);
	\draw[ultra thick, purple, dashed] (1,0) -- (1,2);
	\node[right] at (1,1.5) {$\X(\tau_1)$};
	\end{scope}
	\end{tikzpicture}
\end{array}\quad\begin{array}{c}
\begin{tikzpicture}
\draw[<->] (-0.5,2) -- (-0.5,0) -- (2.5,0);
\node[left] at (0,2) {$\tau$};
\node[below] at (2.5,0) {$x$};
\draw[ultra thick,->] (0.5,1) -- (2,2);
\node[below] at (0.5,1) {$\X(\tau_0)$};
\node[right] at (2,2) {$\X(\tau_1)$};
\end{tikzpicture}
\end{array}
\ee
where the figure on the right shows the $(\tau,x)$-spacetime interpretation of the
$\zeta$ instanton that contributes to the matrix element of $\CQ_{\zeta}$.

The analogous $\zeta$-instanton in $ \CE_{ij}(\mathscr{H}) $ will look like
%
%
\be
 \begin{array}{c}
 	\begin{tikzpicture}
 	\begin{scope}[scale=0.7]
 	\draw[->] (2,-0.5) -- (3,-0.5);
 	\node[right] at (3,-0.5) {$\X(\tau_0)$};
 	\draw[purple, ultra thick, dashed] (3,-1) -- (3,2);
 	\draw (0,0.75) -- (1,0.75) (1,1.2) -- (2,1.2) (1,1.3) -- (2,1.3);
 	\filldraw[white] (0.5,0.75) circle (0.1);
 	\draw (0.5,0.75) circle (0.1);
 	\filldraw[black] (1.5,1.25) circle (0.1);
 	\draw[purple, ultra thick] (0,2) -- (0,0.5) (1,1.5) -- (1,0.5) (2,1.5) -- (2,-1) (-0.1,0.5) -- (1.1,0.5) (0.9,1.5) -- (2.1,1.5);
 	\node[below] at (0,0.5) {$+$}; \node[below] at (1,0.5) {$-$}; \node[below] at (2,-1) {$+$};
 	\node[above] at (0,2) {$-$}; \node[above] at (1,1.5) {$-$}; \node[above] at (2,1.5) {$+$};
 	\node[above] at (0.5,0.85) {$1$};
 	\node[below] at (1.5,1.15) {$2$};
 	\node[left] at (0,0.5) {$x_{\cup}$};
 	\node[right] at (2,1.5) {$x_{\cap}$};
 	\end{scope}
 	\end{tikzpicture}
 \end{array}\mathop{\longrightarrow}^{\CQ_{\zeta}} \begin{array}{c}
\begin{tikzpicture}
\begin{scope}[scale=0.7]
 	\draw[->] (0,2.5) -- (3,2.5);
 	\node[right] at (3,2.5) {$\X(\tau_1)$};
\draw[purple, ultra thick, dashed] (3,0) -- (3,3);
\draw[purple, ultra thick] (0,3) -- (0,0.5) (1,1.5) -- (1,0.5) (2,1.5) -- (2,0) (-0.1,0.5) -- (1.1,0.5) (0.9,1.5) -- (2.1,1.5);
\node[below] at (0,0.5) {$+$}; \node[below] at (1,0.5) {$-$}; \node[below] at (2,0) {$+$};
\node[above] at (0,3) {$-$}; \node[above] at (1,1.5) {$-$}; \node[above] at (2,1.5) {$+$};
	\node[left] at (0,0.5) {$x_{\cup}$};
	\node[right] at (2,1.5) {$x_{\cap}$};
\end{scope}
\end{tikzpicture}
\end{array}\qquad
\begin{array}{c}
\begin{tikzpicture}
\draw[<->] (-1,2) -- (-1,0) -- (3,0);
\node[above] at (-1,2) {$\tau$}; \node[right] at (3,0) {$x$};
\draw[ultra thick, ->] (0.5,0.6) -- (2.5,1.8);
\draw[ultra thick] (1,0) to[out=90,in=230]  (2,1.5) (2,0) -- (2,1.5);
\draw[blue!40!black,ultra thick] (0.75,0) -- (0.75,2) (2.25,0) -- (2.25,2);
\begin{scope}[shift={(2,0)}]
\filldraw[red] (-0.1,-0.1) -- (-0.1,0.1) -- (0.1,0.1) -- (0.1,-0.1) -- (-0.1,-0.1);
\end{scope}
\begin{scope}[shift={(1,0)}]
\filldraw[red] (-0.1,-0.1) -- (-0.1,0.1) -- (0.1,0.1) -- (0.1,-0.1) -- (-0.1,-0.1);
\end{scope}
\begin{scope}[shift={(2,1.5)}]
\filldraw[black] (-0.1,-0.1) -- (-0.1,0.1) -- (0.1,0.1) -- (0.1,-0.1) -- (-0.1,-0.1);
\end{scope}
\node[below] at (1,-0.1) {$1$};
\node[below] at (2,-0.1) {$2$};
\node[left] at (0.5,0.6) {$\X(\tau_0)$};
\node[right] at (2.5,1.8) {$\X(\tau_1)$};
\node (A) at (0.25,-0.5) {$x_{\cup}$}; \node (B) at (2.75,-0.5) {$x_{\cap}$};
\path (A) edge[->] (0.65,-0.1) (B) edge[->] (2.35,-0.1);
\end{tikzpicture}
\end{array}
\ee
where on the right we show the spacetime interpretation of a $\zeta$-instanton that contributes to
the differential on $ \CE_{ij}(\mathscr{H}) $.
In this process the soliton $\cal X$ interacts with operators on the interfaces as indicated by the curved web diagram.  Note that the interaction with the
branes  does not add new moduli since the positions of all the vertices are defined by the curved trajectory for $\cal X$.

The second Riedemeister 0 move involves a ``trident'' link diagram:
\be\label{ex_move_b}
\begin{split}
V_{23}\boxtimes {\cal R}_{12}\sim V_{12}\boxtimes {\cal R}_{23}^{-1}\\
\begin{array}{c}
	\begin{tikzpicture}
	\begin{scope}[scale=0.5]
	\draw[ultra thick] (1,0) -- (1,2) (0,2) to[out=270,in=135] (0.8,1.2) (1.2,0.8) to[out=315,in=180] (1.75,0) to[out=0,in=270] (2,2);
	\end{scope}
	\end{tikzpicture}
\end{array}\sim \begin{array}{c}
\begin{tikzpicture}
\begin{scope}[xscale=-1]
\begin{scope}[scale=0.5]
\draw[ultra thick] (1,0) -- (1,2) (0,2) to[out=270,in=135] (0.8,1.2) (1.2,0.8) to[out=315,in=180] (1.75,0) to[out=0,in=270] (2,2);
\end{scope}
\end{scope}
\end{tikzpicture}
\end{array}
\end{split}
\ee
Let us decompose both sides of this equality diagrammatically:
\be
\begin{split}
	LHS=q\begin{array}{c}
		\begin{tikzpicture}
		\begin{scope}[scale=0.6]
		\draw (1,0.25) -- (2,0.25);
		\filldraw[white] (1.5,0.25) circle (0.1);
		\draw (1.5,0.25) circle (0.1);
		\begin{scope}[shift={(0,-1)}]
		\draw[ultra thick, purple] (0,1) -- (0,1.5) to[out=90,in=210] (0.5,2) to[out=30,in=270] (1,2.5) -- (1,3) (0,3) -- (0,2.5) to[out=270,in=150] (0.4,2.1) (0.6,1.9) to[out=330,in=90] (1,1.5) -- (1,1) (0.75,1) -- (2.25,1) (2,1) -- (2,3);
		\end{scope}
		\node[below] at (0,0) {$-$}; \node[below] at (1,0) {$+$}; \node[below] at (2,0) {$-$};
		\node[above] at (0,2) {$-$}; \node[above] at (1,2) {$-$}; \node[above] at (2,2) {$+$};
		\end{scope}
		\end{tikzpicture}
	\end{array}\oplus q^{-1}\begin{array}{c}
	\begin{tikzpicture}
	\begin{scope}[scale=0.6]
	\begin{scope}[shift={(0,-1)}]
	\draw[ultra thick, purple] (0,1) -- (0,1.5) to[out=90,in=210] (0.5,2) to[out=30,in=270] (1,2.5) -- (1,3) (0,3) -- (0,2.5) to[out=270,in=150] (0.4,2.1) (0.6,1.9) to[out=330,in=90] (1,1.5) -- (1,1) (0.75,1) -- (2.25,1) (2,1) -- (2,3);
	\end{scope}
	\node[below] at (0,0) {$-$}; \node[below] at (1,0) {$+$}; \node[below] at (2,0) {$-$};
	\node[above] at (0,2) {$+$}; \node[above] at (1,2) {$-$}; \node[above] at (2,2) {$-$};
	\end{scope}
	\end{tikzpicture}
\end{array}\oplus \begin{array}{c}
\begin{tikzpicture}
\begin{scope}[scale=0.6]
\draw (1,0.25) -- (2,0.25);
\filldraw[white] (1.5,0.25) circle (0.1);
\draw (1.5,0.25) circle (0.1);
\begin{scope}[shift={(0,-1)}]
\draw[ultra thick, purple] (0,1) -- (0,1.5) to[out=90,in=210] (0.5,2) to[out=30,in=270] (1,2.5) -- (1,3) (0,3) -- (0,2.5) to[out=270,in=150] (0.4,2.1) (0.6,1.9) to[out=330,in=90] (1,1.5) -- (1,1) (0.75,1) -- (2.25,1) (2,1) -- (2,3);
\end{scope}
\node[below] at (0,0) {$+$}; \node[below] at (1,0) {$+$}; \node[below] at (2,0) {$-$};
\node[above] at (0,2) {$-$}; \node[above] at (1,2) {$+$}; \node[above] at (2,2) {$+$};
\end{scope}
\end{tikzpicture}
\end{array}\oplus\\
\oplus q\begin{array}{c}
	\begin{tikzpicture}
	\begin{scope}[scale=0.6]
	\draw (0,0.5) -- (1,0.5) (1,0.25) -- (2,0.25);
	\filldraw[white] (1.5,0.25) circle (0.1);
	\draw (1.5,0.25) circle (0.1);
	\begin{scope}[shift={(0,-1)}]
	\draw[ultra thick, purple] (0,1) -- (0,1.5) to[out=90,in=210] (0.5,2) to[out=30,in=270] (1,2.5) -- (1,3) (0,3) -- (0,2.5) to[out=270,in=150] (0.4,2.1) (0.6,1.9) to[out=330,in=90] (1,1.5) -- (1,1) (0.75,1) -- (2.25,1) (2,1) -- (2,3);
	\end{scope}
	\node[below] at (0,0) {$+$}; \node[below] at (1,0) {$+$}; \node[below] at (2,0) {$-$};
	\node[above] at (0,2) {$+$}; \node[above] at (1,2) {$-$}; \node[above] at (2,2) {$+$};
	\end{scope}
	\end{tikzpicture}
\end{array}\oplus q^{-1}\begin{array}{c}
\begin{tikzpicture}
\begin{scope}[scale=0.6]
\draw (0,0.5) -- (1,0.5) (0,0.6) -- (1,0.6) (1,0.25) -- (2,0.25);
\filldraw (0.5,0.55) circle (0.1);
\filldraw[white] (1.5,0.25) circle (0.1);
\draw (1.5,0.25) circle (0.1);
\begin{scope}[shift={(0,-1)}]
\draw[ultra thick, purple] (0,1) -- (0,1.5) to[out=90,in=210] (0.5,2) to[out=30,in=270] (1,2.5) -- (1,3) (0,3) -- (0,2.5) to[out=270,in=150] (0.4,2.1) (0.6,1.9) to[out=330,in=90] (1,1.5) -- (1,1) (0.75,1) -- (2.25,1) (2,1) -- (2,3);
\end{scope}
\node[below] at (0,0) {$+$}; \node[below] at (1,0) {$+$}; \node[below] at (2,0) {$-$};
\node[above] at (0,2) {$+$}; \node[above] at (1,2) {$-$}; \node[above] at (2,2) {$+$};
\end{scope}
\end{tikzpicture}
\end{array} \oplus \begin{array}{c}
\begin{tikzpicture}
\begin{scope}[scale=0.6]
\begin{scope}[shift={(0,-1)}]
\draw[ultra thick, purple] (0,1) -- (0,1.5) to[out=90,in=210] (0.5,2) to[out=30,in=270] (1,2.5) -- (1,3) (0,3) -- (0,2.5) to[out=270,in=150] (0.4,2.1) (0.6,1.9) to[out=330,in=90] (1,1.5) -- (1,1) (0.75,1) -- (2.25,1) (2,1) -- (2,3);
\end{scope}
\node[below] at (0,0) {$+$}; \node[below] at (1,0) {$+$}; \node[below] at (2,0) {$-$};
\node[above] at (0,2) {$+$}; \node[above] at (1,2) {$+$}; \node[above] at (2,2) {$-$};
\end{scope}
\end{tikzpicture}
\end{array}
\end{split}
\ee

\be
\begin{split}
	RHS=q\begin{array}{c}
		\begin{tikzpicture}
		\begin{scope}[scale=0.6]
		\draw (0,0.25) -- (1,0.25);
		\filldraw[white] (0.5,0.25) circle (0.1);
		\draw (0.5,0.25) circle (0.1);
		\begin{scope}[shift={(2,0)}]
		\begin{scope}[xscale=-1]
		\begin{scope}[shift={(0,-1)}]
		\draw[ultra thick, purple] (0,1) -- (0,1.5) to[out=90,in=210] (0.5,2) to[out=30,in=270] (1,2.5) -- (1,3) (0,3) -- (0,2.5) to[out=270,in=150] (0.4,2.1) (0.6,1.9) to[out=330,in=90] (1,1.5) -- (1,1) (0.75,1) -- (2.25,1) (2,1) -- (2,3);
		\end{scope}
		\end{scope}
		\end{scope}
		\node[below] at (0,0) {$+$}; \node[below] at (1,0) {$-$}; \node[below] at (2,0) {$-$};
		\node[above] at (0,2) {$-$}; \node[above] at (1,2) {$-$}; \node[above] at (2,2) {$+$};
		\end{scope}
		\end{tikzpicture}
	\end{array}\oplus
	q^{-1}\begin{array}{c}
		\begin{tikzpicture}
		\begin{scope}[scale=0.6]
		\begin{scope}[shift={(2,0)}]
		\begin{scope}[xscale=-1]
		\begin{scope}[shift={(0,-1)}]
		\draw[ultra thick, purple] (0,1) -- (0,1.5) to[out=90,in=210] (0.5,2) to[out=30,in=270] (1,2.5) -- (1,3) (0,3) -- (0,2.5) to[out=270,in=150] (0.4,2.1) (0.6,1.9) to[out=330,in=90] (1,1.5) -- (1,1) (0.75,1) -- (2.25,1) (2,1) -- (2,3);
		\end{scope}
		\end{scope}
		\end{scope}
		\node[below] at (0,0) {$+$}; \node[below] at (1,0) {$-$}; \node[below] at (2,0) {$-$};
		\node[above] at (0,2) {$+$}; \node[above] at (1,2) {$-$}; \node[above] at (2,2) {$-$};
		\end{scope}
		\end{tikzpicture}
	\end{array}\oplus
	\begin{array}{c}
		\begin{tikzpicture}
		\begin{scope}[scale=0.6]
		\draw (0,0.25) -- (1,0.25);
		\filldraw[white] (0.5,0.25) circle (0.1);
		\draw (0.5,0.25) circle (0.1);
		\begin{scope}[shift={(2,0)}]
		\begin{scope}[xscale=-1]
		\begin{scope}[shift={(0,-1)}]
		\draw[ultra thick, purple] (0,1) -- (0,1.5) to[out=90,in=210] (0.5,2) to[out=30,in=270] (1,2.5) -- (1,3) (0,3) -- (0,2.5) to[out=270,in=150] (0.4,2.1) (0.6,1.9) to[out=330,in=90] (1,1.5) -- (1,1) (0.75,1) -- (2.25,1) (2,1) -- (2,3);
		\end{scope}
		\end{scope}
		\end{scope}
		\node[below] at (0,0) {$+$}; \node[below] at (1,0) {$-$}; \node[below] at (2,0) {$+$};
		\node[above] at (0,2) {$-$}; \node[above] at (1,2) {$+$}; \node[above] at (2,2) {$+$};
		\end{scope}
		\end{tikzpicture}
	\end{array}\oplus\\
	\oplus
	q\begin{array}{c}
		\begin{tikzpicture}
		\begin{scope}[scale=0.6]
		\draw (1,1.5) -- (2,1.5);
		\filldraw[white] (1.5,1.5) circle (0.1);
		\draw (1.5,1.5) circle (0.1);
		\begin{scope}[shift={(2,0)}]
		\begin{scope}[xscale=-1]
		\begin{scope}[shift={(0,-1)}]
		\draw[ultra thick, purple] (0,1) -- (0,1.5) to[out=90,in=210] (0.5,2) to[out=30,in=270] (1,2.5) -- (1,3) (0,3) -- (0,2.5) to[out=270,in=150] (0.4,2.1) (0.6,1.9) to[out=330,in=90] (1,1.5) -- (1,1) (0.75,1) -- (2.25,1) (2,1) -- (2,3);
		\end{scope}
		\end{scope}
		\end{scope}
		\node[below] at (0,0) {$+$}; \node[below] at (1,0) {$-$}; \node[below] at (2,0) {$+$};
		\node[above] at (0,2) {$+$}; \node[above] at (1,2) {$-$}; \node[above] at (2,2) {$+$};
		\end{scope}
		\end{tikzpicture}
	\end{array} \oplus
	q^{-1}\begin{array}{c}
		\begin{tikzpicture}
		\begin{scope}[scale=0.6]
		\draw (1,1.5) -- (2,1.5) (1,1.6) -- (2,1.6);
		\begin{scope}[shift={(2,0)}]
		\begin{scope}[xscale=-1]
		\begin{scope}[shift={(0,-1)}]
		\draw[ultra thick, purple] (0,1) -- (0,1.5) to[out=90,in=210] (0.5,2) to[out=30,in=270] (1,2.5) -- (1,3) (0,3) -- (0,2.5) to[out=270,in=150] (0.4,2.1) (0.6,1.9) to[out=330,in=90] (1,1.5) -- (1,1) (0.75,1) -- (2.25,1) (2,1) -- (2,3);
		\end{scope}
		\end{scope}
		\end{scope}
		\node[below] at (0,0) {$+$}; \node[below] at (1,0) {$-$}; \node[below] at (2,0) {$+$};
		\node[above] at (0,2) {$+$}; \node[above] at (1,2) {$-$}; \node[above] at (2,2) {$+$};
		\end{scope}
		\end{tikzpicture}
	\end{array} \oplus
	\begin{array}{c}
		\begin{tikzpicture}
		\begin{scope}[scale=0.6]
		\begin{scope}[shift={(2,0)}]
		\begin{scope}[xscale=-1]
		\begin{scope}[shift={(0,-1)}]
		\draw[ultra thick, purple] (0,1) -- (0,1.5) to[out=90,in=210] (0.5,2) to[out=30,in=270] (1,2.5) -- (1,3) (0,3) -- (0,2.5) to[out=270,in=150] (0.4,2.1) (0.6,1.9) to[out=330,in=90] (1,1.5) -- (1,1) (0.75,1) -- (2.25,1) (2,1) -- (2,3);
		\end{scope}
		\end{scope}
		\end{scope}
		\node[below] at (0,0) {$+$}; \node[below] at (1,0) {$-$}; \node[below] at (2,0) {$+$};
		\node[above] at (0,2) {$+$}; \node[above] at (1,2) {$+$}; \node[above] at (2,2) {$-$};
		\end{scope}
		\end{tikzpicture}
	\end{array}
\end{split}
\ee

The $U$-map for the first, second, third and sixth terms is defined by the obvious hovering solution, and the $\sigma\zeta$-instanton defining the map on the fourth and fifth complex involves a trivalent vertex in $(\sigma,x)$-space.

%
%

\subsection{Reidemiester I}\label{subsec:R1}
Now we consider the Reidemeister move I:
\be\label{ReidI}
\begin{split}
\cup_{12}\boxtimes \CR^{-1}_{23}\boxtimes \cap_{12}\sim \left[q^{-\frac{3}{2}}t\right]{\cal I}_{33}\\
\begin{array}{c}
\begin{tikzpicture}
\draw[ultra thick] (0.5,-0.5) -- (-0.5,0.5) to[out=135,in=225] (-0.5,-0.5) -- (-0.1,-0.1) (0.1,0.1) -- (0.5,0.5);
\end{tikzpicture}
\end{array}\sim\left[q^{-\frac{3}{2}}t\right] \begin{array}{c}
\begin{tikzpicture}
\draw[ultra thick] (0,0) -- (0,1);
\end{tikzpicture}
\end{array}
\end{split}
\ee

Let us denote the LHS as $\mathscr{C}$ and expand it in terms of Chan-Paton data:
\be
\CE_{ij}\left(\mathscr{C}\right)=
q^{\frac{1}{2}}\left[
\begin{array}{c}
\begin{tikzpicture}
\begin{scope}[scale=0.8]
\draw (0,0.25) -- (1,0.25);
\filldraw[white] (0.5,0.25) circle (0.1);
\draw (0.5,0.25) circle (0.1);
\draw[purple, ultra thick] (-0.25,0) -- (1.25,0) (-0.25,2) -- (1.25,2) (0,0) -- (0,2) (1,2) -- (1,1.5) to[out=270,in=150] (1.5,1) to[out=330,in=90] (2,0.5) -- (2,0) (1,0) -- (1,0.5) to[out=90,in=210] (1.4,0.9) (1.6,1.1) to[out=30,in=270] (2,1.5) -- (2,2);
\node[below] at (0,0) {$+$}; \node[below] at (1,0) {$-$}; \node[below] at (2,0) {$+$};
\node[above] at (0,2) {$-$}; \node[above] at (1,2) {$+$}; \node[above] at (2,2) {$+$};
\node[left] at (-0.1,0.25) {$x_{\cup}^{s}$};
\end{scope}
\end{tikzpicture}
\end{array}\oplus \begin{array}{c}
\begin{tikzpicture}
\begin{scope}[scale=0.8]
\draw (0,1.7) -- (1,1.7) (0,1.8) -- (1,1.8) (1,1.5) -- (2,1.5);
\filldraw (0.5,1.75) circle (0.1);
\filldraw[white] (1.5,1.5) circle (0.1);
\draw (1.5,1.5) circle (0.1);
\draw[purple, ultra thick] (-0.25,0) -- (1.25,0) (-0.25,2) -- (1.25,2) (0,0) -- (0,2) (1,2) -- (1,1.5) to[out=270,in=150] (1.5,1) to[out=330,in=90] (2,0.5) -- (2,0) (1,0) -- (1,0.5) to[out=90,in=210] (1.4,0.9) (1.6,1.1) to[out=30,in=270] (2,1.5) -- (2,2);
\node[below] at (0,0) {$+$}; \node[below] at (1,0) {$-$}; \node[below] at (2,0) {$+$};
\node[above] at (0,2) {$-$}; \node[above] at (1,2) {$+$}; \node[above] at (2,2) {$+$};
\node[left] at (-0.1,1.75) {$x_{\cap}^{s}$};
\node[right] at (2,1.5) {$x_1$};
\end{scope}
\end{tikzpicture}
\end{array}
\right]\oplus \\ \oplus q^{-\frac{3}{2}}\left[
\begin{array}{c}
	\begin{tikzpicture}
	\begin{scope}[scale=0.8]
	\filldraw (0.5,1.75) circle (0.1);
	\draw (0,1.7) -- (1,1.7) (0,1.8) -- (1,1.8) (1,1.5) -- (2,1.5) (1,1.4) -- (2,1.4);
	\draw[purple, ultra thick] (-0.25,0) -- (1.25,0) (-0.25,2) -- (1.25,2) (0,0) -- (0,2) (1,2) -- (1,1.5) to[out=270,in=150] (1.5,1) to[out=330,in=90] (2,0.5) -- (2,0) (1,0) -- (1,0.5) to[out=90,in=210] (1.4,0.9) (1.6,1.1) to[out=30,in=270] (2,1.5) -- (2,2);
	\node[below] at (0,0) {$+$}; \node[below] at (1,0) {$-$}; \node[below] at (2,0) {$+$};
	\node[above] at (0,2) {$-$}; \node[above] at (1,2) {$+$}; \node[above] at (2,2) {$+$};
	\node[left] at (-0.1,1.75) {$x_{\cap}^{s}$};
	\node[right] at (2,1.4) {$x_2$};
	\end{scope}
	\end{tikzpicture}
\end{array}\oplus
\begin{array}{c}
	\begin{tikzpicture}
	\begin{scope}[scale=0.8]
	\filldraw (0.5,1.75) circle (0.1);
	\draw (0,1.7) -- (1,1.7) (0,1.8) -- (1,1.8);
	\draw[purple, ultra thick] (-0.25,0) -- (1.25,0) (-0.25,2) -- (1.25,2) (0,0) -- (0,2) (1,2) -- (1,1.5) to[out=270,in=150] (1.5,1) to[out=330,in=90] (2,0.5) -- (2,0) (1,0) -- (1,0.5) to[out=90,in=210] (1.4,0.9) (1.6,1.1) to[out=30,in=270] (2,1.5) -- (2,2);
	\node[below] at (0,0) {$+$}; \node[below] at (1,0) {$-$}; \node[below] at (2,0) {$-$};
	\node[above] at (0,2) {$-$}; \node[above] at (1,2) {$+$}; \node[above] at (2,2) {$-$};
	\node[left] at (-0.1,1.75) {$x_{\cap}^{s}$};
	\end{scope}
	\end{tikzpicture}
\end{array}
\right]
\ee

First we use homotopy and concentrate on the part of $\bf P$-degree $\frac{1}{2}$. We start to shrink the loop homotopically. During this action the binding point with coordinate $x_{\cap}^s$ approaches $x_1$. Suppose they meet at some homotopy time $\sigma^*$ at the value $x_0$. Notice that we can define a position of the binding point as a smooth evolution of a root to equation:
\be
{\rm Im}\left[\zeta^{-1}Z(x)\right]=0
\ee
Note further that
\be
{\rm Im}\left[\zeta^{-1}Z_{\cup}^{s}(x_0)\right]={\rm Im}\left[\zeta^{-1}Z_{\cap}^{s}(x_0)\right]+{\rm Im}\left[\zeta^{-1}Z_1(x_0)\right]=0
\ee
Therefore we conclude $x_{\cup}^s(\sigma^*)=x_{\cap}^s(\sigma^*)=x_1(\sigma^*)=x_0$. We can plot corresponding diagrams showing the
 evolution of the binding points with the homotopy flow:
\be\label{eq:hmtpy_ReidI}
\begin{array}{c}
\begin{tikzpicture}
\begin{scope}[scale=0.8]
\draw[<->] (0,2.5) -- (0,0) -- (4,0);
\draw[dashed] (0,1.5) -- (1.5,1.5) (1.5,1.5) -- (1.5,0);
\node[left] at (0,1.5) {$\sigma^*$};
\node[above] at (0,2.5) {$\sigma$}; \node[right] at (4,0) {$x$};
\node[below] at (1.5,0) {$x_0$}; \node[below] at (0.5,-0.1) {$x_{\cup}^s$};
\draw[ultra thick] (0.5,0) to[out=90,in=190] (1.5,1.5);
\begin{scope}[shift={(0.5,0)}]
\filldraw[red] (-0.1,0.1) -- (0.1,0.1) -- (0.1,-0.1) -- (-0.1,-0.1) -- (-0.1,0.1);
\end{scope}
\begin{scope}[shift={(1.5,1.5)}]
\filldraw (-0.1,0.1) -- (0.1,0.1) -- (0.1,-0.1) -- (-0.1,-0.1) -- (-0.1,0.1);
\end{scope}
\node[below] at (2,-1) {(a)};
\end{scope}
\end{tikzpicture}
\end{array}\;
\begin{array}{c}
	\begin{tikzpicture}
	\begin{scope}[scale=0.8]
	\draw[<->] (0,2.5) -- (0,0) -- (4,0);
	\draw[dashed] (0,1.5) -- (1.5,1.5) (1.5,1.5) -- (1.5,0);
	\node[left] at (0,1.5) {$\sigma^*$};
	\node[above] at (0,2.5) {$\sigma$}; \node[right] at (4,0) {$x$};
	\node[below] at (1.5,0) {$x_0$};
	\node[below] at (2,-0.1) {$x_1$}; \node[below] at (3,-0.1) {$x_{\cap}^s$};
	\draw[ultra thick] (2,0) to[out=90,in=270] (1.5,1.5) (3,0) to[out=90,in=350] (1.5,1.5);
	\begin{scope}[shift={(2,0)}]
	\filldraw[red] (-0.1,0.1) -- (0.1,0.1) -- (0.1,-0.1) -- (-0.1,-0.1) -- (-0.1,0.1);
	\end{scope}
	\begin{scope}[shift={(1.5,1.5)}]
	\filldraw (-0.1,0.1) -- (0.1,0.1) -- (0.1,-0.1) -- (-0.1,-0.1) -- (-0.1,0.1);
	\end{scope}
	\begin{scope}[shift={(3,0)}]
	\filldraw[red] (-0.1,0.1) -- (0.1,0.1) -- (0.1,-0.1) -- (-0.1,-0.1) -- (-0.1,0.1);
	\end{scope}
	\node[below] at (2,-1) {(b)};
	\end{scope}
	\end{tikzpicture}
\end{array}\;
\begin{array}{c}
	\begin{tikzpicture}
	\begin{scope}[scale=0.8]
	\draw[<->] (0,2.5) -- (0,0) -- (4,0);
	\draw[dashed] (0,1.5) -- (1.5,1.5) (1.5,1.5) -- (1.5,0);
	\node[left] at (0,1.5) {$\sigma^*$};
	\node[above] at (0,2.5) {$\sigma$}; \node[right] at (4,0) {$x$};
	\node[below] at (1.5,0) {$x_0$};
	\node[below] at (2,-0.1) {$x_1$}; \node[below] at (3,-0.1) {$x_{\cap}^s$};
	\draw[ultra thick] (2,0) to[out=90,in=270] (1.5,1.5) (3,0) to[out=90,in=350] (1.5,1.5);
	\node[below] at (0.5,-0.1) {$x_{\cup}^s$};
	\draw[ultra thick] (0.5,0) to[out=90,in=190] (1.5,1.5);
	\begin{scope}[shift={(2,0)}]
	\filldraw[red] (-0.1,0.1) -- (0.1,0.1) -- (0.1,-0.1) -- (-0.1,-0.1) -- (-0.1,0.1);
	\end{scope}
	\begin{scope}[shift={(1.5,1.5)}]
	\filldraw (-0.1,0.1) -- (0.1,0.1) -- (0.1,-0.1) -- (-0.1,-0.1) -- (-0.1,0.1);
	\end{scope}
	\begin{scope}[shift={(3,0)}]
	\filldraw[red] (-0.1,0.1) -- (0.1,0.1) -- (0.1,-0.1) -- (-0.1,-0.1) -- (-0.1,0.1);
	\end{scope}
	\begin{scope}[shift={(0.5,0)}]
	\filldraw[red] (-0.1,0.1) -- (0.1,0.1) -- (0.1,-0.1) -- (-0.1,-0.1) -- (-0.1,0.1);
	\end{scope}	
	\node[below] at (2,-1) {(c)};
	\end{scope}
	\end{tikzpicture}
\end{array}
\ee
In   plot (a) we depicted the evolution of the first critical point; in   plot (b) we depicted evolution of the second critical point;
in plot (c) we depicted both plots (a) and (b) in the same   $(x,\sigma)$-plane. As we see these two critical points merge through a 3-valent interior amplitude vertex. So far we have shown that the term of $\bf P$-degree $\frac{1}{2}$ is quasi-isomorphic to a null complex.

For terms of $\bf P$-degree $-\frac{3}{2}$ we will show that they are isomorphic to the terms of the identity interface. Here we apply the same \FINDANAME argument as we used in the Reidemeister 0a move since we compare generically non-homotopic theories with different number of strands. First we notice that insertion of LHS interface or RHS interface does not change the number of generators of the complex, all the $\bf P$-degrees are shifted by $-\frac{3}{2}$ and $\bf F$-numbers are shifted by $+1$ if we switch from left to right. Also we should demonstrate that switching between the sides of the relation does not change $\zeta$-instanton solutions. For the identity interface $\CI_{33}$ on the RHS a typical instanton solution is illustrated in \eqref{eq:identity} above. Under $U$ the corresponding $\zeta$-instanton contributing to $\mathscr{C}$ has the following form:
\be
\begin{array}{c}
	\begin{tikzpicture}
	\begin{scope}[scale=0.8]
	\draw[ultra thick, purple, dashed] (3.4,0) -- (3.4,2);
	\draw[->] (2,0.25) -- (3.4,0.25);
	\node[above] at (2.7,0.25) {$\CX(\tau_0)$};
	\filldraw (0.5,1.75) circle (0.1);
	\draw (0,1.7) -- (1,1.7) (0,1.8) -- (1,1.8);
	\draw[purple, ultra thick] (-0.25,0) -- (1.25,0) (-0.25,2) -- (1.25,2) (0,0) -- (0,2) (1,2) -- (1,1.5) to[out=270,in=150] (1.5,1) to[out=330,in=90] (2,0.5) -- (2,0) (1,0) -- (1,0.5) to[out=90,in=210] (1.4,0.9) (1.6,1.1) to[out=30,in=270] (2,1.5) -- (2,2);
	\node[below] at (0,0) {$+$}; \node[below] at (1,0) {$-$}; \node[below] at (2,0) {$+$};
	\node[above] at (0,2) {$-$}; \node[above] at (1,2) {$+$}; \node[above] at (2,2) {$-$};
	\node[left] at (-0.1,1.75) {$x_{\cap}^{s}$};
	\end{scope}
	\end{tikzpicture}
\end{array}\mathop{\longrightarrow}^{\CQ_{\zeta}}
\begin{array}{c}
	\begin{tikzpicture}
	\begin{scope}[scale=0.8]
	\draw[ultra thick, purple, dashed] (3.4,0) -- (3.4,2);
	\draw[->] (2,1.75) -- (3.4,1.75);
	\node[below] at (2.7,1.75) {$\CX(\tau_1)$};
	\filldraw (0.5,1.75) circle (0.1);
	\draw (0,1.7) -- (1,1.7) (0,1.8) -- (1,1.8) (1,1.5) -- (2,1.5) (1,1.4) -- (2,1.4);
	\draw[purple, ultra thick] (-0.25,0) -- (1.25,0) (-0.25,2) -- (1.25,2) (0,0) -- (0,2) (1,2) -- (1,1.5) to[out=270,in=150] (1.5,1) to[out=330,in=90] (2,0.5) -- (2,0) (1,0) -- (1,0.5) to[out=90,in=210] (1.4,0.9) (1.6,1.1) to[out=30,in=270] (2,1.5) -- (2,2);
	\node[below] at (0,0) {$+$}; \node[below] at (1,0) {$-$}; \node[below] at (2,0) {$+$};
	\node[above] at (0,2) {$-$}; \node[above] at (1,2) {$+$}; \node[above] at (2,2) {$-$};
	\node[left] at (-0.1,1.75) {$x_{\cap}^{s}$};
	\node[above] at (1.5,1.5) {$x_2$};
	\end{scope}
	\end{tikzpicture}
\end{array}
\qquad \begin{array}{c}
\begin{tikzpicture}
\begin{scope}[scale=0.8]
\draw[<->] (0,3) -- (0,0) -- (3.5,0);
\node[right] at (3.5,0) {$x$}; \node[above] at (0,3) {$\tau$};
\draw (0,2.5) -- (3.5,2.5);
\draw[ultra thick, ->] (0.5,1) -- (2.75,1.75);
\draw[ultra thick] (2,0) -- (2,2.5) (1,2.5) -- (1,1.16);
\begin{scope}[shift={(2,0)}]
\filldraw[red] (-0.1,0.1) -- (0.1,0.1) -- (0.1,-0.1) -- (-0.1,-0.1) -- (-0.1,0.1);
\end{scope}
\begin{scope}[shift={(2,2.5)}]
\filldraw[red] (-0.1,0.1) -- (0.1,0.1) -- (0.1,-0.1) -- (-0.1,-0.1) -- (-0.1,0.1);
\end{scope}
\begin{scope}[shift={(1,2.5)}]
\filldraw[red] (-0.1,0.1) -- (0.1,0.1) -- (0.1,-0.1) -- (-0.1,-0.1) -- (-0.1,0.1);
\end{scope}
\begin{scope}[shift={(1,1.16)}]
\filldraw (-0.1,0.1) -- (0.1,0.1) -- (0.1,-0.1) -- (-0.1,-0.1) -- (-0.1,0.1);
\end{scope}
\node[below] at (0.7,1) {$\CX(\tau_0)$}; \node[right] at (2.75,1.75) {$\CX(\tau_1)$};
\node[above] at (1,2.6) {$x_2$}; \node[above] at (2,2.6) {$x_{\cap}^s$};
 \node[below] at (2,-0.1) {$x_{\cap}^s$};
\end{scope}
\end{tikzpicture}
\end{array}
\ee
Notice that the vertical trajectory located at $x_{\cap}^s$ intersects the trajectory of the boosted soliton without junction because they correspond to different strands.

Naively, there are more Reidemeister type I moves: we can pair $\CR$- or $\CR^{-1}$-interface with caps or cups. However we can connect other versions of this relation via other types of Reidemeister moves, for example,
\be
\begin{array}{c}
\begin{tikzpicture}
\draw[ultra thick] (-0.5,-0.5) -- (0.5,0.5) to[out=45,in=135] (-0.5,0.5) -- (-0.1,0.1) (0.1,-0.1) -- (0.5,-0.5);
\end{tikzpicture}
\end{array}\mathop{\sim}^{\eqref{ex_move_a}}\begin{array}{c}
\begin{tikzpicture}
\draw[ultra thick] (-0.5,-0.5) -- (0.5,0.5) to[out=45,in=135] (-0.5,0.5) -- (-0.1,0.1) (0.1,-0.1) -- (0.5,-0.5) to[out=315,in=90] (1.25,-0.3) -- (1.25,-0.7);
\end{tikzpicture}
\end{array}\mathop{\sim}^{\eqref{ex_move_b}}\begin{array}{c}
\begin{tikzpicture}
\draw[ultra thick] (0,0) to[out=45,in=315] (0.5,1) -- (0,1.5) to[out=135,in=225] (0,0.5) -- (0.4,0.9) (0.6,1.1) to[out=45,in=90] (1.25,1) -- (1.25,0);
\end{tikzpicture}
\end{array}\mathop{\sim}^{\eqref{ReidI}}  \left[q^{-\frac{3}{2}}t\right]\begin{array}{c}
\begin{tikzpicture}
\draw[ultra thick] (0,0) -- (0,0.5) to[out=90,in=90] (0.7,0.5) -- (0.7,0);
\end{tikzpicture}
\end{array}
\ee

\subsection{Reidemiester II}\label{subsec:R2}

The second Reidemeister move may be depicted as:

\be
\begin{split}
	{\cal R}\boxtimes {\cal R}^{-1}\sim {\cal I}\\
\begin{array}{c}
	\begin{tikzpicture}
	\begin{scope}[scale=0.8]
	\draw[ultra thick] (0,0) to[out=90,in=210] (0.5,0.5) to[out=30,in=270] (1,1) (1,0) to [out=90,in=330] (0.6,0.4) (0.4,0.6) to[out=150,in=270] (0,1);
	\begin{scope}[xscale=-1,shift={(-1,1)}]
	\draw[ultra thick] (0,0) to[out=90,in=210] (0.5,0.5) to[out=30,in=270] (1,1) (1,0) to [out=90,in=330] (0.6,0.4) (0.4,0.6) to[out=150,in=270] (0,1);
	\end{scope}
	\end{scope}
	\end{tikzpicture}
\end{array}\sim\begin{array}{c}
\begin{tikzpicture}
\begin{scope}[scale=0.8]
\draw[ultra thick] (0,0) to[out=70,in=290] (0,2) (1,0) to[out=110,in=250] (1,2);
\end{scope}
\end{tikzpicture}
\end{array}
\end{split}
\ee

The LHS gives the following contributions to the MSW complex:
\be\label{eq:RII-complex}
\begin{split}
	\IM\left(
	\begin{array}{c}
		\begin{tikzpicture}
		\begin{scope}[scale=0.8]
		\draw[ultra thick] (0,0) to[out=90,in=210] (0.5,0.5) to[out=30,in=270] (1,1) (1,0) to [out=90,in=330] (0.6,0.4) (0.4,0.6) to[out=150,in=270] (0,1);
		\begin{scope}[xscale=-1,shift={(-1,1)}]
		\draw[ultra thick] (0,0) to[out=90,in=210] (0.5,0.5) to[out=30,in=270] (1,1) (1,0) to [out=90,in=330] (0.6,0.4) (0.4,0.6) to[out=150,in=270] (0,1);
		\end{scope}
		\end{scope}
		\end{tikzpicture}
	\end{array}\right)=
	\begin{array}{c}
		\begin{tikzpicture}
		\begin{scope}[scale=0.8]
		\draw[ultra thick,purple] (0,-0.5) -- (0,0) to[out=90,in=210] (0.5,0.5) to[out=30,in=270] (1,1) (1,-0.5) -- (1,0) to [out=90,in=330] (0.6,0.4) (0.4,0.6) to[out=150,in=270] (0,1);
		\begin{scope}[yscale=-1,shift={(0,-2.5)}]
		\draw[ultra thick,purple] (0,-0.5) -- (0,0) to[out=90,in=210] (0.5,0.5) to[out=30,in=270] (1,1) (1,-0.5) -- (1,0) to [out=90,in=330] (0.6,0.4) (0.4,0.6) to[out=150,in=270] (0,1);
		\end{scope}
		\node[below] at (0,-0.5) {$-$};\node[below] at (1,-0.5) {$-$};
		\node at (0,1.25) {$-$};\node at (1,1.25) {$-$};
		\node[above] at (0,3) {$-$};\node[above] at (1,3) {$-$};
		\end{scope}
		\end{tikzpicture}
	\end{array}\oplus
	\begin{array}{c}
		\begin{tikzpicture}
		\begin{scope}[scale=0.8]
		\draw[ultra thick,purple] (0,-0.5) -- (0,0) to[out=90,in=210] (0.5,0.5) to[out=30,in=270] (1,1) (1,-0.5) -- (1,0) to [out=90,in=330] (0.6,0.4) (0.4,0.6) to[out=150,in=270] (0,1);
		\begin{scope}[yscale=-1,shift={(0,-2.5)}]
		\draw[ultra thick,purple] (0,-0.5) -- (0,0) to[out=90,in=210] (0.5,0.5) to[out=30,in=270] (1,1) (1,-0.5) -- (1,0) to [out=90,in=330] (0.6,0.4) (0.4,0.6) to[out=150,in=270] (0,1);
		\end{scope}
		\node[below] at (0,-0.5) {$+$};\node[below] at (1,-0.5) {$+$};
		\node at (0,1.25) {$+$};\node at (1,1.25) {$+$};
		\node[above] at (0,3) {$+$};\node[above] at (1,3) {$+$};
		\end{scope}
		\end{tikzpicture}
	\end{array}\oplus
	\begin{array}{c}
		\begin{tikzpicture}
		\begin{scope}[scale=0.8]
		\draw[ultra thick,purple] (0,-0.5) -- (0,0) to[out=90,in=210] (0.5,0.5) to[out=30,in=270] (1,1) (1,-0.5) -- (1,0) to [out=90,in=330] (0.6,0.4) (0.4,0.6) to[out=150,in=270] (0,1);
		\begin{scope}[yscale=-1,shift={(0,-2.5)}]
		\draw[ultra thick,purple] (0,-0.5) -- (0,0) to[out=90,in=210] (0.5,0.5) to[out=30,in=270] (1,1) (1,-0.5) -- (1,0) to [out=90,in=330] (0.6,0.4) (0.4,0.6) to[out=150,in=270] (0,1);
		\end{scope}
		\node[below] at (0,-0.5) {$-$};\node[below] at (1,-0.5) {$+$};
		\node at (0,1.25) {$+$};\node at (1,1.25) {$-$};
		\node[above] at (0,3) {$-$};\node[above] at (1,3) {$+$};
		\end{scope}
		\end{tikzpicture}
	\end{array}\oplus
	\begin{array}{c}
		\begin{tikzpicture}
		\begin{scope}[scale=0.8]
		\draw[ultra thick,purple] (0,-0.5) -- (0,0) to[out=90,in=210] (0.5,0.5) to[out=30,in=270] (1,1) (1,-0.5) -- (1,0) to [out=90,in=330] (0.6,0.4) (0.4,0.6) to[out=150,in=270] (0,1);
		\begin{scope}[yscale=-1,shift={(0,-2.5)}]
		\draw[ultra thick,purple] (0,-0.5) -- (0,0) to[out=90,in=210] (0.5,0.5) to[out=30,in=270] (1,1) (1,-0.5) -- (1,0) to [out=90,in=330] (0.6,0.4) (0.4,0.6) to[out=150,in=270] (0,1);
		\end{scope}
		\node[below] at (0,-0.5) {$+$};\node[below] at (1,-0.5) {$-$};
		\node at (0,1.25) {$-$};\node at (1,1.25) {$+$};
		\node[above] at (0,3) {$+$};\node[above] at (1,3) {$-$};
		\end{scope}
		\end{tikzpicture}
	\end{array}\oplus\\
	\oplus q\left(\begin{array}{c}
		\begin{tikzpicture}
		\begin{scope}[scale=0.8]
		\draw (0,2.5) -- (1,2.5);
		\filldraw[white] (0.5,2.5) circle (0.1);
		\draw (0.5,2.5) circle (0.1);
		\draw[ultra thick,purple] (0,-0.5) -- (0,0) to[out=90,in=210] (0.5,0.5) to[out=30,in=270] (1,1) (1,-0.5) -- (1,0) to [out=90,in=330] (0.6,0.4) (0.4,0.6) to[out=150,in=270] (0,1);
		\begin{scope}[yscale=-1,shift={(0,-2.5)}]
		\draw[ultra thick,purple] (0,-0.5) -- (0,0) to[out=90,in=210] (0.5,0.5) to[out=30,in=270] (1,1) (1,-0.5) -- (1,0) to [out=90,in=330] (0.6,0.4) (0.4,0.6) to[out=150,in=270] (0,1);
		\end{scope}
		\node[below] at (0,-0.5) {$+$};\node[below] at (1,-0.5) {$-$};
		\node at (0,1.25) {$-$};\node at (1,1.25) {$+$};
		\node[above] at (0,3) {$+$};\node[above] at (1,3) {$-$};
		\end{scope}
		\end{tikzpicture}
	\end{array}\oplus \begin{array}{c}
	\begin{tikzpicture}
	\begin{scope}[scale=0.8]
	\draw (0,0) -- (1,0);
	\draw[ultra thick,purple] (0,-0.5) -- (0,0) to[out=90,in=210] (0.5,0.5) to[out=30,in=270] (1,1) (1,-0.5) -- (1,0) to [out=90,in=330] (0.6,0.4) (0.4,0.6) to[out=150,in=270] (0,1);
	\begin{scope}[yscale=-1,shift={(0,-2.5)}]
	\draw[ultra thick,purple] (0,-0.5) -- (0,0) to[out=90,in=210] (0.5,0.5) to[out=30,in=270] (1,1) (1,-0.5) -- (1,0) to [out=90,in=330] (0.6,0.4) (0.4,0.6) to[out=150,in=270] (0,1);
	\end{scope}
	\node[below] at (0,-0.5) {$+$};\node[below] at (1,-0.5) {$-$};
	\node at (0,1.25) {$+$};\node at (1,1.25) {$-$};
	\node[above] at (0,3) {$+$};\node[above] at (1,3) {$-$};
	\end{scope}
	\end{tikzpicture}
\end{array}\right) \oplus q^{-1}\left(\begin{array}{c}
\begin{tikzpicture}
\begin{scope}[scale=0.8]
\draw (0,2.5) -- (1,2.5) (0,2.6) -- (1,2.6);
\draw[ultra thick,purple] (0,-0.5) -- (0,0) to[out=90,in=210] (0.5,0.5) to[out=30,in=270] (1,1) (1,-0.5) -- (1,0) to [out=90,in=330] (0.6,0.4) (0.4,0.6) to[out=150,in=270] (0,1);
\begin{scope}[yscale=-1,shift={(0,-2.5)}]
\draw[ultra thick,purple] (0,-0.5) -- (0,0) to[out=90,in=210] (0.5,0.5) to[out=30,in=270] (1,1) (1,-0.5) -- (1,0) to [out=90,in=330] (0.6,0.4) (0.4,0.6) to[out=150,in=270] (0,1);
\end{scope}
\node[below] at (0,-0.5) {$+$};\node[below] at (1,-0.5) {$-$};
\node at (0,1.25) {$-$};\node at (1,1.25) {$+$};
\node[above] at (0,3) {$+$};\node[above] at (1,3) {$-$};
\end{scope}
\end{tikzpicture}
\end{array}\oplus \begin{array}{c}
\begin{tikzpicture}
\begin{scope}[scale=0.8]
\draw (0,0) -- (1,0) (0,-0.1) -- (1,-0.1);
\filldraw (0.5,-0.05) circle (0.1);
\draw[ultra thick,purple] (0,-0.5) -- (0,0) to[out=90,in=210] (0.5,0.5) to[out=30,in=270] (1,1) (1,-0.5) -- (1,0) to [out=90,in=330] (0.6,0.4) (0.4,0.6) to[out=150,in=270] (0,1);
\begin{scope}[yscale=-1,shift={(0,-2.5)}]
\draw[ultra thick,purple] (0,-0.5) -- (0,0) to[out=90,in=210] (0.5,0.5) to[out=30,in=270] (1,1) (1,-0.5) -- (1,0) to [out=90,in=330] (0.6,0.4) (0.4,0.6) to[out=150,in=270] (0,1);
\end{scope}
\node[below] at (0,-0.5) {$+$};\node[below] at (1,-0.5) {$-$};
\node at (0,1.25) {$+$};\node at (1,1.25) {$-$};
\node[above] at (0,3) {$+$};\node[above] at (1,3) {$-$};
\end{scope}
\end{tikzpicture}
\end{array}\right)
\end{split}
\ee
%
%

The first four terms on the RHS of \eqref{eq:RII-complex} do not contain binding points. Therefore there is a unique, rigid,
interpolation in $\sigma$, satisfying \eqref{Hmtpy}, and transforming one hovering soliton
solution to the other. These terms give us the desired identity interface.

\begin{figure}[h!]
	\begin{center}
		\includegraphics[scale=0.6]{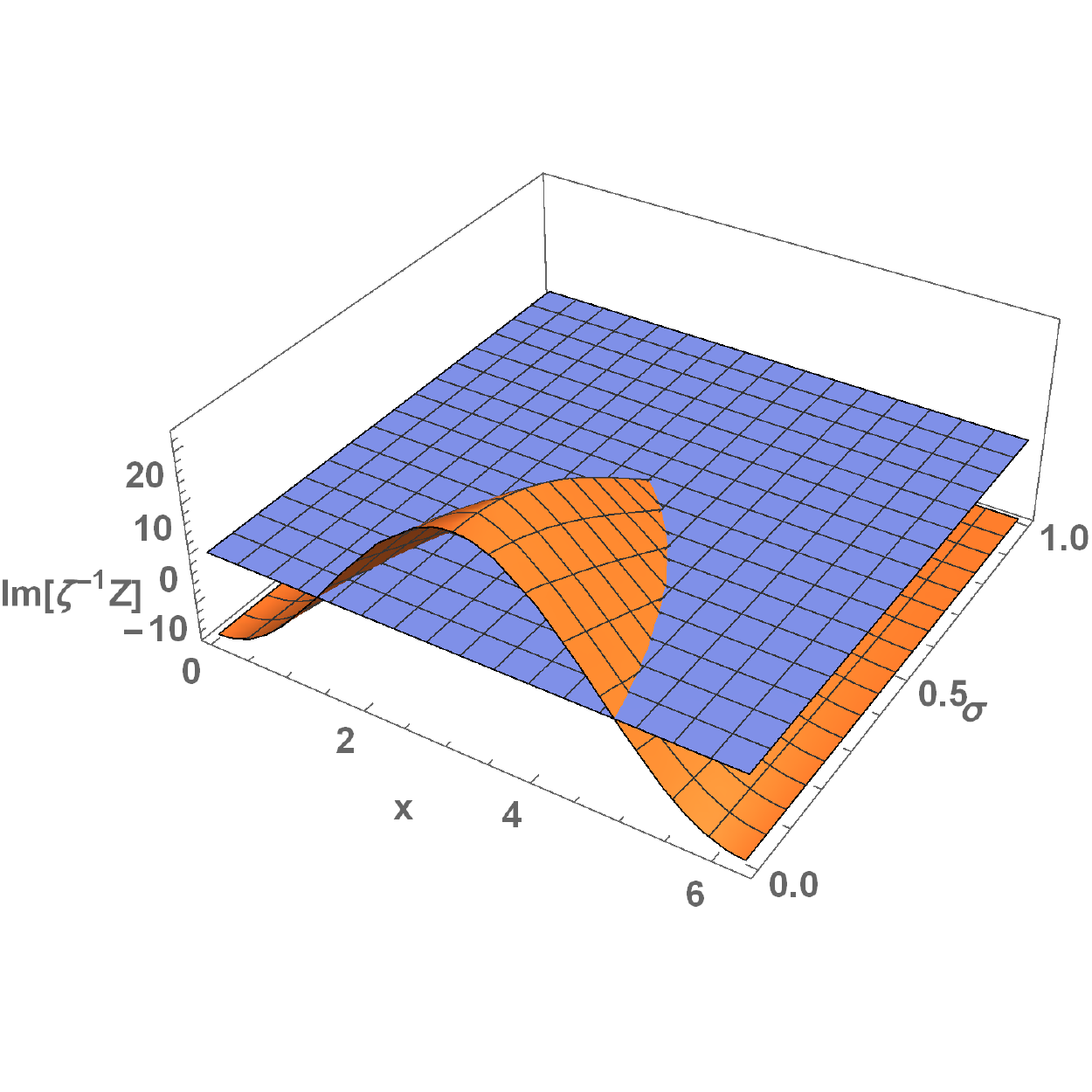}
	\end{center}
	\caption{For the homotopy describing the first Reidemeister move we can plot ${\rm Im}[\zeta^{-1}Z(\phi(x;\sigma))]$ for a forced $\zeta$-soliton.
The horizontal plane on this figure denotes the zero level.
The zeroes of ${\rm Im}[\zeta^{-1}Z]$ correspond to the binding points of the forced $\zeta$-soliton. Note that as $\sigma$ evolves these binding points merge and
annihilate. Thus, we must define $U=0$ on the complexes of \Pdeg-degree $=\pm 1$. Fortunately, these sub-complexes are indeed
quasi-isomorphic to zero.   }\label{fig:Reidemeister_2}
\end{figure}

The remaining four terms on the RHS of \eqref{eq:RII-complex} consist of two  subcomplexes $\IM_1$ and $\IM_{-1}$ of $\Pdeg$-degrees 1 and -1.
They are again quasi-isomorphic to zero.
When solving the equation \eqref{Hmtpy} the solitons associated with the two binding points in the fifth and sixth,
and (separately) in the seventh and eighth terms merge into each other and disappear at a critical
value of $\sigma$. Note that this is possible because the solitons associated with the merging pair of binding points
indeed are homotopic paths in the target space $X$. (In particular, they have the same central charge $Z=\I (W_i - W_j)$.)
As shown in Section 7.4.1 of \cite{Gaiotto:2015aoa} binding points are points where the ``vacuum weights'' $z_{ij}(x)$
become positive imaginary numbers, or alternatively $\zeta^{-1}Z_{ij}(x)$ are positive real numbers.
We can plot, numerically the value of ${\rm Im}[\zeta^{-1}Z]$ as a function of the homotopy parameter $\sigma$ and
see the merging of the binding points in Figure \ref{fig:Reidemeister_2}.

\subsection{Reidemiester III}\label{subsec:R3}

Finally, we come to the third Reidemeister move, depicted as:

\be
\begin{split}
 {\cal R}_{12}\boxtimes {\cal R}_{23}\boxtimes {\cal R}_{12}\sim {\cal R}_{23}\boxtimes {\cal R}_{12}\boxtimes {\cal R}_{23}\\
\begin{array}{c}
	\begin{tikzpicture}
	\begin{scope}[scale=0.4,yscale=-1]
	\draw [ultra thick] (3,-0.5) -- (-1,1.5);
	\draw [ultra thick] (0,-2) -- (0,0.8);
	\draw [ultra thick] (0,1.2) -- (0,2);
	\draw [ultra thick] (-1,-1.5) -- (-0.2,-1.1) (0.2,-0.9) -- (1.8,-0.1);
	\draw [ultra thick] (2.2,0.1) -- (3,0.5);
	\end{scope}
	\end{tikzpicture}
\end{array}\sim
\begin{array}{c}
	\begin{tikzpicture}
	\begin{scope}[scale=0.4,yscale=-1]
	\draw [ultra thick] (1,-1.5) -- (-3,0.5);
	\draw [ultra thick] (0,-0.8) -- (0,2);
	\draw [ultra thick] (0,-1.2) -- (0,-2);
	\draw [ultra thick] (0.2,1.1) -- (1,1.5);
	\draw [ultra thick] (-0.2,0.9) -- (-1.8,0.1);
	\draw [ultra thick] (-2.2,-0.1) -- (-3,-0.5);
	\end{scope}
	\end{tikzpicture}
\end{array}
\end{split}
\ee\label{YB}

The composition of the braiding interfaces leads to many possible boundary conditions, with $8$ possible boundary
conditions at $x=-\infty$ and $8$ at $x=+\infty$. Choosing an ordering on $z_{a_1}, z_{a_2}, z_{a_3}$  we can
denote  $U: \CE_{a,b,c}^{a',b',c'}(LHS) \to \CE_{a,b,c}^{a',b',c'}(RHS)$ by ${\bf Htpy}_{a,b,c}^{a',b',c'}$.

We start with ${\bf Htpy}_{---}^{---}$:
\be
q^{\frac{3}{2}}\begin{array}{c}
	\begin{tikzpicture}
	\begin{scope}[scale=0.7]
	\S;
	\begin{scope}[shift={(1,1.5)}]
	\S;
	\end{scope}
	\begin{scope}[shift={(0,3)}]
	\S;
	\end{scope}
	\draw[ultra thick, purple] (0,1.5) -- (0,2.5);
	\node[below] at (0,0) {$-$};\node[below] at (1,0) {$-$};
	\node at (0,1.25) {$-$}; \node at (1,1.25) {$-$};\node at (2,1.25) {$-$};
	\node at (0,2.75) {$-$}; \node at (1,2.75) {$-$};\node at (2,2.75) {$-$};
	\node[above] at (0,4) {$-$}; \node[above] at (1,4) {$-$};
	\end{scope}
	\end{tikzpicture}
\end{array}\longrightarrow q^{\frac{3}{2}} \begin{array}{c}
\begin{tikzpicture}
\begin{scope}[scale=0.7]
\S;
\begin{scope}[shift={(-1,1.5)}]
\S;
\end{scope}
\begin{scope}[shift={(0,3)}]
\S;
\end{scope}
\draw[ultra thick, purple] (1,1.5) -- (1,2.5);
\node[below] at (0,0) {$-$};\node[below] at (1,0) {$-$};
\node at (-1,1.25) {$-$}; \node at (0,1.25) {$-$};\node at (1,1.25) {$-$};
\node at (-1,2.75) {$-$}; \node at (0,2.75) {$-$};\node at (1,2.75) {$-$};
\node[above] at (0,4) {$-$}; \node[above] at (1,4) {$-$};
\end{scope}
\end{tikzpicture}
\end{array}
\ee
Note that ${\bf Htpy}_{---}^{---}$ has bi-degree zero, and the initial and final complexes
have forced $\zeta$-solitons with no binding points. Therefore under equation \eqref{Hmtpy}
the hovering forced $\zeta$-soliton on the LHS evolves smoothly into the hovering forced $\zeta$-soliton
on the RHS. The $U$-map is naturally an identity. The other cases   ${\bf Htpy}_{--+}^{+--}$, ${\bf Htpy}_{-+-}^{-+-}$, ${\bf Htpy}_{-++}^{++-}$, ${\bf Htpy}_{+--}^{--+}$, ${\bf Htpy}_{+-+}^{+-+}$, ${\bf Htpy}_{++-}^{-++}$ and ${\bf Htpy}_{+++}^{+++}$ are entirely similar.

Now we turn to the next type of case. This is exemplified by ${\bf Htpy}_{-+-}^{+--}$:
%
%
\be
q^{\frac{1}{2}}\begin{array}{c}
	\begin{tikzpicture}
	\begin{scope}[scale=0.7]
	\draw (0,3.1) -- (1,3.1);
	\S;
	\begin{scope}[shift={(1,1.5)}]
	\S;
	\end{scope}
	\begin{scope}[shift={(0,3)}]
	\S;
	\end{scope}
	\draw[ultra thick, purple] (0,1.5) -- (0,2.5);
	\node[below] at (0,0) {$-$};\node[below] at (1,0) {$+$};
	\node at (0,1.25) {$+$}; \node at (1,1.25) {$-$};\node at (2,1.25) {$-$};
	\node at (0,2.75) {$+$}; \node at (1,2.75) {$-$};\node at (2,2.75) {$-$};
	\node[above] at (0,4) {$+$}; \node[above] at (1,4) {$-$};
	\end{scope}
	\end{tikzpicture}
\end{array}\oplus\; q^{-\frac{3}{2}} \begin{array}{c}
\begin{tikzpicture}
\begin{scope}[scale=0.7]
\draw (0.1,3.2) -- (0.9,3.2) (0,3.1) -- (1,3.1);
\filldraw (0.5,3.15) circle (0.1);
\S;
\begin{scope}[shift={(1,1.5)}]
\S;
\end{scope}
\begin{scope}[shift={(0,3)}]
\S;
\end{scope}
\draw[ultra thick, purple] (0,1.5) -- (0,2.5);
\node[below] at (0,0) {$-$};\node[below] at (1,0) {$+$};
\node at (0,1.25) {$+$}; \node at (1,1.25) {$-$};\node at (2,1.25) {$-$};
\node at (0,2.75) {$+$}; \node at (1,2.75) {$-$};\node at (2,2.75) {$-$};
\node[above] at (0,4) {$+$}; \node[above] at (1,4) {$-$};
\end{scope}
\end{tikzpicture}
\end{array}\longrightarrow
q^{\frac{1}{2}} \begin{array}{c}
	\begin{tikzpicture}
	\begin{scope}[scale=0.7]
	\draw (0,0.1) -- (1,0.1);
	\S;
	\begin{scope}[shift={(-1,1.5)}]
	\S;
	\end{scope}
	\begin{scope}[shift={(0,3)}]
	\S;
	\end{scope}
	\draw[ultra thick, purple] (1,1.5) -- (1,2.5);
	\node[below] at (0,0) {$+$};\node[below] at (1,0) {$-$};
	\node at (-1,1.25) {$-$}; \node at (0,1.25) {$+$};\node at (1,1.25) {$-$};
	\node at (-1,2.75) {$+$}; \node at (0,2.75) {$-$};\node at (1,2.75) {$-$};
	\node[above] at (0,4) {$-$}; \node[above] at (1,4) {$-$};
	\end{scope}
	\end{tikzpicture}
\end{array}\oplus\;
q^{-\frac{3}{2}} \begin{array}{c}
	\begin{tikzpicture}
	\begin{scope}[scale=0.7]
	\draw (0,0.1) -- (1,0.1) (0.1,0.2) -- (0.9,0.2);
	\filldraw (0.5,0.15) circle (0.1);
	\S;
	\begin{scope}[shift={(-1,1.5)}]
	\S;
	\end{scope}
	\begin{scope}[shift={(0,3)}]
	\S;
	\end{scope}
	\draw[ultra thick, purple] (1,1.5) -- (1,2.5);
	\node[below] at (0,0) {$+$};\node[below] at (1,0) {$-$};
	\node at (-1,1.25) {$-$}; \node at (0,1.25) {$+$};\node at (1,1.25) {$-$};
	\node at (-1,2.75) {$+$}; \node at (0,2.75) {$-$};\node at (1,2.75) {$-$};
	\node[above] at (0,4) {$-$}; \node[above] at (1,4) {$-$};
	\end{scope}
	\end{tikzpicture}
\end{array}
\ee
This case is similar to the case without binding points. In this case the single binding point evolves smoothly as a function of $\sigma$ from a large
positive value of $x$ to a large negative value of $x$. Again there is a unique rigid $\sigma\zeta$-instanton.
The cases  ${\bf Htpy}_{+--}^{-+-}$, ${\bf Htpy}_{+-+}^{++-}$ and ${\bf Htpy}_{++-}^{+-+}$ work in the same way.

The final, and  most interesting, cases are  ${\bf Htpy}_{+--}^{+--}$ and ${\bf Htpy}_{++-}^{++-}$, which are again quite similar.
(They have the same number of generators and they transform into each other under homotopy by the same mechanisms.)
We will focus on  ${\bf Htpy}_{+--}^{+--}$.  The domain and range complexes for the $U$-map are now:
\be\nn
	LHS({\bf Htpy}_{+--}^{+--})=q^{\frac{3}{2}} \begin{array}{c}
		\begin{tikzpicture}
		\begin{scope}[scale=0.7]
		\draw[blue] (0,0.1) -- (1,0.1);
		\draw[red] (0,3.1) -- (1,3.1);
		\S;
		\begin{scope}[shift={(1,1.5)}]
		\S;
		\end{scope}
		\begin{scope}[shift={(0,3)}]
		\S;
		\end{scope}
		\draw[ultra thick, purple] (0,1.5) -- (0,2.5);
		\node[below] at (0,0) {$+$};\node[below] at (1,0) {$-$};
		\node at (0,1.25) {$+$}; \node at (1,1.25) {$-$};\node at (2,1.25) {$-$};
		\node at (0,2.75) {$+$}; \node at (1,2.75) {$-$};\node at (2,2.75) {$-$};
		\node[above] at (0,4) {$+$}; \node[above] at (1,4) {$-$};
		\end{scope}
		\end{tikzpicture}
	\end{array}\oplus q^{-\frac{5}{2}}\left(\begin{array}{c}
	\begin{tikzpicture}
	\begin{scope}[scale=0.7]
	\draw (1,1.6) -- (2,1.6) (1.1,1.7) -- (1.9,1.7);
	\filldraw (1.5,1.65) circle (0.1);
	\S;
	\begin{scope}[shift={(1,1.5)}]
	\S;
	\end{scope}
	\begin{scope}[shift={(0,3)}]
	\S;
	\end{scope}
	\draw[ultra thick, purple] (0,1.5) -- (0,2.5);
	\node[below] at (0,0) {$+$};\node[below] at (1,0) {$-$};
	\node at (0,1.25) {$-$}; \node at (1,1.25) {$+$};\node at (2,1.25) {$-$};
	\node at (0,2.75) {$-$}; \node at (1,2.75) {$+$};\node at (2,2.75) {$-$};
	\node[above] at (0,4) {$+$}; \node[above] at (1,4) {$-$};
	\end{scope}
	\end{tikzpicture}
\end{array}\oplus \begin{array}{c}
\begin{tikzpicture}
\begin{scope}[scale=0.7]
\draw (0,0.1) -- (1,0.1) (0.1,0.2) -- (0.9,0.2);
\filldraw (0.5,0.15) circle (0.1);
\draw (0,3.1) -- (1,3.1) (0.1,3.2) -- (0.9,3.2);
\filldraw (0.5,3.15) circle (0.1);
\S;
\begin{scope}[shift={(1,1.5)}]
\S;
\end{scope}
\begin{scope}[shift={(0,3)}]
\S;
\end{scope}
\draw[ultra thick, purple] (0,1.5) -- (0,2.5);
\node[below] at (0,0) {$+$};\node[below] at (1,0) {$-$};
\node at (0,1.25) {$+$}; \node at (1,1.25) {$-$};\node at (2,1.25) {$-$};
\node at (0,2.75) {$+$}; \node at (1,2.75) {$-$};\node at (2,2.75) {$-$};
\node[above] at (0,4) {$+$}; \node[above] at (1,4) {$-$};
\end{scope}
\end{tikzpicture}
\end{array}\right)\oplus\ee
\be
\oplus q^{-\frac{1}{2}}\left(\begin{array}{c}
	\begin{tikzpicture}
	\begin{scope}[scale=0.7]
	\draw[orange] (1,1.6) -- (2,1.6);
	\S;
	\begin{scope}[shift={(1,1.5)}]
	\S;
	\end{scope}
	\begin{scope}[shift={(0,3)}]
	\S;
	\end{scope}
	\draw[ultra thick, purple] (0,1.5) -- (0,2.5);
	\node[below] at (0,0) {$+$};\node[below] at (1,0) {$-$};
	\node at (0,1.25) {$-$}; \node at (1,1.25) {$+$};\node at (2,1.25) {$-$};
	\node at (0,2.75) {$-$}; \node at (1,2.75) {$+$};\node at (2,2.75) {$-$};
	\node[above] at (0,4) {$+$}; \node[above] at (1,4) {$-$};
	\end{scope}
	\end{tikzpicture}
\end{array}_{\Psi_1}\oplus \begin{array}{c}
\begin{tikzpicture}
\begin{scope}[scale=0.7]
\draw[blue] (0,0.1) -- (1,0.1) (0.1,0.2) -- (0.9,0.2);
\filldraw (0.5,0.15) circle (0.1);
\draw[red] (0,3.1) -- (1,3.1);
\S;
\begin{scope}[shift={(1,1.5)}]
\S;
\end{scope}
\begin{scope}[shift={(0,3)}]
\S;
\end{scope}
\draw[ultra thick, purple] (0,1.5) -- (0,2.5);
\node[below] at (0,0) {$+$};\node[below] at (1,0) {$-$};
\node at (0,1.25) {$+$}; \node at (1,1.25) {$-$};\node at (2,1.25) {$-$};
\node at (0,2.75) {$+$}; \node at (1,2.75) {$-$};\node at (2,2.75) {$-$};
\node[above] at (0,4) {$+$}; \node[above] at (1,4) {$-$};
\end{scope}
\end{tikzpicture}
\end{array}_{\Psi_2}\oplus \begin{array}{c}
\begin{tikzpicture}
\begin{scope}[scale=0.7]
\draw[black!40!green] (0,0.1) -- (1,0.1);
\draw[magenta] (0,3.1) -- (1,3.1) (0.1,3.2) -- (0.9,3.2);
\filldraw (0.5,3.15) circle (0.1);
\S;
\begin{scope}[shift={(1,1.5)}]
\S;
\end{scope}
\begin{scope}[shift={(0,3)}]
\S;
\end{scope}
\draw[ultra thick, purple] (0,1.5) -- (0,2.5);
\node[below] at (0,0) {$+$};\node[below] at (1,0) {$-$};
\node at (0,1.25) {$+$}; \node at (1,1.25) {$-$};\node at (2,1.25) {$-$};
\node at (0,2.75) {$+$}; \node at (1,2.75) {$-$};\node at (2,2.75) {$-$};
\node[above] at (0,4) {$+$}; \node[above] at (1,4) {$-$};
\end{scope}
\end{tikzpicture}
\end{array}_{\Psi_3}\right)\label{eq:RIII-Hard}
\ee
\be\nn
RHS({\bf Htpy}_{+--}^{+--})= q^{\frac{3}{2}} \begin{array}{c}
	\begin{tikzpicture}
	\begin{scope}[scale=0.7]
	\draw[gray] (-1,1.6) -- (0,1.6);
	\S;
	\begin{scope}[shift={(-1,1.5)}]
	\S;
	\end{scope}
	\begin{scope}[shift={(0,3)}]
	\S;
	\end{scope}
	\draw[ultra thick, purple] (1,1.5) -- (1,2.5);
	\node[below] at (0,0) {$-$};\node[below] at (1,0) {$-$};
	\node at (-1,1.25) {$+$}; \node at (0,1.25) {$-$};\node at (1,1.25) {$-$};
	\node at (-1,2.75) {$+$}; \node at (0,2.75) {$-$};\node at (1,2.75) {$-$};
	\node[above] at (0,4) {$-$}; \node[above] at (1,4) {$-$};
	\end{scope}
	\end{tikzpicture}
\end{array}\oplus q^{-\frac{1}{2}}\begin{array}{c}
\begin{tikzpicture}
\begin{scope}[scale=0.7]
\draw[gray] (-1,1.6) -- (0,1.6) (-1,1.7) -- (0,1.7);
\filldraw (-0.5,1.65) circle (0.1);
\S;
\begin{scope}[shift={(-1,1.5)}]
\S;
\end{scope}
\begin{scope}[shift={(0,3)}]
\S;
\end{scope}
\draw[ultra thick, purple] (1,1.5) -- (1,2.5);
\node[below] at (0,0) {$-$};\node[below] at (1,0) {$-$};
\node at (-1,1.25) {$+$}; \node at (0,1.25) {$-$};\node at (1,1.25) {$-$};
\node at (-1,2.75) {$+$}; \node at (0,2.75) {$-$};\node at (1,2.75) {$-$};
\node[above] at (0,4) {$-$}; \node[above] at (1,4) {$-$};
\end{scope}
\end{tikzpicture}
\end{array}_{\Psi_4}
\ee

The $\sigma\zeta$ instanton mapping the complexes of $\Pdeg$-degree $+\frac{3}{2}$ involve a curved web in $(\sigma, x)$ space that
makes use of a trivalent vertex:
\be\label{3-valent_ReidIII}
\begin{array}{c}
\begin{tikzpicture}
\draw[dashed] (0,1) -- (1.5,1);
\node[left] at (0,1) {$\sigma_0$};
\draw[<->] (0,2.5) -- (0,0) -- (3,0);
\node[below] at (3,0) {$x$};
\node[left] at (0,2.5) {$\sigma$};
\draw (0,2) -- (3,2);
\draw[ultra thick] (1,0) -- (1.5,1) -- (2,0) (1.5,1) -- (1.5,2);
\begin{scope}[shift={(1.5,1)}]
\filldraw (-0.1,0.1) -- (0.1,0.1) -- (0.1,-0.1) -- (-0.1,-0.1);
\end{scope}
\begin{scope}[shift={(1,0)}]
\filldraw[blue] (-0.1,0.1) -- (0.1,0.1) -- (0.1,-0.1) -- (-0.1,-0.1);
\end{scope}
\begin{scope}[shift={(2,0)}]
\filldraw[red] (-0.1,0.1) -- (0.1,0.1) -- (0.1,-0.1) -- (-0.1,-0.1);
\end{scope}
\begin{scope}[shift={(1.5,2)}]
\filldraw[gray] (-0.1,0.1) -- (0.1,0.1) -- (0.1,-0.1) -- (-0.1,-0.1);
\end{scope}
\end{tikzpicture}
\end{array}
\ee

The complex of  $\Pdeg$-degree $-\frac{5}{2}$ on the LHS of \eqref{eq:RIII-Hard} is mapped by $U$ to zero. The reason is that the evolution in
$\sigma$ given by \eqref{Hmtpy} makes use of a trivalent graph terminating the curved web at a critical value of $\sigma$ analogously to the case we have encountered when discussed the Reidemeister move I \eqref{eq:hmtpy_ReidI}.







%
%

Finally, the most nontrivial part comes from the map of subcomplexes of $\Pdeg$-degree $-\frac{1}{2}$.
Let us draw the homotopy classes of the soliton paths illustrated in the complexes with degree $q^{-1/2}$
in \eqref{eq:RIII-Hard}. Noting the color code of the soltions in \eqref{eq:RIII-Hard} we obtain the field configurations
(with all the solitons flowing from left to right):
$$
\begin{array}{c}
\begin{tikzpicture}
\draw[thick, black!40!green] (0,0) to[out=45,in=135] (2,0);
\draw[thick, blue] (0,0) to[out=315,in=225] (2,0);
\draw[thick, red] (2,0) to[out=45,in=135] (3.5,0);
\draw[thick, magenta] (2,0) to[out=315,in=225] (3.5,0);
\draw[thick,orange] (0,0) to [out=90,in=180] (1,0.7) to[out=0,in=180] (2.3,-0.7) to[out=0,in=270] (3.5,0);
\begin{scope}[yscale=-1]
\draw[thick,gray] (0,0) to [out=90,in=180] (1,0.7) to[out=0,in=180] (2.3,-0.7) to[out=0,in=270] (3.5,0);
\end{scope}
\draw (0.5,0) circle (0.1) (2.5,0) circle (0.1) (4,0) circle (0.1);
\begin{scope}
\draw[purple, ultra thick] (-0.1,-0.1) -- (0.1,0.1) (0.1,-0.1) -- (-0.1,0.1);
\end{scope}
\begin{scope}[shift={(2,0)}]
\draw[purple, ultra thick] (-0.1,-0.1) -- (0.1,0.1) (0.1,-0.1) -- (-0.1,0.1);
\end{scope}
\begin{scope}[shift={(3.5,0)}]
\draw[purple, ultra thick] (-0.1,-0.1) -- (0.1,0.1) (0.1,-0.1) -- (-0.1,0.1);
\end{scope}
\end{tikzpicture}
\end{array}
$$
Note that since we are considering  boundary conditions $+--$, with a single $+$, there is a single
``moving'' LG field $w_i$, illustrated above, while all other $w_j(x)$ are approximately constant.
To understand the topology of the  orange and grey curves it helps to push the soliton path in
\eqref{eq:RIII-Hard} to the bottom of the three strands.

Now we depict the migration of all the binding points with the homotopy time $\sigma$ taking us from the LHS to the RHS of \eqref{eq:RIII-Hard}:
\be
\begin{array}{c}
	\begin{tikzpicture}
	\begin{scope}[scale=0.6]
	\draw[<->] (0,3.5) -- (0,0) -- (5.5,0);
	\draw (0,3) -- (5,3);
	\node[right] at (5.5,0) {$x$};
	\node[left] at (0,3.5) {$\sigma$};
	\draw[dashed] (2,0) -- (2,3);
	\draw[ultra thick] (2,0) -- (2,1);
	\begin{scope}[shift={(2,0)}]
	\filldraw[orange] (-0.1,0.1) -- (0.1,0.1) -- (0.1,-0.1) -- (-0.1,-0.1);
	\end{scope}
	\begin{scope}[shift={(2,1)}]
	\filldraw[black] (-0.1,0.1) -- (0.1,0.1) -- (0.1,-0.1) -- (-0.1,-0.1);
	\end{scope}
	\node at (2,-0.5) {$\Psi_1$};
	\node[above right] at (2,1) {$1$};
	\end{scope}
	\end{tikzpicture}
\end{array}\;
\begin{array}{c}
	\begin{tikzpicture}
	\begin{scope}[scale=0.6]
	\draw[<->] (0,3.5) -- (0,0) -- (5.5,0);
	\draw (0,3) -- (5,3);
	\node[right] at (5.5,0) {$x$};
	\node[left] at (0,3.5) {$\sigma$};
	\draw[dashed] (2,0) -- (2,3);
	\draw[ultra thick] (1,0) to[out=90,in=190] (2,1) (3.5,0) to[out=90,in=350] (2,1);
	\begin{scope}[shift={(1,0)}]
	\filldraw[green] (-0.1,0.1) -- (0.1,0.1) -- (0.1,-0.1) -- (-0.1,-0.1);
	\end{scope}
	\begin{scope}[shift={(3.5,0)}]
	\filldraw[purple] (-0.1,0.1) -- (0.1,0.1) -- (0.1,-0.1) -- (-0.1,-0.1);
	\end{scope}
	\begin{scope}[shift={(2,1)}]
	\filldraw[black] (-0.1,0.1) -- (0.1,0.1) -- (0.1,-0.1) -- (-0.1,-0.1);
	\end{scope}
	\node at (2,-0.5) {$\Psi_3$};
	\node[above right] at (2,1) {$1$};
	\end{scope}
	\end{tikzpicture}
\end{array}\;
\begin{array}{c}
	\begin{tikzpicture}
	\begin{scope}[scale=0.6]
	\draw[<->] (0,3.5) -- (0,0) -- (5.5,0);
	\draw (0,3) -- (5,3);
	\node[right] at (5.5,0) {$x$};
	\node[left] at (0,3.5) {$\sigma$};
	\draw[dashed] (2,0) -- (2,3);
	\draw[ultra thick] (0.5,0) to[out=90,in=200] (2,2) (4.5,0) to[out=90,in=340] (2,2) (2,2) -- (2,3);
	\begin{scope}[shift={(0.5,0)}]
	\filldraw[blue] (-0.1,0.1) -- (0.1,0.1) -- (0.1,-0.1) -- (-0.1,-0.1);
	\end{scope}
	\begin{scope}[shift={(4.5,0)}]
	\filldraw[red] (-0.1,0.1) -- (0.1,0.1) -- (0.1,-0.1) -- (-0.1,-0.1);
	\end{scope}
	\begin{scope}[shift={(2,2)}]
	\filldraw[black] (-0.1,0.1) -- (0.1,0.1) -- (0.1,-0.1) -- (-0.1,-0.1);
	\end{scope}
	\begin{scope}[shift={(2,3)}]
	\filldraw[gray] (-0.1,0.1) -- (0.1,0.1) -- (0.1,-0.1) -- (-0.1,-0.1);
	\end{scope}
	\node at (2,-0.5) {$\Psi_2$};
	\node[above] at (2,3) {$\Psi_4$};
	\node[above right] at (2,2) {$2$};
	\end{scope}
	\end{tikzpicture}
\end{array}\label{eq:ReidIII_diag}
\ee

Note that the orange curve is the composition of the green and purple curves. Therefore the
 critical points $\Psi_1$ and $\Psi_3$ of $h$ merge into each other during the homotopy while the critical point $\Psi_2$ transforms to the point $\Psi_4$ via a 3-valent interior amplitude vertex. Moreover, the solitons marked by orange and gray binding points have \underline{identical}  central charges.
(To derive this, note that the difference of the central charges is the period of $dW$ with the Landau-Ginzburg field $w$ moving around the figure eight
described by the composition of the orange and (inverse) grey curves.  The winding around the second singularity cancels the winding around the first singularity.)
Since the central charges are identical the curved webs will be
 subspaces of a single curve in the $(x,\sigma)$ plane, illustrated by the dashed vertical curve in diagram \eqref{eq:ReidIII_diag}. The two black bulk vertices $1$ and $2$ that are lying on this dashed line could in principle have been at the same value of $\sigma$, leading to a complicated multivalent vertex. Nevertheless this does not happen since these solitons are not present in the spectrum at the same point of the $\sigma$-parameter space. In fact, chambers where gray and orange solitons are present in the spectrum of the theory are separated by another chamber where there are no solitons for  this central charge. We can observe this explicitly by observing how the
 steepest descent paths evolve as a function of $\sigma$:
$$
\begin{array}{c}
\begin{tikzpicture}
\draw (3.5,1) -- (3.5,-1) (7.5,1) -- (7.5,-1);
\draw[orange, thick] (0,0) to[out=90,in=180] (1,0.5) to[out=0,in=180] (2.5,0);
\draw[thick] (0,0) to[out=270, in=180] (1,-0.5) -- (3,-0.5);
\draw (0.5,0) circle (0.1) (2.2,0.5) circle (0.1) (3,0) circle (0.1);
\begin{scope}
\draw[purple, ultra thick] (-0.1,-0.1) -- (0.1,0.1) (0.1,-0.1) -- (-0.1,0.1);
\end{scope}
\begin{scope}[shift={(1.7,0.5)}]
\draw[purple, ultra thick] (-0.1,-0.1) -- (0.1,0.1) (0.1,-0.1) -- (-0.1,0.1);
\end{scope}
\begin{scope}[shift={(2.5,0)}]
\draw[purple, ultra thick] (-0.1,-0.1) -- (0.1,0.1) (0.1,-0.1) -- (-0.1,0.1);
\end{scope}
\begin{scope}[shift={(4,0)}]
\begin{scope}[yscale=-1]
\draw[thick] (0,0) to[out=270, in=180] (1,-0.5) -- (3,-0.5);
\end{scope}
\draw[thick] (0,0) to[out=270, in=180] (1,-0.5) -- (3,-0.5);
\draw (0.5,0) circle (0.1) (2,0) circle (0.1) (3,0) circle (0.1);
\begin{scope}
\draw[purple, ultra thick] (-0.1,-0.1) -- (0.1,0.1) (0.1,-0.1) -- (-0.1,0.1);
\end{scope}
\begin{scope}[shift={(1.5,0)}]
\draw[purple, ultra thick] (-0.1,-0.1) -- (0.1,0.1) (0.1,-0.1) -- (-0.1,0.1);
\end{scope}
\begin{scope}[shift={(2.5,0)}]
\draw[purple, ultra thick] (-0.1,-0.1) -- (0.1,0.1) (0.1,-0.1) -- (-0.1,0.1);
\end{scope}
\end{scope}
\begin{scope}[shift={(8,0)}]
\begin{scope}[yscale=-1]
\draw[gray, thick] (0,0) to[out=90,in=180] (1,0.5) to[out=0,in=180] (2.5,0);
\draw[thick] (0,0) to[out=270, in=180] (1,-0.5) -- (3,-0.5);
\end{scope}
\draw (0.5,0) circle (0.1) (2.2,-0.5) circle (0.1) (3,0) circle (0.1);
\begin{scope}
\draw[purple, ultra thick] (-0.1,-0.1) -- (0.1,0.1) (0.1,-0.1) -- (-0.1,0.1);
\end{scope}
\begin{scope}[shift={(1.7,-0.5)}]
\draw[purple, ultra thick] (-0.1,-0.1) -- (0.1,0.1) (0.1,-0.1) -- (-0.1,0.1);
\end{scope}
\begin{scope}[shift={(2.5,0)}]
\draw[purple, ultra thick] (-0.1,-0.1) -- (0.1,0.1) (0.1,-0.1) -- (-0.1,0.1);
\end{scope}
\end{scope}
\end{tikzpicture}
\end{array}
$$
An even simpler proof proceeds by noting that $x_{\rm blue}(\sigma)<x_{\rm green}(\sigma)$ and $x_{\rm purple}(\sigma)<x_{\rm red}(\sigma)$ for all values $\sigma$.
The value $\sigma_0$ where the green and purple solitons merge into each other can be defined  from the condition $x_{\rm green}(\sigma_0)=x_{\rm purple}(\sigma_0)$. Therefore we conclude $x_{\rm blue}(\sigma_0)<x_{\rm red}(\sigma_0)$. Hence  when the green and purple solitons merge, the blue and red ones are well separated, as depicted in our diagrams.

In the same way we check ${\rm \bf Hpty}_{++-}^{++-}$. All the other possible groups are empty so we have demonstrated homotopy invariance of the cohomology under Reidemeister move III.

\section{Obstructions To Existence Of $\zeta$-Instantons}\label{sec:Obstruction}

In Section \ref{sec:R-Invariance} we demonstrated the invariance of link homology under Reidemeister moves
making use of $\sigma\zeta$-instantons and the instanton transplant argument. We hope the arguments were reasonably persuasive.
(They were never mathematically rigorous.) In some of the arguments we simply assumed that
the required instantons exist. Closer inspection shows that there can be obstructions to the existence
of these instantons. In this section we will argue that obstructions really do exist in the YYLG model
and in the next section \ref{sec:ResolveObstruction} we will argue that they are resolved in the MLG model.
 We will discuss two kinds of obstructions:

The first obstruction appears when proving the Reidemeister I move. It turns out, in the YYLG model, that an
 approximate solution described by a curved web trajectory can not be continued in $\sigma$
  since it intersects a wall of marginal stability. Across this wall the boosted soliton solution stops existing.
 We explain this problem in more detail in Section \ref{subsec:Problem-RI} below.

The second obstruction appears when proving the Reidemeister III move.   Suppose that the initial and final positions of the points $z_a$ is the same and consider a
$\zeta$ instanton interpolating between $\phi_{\rm in}(x)$ and $\phi_{\rm out}(x)$. These two field configurations define open paths $\wp_1$ and $\wp_2$ in the
target space $X$ between vacua $\phi_i$ and $\phi_j$. Therefore $\wp_1\circ \wp_2^{-1}$ defines a closed loop in $X$. If that closed loop is homotopically
nontrivial then a $\zeta$-instanton cannot exist, since such an instanton would be a smooth field configuration on all of $\IR^2$. On the circle
at infinity such an instanton would describe  the
loop $\wp_1\circ \wp_2^{-1}$ and hence the instanton would define a null homotopy of that loop.

\subsection{An Obstruction To Proving RI}\label{subsec:Problem-RI}

To illustrate the first obstruction consider the Reidemeister I relation:
\be\label{eq:ReidI-obs}
\begin{array}{c}
	\begin{tikzpicture}
	\begin{scope}[yscale=-1]
	\draw[ultra thick] (0,0) -- (1,1) to [out=45,in=135] (0,1) -- (0.4,0.6) (0.6,0.4) -- (1,0);
	\end{scope}
	\end{tikzpicture}
\end{array}=q^{\frac{3}{2}}t^{-1}\begin{array}{c}
\begin{tikzpicture}
\begin{scope}[yscale=-1]
\draw[ultra thick] (0,0) -- (0,0.8) to [out=90,in=90] (1,0.8) -- (1,0);
\end{scope}
\end{tikzpicture}
\end{array}\oplus\left(
\begin{array}{c}
	\begin{tikzpicture}
	\begin{scope}[scale=0.7]
	\draw[purple, ultra thick] (-0.25,0) -- (1.25,0) (0,0) -- (0,0.5) to[out=90,in=210] (0.4,0.9) (0.6,1.1) to[out=30,in=270] (1,1.5) -- (1,2) (1,0) -- (1,0.5) to[out=90,in=330] (0.5,1) to[out=150,in=270] (0,1.5) -- (0,2);
	\node[below] at (0,0) {$+$}; \node[below] at (1,0) {$-$};
	\node[above] at (0,2) {$-$}; \node[above] at (1,2) {$+$};
	\end{scope}
	\end{tikzpicture}
\end{array}\oplus
\begin{array}{c}
	\begin{tikzpicture}
	\begin{scope}[scale=0.7]
	\draw (0,0.25) -- (1,0.25) (0,1.4) -- (1,1.4) (0,1.5) -- (1,1.5);
	\node[right] at (1.1,0.25) {$x_{\cup}^s$};
	\node[right] at (1.1,1.5) {$x_1$};
	\filldraw[white] (0.5,0.25) circle (0.1);
	\draw (0.5,0.25) circle (0.1);
	\draw[purple, ultra thick] (-0.25,0) -- (1.25,0) (0,0) -- (0,0.5) to[out=90,in=210] (0.4,0.9) (0.6,1.1) to[out=30,in=270] (1,1.5) -- (1,2) (1,0) -- (1,0.5) to[out=90,in=330] (0.5,1) to[out=150,in=270] (0,1.5) -- (0,2);
	\node[below] at (0,0) {$+$}; \node[below] at (1,0) {$-$};
	\node[above] at (0,2) {$-$}; \node[above] at (1,2) {$+$};
	\end{scope}
	\end{tikzpicture}
\end{array}
\right)\label{RI_problem}
\ee
The second and third terms on the RHS should be equivalent to zero. As we have discussed,
 a necessary condition for this to take place is an existence of a $\zeta$-instanton.
Suppose we denote punctures as $z_a$ and $z_b$ at the bottom of the diagram, then the soliton at $x_{\cup}^s$ carries LG field $w$ from $z_a$ to $z_b$, and the soliton at $x_1$ carries LG field $w$ back from $z_b$ to $z_a$. Moreover these solitons have opposite values of $\int dW$.
%
%
Paths corresponding to these solitons on the cyclic cover $X$ therefore form a closed loop, hence there is a curved trajectory satisfying (\ref{domain_wall}) and connecting these solitons in a curved web:
\begin{center}
\begin{tikzpicture}
\begin{scope}[scale=0.7]
\draw[->] (0,-0.2) -- (0,3);
\draw[->] (-0.2,0) -- (5,0);
\node[left] at (0,3) {$\tau$};
\node[right] at (5,0) {$x$};
\draw[ultra thick] (0.5,0) to[out=90,in=180] (2,1.9) to[out=0,in=90] (3.5,0);
\draw[dashed] (2,1.9) -- (2,0);
\begin{scope}[shift={(3.5,0)}]
\filldraw[red] (-0.1,0.1) -- (0.1,0.1) -- (0.1,-0.1) -- (-0.1,-0.1);
\end{scope}
\begin{scope}[shift={(0.5,0)}]
\filldraw[red] (-0.1,0.1) -- (0.1,0.1) -- (0.1,-0.1) -- (-0.1,-0.1);
\end{scope}
\node[below] at (0.5,-0.1) {$x_{\cup}^{s}$};
\node[below] at (3.5,-0.1) {$x_1$};
\node[below] at (2,0) {$x_0$};
\end{scope}
\end{tikzpicture}
\end{center}
Indeed these two solitons behave like a soliton-anti-soliton pair.

\begin{figure}[h!]
\begin{center}
	\begin{tikzpicture}
	\begin{scope}[scale=0.7]
	\draw[->] (-4.5,-0.8) -- (5.5,-0.8);
	\node[right] at (5.5,-0.8) {$x$};
	\draw[ultra thick] (0.5,-0.7) -- (0.5,-0.9) (-3.25,-0.7) -- (-3.25,-0.9) (4.25,-0.7) -- (4.25,-0.9);
	\node[below] at (-3.25,-0.9) {$x_{\cup}^s$}; \node[below] at (0.5,-0.9) {$x_0$}; \node[below] at (4.25,-0.9) {$x_1$};
	\draw[->] (-2.5,0.25) -- (-0.5,0.25);
	\node[below] at (-1.5,0.25) {$\circlearrowright$};
	\draw[->] (3.5,0.25) -- (1.5,0.25);
	\node[below] at (2.5,0.25) {$\circlearrowleft$};
	\begin{scope}[shift={(-4.5,1)}]
	\begin{scope}[yscale=-1]
	\draw (0,0) to[out=270,in=180] (0.5,-0.5) to[out=0,in=210] (2,0);
	\draw[->] (0,0) to[out=270,in=180] (0.5,-0.5);
	\begin{scope}
	\draw[purple, ultra thick] (-0.1,-0.1) -- (0.1,0.1) (-0.1,0.1) -- (0.1,-0.1);
	\end{scope}
	\begin{scope}[shift={(2,0)}]
	\draw[purple, ultra thick] (-0.1,-0.1) -- (0.1,0.1) (-0.1,0.1) -- (0.1,-0.1);
	\end{scope}
	\draw (0.5,0) circle (0.1) (2.5,0) circle (0.1);
	\end{scope}
	\end{scope}
	\begin{scope}[shift={(3,1)}]
	\begin{scope}
	\draw (0,0) to[out=270,in=180] (0.5,-0.5) to[out=0,in=210] (2,0);
	\draw[->] (0,0) to[out=270,in=180] (0.5,-0.5);
	\end{scope}
	\begin{scope}
	\draw[purple, ultra thick] (-0.1,-0.1) -- (0.1,0.1) (-0.1,0.1) -- (0.1,-0.1);
	\end{scope}
	\begin{scope}[shift={(2,0)}]
	\draw[purple, ultra thick] (-0.1,-0.1) -- (0.1,0.1) (-0.1,0.1) -- (0.1,-0.1);
	\end{scope}
	\draw (0.5,0) circle (0.1) (2.5,0) circle (0.1);
	\end{scope}
	\begin{scope}
	\draw[->] (0,0) to [out=270,in=180] (0.5,-0.5) to [out=0,in=270] (1,0);
	\draw (1,-0.1) to[out=90,in=315] (0,2) (0,0) to[out=45,in=270] (1,2);
	\draw [->] (0,2) to[out=90,in=180] (0.5,2.5) to[out=0,in=90] (1,2);
	\draw (0.5,0) circle (0.1) (0.5,2) circle (0.1);
	\node[right] at (1,0) {$\phi_{x_1}$};
	\node[right] at (1,2) {$\phi_{x_{\cup}^s}$};
	\begin{scope}
	\draw[purple, ultra thick] (-0.1,-0.1) -- (0.1,0.1) (-0.1,0.1) -- (0.1,-0.1);
	\end{scope}
	\begin{scope}[shift={(0,2)}]
	\draw[purple, ultra thick] (-0.1,-0.1) -- (0.1,0.1) (-0.1,0.1) -- (0.1,-0.1);
	\end{scope}
	\end{scope}
	\end{scope}
	\end{tikzpicture}
\end{center}
\caption{In order to compare the homotopy classes of the solitons shown in equation \eqref{eq:ReidI-obs} we continuously evolve
the solitons as we increase
$x^s_{\cup}$ and decrease $x_1$. This figure shows how they evolve to the common point $x_0$, where it they are clearly not homotopic.}\label{fig:RI_problem}
\end{figure}
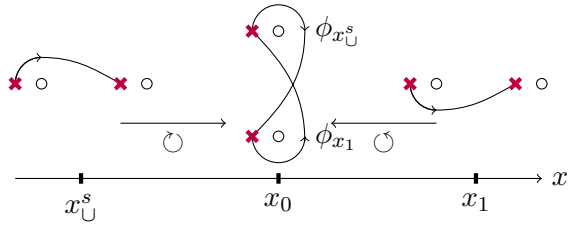

Although there is a suitable curved web we should ask if  there is really a corresponding $\zeta$-instanton.
 We claim there is \underline{not} because these  solitons belong to different homotopy classes in $X$.
In order to understand this we order $z_a$ (thus working on $P_m(\{z_a\})$) and rotate $z_a$, $z_b$ so that we can compare solitons in a common theory. For the soliton $\phi_{\cup}^s$ we rotate $z_a$, $z_b$ by $\frac{\pi}{2}$ clockwise and for the soliton $\phi_{x_1}$ we rotate by  $\frac{\pi}{2}$ counter-clockwise. (In terms of the braiding in equation \eqref{RI_problem} we decrease the value $x_1$ and increase the value of $x_{\cup}^s$ to a common value $x_0$.) The continuation of the solitons $\phi_{\cup}^s$ and $\phi_{x_1}$ is shown in Figure \ref{fig:RI_problem}.
It is now evident that these field configurations are not homotopic in the YYLG model.

Note that if we continue to increase $x_{\cup}^s$ past $x_0$ then $\phi_{x_\cup^s}$ decays into a pair of a ``purely flavour soliton'' and another soliton as in:
\begin{center}
	\begin{tikzpicture}
		\begin{scope}[scale=0.7]
		\draw[ultra thick] (0,2) to[out=90,in=180] (0.5,2.5) to[out=0,in=90] (1,2) to[out=270,in=90] (1,0);
		\begin{scope}[shift={(1,0)}]
		\draw (0.5,0) circle (0.1);
		\draw[purple, ultra thick] (-0.1,-0.1) -- (0.1,0.1) (-0.1,0.1) -- (0.1,-0.1);
		\end{scope}
		\begin{scope}[shift={(0,2)}]
		\draw (0.5,0) circle (0.1);
		\draw[purple, ultra thick] (-0.1,-0.1) -- (0.1,0.1) (-0.1,0.1) -- (0.1,-0.1);
		\end{scope}
		\end{scope}
	\end{tikzpicture} \qquad
	\begin{tikzpicture}
	\begin{scope}[scale=0.7]
	\draw[ultra thick] (0,2) to[out=90,in=180] (0.5,2.5) to[out=0,in=90] (0.8,2) to[out=270,in=0] (0.3,1.7) to[out=270,in=90] (0.3,0);
	\begin{scope}[shift={(0.3,0)}]
	\draw (0.5,0) circle (0.1);
	\draw[purple, ultra thick] (-0.1,-0.1) -- (0.1,0.1) (-0.1,0.1) -- (0.1,-0.1);
	\end{scope}
	\begin{scope}[shift={(0,2)}]
	\draw (0.5,0) circle (0.1);
	\draw[purple, ultra thick] (-0.1,-0.1) -- (0.1,0.1) (-0.1,0.1) -- (0.1,-0.1);
	\end{scope}
	\end{scope}
	\end{tikzpicture} \qquad
	\begin{tikzpicture}
	\begin{scope}[scale=0.7]
	\draw[ultra thick] (0,2) to[out=90,in=180] (0.5,2.3) to[out=0,in=90] (0.8,2) to[out=270,in=0] (0.5,1.7) to[out=180,in=270] (0,2) -- (0,0);
	\begin{scope}[shift={(0,0)}]
	\draw (0.5,0) circle (0.1);
	\draw[purple, ultra thick] (-0.1,-0.1) -- (0.1,0.1) (-0.1,0.1) -- (0.1,-0.1);
	\end{scope}
	\begin{scope}[shift={(0,2)}]
	\draw (0.5,0) circle (0.1);
	\draw[purple, ultra thick] (-0.1,-0.1) -- (0.1,0.1) (-0.1,0.1) -- (0.1,-0.1);
	\end{scope}
	\end{scope}
	\end{tikzpicture}
\end{center}
Similarly if we continue to decrease $x_1$ past $x_0$ then $\phi_{x_1}$ decays into a pair of a ``purely flavour soliton'' and another soliton as in:
\begin{center}
	\begin{tikzpicture}
	\begin{scope}[scale=0.7,yscale=-1]
	\draw[ultra thick] (0,2) to[out=90,in=180] (0.5,2.5) to[out=0,in=90] (1,2) to[out=270,in=90] (1,0);
	\begin{scope}[shift={(1,0)}]
	\draw (0.5,0) circle (0.1);
	\draw[purple, ultra thick] (-0.1,-0.1) -- (0.1,0.1) (-0.1,0.1) -- (0.1,-0.1);
	\end{scope}
	\begin{scope}[shift={(0,2)}]
	\draw (0.5,0) circle (0.1);
	\draw[purple, ultra thick] (-0.1,-0.1) -- (0.1,0.1) (-0.1,0.1) -- (0.1,-0.1);
	\end{scope}
	\end{scope}
	\end{tikzpicture} \qquad
	\begin{tikzpicture}
	\begin{scope}[scale=0.7,yscale=-1]
	\draw[ultra thick] (0,2) to[out=90,in=180] (0.5,2.5) to[out=0,in=90] (0.8,2) to[out=270,in=0] (0.3,1.7) to[out=270,in=90] (0.3,0);
	\begin{scope}[shift={(0.3,0)}]
	\draw (0.5,0) circle (0.1);
	\draw[purple, ultra thick] (-0.1,-0.1) -- (0.1,0.1) (-0.1,0.1) -- (0.1,-0.1);
	\end{scope}
	\begin{scope}[shift={(0,2)}]
	\draw (0.5,0) circle (0.1);
	\draw[purple, ultra thick] (-0.1,-0.1) -- (0.1,0.1) (-0.1,0.1) -- (0.1,-0.1);
	\end{scope}
	\end{scope}
	\end{tikzpicture} \qquad
	\begin{tikzpicture}
	\begin{scope}[scale=0.7,yscale=-1]
	\draw[ultra thick] (0,2) to[out=90,in=180] (0.5,2.3) to[out=0,in=90] (0.8,2) to[out=270,in=0] (0.5,1.7) to[out=180,in=270] (0,2) -- (0,0);
	\begin{scope}[shift={(0,0)}]
	\draw (0.5,0) circle (0.1);
	\draw[purple, ultra thick] (-0.1,-0.1) -- (0.1,0.1) (-0.1,0.1) -- (0.1,-0.1);
	\end{scope}
	\begin{scope}[shift={(0,2)}]
	\draw (0.5,0) circle (0.1);
	\draw[purple, ultra thick] (-0.1,-0.1) -- (0.1,0.1) (-0.1,0.1) -- (0.1,-0.1);
	\end{scope}
	\end{scope}
	\end{tikzpicture}
\end{center}
In other words $x_0$ defines a wall of marginal stability since two purely flavour solitons have different flavour charge: one winds around puncture $z_a$, another winds around $z_b$. Pictorially we can illustrate this phenomenon by the following plot of central charges of solitons as functions of $x$:
\begin{center}
\begin{tikzpicture}
\begin{scope}[scale=0.6]
\draw[red, dashed, ultra thick] (2,0) -- (2,4);
\draw[ultra thick] (0,2) -- (4,2) (0,4) -- (1,4) to[out=0,in=110] (2,2) to[out=290,in=180] (3,0) -- (4,0) (0,3) -- (1,3) to[out=0,in=130] (2,2) to[out=310,in=180] (3,1) -- (4,1);
\filldraw (2,2) circle (0.1);
\node[left] at (0,2) {$\CB_a,\; \CB_b$}; \node[right] at (4,2) {$\CB_a,\; \CB_b$};
\node[left] at (0,3) {$\mathscr{X}$}; \node[right] at (4,1) {$\tilde{\mathscr{X}}$};
\node[left] at (0,4) {$\mathscr{Y}$}; \node[right] at (4,0) {$\mathscr{Y}$};
\draw[<->] (-2.5,4) -- (-2.5,-0.5) -- (5,-0.5);
\node[right] at (5,-0.5) {$x$}; \node[below] at (2,-0.5) {$x_0$}; \node [left] at (-2.5,4) {$Z$};
\node[right] (A) at (3,3) {MS-wall}; \path (A) edge[->] (2,3.5);
\end{scope}
\end{tikzpicture}
\end{center}
In this diagram we denote flavour solitons winding around $z_a$ and $z_b$ by $\CB_a$ and $\CB_b$ respectively. Here $\mathscr{Y}$ is a soliton homotopic to the double line soliton in $x_{\cup}^s$; this soliton varies smoothly through the wall of marginal stability. $\mathscr{X}$ and $\tilde{\mathscr{X}}$ are the soliton $\phi_{\cup}^s$ and the anti-soliton of $\phi_{x_1}$, respectively. The central charges of these solitons are related by:
\be
Z(\CB_a)=Z(\CB_b)=2\pi,\quad Z(\mathscr{X})=Z(\tilde{\mathscr{X}})=Z(\mathscr{Y})+Z(\CB_a)
\ee

A key point is that the soliton $\mathscr{X}$ does not exist after the wall of marginal stability. Rather it decays  into $\mathscr{Y}$ and $\CB_a$.
Similarly, the  soliton $\tilde{\mathscr{X}}$ does not exist before the wall, and after the wall it is formed as a bound state of $\mathscr{Y}$ and $\CB_b$.
This is really the origin of the problem with RI-invariance, for had there not
been the wall of marginal stability we could have continued one of the two solitons to obtain suitable complexes.
In this sense, walls of marginal stability with purely flavour solitons can obstruct the existence of $\zeta$-instantons.

We would like to stress that the obstruction we have presented is  invisible in the construction of Jones polynomials. That is, it is
invisible when we switch from complexes to their Euler characteristics. In particular, across walls of marginal stability   associated with
the obstruction the spectrum of the solitons changes in a way so that neither the number of solitons nor their topological charge changes.
Recall   that the topological charge is determined by the vacua at $x \to \pm \infty$, so the central charges do not change either.
In order to describe the change of the spectrum one needs to take into account $\pi_1(X)$, and this can be interpreted as a change of
\underline{flavour} charge.  Indeed  in this sense  $\CB_a$ and $\CB_b$ have different flavour charges.
The standard wall-crossing formula of \cite{Cecotti:1992rm} does not take into account these flavor charges.  But one uses
precisely these  flavour-blind  wall-crossing formulae to prove   flatness of the decategorified parallel transport on the parameter space of LG potentials.
This flatness, in turn,
is an essential ingredient in the proof of invariance under Reidemeister moves of the construction of the Jones polynomial and related Chern-Simons link
polynomials.  That is why this obstruction was not be detected in the application of the YYLG formalism to the Jones polynomial \cite{Bigelow,Gaiotto:2011nm}.

\subsection{An Obstruction To Proving RIII}\label{subsec:Problem-RIII}

In order to illustrate the second obstruction let us revisit the proof of invariance of link homology
under the RIII move. We expand both sides of the Yang-Baxter relation as:
\be
\begin{array}{c}
	\begin{tikzpicture}
	\begin{scope}[scale=0.4,yscale=-1]
	\draw [ultra thick] (3,-0.5) -- (-1,1.5);
	\draw [ultra thick] (0,-2) -- (0,0.8);
	\draw [ultra thick] (0,1.2) -- (0,2);
	\draw [ultra thick] (-1,-1.5) -- (-0.2,-1.1) (0.2,-0.9) -- (1.8,-0.1);
	\draw [ultra thick] (2.2,0.1) -- (3,0.5);
	\end{scope}
	\end{tikzpicture}
\end{array}=:\bigoplus_{I,J,n}q^n{\CE}_I^J[n],\\
\begin{array}{c}
	\begin{tikzpicture}
	\begin{scope}[scale=0.4,yscale=-1]
	\draw [ultra thick] (1,-1.5) -- (-3,0.5);
	\draw [ultra thick] (0,-0.8) -- (0,2);
	\draw [ultra thick] (0,-1.2) -- (0,-2);
	\draw [ultra thick] (0.2,1.1) -- (1,1.5);
	\draw [ultra thick] (-0.2,0.9) -- (-1.8,0.1);
	\draw [ultra thick] (-2.2,-0.1) -- (-3,-0.5);
	\end{scope}
	\end{tikzpicture}
\end{array}=:\bigoplus_{I,J,n}q^n \tilde{\CE}_I^J[n]
\ee
Where $I$ and $J$ are vacua at spatial infinities, and $n$ is $\Pdeg$-degree.

Let us focus on the matrix element:
\be\label{problem}
\CE_{(++-)}^{(++-)}[2]=\begin{array}{c}
\begin{tikzpicture}
\begin{scope}[scale=0.7]
\draw (1.1,1.7) -- (1.9,1.7);
\node[below] at (0,0) {$+$};\node[below] at (1,0) {$+$};
\node at (0,1.25) {$+$};\node at (1,1.25) {$+$}; \node at (2,1.25) {$-$};
\node at (0,2.75) {$+$};\node at (1,2.75) {$+$}; \node at (2,2.75) {$-$};
\node[above] at (0,4) {$+$};\node[above] at (1,4) {$+$};
\S
\begin{scope}[shift={(1,1.5)}]
\S
\end{scope}
\begin{scope}[shift={(0,3)}]
\S
\end{scope}
\draw[ultra thick,purple] (0,1.5) -- (0,2.5) (2,0) -- (2,1) (2,3) -- (2,4);
\node[below] at (2,0) {$-$}; \node[above] at (2,4) {$-$};
\node[below] at (0,-0.5) {$v_1$}; \node[below] at (1,-0.5) {$v_2$}; \node[below] at (2,-0.5) {$v_3$};
\end{scope}
\end{tikzpicture}
\end{array},\quad \tilde\CE_{(++-)}^{(++-)}[2]=\begin{array}{c}
\begin{tikzpicture}
\begin{scope}[scale=0.7]
\draw (1.1,0.2) -- (1.9,0.2) (1.1,3.2) -- (1.9,3.2);
\node[below] at (1,0) {$+$};\node[below] at (2,0) {$-$};
\node at (0,1.25) {$+$};\node at (1,1.25) {$+$}; \node at (2,1.25) {$-$};
\node at (0,2.75) {$+$};\node at (1,2.75) {$+$}; \node at (2,2.75) {$-$};
\node[above] at (1,4) {$+$};\node[above] at (2,4) {$-$};
\begin{scope}[shift={(0,1.5)}]
\S
\end{scope}
\begin{scope}[shift={(1,0)}]
\S
\end{scope}
\begin{scope}[shift={(1,3)}]
\S
\end{scope}
\draw[ultra thick,purple] (2,1.5) -- (2,2.5) (0,0) -- (0,1) (0,3) -- (0,4);
\node[below] at (0,0) {$+$}; \node[above] at (0,4) {$+$};
\node[below] at (0,-0.5) {$v_1$}; \node[below] at (1,-0.5) {$v_2$}; \node[below] at (2,-0.5) {$v_3$};
\end{scope}
\end{tikzpicture}
\end{array}
\ee

We search for a suitable map $U:\CE \to \tilde{\CE}$ associated to a $\sigma\zeta$-instanton.

The curved web for this homotopy process is expected to include a 3-valent interior amplitude (see \eqref{3-valent_ReidIII}), however for this solution to exist one needs the corresponding boundary conditions to be homotopically trivial in the target space as explained above.

Both $\CE_{(++-)}^{(++-)}[2]$ and $\tilde\CE_{(++-)}^{(++-)}[2]$ are one-dimensional complexes. Both are generated by wave-functions of interfaces corresponding to certain  paths in the field space $X$, we denote these paths as $p_1$ and $p_2$. In order to compare them we should drag solitons smoothly to some $x$ where theories we are comparing are the same. We choose this $x$ at the very bottom of interfaces:
\be
\CE_{(++-)}^{(++-)}[2]\rightsquigarrow p_1,\quad \tilde{\CE}_{(++-)}^{(++-)}[2]\rightsquigarrow p_2
\ee
Or, explicitly,
\be\label{eq:p1p2}
\begin{array}{c|c}
	p_1: & p_2:\\
	\left.\begin{array}{c}
		{\color{red}w_1}:v_1\to v_3,\; {\color{blue}w_2}=v_2
	\end{array}\right.& \left\{\begin{array}{rl}
	x<x_1:& {\color{red}w_1}=v_1,\; {\color{blue}w_2}:v_2\to v_3\\
	x_1<x<x_2:& {\color{red}w_1}=v_1,\; {\color{blue}w_2}= v_3\\
	x>x_2:&  {\color{red}w_1}=v_1\to v_2,\; {\color{blue}w_2}= v_3\\
\end{array}\right.\\
\begin{tikzpicture}
\draw (0,0) circle (0.1) (-1.5,0) circle (0.1) (1.5,0) circle (0.1);
\node[below] at (-1.5,-0.1) {$v_1$};
\node[above] at (0,0.1) {$v_2$};
\node[below] at (1.5,-0.1) {$v_3$};
\begin{scope}[shift={(-0.5,0)}]
\draw[blue,thick] (-0.1,-0.1) -- (0.1,0.1) (-0.1,0.1) -- (0.1,-0.1);
\end{scope}
\draw[red,thick,->] (-2,0) to[out=90,in=180] (-1.5,0.5) to[out=0,in=90] (-1,0) to[out=270,in=180] (0,-0.5) to[out=0,in=225] (1,0);
\end{tikzpicture} & \begin{tikzpicture}
\draw (0,0) circle (0.1) (-1.5,0) circle (0.1) (1.5,0) circle (0.1);
\node[below] at (-1.5,-0.1) {$v_1$};
\node[below] at (0,-0.1) {$v_2$};
\node[below] at (1.5,-0.1) {$v_3$};
\draw[blue,thick,->] (-0.5,0) to[out=90,in=180] (0,0.5) to[out=0,in=135] (1,0);
\draw[red,thick,->] (-2,0) to[out=90,in=180] (-1.5,0.5) to[out=0,in=135] (-0.5,0);
\end{tikzpicture}\\
\end{array}
\ee
The path $p_1$ illustrates a transition of the field $w_1$ from the vacuum near $v_1$ to the vacuum near $v_3$ along a trajectory depicted by the red curve in this diagram.
Meanwhile the  field $w_2$ remains nearly at rest in a vacuum near $v_2 $. The path $p_2$ can be divided in two stages. In the first stage the field $w_2$ makes a transition from the vacuum near $v_2$ to the vacuum near $v_3$, in this way the vacuum near $v_2$ becomes vacant. Further on in $x$, in the second stage the field $w_1$ makes a transition from the vacuum near $v_1$ to the vacuum near $v_2$. Notice that final field values $w_1$ and $w_2$ are permuted for $p_1$ and $p_2$, however on $X$ this is the same point.
Thus, the  boundary condition on the $\sigma\zeta$-soliton at infinity define a closed loop in $X$ homotopic to $p_1\circ p_2^{-1}$.

While it is a little complicated to understand the loop $p_1\circ p_2^{-1}$ it is rather easy to see that
\be\label{eq:R_III_Problem}
p_1\circ p_2^{-1}\circ p_1\circ p_2^{-1}=\begin{array}{c}
		\begin{tikzpicture}
		\draw (0,0) circle (0.1);
		\draw[blue,->, ultra thick] (0,0.5) to[out=225,in=90] (-0.25,0) to[out=270,in=180] (0,-0.25);
		\draw[blue, ultra thick] (0,-0.25) to[out=0,in=270] (0.25,0) to[out=90,in=315] (0,0.5);
		\begin{scope}[scale=2]
		\draw[red,->, ultra thick] (0,0.5) to[out=225,in=90] (-0.25,0) to[out=270,in=180] (0,-0.25);
		\draw[red, ultra thick] (0,-0.25) to[out=0,in=270] (0.25,0) to[out=90,in=315] (0,0.5);
		\node[below] at (0,-0.27) {$v_2$};
		\end{scope}
		\end{tikzpicture}
	\end{array}
\ee
By this we mean that the projection into $P_2(\{ z_a \})$ is a loop  $(w_1(s), w_2(s))$ with $w_1(s)$ and $w_2(s)$ shown as above.
Therefore $p_1\circ p_2^{-1}\circ p_1\circ p_2^{-1}$ is clearly a nonzero element of $\pi_1(X)$.
It follows that there is an obstruction in the YYLG to the existence of the required $\sigma\zeta$-instanton for Reidemeister III invariance.

\section{Resolving The Obstruction}\label{sec:ResolveObstruction}

The obstructions described in the previous section can be traced to the singularities in the YYLG when $w_i = z_a$ or
when $w_i = w_j $ for $i\not=j$.
This suggests that by working in the monopole LG model, where there are no such singularities, the obstructions will vanish.
In this section we demonstrate that this is correct in a little more detail.
We first check that the obstruction to RI vanishes by   showing numerically that in the NMLG and MLG models
the  crucial solutions to the forced $\zeta$-soliton equation continue to exist through critical values of the homotopy parameter
and hence vary smoothly across marginal stability walls of the YYLG model.  The second obstruction vanishes for the MLG model
because the target space $X$ is the simply connected universal cover of monopole moduli space.

\subsection{Resolution Of The Problem With RI}\label{sec:Res_RI}

Let us compare the soliton equations in the YYLG and the NMLG models. Assuming
 the  K\"ahler metric is the Euclidean metric the equation in the YYLG model can
 be written in the form:
\be
\frac{dw_i}{dx}=\frac{\I\zeta}{2}\overline{\frac{\p W_{\rm NMLG}(w,Y)}{\p w_i}},\quad 0=\frac{\I\zeta}{2}\overline{\frac{\p W_{\rm NMLG}(w,Y)}{\p Y_i}}
\ee
Indeed making the change of variables  \eqref{eq:subs},   the second equation becomes an algebraic equation for $\tilde Y_i$, and then the first equation becomes the forced $\zeta$-soliton equation in the YYLG model. This results in the difficulty encountered in Section \ref{sec:Obstruction} above. In the NMLG model (or the MLG model when
the monopoles are widely separated) we solve a differential equation in both fields $w_i$ and $Y_i$:
\be\label{eq:NMLG}
\frac{dw_i}{dx}=\frac{\I\zeta}{2}\overline{\frac{\p W_{\rm NMLG}(w,Y)}{\p w_i}},\quad \frac{dY_i}{dx}=\frac{\I\zeta}{2}\overline{\frac{\p W_{\rm NMLG}(w,Y)}{\p Y_i}}
\ee
By introducing extra dynamical fields $Y_i$ and working in the NMLG or MLG models the solitons evolve smoothly through the point $x_0$.

\begin{figure}[h!]
\begin{center}
\begin{tikzpicture}
	\draw (0,-0.5) -- (0,2.5);
	\node[below] at (0,-0.5) {$x_0$};
	\begin{scope}[shift={(3,0)}]
		\draw[red,ultra thick] (0,2) to[out=250,in=90] (-0.2,1) to[out=270,in=110] (0,0);
		\draw[blue,ultra thick] (0,2) to[out=45,in=90] (0.8,2) to[out=270,in=315] (0.1,1.9) -- (0,0);
		\draw[orange,ultra thick] (0,2) to[out=180,in=90] (-0.8,1) to[out=270,in=180] (0,0);
		\draw[violet,ultra thick] (0,2) to[out=0,in=90] (0.8,1) to[out=270,in=0] (0,0);
		\node[right] at (0.8,2) {$z_1$}; \node[right] at (0.8,0) {$z_2$};
		\draw (0.5,0) circle (0.1) (0.5,2) circle (0.1);
		\draw[purple,ultra thick] (-0.1,2.1) -- (0.1,1.9) (0.1,2.1) -- (-0.1,1.9) (-0.1,0.1) -- (0.1,-0.1) (0.1,0.1) -- (-0.1,-0.1);
		\node[below] at (0,-0.5) {$x_0+\epsilon$};
	\end{scope}
	\begin{scope}[shift={(-3,0)}]
		\begin{scope}[yscale=-1, shift={(0,-2)}]
		\draw[red,ultra thick] (0,2) to[out=250,in=90] (-0.2,1) to[out=270,in=110] (0,0);
		\draw[blue,ultra thick] (0,2) to[out=45,in=90] (0.8,2) to[out=270,in=315] (0.1,1.9) -- (0,0);
		\draw[orange,ultra thick] (0,2) to[out=180,in=90] (-0.8,1) to[out=270,in=180] (0,0);
		\draw[violet,ultra thick] (0,2) to[out=0,in=90] (0.8,1) to[out=270,in=0] (0,0);
		\node[right] at (0.8,2) {$z_2$}; \node[right] at (0.8,0) {$z_1$};
		\end{scope}
		\draw (0.5,0) circle (0.1) (0.5,2) circle (0.1);
		\draw[purple,ultra thick] (-0.1,2.1) -- (0.1,1.9) (0.1,2.1) -- (-0.1,1.9) (-0.1,0.1) -- (0.1,-0.1) (0.1,0.1) -- (-0.1,-0.1);
		\node[below] at (0,-0.5) {$x_0-\epsilon$};
	\end{scope}
\end{tikzpicture}
\end{center}
\caption{This figure compares the solitons in the YYLG and NMLG models on the left- and right-hand side of the marginal stability wall.
The blue ({\color{blue} $\blacksquare$}) curves
are solitons $\phi_{x_{\cup}^s}$ and $\phi_1$ in the YYLG model. Note that they get pinched indicating
that they decay across the wall of marginal stability.
By contrast, the violet ({\color{violet} $\blacksquare$}) curves are the
corresponding solitons $\phi_{x_{\cup}^s}$ and $\phi_1$ in the NMLG model.
Note that these evolve  smoothly across the wall. Recall that there are two solitons between pairs of  the second solution to $\zeta$-soliton equation evolves smoothly in both YYLG ({\color{red} $\blacksquare$}) and NMLG ({\color{orange} $\blacksquare$}) models.
}\label{fig:Res_RI}
\end{figure}
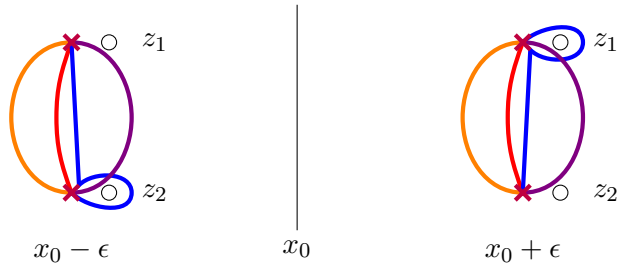

We could claim this on general grounds by analysing the fundamental group of the target space $X$ in the corresponding models. This argument can be checked by numerical evaluation of solutions to equation \eqref{eq:NMLG} in the neighbourhood of the (YYLG) marginal stability point $x_0$.\footnote{
	Technically we apply a variational principal to an ansatz for a solution. We discretize an interval in the real line, make an ansatz for the field configuration and minimize the energy numerically using Mathematica.
} The result is shown in Figure \ref{fig:Res_RI}. In the NMLG and MLG models the two solutions $\phi_{x_\cup^s}$ and $\phi_{x_1}$ evolve smoothly into each other.

\subsection{Resolution Of The Problem With RIII}

Let us return to the problematic path \eqref{eq:R_III_Problem}.  For the
 NMLG model, since the $w_i$ can pass through $z_a$ the path $(p_1\circ p_2^{-1})^2$  can be simplified to:
\begin{center}
\begin{tikzpicture}
\begin{scope}[shift={(0,1)}]
\draw[blue,ultra thick] (-0.1,-0.1) -- (0.1,0.1) (-0.1,0.1) -- (0.1,-0.1);
\end{scope}
\draw (0,0) circle (0.1);
\node[right] at (0.1,0) {$v_2$}; \node [left] at (-0.1,1) {$w_2$};
\draw[ultra thick, red, ->] (0,0.5) to[out=0,in=270] (0.8,1) to[out=90,in=330] (0,1.7) to[out=210,in=90] (-0.8,1) to[out=270,in=180] (0,0.5);
\node[right] at (0.8,1) {$w_1$};
\end{tikzpicture}
\end{center}
But if $w_i\neq w_j$ for $i\neq j$ this is still homotopically non-trivial.

\begin{figure}[h!]
\begin{center}
\includegraphics[scale=0.5]{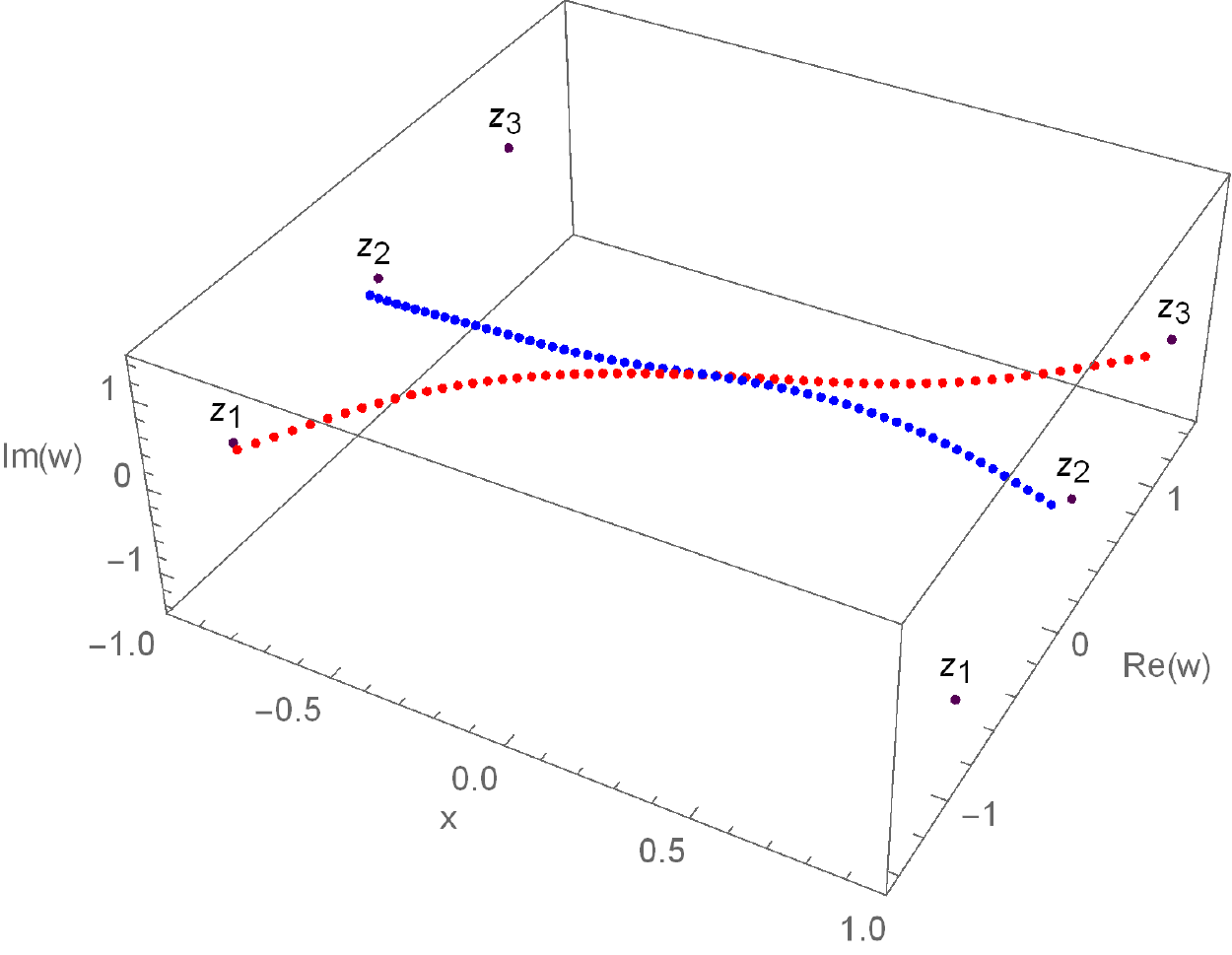}\quad \includegraphics[scale=0.5]{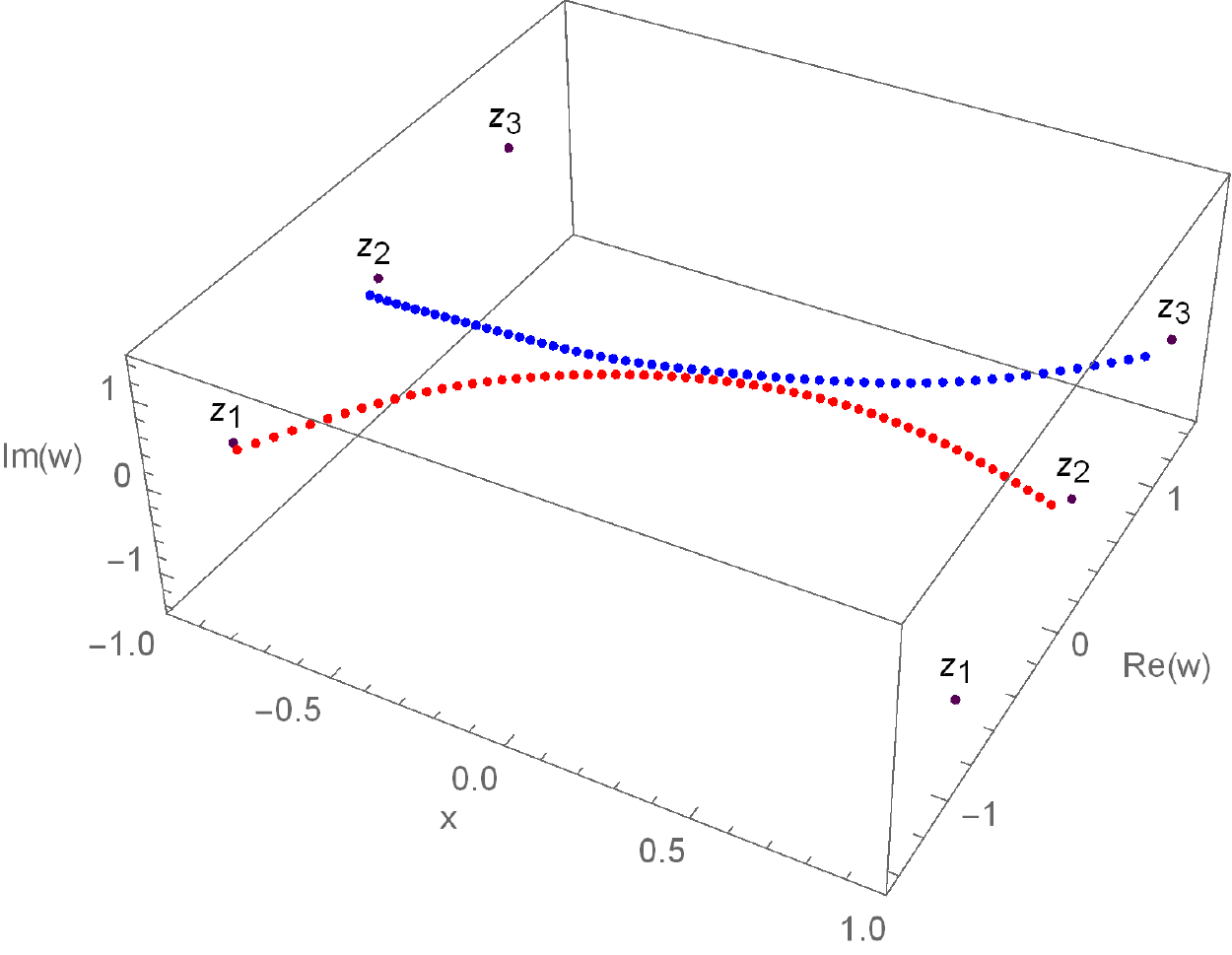}
\end{center}
\caption{Numerically constructed paths $p_1$ and $p_2$ in the MLG model.}\label{fig:p1p2}
\end{figure}

On the other hand, for the MLG model $X$ and $W$ is non-singular so the problem disappears.
As a check, in Figure \ref{fig:p1p2} we have numerically constructed approximations to the soliton paths analogous  to $p_1$ and $p_2$ of \eqref{eq:p1p2} in the MLG model.
We work at a  point of the parameter space close to the critical value $\sigma_0$ (for example, $\sigma_0$ in diagram \ref{3-valent_ReidIII}) when two solitons in $\tilde\CE_{(++-)}^{(++-)}[2]$ of \eqref{problem}  merge to give a single soliton in $\CE_{(++-)}^{(++-)}[2]$.
As shown in Figure \ref{fig:p1p2} in the MLG model the corresponding paths are homotopic to each other since monopoles are indistinguishable. In  the left plot the trajectories of two migrating monopoles in the $w$-plane intersect, so the coordinates $w_1$ and $w_2$ get interchanged in the intersection point, and we get two trajectories depicted in the right plot.

\section{Future Directions}\label{sec:FutureDirections}

It will be apparent to any reader who has tried to work through Sections \ref{sec:Examples} and \ref{sec:R-Invariance}
that an important weak point in our presentation has been the precise determination of the differential on the link
complex. We have used tricks and self-consistency to determine it in the cases we have discussed. There are two
aspects which should be improved. First, one needs a more systematic way to determine the   interior amplitudes $\beta$
associated to the vertices appearing in curved webs. These interior amplitudes define solutions of an $L_\infty$
Maurer-Cartan equation and have a path-integral definition \cite{Gaiotto:2015aoa} but a really satisfying physical
interpretation remains to be given. Certainly, there is no systematic approach for computing them. Second, it would
be good to have a more algorithmic approach to determining the signs of the matrix elements of $\CQ_{\zeta}$
than the procedure outlined in Appendix \ref{app:Signs}.

There are, of course, many other directions in which it would be nice to generalize our computations. One
direction is to include higher spin representations, that is, to take the case where some $k_a > 1$.
(Some discussion of this generalization can be found in \cite{GalakhovPhD}.) It seems challenging, but
possible, to generalize the treatment to link homologies following from 5D SYM with gauge groups of
rank larger than one.

We have shown in examples that the Poincar\'e polynomials of the link cohomologies derived from the LG model coincide with the corresponding Khovanov polynomial
(up to a simple redefinition of variables) for a few simple knots and links. It is expected that this equivalence will hold for all links, and it would be
nice to have a systematic proof that this is the case.   In particular, there are actually \underline{two} versions of Khovanov homology, called
even and odd  Khovanov homology \cite{Odd}. How would that difference arise in the framework of this paper?  It is natural to suspect that there are
inequivalent sign choices for the differential, i.e. for the matrix elements $\langle \Psi_1 \vert \CQ_{\zeta} \vert \Psi_2 \rangle$. Such
a sign change can in turn arise naturally from coupling  the supersymmetric quantum mechanics of Morse theory to a nontrivial flat connection that is
torsion of order two. Since the target space is a space of maps of one-dimensional space into the moduli space of monopoles such an order two connection
would be induced by a torsion $B$-field on the monopole moduli space.
\footnote{The stable homotopy type of the moduli space is the the moduli space of all continuous maps from $S^2 \rightarrow S^2$. We are interested
in $\pi_2$ of this space and $[S^2, {\rm Map}(S^2,S^2)]$ receives a map from  $\pi_4(S^2) \cong \IZ_2$. So there is room for the required $B$-field. It was suggested to GM some years ago by E. Witten that the difference
between even and odd Khovanov theory could be attributed to a $B$-field on monopole moduli space.}
Unfortunately, the first knot where the difference between even and odd
Khovanov homology appears is  $8_{19}$, and this is far beyond the examples we have presented here.





\appendix

\section{Fermion Number Grading And  $tt^*$-geometry}\label{app:FermionTTstar}


In this section we describe a formula that allows us to compute the (relative) fermion numbers of
the quantum groundstates associated to forced LG solitons. To do this we consider a slight
generalization of the famous CFIV index \cite{Cecotti:1992qh}.

Consider a family of 2d N=(2,2) quantum field theories parametrized by deformations of the
actions by chiral operators. This is usually written $S \to S + \sum_a (\int d^2\theta t_a \; \Phi_a + c.c.) $
where $t_a$ are coordinates on the chiral ring relative to a basis of chiral operators $\Phi_a$.
Now consider a path $t_a(x)$ in deformation space. We will initially consider it as a function of a real
variable $x$, but it will be important to assume that it admits analytic continuation to a holomorphic
function of $x$ in the complex plane.

Consider the partition function of a LG interface on  the cylinder $\IR \times S^1$  with the $S^1$ factor identified with a thermal
circle of radius $\beta$ and $\IR$ is viewed as a spatial coordinate. We must choose boundary conditions for LG fields and we stipulate
that the fields approach the  vacuum $\phi_i$
for $x \to - \infty$ and $\phi_{j'}$ for $x\to + \infty$.  Thus, we really get a matrix of partition functions:
\be
\Psi_{ij'}[t_a(x)]:=\Tr_{\CH_{ij'}}\; e^{-\beta H(t_a(x),\zeta)}e^{\I \pi  {\bf F}}
\ee
Here $H(t_a(x),\zeta)$ is the Hamiltonian of the $1+1$ dimensional QFT following from \eqref{eq:InterfaceMorseFunction}.

Now we consider the asymptotic behavior of $\Psi_{ij'}$ as $\beta \to + \infty$. The leading approximation should be given by a sum over forced LG
solitons:
\be
\Psi_{ij'}[t_a(x)]\simeq\sum\lm_{s\in \fS_{ij'}}\Delta_s e^{-\half (\zeta^{-1}\beta Z_s+\zeta\beta Z_s)}E_{ij'}(s)+\mbox{multi-soliton~ contributions}\label{expansion}
\ee
Here $\fS_{ij'}$ is the set of solutions to the forced $\zeta$-soliton equation \eqref{forced_soliton}
approaching $i$ and $j'$ vacuum solutions at spatial infinities. The factor $\Delta_s=e^{\pi \I {\bf F}_s}$ a product of  1-loop determinants and
is the phase of the Dirac operator computed around the forced $\zeta$-soliton. The fermion number ${\bf F}_s$ is given by the $\eta$-invariant
of the corresponding Dirac operator in the soliton background.  $E_{ij'}(s)$ is a simply
\be
E_{ij'}(s)=\left\{\begin{array}{l}
	1,\quad {\rm if}\; \phi(x|s)\to\left\{\begin{array}{l}
		\phi_i,\;x\to-\infty\\
		\phi_{j'},\;x\to+\infty
	\end{array}\right.\\
	0,\quad{\rm otherwise}
\end{array}\right.
\ee

We now want to consider a family of paths $\wp_x$ in the space $C$ of deformation parameters.
The path $\wp_x: \IR \to C$ runs the original path $t_a(u)$ in $C$  just for $u\leq x$ and
then for $u\geq x$ the deformation parameter $\wp_x(u) = t_a(x)$ is held constant.   Accordingly define the function $z_a(u;x)$:
\be
\begin{array}{c}
\begin{tikzpicture}
\draw[<->] (0,1.5) -- (0,0) -- (5,0);
\node[right] at (5,0) {$u$};
\node[above] at (0,1.5) {$t_a(u)$};
\draw[gray] (3,1) to[out=45,in=135] (4,1) -- (4.5,0.5);
\draw[thick] (0.5,0.5) -- (1,1) to[out=45,in=135] (2,1) to[out=315,in=225] (3,1) -- (4.5,1);
\draw[dashed] (3,1) -- (3,0);
\node[below] at (3,0) {$x$};
\end{tikzpicture}
\end{array}\qquad
z_a(u;x):=\left\{\begin{array}{l}
t_a(u),\quad u<x,\\
t_a(x),\quad u\geq x;
\end{array}\right.
\ee
The matrix of partition functions  $\Psi[z_a(u;x)]$ will now be denoted as $\Psi(x)$. It can be shown
(see \cite{GalakhovPhD}) to be  a \underline{flat} section of the $tt^*$ connection
of \cite{Cecotti:1992rm}:
\be
{\cal A}=\frac{\I}{2}\zeta^{-1}\beta C_a dz_a+A_a dz_a+\bar A_a d\bar z_a-\frac{\I}{2}\zeta \beta \bar C_a d\bar z_a
\ee
where $z_a=z_a(x;\infty)$, and $C_a$ are chiral ring structure constants:
\be
\Phi_a \Phi_b=:\sum\lm_c (C_a)_b^c\; \Phi_c
\ee
Following \cite{Cecotti:1992rm} we choose a suitable gauge:
\be
A_a\to g^{-1}\p_a g+A_a,\quad \bar A_a\to 0
\ee
where $g$ is the inverse Zamolodchikov metric, (that is, the $tt^*$ metric)
and $A$ is purely holomorphic. In \cite{Cecotti:1992rm} authors use a canonical basis where $A$ vanishes after this gauge transformation but
we find it useful to choose a slightly different basis where it is pure gauge, but nonzero.
We would like to choose chiral ring generators $\Phi_a:=\p_{z_a}W(\phi^I|z_a)$ depending as functions on $z_a$, so we can calculate the remnant connection $A_a$ as:
\be
\p_{z_a} \Phi_b=:-\sum\lm_c (A_a)_b^c\; \Phi_c
\ee

Now the partition function $\Psi_{ij'}[t_a(u;x)]$, as a function of $x$,  has a holomorphic extension to the complex $x$-plane
compatible with the flatness of the $tt*$ connection. The pullback of the flatness equation
for $\Psi$ to the complex $x$ plane has a holomorphic part given by:
\be\label{eq:tt^*}
\p\Psi(x)+\frac{\I}{2}\zeta^{-1}\beta C\Psi(x)+\left(g^{-1}\p g+A\right)\Psi(x)=0
\ee
where $C=C_a dz_a$. The  $\bar\p$-part of the flatness equation involves $\bar C$.

Let us now return to the asymptotics of these partition functions   for $\beta\to \infty$. According to \cite{SN,GLM,Gabella}
they should have the asymptotic form:
\be\label{eq:approx}
\Psi_{ij'}(x)\mathop{\sim}\lm_{\beta\to \infty}\sum\lm_{\gamma_{ij'}\in\CD_{ij'}} e^{-\frac{\I}{2}\zeta^{-1}\beta\int\lm_{\gamma_{ij'}}\lambda+\pi \I \int\lm_{\gamma_{ij'}}\omega}\left(\psi_{\gamma_{ij'}} (x)+\zeta \beta^{-1}\chi_{\gamma_{ij'}}(x)+\ldots\right)
\ee
Here $\CD_{ij'}$ is a set of ``detours'' -- namely open paths on a certain spectral cover $\Sigma$ defined in equation \eqref{eq:spec_cov} below, these paths begin on sheet $i$ over some initial point $x_0$ and end on sheet $j'$ over point $x$.
The functions $\psi_{\gamma_{ij'}} (x)$ and $\chi_{\gamma_{ij'}}(x)$  are independent of $\zeta$ and define matrices whose rows and columns
are labeled by vacua $i$ and $j'$.  The one-form
$\lambda$ is the canonical Liouville one-form on $T^*\CC$ where $\CC$ is the complex $x$-plane (pulled back to $\Sigma$) and $\omega$ is another
one-form on $\Sigma$, derived below. Both forms are integrated over $\gamma_{ij'}$.

Substituting this expansion into equation \eqref{eq:tt^*} and writing the first two  orders in the expansion in powers of $\beta$
gives the leading conditions (we drop the subscript $\gamma_{ij'}$ on $\psi$ and $\chi$ here):
\begin{subequations}
\be
\label{eq:asmpt_a} (\lambda \mathds{1}-C)\psi=0\\
\pi \I \omega \psi-\frac{\I}{2}\left(\lambda -C\right)\chi+ (\p+A+\left(g^{-1}\p g\right))\psi=0 \label{eq:asmpt_b}
\ee
\end{subequations}
The first equation \eqref{eq:asmpt_a} has a solution iff
\be\label{eq:spec_cov}
{\rm det}\; \left(\lambda \mathds{1} -C\right)=0
\ee
This equation defines a spectral cover $\pi:\Sigma \to \IC$ of the $x$-plane. Let us denote distinct roots of this covering equation by $\lambda^{(i)}$. For each $i$ we have a one-dimensional null space. Indeed, in a  a massive theory the index $i$ should run over $1, \dots, N_{\rm vac}$.  Choose a vector $\psi_i$ in the null space for  $\lambda^{(i)}$.
These vectors for different $i=1,\ldots,N_{\rm vac}$ form a basis of sections for a vector bundle $\CV$ over the $x$-plane of dimension $N_{\rm vac}$.
Take the leading matrix $\psi(x)$ in equation \eqref{eq:approx} to be the square matrix with column vectors $\psi_i$.
Any other section of $\CV$, in particular the $i^{th}$ column $\chi_i(x)$ of the subleading matrix $\chi(x)$ in
\eqref{eq:approx}, can be expanded as
\be
\chi_i(x)=\sum\lm_j c^j_i(x) \psi_j(x)
\ee
where $c^j_i(x)$ are holomorphic functions of $x$.
Substituting this expansion into \eqref{eq:asmpt_b} we get:
\be\label{eq:diff-for-psi1}
\pi \I \omega \psi_i-\frac{\I}{2}\sum\lm_j c^j_i(\lambda^{(i)}-\lambda^{(j)})\psi_j+ ( \p  +A +\left(g^{-1}\p g\right))\psi_i=0
\ee
Making a gauge transform with $g$ (recall $g$ is the $tt^*$ metric) this equation can equally well be written as
\be\label{eq:diff-for-psi2}
\pi \I \omega \psi_i^{(g)} -\frac{\I}{2}\sum\lm_j c^j_i(\lambda^{(i)}-\lambda^{(j)})\psi_j^{(g)} + ( \p  +A^{(g)} )\psi_i^{(g)} =0
\ee
where a superscript $*^{(g)}$ implies a suitable action by the gauge transformation $g$.

As we can see from this equation $\omega$ depends on the choice of the eigenvector $\psi_i$, in other words on the choice of the sheet $i$ of the cover $\Sigma$, therefore  $\omega$ should be interpreted as a differential form on  $\Sigma$.

Notice that the operation
\be
{\rm det}\left[\psi_1^{(g)},\ldots,\psi_{i-1}^{(g)},\eta,\psi_{i+1}^{(g)},\ldots,\psi_{N_{\rm vac}}^{(g)}\right]
\ee
as applied to a vector $\eta$ acts as a linear map to the complex numbers. It is non-zero only if $\eta$ and $\psi_i^{(g)}$ are collinear. We apply this linear map to equation \eqref{eq:diff-for-psi2} and derive the following expression for the value of the form $\omega$ on the $i^{\rm th}$ sheet:
\be
\omega^{(i)}=-\frac{1}{\pi \I}\frac{{\rm det}\left[\psi_1^{(g)},\ldots,\left(\p+A^{(g)}\right)\psi_i^{(g)},\ldots,\psi_{N_{\rm vac}}^{(g)}\right]}{{\rm det}\left[\psi_1^{(g)},\ldots,\psi_{N_{\rm vac}}^{(g)}\right]}
\ee

As we circle around a simple branch point of type  $ij$ the  roots the of the spectral cover equation get permuted. The root $\lambda^{(i)}$ becomes $\lambda^{(j)}$
and simultaneously eigenvector $\psi_i$ becomes $\psi_j$. Therefore $\omega^{(i)}$ becomes $\omega^{(j)}$. Hence we conclude that $\omega$ is a well-defined
one-form on $\Sigma$.

Now we use the fact that  the metric $g$ is flat modulo exponential corrections in the limit $\beta \to\infty$  \cite[eq.(4.6)]{Cecotti:1992rm}:
\be
g^{\bar j i}\simeq \delta^{ji}+\frac{\I}{\pi}\mu_{ij}K_0\left(\beta |Z_{ij}|\right)+\ldots
\ee
Therefore neglecting these corrections we derive:
\be
\omega^{(i)}=-\frac{1}{\pi \I}\frac{{\rm det}\left[\psi_1,\ldots,\left(\p+A\right)\psi_i,\ldots,\psi_{N_{\rm vac}}\right]}{{\rm det}\left[\psi_1,\ldots,\psi_{N_{\rm vac}}\right]}+O\left(e^{-\beta |Z_{ij}|}\right)
\ee
%
%

As we have mentioned, the sheets of the spectral cover \eqref{eq:spec_cov} are in 1-to-1 correspondence with the vacua in $\IV$.
We would like to compare the expansions \eqref{expansion} and \eqref{eq:approx} in the limit $\beta\to \infty$.
%
%
%
Then we have two expansions for the same quantity. Comparing these expansions  we see that the
detours $\CD_{ij'}$ are in 1-to-1 correspondence with solutions $\fS_{ij'}$ to the forced $\zeta$-soliton equation.
 This gives an invertible map
\be
 \CP:\fS_{ij'}\to \CD_{ij'},\quad \CP^{-1}:\;\CD_{ij'}\to \fS_{ij'}
\ee
and under this correspondence we have:
\be
Z_s=-\frac{\I}{2}\int\lm_{\CP(s)}\lambda,\quad {\bf F}_s=\int\lm_{\CP(s)}\omega .
\ee

\bigskip
\noindent
{\bf Remarks:}

\begin{enumerate}

\item Note that we have assumed that we can choose a basis $\psi_i$ so that the leading term coincides with $E_{ij'}$. A change of basis,
in particular a change of phase of the basis leads to a gauge transform  $\omega\to \omega + d\varphi$ and a shift of Fermion number
${\bf F}_{\gamma_{ij'}} \to {\bf F}_{\gamma_{ij'}} + \varphi(j') - \varphi(i)$. We will in fact make use of this freedom below.

\item Notice we managed to suppress the contribution of the anti-holomorphic part of the $tt^*$-equations by working in the complex $x$-plane. If one assumes an opposite limit $\beta\to0$ from the very beginning the partition function can be approximated by an integral of $\exp(-\frac{\I}{2}\zeta^{-1}\beta W)$ (see for example \cite{Hori:2000ck}). These integrals satisfy certain differential equations similar to the equations for a flat section of the $tt^*$
     connection, with the important difference that there is no   anti-holomorphic part from the very beginning.
     In this case quantities like $e^{\pi \I\int \omega}$ play a  role determining
the  phases of Stokes coefficients in the WKB expansion of solutions to these equations.\footnote{This derivation relating fermion numbers of solitons to Stokes coefficients is analogous to one using the IR asymptotic behavior of the $tt^*$-connection in \cite[Sec.4.4]{Cecotti:1992rm}.}

\end{enumerate}

\subsection{Example}

As an application of the above discussion
 we now calculate the fermion numbers of the chain complexes appearing
  in the Chan-Paton data for the interface corresponding to the $\cal R$-interface \eqref{rules_a}.

Consider the YYLG theory with one LG field $w$ and two punctures $z_a$ and $z_b$:
\be
W=\log(w-z_a)+\log(w-z_b)+cw
\ee
This theory supports the two solitons illustrated in Figure \ref{fig:2_solitons}. In this model
there are two vacua and we find it convenient to use a basis of chiral fields   $\Phi_a=\frac{1}{z_a-w}$ and $\Phi_b=\frac{1}{z_b-w}$.
The corresponding chiral ring structure constants and the connection
in this basis  are
\be
C_a=\left(\begin{array}{cc}
	c+\frac{1}{z_a-z_b} & -\frac{1}{z_a-z_b}\\
	-\frac{1}{z_a-z_b} & \frac{1}{z_a-z_b}\\
\end{array}\right), \quad C_b=\left(\begin{array}{cc}
\frac{1}{z_b-z_a} & -\frac{1}{z_b-z_a}\\
-\frac{1}{z_b-z_a} & c+\frac{1}{z_b-z_a}\\
\end{array}\right)\\
A_a=A_b=0
\label{eq:A-connect}
\ee
so the spectral curve is
\be
{\rm det}\; \left(\lambda \mathds{1}+C_a d z_a+C_b d z_b \right)=0
\ee
This equation has two roots, and in the large $c$ limit they have a simple form:
\be
\lambda_{(+-)}=cd z_a+O(c^0),\quad \lambda_{(-+)}=cd z_b+O(c^0)
\ee
illustrating the correspondence to the vacua  mentioned above. 

Without loss of generality we can assume $z_a=e^{\I x}$, $z_b=-e^{\I x}$. Then the  spectral curve reads:
\be
\lambda^2+(1+c^2 e^{2\I x})dx^2=0
\ee
And we easily derive the form $\omega$:
%
%
%
%
%
%
%
%
\be
\omega=-\left[\frac{\left(ce^{ix}\right)^2}{1+\left(ce^{ix}\right)^2}+\frac{ce^{ix}}{\sqrt{1+\left(ce^{ix}\right)^2}}\right]\frac{dx}{2\pi}
\ee

The corresponding detours are depicted in Figure \ref{fig:spec_net}. The two solitons depicted in Figure \ref{fig:2_solitons} correspond to the detours $d_1$ and $d_2$.
In the $x$-evolution the detour $d_1$ appears before the detour $d_2$ so it corresponds to the soliton denoted by the double line in Figure \ref{fig:2_solitons}, while the
detour $d_2$ corresponds to the soliton depicted by the single line. The concatenation $d_1\circ d_2^{-1}=:\gamma$ is a closed cycle. We assume that the detours start from the sheet where the square root in the large $c$ limit behaves as
\be
\sqrt{1+c^2 e^{2\I x}}\sim c e^{\I x}+\ldots
\ee
\begin{figure}[h!]
	\begin{center}
		\begin{tikzpicture}
		\draw[->] (-1,0) -- (3,0);
		\node[right] at (3,0) {$x$};
		\node[below] at (0,0) {$0$};
		\node[below] at (2,0) {$\pi$};
		\draw[dashed] (0,1) -- (0,2) (2,1) -- (2,2) (0,1) to[out=315,in=90] (1,-1) (0,1) to[out=225,in=90] (-1,-1) (2,1) to[out=315,in=90] (3,-1) (2,1) to[out=225,in=90] (1,-1);
		\draw[red] (0,0) to[out=90,in=270] (-0.5,1) to[out=90,in=180] (0,1.5) to[out=0,in=150] (2,0);
		\begin{scope}[xscale=-1,shift={(-2,0)}]
		\draw[blue] (0,0) to[out=90,in=270] (-0.5,1) to[out=90,in=180] (0,1.5) to[out=0,in=150] (2,0);
		\end{scope}
		\draw[violet] (0,0.7) to[out=180,in=180] (0,1.3) to[out=0,in=150] (1,2) to[out=330,in=180] (2,0.7) to[out=0,in=0] (2,1.3) to[out=180,in=30] (1,2) to[out=210,in=0] (0,0.7);
		\begin{scope}[shift={(0,1)}]
		\draw[orange, ultra thick] (-0.1,-0.1) -- (0.1,0.1) (-0.1,0.1) -- (0.1,-0.1);
		\end{scope}
		\begin{scope}[shift={(2,1)}]
		\draw[orange, ultra thick] (-0.1,-0.1) -- (0.1,0.1) (-0.1,0.1) -- (0.1,-0.1);
		\end{scope}
		\draw[ultra thick] (0,0.1) -- (0,-0.1) (2,0.1) -- (2,-0.1);
		\node[left] at (-0.5,1) {$\I\; \log\; c-\frac{\pi}{2}$};
		\node[right] at (2.5,1) {$\I\; \log\; c+\frac{\pi}{2}$};
		\node[above] at (1,2) {$\color{violet} \gamma$};
		\node[above left] at (0,1.5) {$\color{red} d_1$};
		\node[above right] at (2,1.5) {$\color{blue} d_2$};
		\end{tikzpicture}
	\end{center}
	\caption{Detours for $\CR$-interface}\label{fig:spec_net}
\end{figure}

Thus we calculate the difference of fermion numbers:
\be
{\bf F}_1-{\bf F}_2=\int\lm_{d_1}\omega-\int\lm_{d_2}\omega=\oint\lm_{\gamma}\omega=1
\ee


Now, displaying the fermion number of each component of \eqref{rules_a}  in square brackets we have:
\be
	\CE\left(\begin{array}{c}
		\begin{tikzpicture}
		\begin{scope}[scale=0.8]
		\draw[ultra thick] (0,0) -- (0,0.5) to[out=90,in=210] (0.5,1) to[out=30,in=270] (1,1.5) -- (1,2) (1,0) -- (1,0.5) to[out=90,in=330] (0.6,0.9) (0.4,1.1) to[out=150,in=270] (0,1.5) -- (0,2);
		\end{scope}
		\end{tikzpicture}
	\end{array}\right)= q^{\frac{1}{2}}[f_1+\varphi_{(++)}] \begin{array}{c}
	\begin{tikzpicture}
	\begin{scope}[scale=0.8]
	\node[above] at (0,2) {$+$};
	\node[above] at (1,2) {$+$};
	\node[below] at (0,0) {$+$};
	\node[below] at (1,0) {$+$};
	\draw[ultra thick, purple] (0,0) -- (0,0.5) to[out=90,in=210] (0.5,1) to[out=30,in=270] (1,1.5) -- (1,2) (1,0) -- (1,0.5) to[out=90,in=330] (0.6,0.9) (0.4,1.1) to[out=150,in=270] (0,1.5) -- (0,2);
	\end{scope}
	\end{tikzpicture}
\end{array}\oplus q^{\frac{1}{2}}[f_2+\varphi_{(--)}]\begin{array}{c}
\begin{tikzpicture}
\begin{scope}[scale=0.8]
\node[above] at (0,2) {$-$};
\node[above] at (1,2) {$-$};
\node[below] at (0,0) {$-$};
\node[below] at (1,0) {$-$};
\draw[ultra thick, purple] (0,0) -- (0,0.5) to[out=90,in=210] (0.5,1) to[out=30,in=270] (1,1.5) -- (1,2) (1,0) -- (1,0.5) to[out=90,in=330] (0.6,0.9) (0.4,1.1) to[out=150,in=270] (0,1.5) -- (0,2);
\end{scope}
\end{tikzpicture}
\end{array}\oplus \nn\ee 
\be
\oplus q^{-\frac{1}{2}}[f_3+\varphi_{(-+)}]\begin{array}{c}
\begin{tikzpicture}
\begin{scope}[scale=0.8]
\node[above] at (0,2) {$-$};
\node[above] at (1,2) {$+$};
\node[below] at (0,0) {$+$};
\node[below] at (1,0) {$-$};
\draw[ultra thick, purple] (0,0) -- (0,0.5) to[out=90,in=210] (0.5,1) to[out=30,in=270] (1,1.5) -- (1,2) (1,0) -- (1,0.5) to[out=90,in=330] (0.6,0.9) (0.4,1.1) to[out=150,in=270] (0,1.5) -- (0,2);
\end{scope}
\end{tikzpicture}
\end{array}\oplus q^{-\frac{1}{2}}[f_4+\varphi_{(+-)}]\begin{array}{c}
\begin{tikzpicture}
\begin{scope}[scale=0.8]
\node[above] at (0,2) {$+$};
\node[above] at (1,2) {$-$};
\node[below] at (0,0) {$-$};
\node[below] at (1,0) {$+$};
\draw[ultra thick, purple] (0,0) -- (0,0.5) to[out=90,in=210] (0.5,1) to[out=30,in=270] (1,1.5) -- (1,2) (1,0) -- (1,0.5) to[out=90,in=330] (0.6,0.9) (0.4,1.1) to[out=150,in=270] (0,1.5) -- (0,2);
\end{scope}
\end{tikzpicture}
\end{array}\oplus\ee
\be\nn \oplus q^{\frac{1}{2}}[f_5+\varphi_{(+-)}]\begin{array}{c}
\begin{tikzpicture}
\begin{scope}[scale=0.8]
\node[above] at (0,2) {$+$};
\node[above] at (1,2) {$-$};
\node[below] at (0,0) {$+$};
\node[below] at (1,0) {$-$};
\draw (0,0.45) -- (1,0.45);
\draw[ultra thick, purple] (0,0) -- (0,0.5) to[out=90,in=210] (0.5,1) to[out=30,in=270] (1,1.5) -- (1,2) (1,0) -- (1,0.5) to[out=90,in=330] (0.6,0.9) (0.4,1.1) to[out=150,in=270] (0,1.5) -- (0,2);
\end{scope}
\end{tikzpicture}
\end{array}\oplus q^{-\frac{3}{2}}[f_5+\varphi_{(+-)}+1]\begin{array}{c}
\begin{tikzpicture}
\begin{scope}[scale=0.8]
\node[above] at (0,2) {$+$};
\node[above] at (1,2) {$-$};
\node[below] at (0,0) {$+$};
\node[below] at (1,0) {$-$};
\draw (0,0.5) -- (1,0.5) (0,0.4) -- (1,0.4);
\filldraw[black] (0.5,0.45) circle (0.1);
\draw[ultra thick, purple] (0,0) -- (0,0.5) to[out=90,in=210] (0.5,1) to[out=30,in=270] (1,1.5) -- (1,2) (1,0) -- (1,0.5) to[out=90,in=330] (0.6,0.9) (0.4,1.1) to[out=150,in=270] (0,1.5) -- (0,2);
\end{scope}
\end{tikzpicture}
\end{array}
\ee
%
%
%
Here we suppose that all the values of the gauge potential $\varphi$ are fixed in the initial vacua, so the only free parameters are the values of $\varphi$ in the final vacuum.
Now we fix
\be
\varphi_{(++)}=-f_1,\quad \varphi_{(--)}=-f_2,\quad \varphi_{(-+)}=-f_3,\quad \varphi_{(+-)}=-f_4
\ee
thus setting the fermion numbers of the first four subcomplexes to be zero. Therefore,
denoting  $f_0=f_5-f_4$ we can write:
\be
\begin{split}
	\CE\left(\begin{array}{c}
		\begin{tikzpicture}
		\begin{scope}[scale=0.8]
		\draw[ultra thick] (0,0) -- (0,0.5) to[out=90,in=210] (0.5,1) to[out=30,in=270] (1,1.5) -- (1,2) (1,0) -- (1,0.5) to[out=90,in=330] (0.6,0.9) (0.4,1.1) to[out=150,in=270] (0,1.5) -- (0,2);
		\end{scope}
		\end{tikzpicture}
	\end{array}\right)= q^{\frac{1}{2}}[0] \begin{array}{c}
	\begin{tikzpicture}
	\begin{scope}[scale=0.8]
	\node[above] at (0,2) {$+$};
	\node[above] at (1,2) {$+$};
	\node[below] at (0,0) {$+$};
	\node[below] at (1,0) {$+$};
	\draw[ultra thick, purple] (0,0) -- (0,0.5) to[out=90,in=210] (0.5,1) to[out=30,in=270] (1,1.5) -- (1,2) (1,0) -- (1,0.5) to[out=90,in=330] (0.6,0.9) (0.4,1.1) to[out=150,in=270] (0,1.5) -- (0,2);
	\end{scope}
	\end{tikzpicture}
\end{array}\oplus q^{\frac{1}{2}}[0]\begin{array}{c}
\begin{tikzpicture}
\begin{scope}[scale=0.8]
\node[above] at (0,2) {$-$};
\node[above] at (1,2) {$-$};
\node[below] at (0,0) {$-$};
\node[below] at (1,0) {$-$};
\draw[ultra thick, purple] (0,0) -- (0,0.5) to[out=90,in=210] (0.5,1) to[out=30,in=270] (1,1.5) -- (1,2) (1,0) -- (1,0.5) to[out=90,in=330] (0.6,0.9) (0.4,1.1) to[out=150,in=270] (0,1.5) -- (0,2);
\end{scope}
\end{tikzpicture}
\end{array}\oplus  q^{-\frac{1}{2}}[0]\begin{array}{c}
\begin{tikzpicture}
\begin{scope}[scale=0.8]
\node[above] at (0,2) {$-$};
\node[above] at (1,2) {$+$};
\node[below] at (0,0) {$+$};
\node[below] at (1,0) {$-$};
\draw[ultra thick, purple] (0,0) -- (0,0.5) to[out=90,in=210] (0.5,1) to[out=30,in=270] (1,1.5) -- (1,2) (1,0) -- (1,0.5) to[out=90,in=330] (0.6,0.9) (0.4,1.1) to[out=150,in=270] (0,1.5) -- (0,2);
\end{scope}
\end{tikzpicture}
\end{array}\oplus\\ \oplus q^{-\frac{1}{2}}[0]\begin{array}{c}
\begin{tikzpicture}
\begin{scope}[scale=0.8]
\node[above] at (0,2) {$+$};
\node[above] at (1,2) {$-$};
\node[below] at (0,0) {$-$};
\node[below] at (1,0) {$+$};
\draw[ultra thick, purple] (0,0) -- (0,0.5) to[out=90,in=210] (0.5,1) to[out=30,in=270] (1,1.5) -- (1,2) (1,0) -- (1,0.5) to[out=90,in=330] (0.6,0.9) (0.4,1.1) to[out=150,in=270] (0,1.5) -- (0,2);
\end{scope}
\end{tikzpicture}
\end{array} \oplus q^{\frac{1}{2}}[f_0]\begin{array}{c}
\begin{tikzpicture}
\begin{scope}[scale=0.8]
\node[above] at (0,2) {$+$};
\node[above] at (1,2) {$-$};
\node[below] at (0,0) {$+$};
\node[below] at (1,0) {$-$};
\draw (0,0.45) -- (1,0.45);
\draw[ultra thick, purple] (0,0) -- (0,0.5) to[out=90,in=210] (0.5,1) to[out=30,in=270] (1,1.5) -- (1,2) (1,0) -- (1,0.5) to[out=90,in=330] (0.6,0.9) (0.4,1.1) to[out=150,in=270] (0,1.5) -- (0,2);
\end{scope}
\end{tikzpicture}
\end{array}\oplus q^{-\frac{3}{2}}[f_0+1]\begin{array}{c}
\begin{tikzpicture}
\begin{scope}[scale=0.8]
\node[above] at (0,2) {$+$};
\node[above] at (1,2) {$-$};
\node[below] at (0,0) {$+$};
\node[below] at (1,0) {$-$};
\draw (0,0.5) -- (1,0.5) (0,0.4) -- (1,0.4);
\filldraw[black] (0.5,0.45) circle (0.1);
\draw[ultra thick, purple] (0,0) -- (0,0.5) to[out=90,in=210] (0.5,1) to[out=30,in=270] (1,1.5) -- (1,2) (1,0) -- (1,0.5) to[out=90,in=330] (0.6,0.9) (0.4,1.1) to[out=150,in=270] (0,1.5) -- (0,2);
\end{scope}
\end{tikzpicture}
\end{array}
\end{split}\ee

The fermion number $f_0$ cannot be fixed by gauge transformations.
Nevertheless we can calculate it considering the generators $\Psi_1$ and $\Psi_2$ of the following two complexes:
\be
\CE\left[\Psi_1\right]=\begin{array}{c}
\begin{tikzpicture}
\draw[ultra thick, purple] (0,0) to[out=90,in=210] (0.5,0.5) to[out=30,in=270] (1,1) (1,0) to[out=90,in=330] (0.6,0.4) (0.4,0.6) to[out=150,in=270] (0,1);
\begin{scope}[shift={(0,1)}]
\draw[ultra thick, purple] (0,0) to[out=90,in=210] (0.5,0.5) to[out=30,in=270] (1,1) (1,0) to[out=90,in=330] (0.6,0.4) (0.4,0.6) to[out=150,in=270] (0,1);
\end{scope}
\node[below] at (0,0) {$+$}; \node[below] at (1,0) {$-$};
\node[above] at (0,2) {$+$}; \node[above] at (1,2) {$-$};
\end{tikzpicture}
\end{array},\quad
\CE\left[\Psi_2\right]=\begin{array}{c}
	\begin{tikzpicture}
	\begin{scope}[shift={(0,1)}]
	\draw (0.1,0.2) -- (0.9,0.2) (0.2,0.3) -- (0.8,0.3);
	\filldraw[black] (0.5,0.25) circle (0.1);
	\end{scope}
	\begin{scope}[shift={(0,-1)}]
	\draw (0.1,1.25) -- (0.9,1.25);
		\end{scope}
	\draw[ultra thick, purple] (0,0) to[out=90,in=210] (0.5,0.5) to[out=30,in=270] (1,1) (1,0) to[out=90,in=330] (0.6,0.4) (0.4,0.6) to[out=150,in=270] (0,1);
	\begin{scope}[shift={(0,1)}]
	\draw[ultra thick, purple] (0,0) to[out=90,in=210] (0.5,0.5) to[out=30,in=270] (1,1) (1,0) to[out=90,in=330] (0.6,0.4) (0.4,0.6) to[out=150,in=270] (0,1);
	\end{scope}
	\node[below] at (0,0) {$+$}; \node[below] at (1,0) {$-$};
	\node[above] at (0,2) {$+$}; \node[above] at (1,2) {$-$};
	\end{tikzpicture}
\end{array}
\ee
These two generators connect the same vacua and have the same $\Pdeg$-degree, therefore we expect that paths on $\Sigma$ corresponding to these two generators connect the same points, and therefore  integrals of $\omega$ along these paths can be easily compared, and their difference is independent of the gauge transform. In the following diagram we construct paths $\tilde d_1$ and $\tilde d_2$ corresponding $\CE[\Psi_1]$ and $\CE[\Psi_2]$ respectively:
\begin{center}
\begin{tikzpicture}
\draw[blue] (-2,0) to[out=350,in=190] (2,0);
\draw[red] (-2,0) to[out=20,in=150] (0,1.3) to[out=330,in=0] (0,0) to[out=180,in=210] (0,1.3) to[out=30,in=160] (2,0);
\draw[->] (-3,0) -- (3,0);
\draw[ultra thick] (-2,-0.1) -- (-2,0.1) (0,-0.1) -- (0,0.1) (2,-0.1) -- (2,0.1);
\node[right] at (3,0) {$x$};
\node[below] at (-2,-0.1) {$-\pi$};
\node[below] at (0,-0.1) {$0$};
\node[below] at (2,-0.1) {$\pi$};
\begin{scope}[shift={(0,1)}]
\draw[ultra thick, orange] (-0.1,-0.1) -- (0.1,0.1) (-0.1,0.1) -- (0.1,-0.1);
\end{scope}
\node[above] at (0,1.4) {$\I\log c+\frac{\pi}{2}$};
\node[above right] at (1.2,0.5) {$\color{red} \tilde d_2$};
\node[below] at (1,-0.1) {$\color{blue} \tilde d_1$};
\end{tikzpicture}
\end{center}

Notice that $\tilde \gamma:=\tilde d_2\circ \tilde d_1^{-1}$ is indeed a closed cycle. Thus we conclude:
\be
{\bf F}(\Psi_2)-{\bf F}(\Psi_1)=2f_0+1=\int\lm_{\tilde d_2}\omega-\int\lm_{\tilde d_1}\omega=\int\lm_{\tilde \gamma}\omega=1
\ee
and from this condition we conclude $f_0=0$. Thus we have recovered the fermion grading stated in Section \ref{sec:Brading}.

\section{Deriving Signs For the Matrix Elements Of $\CQ_{\zeta}$ }\label{app:Signs}

In this section we discuss a method for finding the signs of the  matrix elements $\langle \Psi_2|\IQ_{\zeta}|\Psi_1\rangle$
where $\Psi_1, \Psi_2$ are normalized perturbative groundstates of a Landau-Ginzburg interface defined by a path of
superpotentials.
These are important for determining the differential on the link homology complex constructed using the web formalism.
In general the signs in the MSW complex are rather delicate. A very careful theoretical treatment of these signs is given in Appendix F
of \cite{Gaiotto:2015aoa}. Here we will take a more pragmatic point of view, thinking of the sign as
the ratio of   the top form $\Psi_2\wedge \star \IQ_{\zeta}\Psi_1$ to the volume form on the field space \cite{Mirror}.
Both are forms on an  infinite dimensional space so we will need to use some physical arguments to reduce the problem to manipulation of
forms on a finite-dimensional space. While the matrix elements can in principle be any integer, generically
the matrix element will simply be $\pm 1$ and the only issue is to determine the sign.
We will henceforth assume we are in this generic situation.

Let us begin with some preliminary remarks concerning how the choices of generators for the complex lead to choices
in the signs of some matrix elements.
Given a collection of generators of the complex we can draw an oriented graph. The nodes of the graph correspond
to the generators.   Two nodes of the graph corresponding, say, to generators   $\psi$ and  $\chi$,  of fermion numbers
$f$ and $f+1$ respectively, are connected by an oriented edge if $\langle \chi|\IQ_{\zeta}|\psi\rangle\neq 0$. We orient
the edge from the generator of fermion number $f$ to the generator of fermion number $f+1$ so edges look like
\be
\begin{array}{c}
	\begin{tikzpicture}
	\draw (0,0) circle (0.1);
	\draw (2,0) circle (0.1);
	\node[above] at (0,0.1) {$\psi$};
	\node[above] at (2,0.1) {$\chi$};
	\draw[->] (0.2,0) -- (1.8,0);
	\end{tikzpicture}
\end{array}
\ee
and an example of an oriented graph associated to a complex with a choice of generators is:
$$
\begin{array}{c}
\begin{tikzpicture}
\draw (0,0) circle (0.1) (-1,0.5) circle (0.1) (-1,-0.5) circle (0.1) (1,0.5) circle (0.1) (1,-0.5) circle (0.1) (2,0) circle (0.1) (3,0) circle (0.1);
\draw[->] (-0.8,0.4) -- (-0.2,0.1);
\draw[->] (-0.8,-0.4) -- (-0.2,-0.1);
\draw[->] (1.2,0.4) -- (1.8,0.1);
\draw[->] (1.2,-0.4) -- (1.8,-0.1);
\draw[->] (0.2,0.1) -- (0.8,0.4);
\draw[->] (0.2,-0.1) -- (0.8,-0.4);
\draw[->] (2.2,0) -- (2.8,0);
\end{tikzpicture}
\end{array}
$$

Now, if the graph  is a tree then starting from any root we can choose the orientation of the generator at each node in such a way that all the edges correspond to a positive
matrix element.

On the other hand, if the graph has loops then the signs of some matrix elements are constrained and cannot all be taken to be $+1$.
As an example consider   the simplest loop:
\be
\begin{array}{c}
	\begin{tikzpicture}
	\draw (0,0) circle (0.1) (1,0.5) circle (0.1) (1,-0.5) circle (0.1) (2,0) circle (0.1);
	\draw[->] (1.2,0.4) -- (1.8,0.1);
	\draw[->] (1.2,-0.4) -- (1.8,-0.1);
	\draw[->] (0.2,0.1) -- (0.8,0.4);
	\draw[->] (0.2,-0.1) -- (0.8,-0.4);
	\node[left] at (-0.1,0) {$\alpha$};
	\node[right] at (2.1,0) {$\gamma$};
	\node[above] at (1,0.6) {$\beta$};
	\node[below] at (1,-0.6) {$\delta$};
	\node[above left] at (0.5,0.25) {$+1$};
	\node[above right] at (1.5,0.25) {$+1$};
	\node[below left] at (0.5,-0.25) {$+1$};
	\node[below right] at (1.5,-0.25) {$q$};
	\end{tikzpicture}
\end{array}\label{loop}
\ee
We can start with  $\alpha$, then fix matrix elements along edges going to node $\beta$, then $\gamma$, and from $\alpha$ to $\delta$. At this point
the   orientations of all the nodes are fixed. So the matrix element $q$ is automatically determined by these data. The complex is:
\be
\IM=\left(0\to \IF[\alpha]\mathop{\longrightarrow}\lm^{\IQ_{\zeta}^{(21)}} \IF[\beta]\oplus\IF[\delta]\mathop{\longrightarrow}\lm^{\IQ_{\zeta}^{(32)}}\IF[\gamma]\to 0\right)
\ee
The differential relative to this basis is of the form:
\be
\IQ_{\zeta}^{(21)}=\left(\begin{array}{c}
	1\\ 1\\
\end{array}\right),\quad \IQ_{\zeta}^{(32)}=\left(\begin{array}{cc}
1 & q
\end{array}\right)
\ee
Then condition $\IQ_{\zeta}^2=\IQ_{\zeta}^{(32)}\IQ_{\zeta}^{(21)}=1+q=0$ implies $q=-1$. We can use this kind of argument when the graph has a simple
structure, as indeed we did in the example of the Hopf link in Section \ref{sec:Hopf}. Nevertheless, it is useful to reproduce the above sign $q=-1$ from a field-theoretic
argument since that will teach us how to handle more complicated cases.  Below we will calculate explicitly how the sign of the ratio of infinite-dimensional
differential forms differs for the two paths $\alpha\to\beta\to \gamma$ and $\alpha\to\delta\to \gamma$.

We can easily encounter cases where the above simple logic fails to determine the sign uniquely. For  example, consider this graph:
\be
\begin{array}{c}
	\begin{tikzpicture}
	\node (A) at (0,0) {\begin{tikzpicture} \draw circle (0.1);\end{tikzpicture}};
	\node (B) at (0,1) {\begin{tikzpicture} \draw circle (0.1);\end{tikzpicture}};
	\node (C) at (2,0) {\begin{tikzpicture} \draw circle (0.1);\end{tikzpicture}};
	\node (D) at (2,1) {\begin{tikzpicture} \draw circle (0.1);\end{tikzpicture}};
	\path (A) edge[->] (C) (B) edge[->] (C) (B) edge[->] (D) (A) edge[->] (D);
	\node[left] at (-0.1,0) {$\alpha$}; \node[left] at (-0.1,1) {$\beta$};
	\node[right] at (2.1,0) {$\gamma$}; \node[right] at (2.1,1) {$\delta$};
\end{tikzpicture}
\end{array}
\ee
In this case the complex has the following form:
\be
\IM=\left(0\to  \IF[\alpha]\oplus \IF[\beta]\mathop{\longrightarrow}\lm^{\IQ_{\zeta}}\IF[\gamma]\oplus \IF[\delta]\to 0\right)
\ee
In this case we also can fix three of four matrix elements of the supercharge by adjusting orientation of corresponding forms in nodes of this graph:
\be
\CQ_{\zeta}=\left(\begin{array}{cc}
	1 & 1\\
	1 & \tilde q\\
\end{array}\right)
\ee
In this example there are no self-consistency conditions to fix $\tilde q$ and depending on the sign of $\tilde q$ cohomologies of this toy complex are quite different:
\be
H^{\bullet}\left[\CE(\tilde q=1)\right]=\IZ[\alpha-\beta]\oplus \IZ[\gamma-\delta],\qquad H^{\bullet}\left[\CE(\tilde q=-1)\right]= \emptyset
\ee

\subsection{Signs From Nearly Zero-Frequency Modes}

We now give a general discussion of how quantum field theory arguments can, in principle, determine the signs. We illustrate the discussion by
determining the sign for the example of the loop cohomology graph as in the Hopf link.
In the examples we have not encountered more complicated cohomology graph topologies, however more involved topologies may appear for more complicated knots and links, hence more elaborated sign counting technique may be required. As we mention in Section \ref{sec:FutureDirections} different sign choice rules may lead to inequivalent link homology theories.
%
%

Let us begin with discussion of the relevant  infinite-degree forms on field space.
Using the notation of \cite{Mirror} the supercharges defined in \eqref{eq:su_charges} read (for the sake of brevity we assume the K\"ahler metric is Euclidean $g_{I\bar J}=\delta_{IJ}$):
\be
\begin{split}
	\CQ_{\zeta}=-\int dx\left\{ \bar \psi_-^I\left[(\p_x-\p_t)\phi^I-\frac{\I\zeta}{2}\overline{\p_I W}\right]+\zeta\psi_+^I \left[(\p_x+\p_t)\bar \phi^I+\frac{\I\zeta^{-1}}{2}\p_I W\right] \right\}\\\
	\bar \CQ_{\zeta}=-\int dx\left\{ \zeta^{-1}\bar \psi_+^I  \left[(\p_x+\p_t)\phi^I-\frac{\I\zeta}{2}\overline{\p_I W}\right]+\psi_-^I \left[(\p_x-\p_t)\bar \phi^I+\frac{\I\zeta^{-1}}{2}\p_I W\right] \right\}
\end{split}
\ee
Let us study these operators in the linear approximation around a critical point, namely,  a solution to the forced $\zeta$-soliton equation. The tangent space in the vicinity of this critical point can be described as a span of normalized eigenmodes $(v_n^I(x),u_n^I(x))$ of the Dirac operator in the forced soliton background:
\be
\begin{split}
	\p_x v^I_n(x)-\frac{\I }{2}\overline{\p^2_{IJ}W} u_n^J(x)=\I\kappa_n v_n^I(x)\\
	\p_x u^I_n(x)+\frac{\I }{2}\p^2_{IJ}Wv_n^J(x)=-\I\kappa_n u_n^I(x)
\end{split}
\ee
In principle this equation has another solution $(\bar u_n^I(x),\bar v_n^I(x))$. In the case of zero eigenmode corresponding to the soliton modulus this is just the same solution. Mixing these two solutions together we can achieve a normalization condition $\int dx\;  u_n^I v_n^I=0$.
We decompose the tangent directions in the field space and fermion modes over these eigenmodes:
\be
\begin{split}
	\delta \phi_n^I(x)=\zeta^{\frac{1}{2}}\sum\lm_n \left(f_n v_n^I(x)+\bar f_n \bar u_n^I(x) \right),\quad 	\delta\bar \phi_n^I(x)=\zeta^{-\frac{1}{2}}\sum\lm_n \left(\bar f_n \bar v_n^I(x)+f_n u_n^I(x) \right),\\
	\psi_-^I(x)=\sum\lm_n\left( \chi_-^n v_n^I(x)+\bar\chi_+^n \bar u_n^I(x)\right),\quad 	\bar \psi_-^I(x)=\sum\lm_n \left( \bar\chi_-^n \bar v_n^I(x)+\chi_+^n u_n^I(x)\right),\\
	\psi_+^I(x)=\sum\lm_n\left( \chi_+^n v_n^I(x)+\bar\chi_-^n \bar u_n^I(x)\right),\quad 	\bar \psi_+^I(x)=\sum\lm_n \left( \bar\chi_+^n \bar v_n^I(x)+\chi_-^n u_n^I(x)\right)
\end{split}
\ee
In these coordinates the supercharges take the following form:
\be\label{eq:pert_Q}
\begin{split}
	\CQ_{\zeta}=-\I \zeta^{\frac{1}{2}}\sum\lm_n\left[||v_n||^2-||u_n||^2\right]\left\{\bar\chi_-^n\left(\kappa_n f_n+\frac{\p}{\p \bar f_n}\right)-\chi_+^n\left(\kappa_n \bar f_n+\frac{\p}{\p  f_n}\right)\right\}\\
	\bar \CQ_{\zeta}=\I \zeta^{\frac{1}{2}}\sum\lm_n\left[||v_n||^2-||u_n||^2\right]\left\{\chi_-^n\left(\kappa_n \bar f_n-\frac{\p}{\p  f_n}\right)-\bar \chi_+^n\left(\kappa_n  f_n-\frac{\p}{\p  \bar f_n}\right)\right\}\\
\end{split}
\ee
Here we are being rather sloppy. The sum  over eigenvalues is meant to indicate integral over the continuous spectrum as well. Defining a  Clifford vacuum $|0\rangle$
to be one annihilated by $\chi_{\pm}^n$,   the approximate ground state wave function reads (here we just generalize \cite[eq.(10.171)]{Mirror}):
\be\label{eq:Appx-GndState-1}
\Psi=e^{-\sum\lm_n |\kappa_n||f_n|^2}\prod\lm_{n:\;\kappa_n<0}\bar\chi_{+}^n \prod\lm_{n:\;\kappa_n>0}\bar\chi_{-}^n|0\rangle
\ee
Notice there are no zero modes. Such zero modes represent translation moduli but the (generic) interface breaks translation invariance in the $x$-direction.

We expect that a major role will be played by nearly zero frequency modes that are associated to moduli of solitons bound to the interface. Indeed as just mentioned,  these bound solitons do not have moduli, so the corresponding zero eigenvalue of the Dirac operator is shifted from zero. The shift can be estimated from first order perturbation theory. Suppose some soliton is bound to some point $x_0$; the zero mode $\delta\phi^i_0(x)$ is localized near $x_0$. The correction to its eigenvalue reads:
\be
\Delta\kappa_0=\frac{1}{||\delta \phi_0||^2}{\rm Re}\left[\zeta^{-1}\int dx\; \p^2_{ij}\left(W(z(x))-W(z(x_0)))\right)\delta \phi_0^i(x) \delta \phi_0^j(x)\right]
\ee
Since the  zero mode is well-localized in the $x$-direction we can estimate this contribution as:
\be
\Delta\kappa_0\sim |W|\left|\frac{dz}{dx}\right|\ell_W
\ee
(See equation (16.16) of \cite{Gaiotto:2015aoa} for a related argument.)
Now assume that the parameters of the superpotential are varying adiabatically across the interface so that translational invariance is only
weakly broken. Thus we assume that $|dz/dx|\ll \ell_{W}^{-1}$ and we may also assume that $|dz/dx|\sim\CO(\ell_W^0)$ as well. We can easily rescale the soliton width $\ell_W$ by an overall rescaling of the superpotential: under this rescaling the non-zero eigenvalues behave as $|\kappa_n|\sim \CO(\ell_W^{-1})$, while the shifted zero eigenvalue is not rescaled  $|\kappa_0|\sim \CO(\ell_W^{0})$. These kinds of considerations
lead to the following picture of the  spectrum:   there is some continuum of positive and negative eigenvalues and there are some
low-frequency modes separated by a gap. (The low-frequency modes are descendents of the translational zero-modes of the vanilla solitons at the binding points):
\be
\begin{array}{c}
	\begin{tikzpicture}
	\begin{scope}[shift={(0,1)}]
	\begin{scope}[scale=0.5]
	\foreach \i in {0,...,15}
	{	\draw (0.25*\i,0) -- (0.25+0.25*\i,2);}
	\end{scope}
	\end{scope}
	\begin{scope}[shift={(0,-2)}]
	\begin{scope}[scale=0.5]
	\foreach \i in {0,...,15}
	{	\draw (0.25*\i,0) -- (0.25+0.25*\i,2);}
	\end{scope}
	\end{scope}
	\draw[ultra thick] (0,-2) -- (0,-1) -- (2,-1) -- (2,-2) (0,2) -- (0,1) -- (2,1) -- (2,2) (-0.25,0) -- (2,0);
	\draw[red, thick] (0,0.25) -- (2,0.25) (0,-0.25) -- (2,-0.25);
	\draw[->] (-0.25,-2) -- (-0.25,2); \node[above] at (-0.25,2) {$\kappa$}; \node[left] at (-0.25,0) {$0$};
	\draw[<->] (1.75,0) -- (1.75,1); \node[right] at (1.75,0.5) {$\sim \ell_W^{-1}$};
	\end{tikzpicture}
\end{array}\label{eq:spectrum}
\ee
As we have seen in Figure \ref{fig:2_solitons} we have two types of solitons. In this section let us refer to double-line binding points as binding points of type $a$, and to single line binding points as binding points of type $b$. The low-frequency fermion eigenmodes are localized near binding points $x_i$, so to binding points we associate corresponding representations of Clifford algebras generated by $a(x_i)$, $a^{\dag}(x_i)$, $b(x_i)$, $b^{\dag}(x_i)$:
\be
\{a(x_i),a^{\dag}(x_j) \}=\delta_{ij},\quad  \{b(x_i),b^{\dag}(x_j) \}=\delta_{ij}
\ee
So, generically, the approximate groundstate \eqref{eq:Appx-GndState-1}  associated with a forced soliton
containing a set $\CB_a$ of binding points of type $a$ and a set $\CB_b$ of binding points of type $b$
can be written as:
\be
\Psi= : \prod\lm_{x_i\in\CB_a}a^{\dag}(x_i) \prod\lm_{x_j\in\CB_b}b^{\dag}(x_j) : \wedge \Omega |0\rangle\label{eq:forms}
\ee
In this representation we will denote by a wedge a separation of the modes into higher eigenvalue modes and nearly
zero eigenvalue modes. So by $\Omega$ here we mean the bulk of higher eigenvalue modes. The product over the nearly-zero frequency
creation modes $a^\dagger$ and $b^\dagger$ is  ordered by the ordering of the  corresponding binding points $x_i$ alon the $x$-axis.
This ordering is indicated by the colons.

Now we are able to define the supercharge: remember that solitons of type $a$ in the $\CR$ interface and the cap interface
have soliton number $+1$, while solitons of type $b$ in the $\CR^{-1}$ interface and the cup interface have soliton number $-1$.
So we construct our supercharge as an operator of fermion number $+1$ in analogy to expression \eqref{eq:pert_Q}:
\be\label{eq:su_charge}
\CQ_{\zeta}=\sum\lm_{i\in \CR,\cap}a^{\dag}(x_i)+ \sum\lm_{i\in \CR^{-1},\cup}b(x_i)
\ee

As we have seen in examples the number of bound solitons can change in a complicated ways. So during the parallel transition of the form $\CQ_{\zeta}\Psi$ along the flow
in fieldspace corresponding to a  $\zeta$-instanton,  some of the modes from the large eigenvalues, initially contributing to $\Omega$,
can become nearly zero-frequency modes. Of course the inverse process can also occur:
%
%
\begin{center}
	\begin{tikzpicture}
	\begin{scope}[shift={(0,1)}]
	\begin{scope}[scale=0.5]
	\foreach \i in {0,...,15}
	{	\draw (0.25*\i,0) -- (0.25+0.25*\i,2);}
	\end{scope}
	\end{scope}
	\begin{scope}[shift={(0,-2)}]
	\begin{scope}[scale=0.5]
	\foreach \i in {0,...,15}
	{	\draw (0.25*\i,0) -- (0.25+0.25*\i,2);}
	\end{scope}
	\end{scope}
	\draw[ultra thick] (0,-2) -- (0,-1) -- (2,-1) -- (2,-2) (0,2) -- (0,1) -- (2,1) -- (2,2) (-0.25,0) -- (2,0);
	\draw[red, thick] (0,0.25) -- (2,0.25) (0,-0.25) -- (2,-0.25);
	\draw[->] (-0.25,-2) -- (-0.25,2); \node[above] at (-0.25,2) {$\kappa$}; \node[left] at (-0.25,0) {$0$};
	\filldraw[blue] (1,1.5) circle (0.1); \filldraw[red] (1.5,0.12) circle (0.1);
	\draw[blue,->, ultra thick] (1,1.5) to[out=300,in=60] (1,0.12);
	\draw[red,->, ultra thick] (1.5,0.12) to[out=60,in=300] (1.5,1.5);
	\end{tikzpicture}
\end{center}
So we have to take into account this aspect of spectral flow as well.

\paragraph{Example: a loop in the Hopf link.} As the first simple example of a loop in the graph associated to a link homology complex let us consider the first four states $\CE_1({\rm Hopf})$ (see equation \eqref{eq:E_Hopf}). We neglect the fifth state since its analogous to the fourth one, so the diagram diamond will be equivalent. Here we adopt the following notations: a binding point without solitons we denote by a circle ($\begin{array}{c}
\begin{tikzpicture}
\crl
\end{tikzpicture}
\end{array}$), a soliton of type $a$ we denote by a diamond ($\begin{array}{c}
\begin{tikzpicture}
\dmd
\end{tikzpicture}
\end{array}$), a soliton of type $b$ we denote by a square ($\begin{array}{c}
\begin{tikzpicture}
\sqr
\end{tikzpicture}
\end{array}$).

In this way we depict the states as:
\be
\begin{split}
	\Psi_1=\begin{array}{c}
		\begin{tikzpicture}
		\draw[ultra thick] (-0.5,0) -- (3.5,0);
		\begin{scope}
		\sqr
		\end{scope}
		\begin{scope}[shift={(1,0)}]
		\crl
		\end{scope}
		\begin{scope}[shift={(2,0)}]
		\crl
		\end{scope}
		\begin{scope}[shift={(3,0)}]
		\crl
		\end{scope}
		\end{tikzpicture}
	\end{array}\\
	\Psi_2=\begin{array}{c}
		\begin{tikzpicture}
		\draw[ultra thick] (-0.5,0) -- (3.5,0);
		\begin{scope}
		\sqr
		\end{scope}
		\begin{scope}[shift={(1,0)}]
		\dmd
		\end{scope}
		\begin{scope}[shift={(2,0)}]
		\sqr
		\end{scope}
		\begin{scope}[shift={(3,0)}]
		\crl
		\end{scope}
		\end{tikzpicture}
	\end{array}\\
	\Psi_3=\begin{array}{c}
		\begin{tikzpicture}
		\draw[ultra thick] (-0.5,0) -- (3.5,0);
		\begin{scope}
		\sqr
		\end{scope}
		\begin{scope}[shift={(1,0)}]
		\sqr
		\end{scope}
		\begin{scope}[shift={(2,0)}]
		\dmd
		\end{scope}
		\begin{scope}[shift={(3,0)}]
		\crl
		\end{scope}
		\end{tikzpicture}
	\end{array}\\
	\Psi_4=\begin{array}{c}
		\begin{tikzpicture}
		\draw[ultra thick] (-0.5,0) -- (3.5,0);
		\begin{scope}
		\crl
		\end{scope}
		\begin{scope}[shift={(1,0)}]
		\crl
		\end{scope}
		\begin{scope}[shift={(2,0)}]
		\crl
		\end{scope}
		\begin{scope}[shift={(3,0)}]
		\dmd
		\end{scope}
		\end{tikzpicture}
	\end{array}
\end{split}
\ee
The corresponding perturbative groundstates can be written in the form \eqref{eq:forms} as:
\be\label{eq:B.15}
\begin{split}
	\Psi_1=b^{\dag}(x_1)\wedge \Omega_1|0\rangle\\
	\Psi_2=b^{\dag}(x_1)a^{\dag}(x_2) b^{\dag}(x_3)\wedge \Omega_2|0\rangle\\
	\Psi_3=b^{\dag}(x_1)b^{\dag}(x_2) a^{\dag}(x_3)\wedge \Omega_3|0\rangle\\
	\Psi_4=a^{\dag}(x_4)\wedge \Omega_4|0\rangle
\end{split}
\ee
where the $\Omega_i$ are the contributions of the high-frequency modes to the perturbative groundstate around the four forced solitions.
Now we need to see how the forms $\Omega_i$ are related to each other under the flow defined by a $\zeta$-instanton.  For example, if we consider the matrix element $\langle \Psi_2|\CQ_{\zeta}|\Psi_1\rangle$ and use the formula for the supercharge $\CQ_{\zeta}$ given in \eqref{eq:su_charge} then this operator ``creates'' a new zero mode $a^{\dag}(x_2)$. But $\CQ_\zeta$ cannot create the mode $b^{\dag}(x_3)$. Thus, this mode should come from the high-frequency modes under spectral flow.
It is therefore ``already stored'' in $\Omega_1$ and so we write $\Omega_1=b^{\dag}(x_3)\Omega_2$ (where parallel transport under the $\zeta$-instanton flow
of the LHS is understood).  Following this logic we rewrite (the suitable parallel transport of) all the forms $\Omega_i$ in terms of one common infinite-dimensional
differential form $\Omega \vert 0 \rangle$:
\be
\Psi_1=b^{\dag}(x_1)\wedge b^{\dag}(x_2) b^{\dag}(x_3) a^{\dag}(x_4)\Omega|0\rangle\\
\Psi_2=b^{\dag}(x_1)a^{\dag}(x_2) b^{\dag}(x_3)\wedge b^{\dag}(x_2) a^{\dag}(x_4)\Omega|0\rangle\\
\Psi_3=b^{\dag}(x_1)b^{\dag}(x_2) a^{\dag}(x_3)\wedge b^{\dag}(x_3) a^{\dag}(x_4)\Omega|0\rangle\\
\Psi_4=a^{\dag}(x_4)\wedge b^{\dag}(x_1)b^{\dag}(x_2)b^{\dag}(x_3) a^{\dag}(x_2) a^{\dag}(x_3) \Omega|0\rangle
\ee
Now we can forget about separation of the modes into higher eigenvalue modes and nearly zero eigenvalue modes and calculate matrix elements using effective supercharge $\CQ_{\zeta}=a^{\dagger}(x_2)+a^{\dagger}(x_3)$:
\begin{center}
	\begin{tikzpicture}
	\begin{scope}[scale=0.8]
	\node(A) at (-2,0) {$\Psi_1$};
	\node(B) at (0,1) {$\Psi_2$};
	\node(C) at (0,-1) {$\Psi_3$};
	\node(D) at (2,0) {$\Psi_4$};
	\path (A) edge[->] (B) (B) edge[->] (D) (A) edge[->] (C) (C) edge[->, dashed] (D);
	\end{scope}
	\end{tikzpicture}
\end{center}
and in this way reproduce the sign $q=-1$ determined by self-consistency above.


\begin{thebibliography}{99}

\bibitem{Aganagic:2011sg}
M.~Aganagic and S.~Shakirov,
``Knot Homology and Refined Chern-Simons Index,''
Commun.\ Math.\ Phys.\  {\bf 333}, no. 1, 187 (2015),
arXiv:1105.5117 [hep-th].

\bibitem{Anokhina:2014hha}
A.~Anokhina and A.~Morozov,
``Towards R-matrix construction of Khovanov-Rozansky polynomials. I. Primary $T$-deformation of HOMFLY,''
JHEP {\bf 1407}, 063 (2014),
arXiv:1403.8087 [hep-th].

\bibitem{Atiyah:1988jp}
  M.~F.~Atiyah and N.~J.~Hitchin,
  ``The Geometry And Dynamics Of Magnetic Monopoles. M.B. Porter Lectures,''
  Princeton University Press, 1988.

\bibitem{Arthamonov:2015rha}
S.~Arthamonov and S.~Shakirov,
``Refined Chern-Simons Theory in Genus Two,''
arXiv:1504.02620 [hep-th].

\bibitem{BarNatan}
D.~Bar-Natan,
``On Khovanov's categorification of the Jones polynomial'',
Algebr.\ Geom.\ Topol.\ {\bf 2},
337-370 (2002) [arXiv:math/0201043].


\bibitem{Bigelow}
S. Bigelow, ``A homological definition of the Jones polynomial,''
Geometry \& Topology Monographs
Volume 4: Invariants of knots and 3-manifolds (Kyoto 2001)
Pages 29–41; 	arXiv:math/0201221 [math.GT].


\bibitem{Braverman:2014ysa}
  A.~Braverman, G.~Dobrovolska and M.~Finkelberg,
 ``Gaiotto-Witten superpotential and Whittaker D-modules on monopoles,''
  arXiv:1406.6671 [math.AG].


\bibitem{Cecotti:1992qh}
  S.~Cecotti, P.~Fendley, K.~A.~Intriligator and C.~Vafa,
  ``A New supersymmetric index,''
  Nucl.\ Phys.\ B {\bf 386}, 405 (1992)
  [arXiv:hep-th/9204102].


\bibitem{Cecotti:1992rm}
  S.~Cecotti and C.~Vafa,
``On classification of $N=2$ supersymmetric theories,''
  Commun.\ Math.\ Phys.\  {\bf 158}, 569 (1993)
  [arXiv:hep-th/9211097].


\bibitem{Cheng:2010yw}
  M.~C.~N.~Cheng, R.~Dijkgraaf and C.~Vafa,
   ``Non-Perturbative Topological Strings And Conformal Blocks,''
  JHEP {\bf 1109}, 022 (2011)
  arXiv:1010.4573 [hep-th].

\bibitem{DF}
V.S.~Dotsenko and V.A.~Fateev,
``Conformal algebra and multipoint correlation functions in 2D statistical models'',
Nucl.\ Phys.\ B\ {\bf  240}, 312-348 (1984).

\bibitem{Dolotin:2012sw}
V.~Dolotin and A.~Morozov,
``Introduction to Khovanov Homologies. I. Unreduced Jones superpolynomial,''
JHEP {\bf 1301}, 065 (2013),
arXiv:1208.4994 [hep-th];
%
V.~Dolotin and A.~Morozov,
``Introduction to Khovanov Homologies. II. Reduced Jones superpolynomial,''
J.\ Phys.\ Conf.\ Ser.\  {\bf 411}, 012013 (2013),
arXiv:1209.5109 [hep-th];
%
 V.~Dolotin and A.~Morozov,
 ``Introduction to Khovanov Homologies. III. A new and simple tensor-algebra construction of Khovanov-Rozansky invariants,''
 Nucl.\ Phys.\ B {\bf 878}, 12 (2014)
 arXiv:1308.5759 [hep-th].

\bibitem{Donaldson} S.K.~Donaldson, ``Nahm's equations and the classification of monopoles,''
 Commun. Math. Phys. {\bf 96}, 387-408 (1984).

\bibitem{Drinfeld} V.G.~Drinfeld,
 ``Quasi-Hopf algebras and Knizhnik-Zamolodchikov equations'', Problems of Modern Quantum Field Theory, 1-13 (1989).

\bibitem{Elitzur:1989nr}
  S.~Elitzur, G.~W.~Moore, A.~Schwimmer and N.~Seiberg,
   ``Remarks on the Canonical Quantization of the Chern-Simons-Witten Theory,''
  Nucl.\ Phys.\ B {\bf 326}, 108 (1989).



\bibitem{FY} P.J.~Freyd and D.N.~Yetter,
``Braided compact closed categories with applications to low dimensional topology,'' Advances in Mathematics {\bf 77.2}, 156-182 (1989)

\bibitem{Gabella}
M.~Gabella,
``Quantum Holonomies from Spectral Networks and Framed BPS States,''
arXiv:1603.05258 [hep-th].

\bibitem{Gaiotto:2011nm}
  D.~Gaiotto and E.~Witten,
  ``Knot Invariants from Four-Dimensional Gauge Theory,''
  arXiv:1106.4789 [hep-th].

\bibitem{Gaiotto:2015zna}
  D.~Gaiotto, G.~W.~Moore and E.~Witten,
   ``An Introduction To The Web-Based Formalism,''
  arXiv:1506.04086 [hep-th].


\bibitem{Gaiotto:2015aoa}
  D.~Gaiotto, G.~W.~Moore and E.~Witten,
  ``Algebra of the Infrared: String Field Theoretic Structures in Massive ${\cal N}=(2,2)$ Field Theory In Two Dimensions,''
  arXiv:1506.04087 [hep-th].

\bibitem{SN}
D.~Gaiotto, G.~W.~Moore and A.~Neitzke,
``Spectral networks,''
Annales Henri Poincar\'e {\bf 14}, 1643 (2013)
arXiv:1204.4824 [hep-th].

\bibitem{GLM}
D.~Galakhov, P.~Longhi and G.~W.~Moore,
``Spectral Networks with Spin,''
Commun.\ Math.\ Phys.\  {\bf 340}, no. 1, 171 (2015)
arXiv:1408.0207 [hep-th].

\bibitem{GalakhovPhD} D. Galakhov, ``INTERFACES IN SUPERSYMMETRIC FIELD
THEORIES,''  Rutgers University PhD, May, 2016

\bibitem{GMOMS} A.~Gerasimov, A.~Morozov, M.~Olshanetsky, A.~Marshakov and S.~Shatashvili, ``Wess-Zumino-Witten model as a theory of free fields'',
Int.\ J.\ Mod.\ Phys.  A {\bf 5(13)}, 2495-2589 (1990)

\bibitem{Gukov:2016gkn}
S.~Gukov, P.~Putrov and C.~Vafa,
``Fivebranes and 3-manifold homology,''
arXiv:1602.05302 [hep-th].

\bibitem{Gukov:2004hz}
S.~Gukov, A.~S.~Schwarz and C.~Vafa,
``Khovanov-Rozansky homology and topological strings,''
Lett.\ Math.\ Phys.\  {\bf 74}, 53 (2005)
[arXiv:hep-th/0412243].


\bibitem{Haydys}
  A.~Haydys, ``Seidel-Fukaya Category And Gauge Theory,'' Symplectic Geom. {\bf 13}, no. 1, 151-207 (2015), arXiv:1010.2353 [math.SG].

\bibitem{achiral} J.~Hoste, M.~Thistlethwaite and J.~Weeks,
``The first 1,701,936 knots'',
 The Mathematical Intelligencer {\bf 20(4)}, 33-48 (1998)

\bibitem{Hori:2000ck}
K.~Hori, A.~Iqbal and C.~Vafa,
``D-branes and mirror symmetry,''
[arXiv:hep-th/0005247].

\bibitem{Mirror}
K.~Hori, S.~Katz, A.~Klemm, R.~Pandharipande, R.~Thomas, C.~Vafa, R.Vakil, E.~Zaslow,
``Mirror symmetry'', Clay mathematics monographs, vol. 1, AMS, Providence, USA, 2003

\bibitem{Hurtubise} J. Hurtubise,
``Monopoles and Rational Maps: A Note on a
Theorem of Donaldson,'' Commun. Math. Phys. {\bf 100}, 191-196 (1985);
J. Hurtubise, ``The Classification of Monopoles for the Classical Groups,''
Commun. Math. Phys. {\bf 120}, 613-641 (1989); J. Hurtubise and M. Murray,
``On The Construction of Monopoles for the Classical Groups,''
Commun. Math. Phys. {\bf 122}, 35-89 (1989)


\bibitem{Kapustin:2006pk}
  A.~Kapustin and E.~Witten,
  ``Electric-Magnetic Duality And The Geometric Langlands Program,''
  Commun.\ Num.\ Theor.\ Phys.\  {\bf 1}, 1 (2007)
  [arXiv:hep-th/0604151].

\bibitem{Khovanov}
M.~Khovanov,
``A categorification of the Jones polynomial'',
Duke\ Math.\ J.\ {\bf 101}, 359-426 (2000) [arXiv:math/9908171]



\bibitem{Ruth_Lawrence}
R. J. Lawrence,
``Homological representations of the Hecke algebra,"
Commun.\ Math.\ Phys. {\bf 135 no. 1},  141-191 (1990)

\bibitem{Mazzeo:2013zga}
  R.~Mazzeo and E.~Witten,
  ``The Nahm Pole Boundary Condition,''
 arXiv:1311.3167 [math.DG].

\bibitem{Moore:1989vd}
  G.~W.~Moore and N.~Seiberg,
   ``Lectures On Rcft,''
  RU-89-32, YCTP-P13-89, C89-08-14.

\bibitem{Morozov:2015iha}
A.~Morozov, An.~Morozov and A.~Popolitov,
``Matrix model and dimensions at hypercube vertices,''
arXiv:1508.01957 [hep-th].

\bibitem{Odd}
P.S.~Ozsv\'ath, J.~Rasmussen,  and Z.~Szab\'o,
`` Odd Khovanov homology,''
Algebraic \& Geometric Topology, {\bf 13(3)}, 1465-1488 (2013)



\bibitem{RT}
N.Yu.~Reshetikhin and V.G.~Turaev,
``Ribbon graphs and their invariants derived from quantum groups,''
Commun.\ Math.\ Phys.\ {\bf 127}, 1-26 (1990)


\bibitem{Rolfsen}\url{http://katlas.math.toronto.edu/wiki/The_Rolfsen_Knot_Table}



\bibitem{Witten:1982im}
  E.~Witten,
  ``Supersymmetry and Morse theory,''
  J.\ Diff.\ Geom.\  {\bf 17}, 661 (1982).


\bibitem{Witten:1978bc}
  E.~Witten,
  ``Instantons, the Quark Model, and the 1/n Expansion,''
  Nucl.\ Phys.\ B {\bf 149}, 285 (1979).

\bibitem{Witten:1988hf}
  E.~Witten,
 ``Quantum Field Theory and the Jones Polynomial,''
  Commun.\ Math.\ Phys.\  {\bf 121}, 351 (1989).



\bibitem{Witten:1993yc}
  E.~Witten,
  ``Phases of N=2 theories in two-dimensions,''
  Nucl.\ Phys.\ B {\bf 403}, 159 (1993)
  [hep-th/9301042].




\bibitem{Witten:2010cx}
  E.~Witten,
   ``Analytic Continuation Of Chern-Simons Theory,''
  arXiv:1001.2933 [hep-th].


  \bibitem{NewLook}
  E. Witten, ``A New Look At The Path Integral Of Quantum Mechanics,''
  arXiv:1009.6032 [hep-th].


  \bibitem{Witten:2011zz}
  E.~Witten,
  ``Fivebranes and Knots,''
 arXiv:1101.3216 [hep-th].




\bibitem{WItten:2011pz}
  E.~Witten,
  ``Khovanov Homology And Gauge Theory,''
  arXiv:1108.3103 [math.GT].

\bibitem{Witten:2014xwa}
  E.~Witten,
   ``Two Lectures On The Jones Polynomial And Khovanov Homology,''
  arXiv:1401.6996 [math.GT].


\end{thebibliography}
\end{document}